\documentclass[10pt]{article}
\usepackage{lineno}
\usepackage{scicite}
\usepackage[letterpaper, margin=1in]{geometry}
\usepackage{graphicx}
\usepackage{subcaption}
\usepackage[range-phrase={--},per-mode = single-symbol]{siunitx}
\usepackage{chemformula}
\usepackage[version=4]{mhchem}
\usepackage{hyperref}
\usepackage{cleveref}
\usepackage{authblk}
\usepackage{varwidth}
\usepackage{xcolor}
\usepackage[utf8]{inputenc}
\usepackage[T1]{fontenc}
\usepackage[normalem]{ulem}
\usepackage{adjustbox}

\Crefname{figure}{Figure}{Figures}
\crefname{figure}{Fig.}{Figs.}

\newcommand{\MPz}{MACE-MP-0} %
\newcommand{\MPzbt}{MACE-MP-0b3} %
\newcommand{\MPAz}{MACE-MPA-0} %

\newenvironment{sciabstract}{%
\begin{quote} \bf}
{\end{quote}}

\DeclareSIUnit{\atom}{at}
\DeclareSIUnit\Angstrom{\textup{\AA}}
\newcommand\longvar[1]{\mathchardef\UrlBreakPenalty=100\mathchardef\UrlBigBreakPenalty=100\url{#1}}

\begin{document}

\title{A foundation model for atomistic materials chemistry}

\renewcommand*{\thefootnote}{\fnsymbol{footnote}}
\footnotetext[2]{\normalsize These authors, ordered alphabetically, contributed equally. All others, except for the corresponding author, are also ordered alphabetically.}
\footnotetext[1]{\normalsize Corresponding author: \href{mailto:gc121@cam.ac.uk}{gc121@cam.ac.uk}}
\renewcommand*{\thefootnote}{\arabic{footnote}}
\setcounter{footnote}{0}

\author[1]{Ilyes Batatia$^\dagger$}
\author[2]{Philipp Benner$^\dagger$}
\author[3,4]{Yuan Chiang$^\dagger$}
\author[17]{Alin M. Elena$^\dagger$} 
\author[1]{D\'avid P. Kov\'acs$^\dagger$}
\author[4,13]{Janosh Riebesell$^\dagger$}
\author[12,13]{Xavier R. Advincula}
\author[3,4]{Mark Asta}
\author[30]{Matthew Avaylon}
\author[1]{William J. Baldwin}
\author[12]{Fabian Berger} 
\author[11]{Noam Bernstein}
\author[25]{Arghya Bhowmik}
\author[32]{Filippo Bigi}
\author[10]{Samuel M. Blau}
\author[1,13]{Vlad C\u{a}rare}
\author[32]{Michele Ceriotti}
\author[32]{Sanggyu Chong}
\author[1]{James P. Darby}
\author[18]{Sandip De}
\author[12]{Flaviano Della Pia}
\author[16]{Volker L. Deringer}
\author[1]{Rokas Elijošius}
\author[16]{Zakariya El-Machachi}
\author[31]{Fabio Falcioni}
\author[18]{Edvin Fako}
\author[26]{Andrea C. Ferrari}
\author[16]{John L. A. Gardner}
\author[38]{Miko{\l}aj J. Gawkowski} 
\author[12]{Annalena Genreith-Schriever}
\author[2,6]{Janine George}
\author[15]{Rhys E. A. Goodall} 
\author[6,2]{Jonas Grandel}
\author[12]{Clare P. Grey}
\author[27,36]{Petr Grigorev}
\author[18]{Shuang Han}
\author[13,19]{Will Handley}
\author[9]{Hendrik H. Heenen}
\author[23]{Kersti Hermansson}
\author[22]{Christian Holm}
\author[5]{Cheuk Hin Ho}
\author[1]{Stephan Hofmann}
\author[1]{Jad Jaafar}
\author[9]{Konstantin S. Jakob}
\author[9]{Hyunwook Jung}
\author[12, 38]{Venkat Kapil}
\author[4]{Aaron D. Kaplan}
\author[20]{Nima Karimitari}
\author[28]{James R. Kermode}
\author[24]{Panagiotis Kourtis}
\author[13,19,1]{Namu Kroupa}
\author[23]{Jolla Kullgren}
\author[3,4]{Matthew C. Kuner}
\author[12]{Domantas Kuryla}
\author[1,26]{Guoda Liepuoniute}
\author[1,35]{Chen Lin}
\author[8]{Johannes T. Margraf}
\author[24]{Ioan-Bogdan Magd\u{a}u}
\author[12]{Angelos Michaelides}
\author[1]{J. Harry Moore}
\author[2,6]{Aakash A. Naik}
\author[12]{Samuel P. Niblett}
\author[25]{Sam Walton Norwood}
\author[12,13]{Niamh O'Neill}
\author[5]{Christoph Ortner}
\author[3,4,7]{Kristin A. Persson}
\author[9]{Karsten Reuter}
\author[33]{Andrew S. Rosen} 
\author[16]{Louise A. M. Rosset}
\author[1]{Lars L. Schaaf}
\author[13]{Christoph Schran}
\author[12]{Benjamin X. Shi} 
\author[10]{Eric Sivonxay}
\author[1]{Tam\'as K. Stenczel}
\author[23]{Viktor Svahn}
\author[20]{Christopher Sutton}
\author[27,37]{Thomas D. Swinburne}
\author[31]{Jules Tilly}
\author[1]{Cas van der Oord}
\author[29]{Santiago Vargas}
\author[1]{Eszter Varga-Umbrich}
\author[25]{Tejs Vegge}
\author[8,9]{Martin Vondrák}
\author[5]{Yangshuai Wang}
\author[14]{William C. Witt}
\author[34]{Thomas Wolf}
\author[22]{Fabian Zills}
\author[1]{G\'abor Cs\'anyi$^*$}

\affil[1]{Engineering Laboratory, University of Cambridge, Trumpington St and JJ Thomson Ave, Cambridge, UK}
\affil[2]{Federal Institute of Materials Research and Testing (BAM), Berlin, Germany}
\affil[3]{Department of Materials Science and Engineering, University of California, Berkeley, CA 94720, USA}
\affil[4]{Materials Sciences Division, Lawrence Berkeley National Laboratory, Berkeley, CA 94720, USA}
\affil[5]{Mathematics Department, University of British Columbia, 1984 Mathematics Rd, Vancouver, BC V6T 1Z2, Canada}
\affil[6]{Institute of Condensed Matter Theory and Solid State Optics, Friedrich Schiller University Jena, Germany}
\affil[7]{Molecular Foundry, Lawrence Berkeley National Laboratory, Berkeley, California 94720, USA}
\affil[8]{University of Bayreuth, Bavarian Center for Battery Technology (BayBatt), Bayreuth, Germany}
\affil[9]{Fritz-Haber-Institute of the Max-Planck-Society, Berlin, Germany}
\affil[10]{Energy Technologies Area, Lawrence Berkeley National Laboratory, Berkeley, CA 94720, USA}
\affil[11]{U.~S. Naval Research Laboratory, Washington DC 20375, USA}
\affil[12]{Yusuf Hamied Department of Chemistry, University of Cambridge, Lensfield Road, Cambridge, UK}
\affil[13]{Cavendish Laboratory, University of Cambridge, J. J. Thomson Ave, Cambridge, UK}
\affil[14]{Department of Materials Science and Metallurgy, University of Cambridge, 27 Charles Babbage Road, CB3 0FS, Cambridge, United Kingdom}
\affil[15]{Chemix, Inc., Sunnyvale, CA 94085, USA}
\affil[16]{Inorganic Chemistry Laboratory, Department of Chemistry, University of Oxford, Oxford OX1 3QR, UK}
\affil[17]{Scientific Computing Department, Science and Technology Facilities Council, Daresbury Laboratory, Keckwick Lane, Daresbury WA4 4AD, UK}
\affil[18]{BASF SE, Carl-Bosch-Stra{\ss}e 38, 67056 Ludwigshafen, Germany}
\affil[19]{Kavli Institute for Cosmology, University of Cambridge, Madingley Road, Cambridge CB3 0HA, UK}
\affil[20]{Department of Chemistry and Biochemistry, University of South Carolina, South Carolina 29208, USA}
\affil[22]{Institute for Computational Physics, University of Stuttgart, 70569 Stuttgart, Germany}
\affil[23]{Department of Chemistry--Ångström, Uppsala University, Box 538, S-751 21, Uppsala, Sweden}
\affil[24]{School of Natural and Environmental Science, Newcastle University, Newcastle upon Tyne, NE1 7RU, UK}
\affil[25]{Department of Energy Conversion and Storage, Technical University of Denmark, Anker Engelunds Vej 301, 2800 Kgs. Lyngby, Denmark}
\affil[26]{Cambridge Graphene Centre, University of Cambridge, Cambridge, CB3 0FA, UK}
\affil[27]{Aix-Marseille Universit\'{e}, CNRS, CINaM UMR 7325, Campus de Luminy, 13288 Marseille, France}
\affil[28]{Warwick Centre for Predictive Modelling, School of Engineering, University of Warwick, Coventry CV4 7AL, United Kingdom}
\affil[29]{Department of Chemistry and Biochemistry, University of California -- Los Angeles, 607 Charles E. Young Drive East, Los Angeles, CA, 90095 USA}
\affil[30]{Computing Sciences Area, Lawrence Berkeley National Laboratory, Berkeley, CA 94720, USA}
\affil[31]{InstaDeep, London, W2 1AY, United Kingdom}
\affil[32]{Laboratory of Computational Science and Modeling, Institute of Materials, \'Ecole Polytechnique F\'ed\'erale de Lausanne, 1015 Lausanne, Switzerland}
\affil[33]{Department of Chemical and Biological Engineering, Princeton University, Princeton, NJ 08544, USA}
\affil[34]{Hugging Face Inc., Brooklyn, NY 11201, USA}
\affil[35]{Information Engineering, University of Oxford, Oxford, OX1 3PA, UK}
\affil[36]{CNRS, INSA Lyon, Universite Claude Bernard Lyon 1, MATEIS, UMR5510, 69621 Villeurbanne, France}
\affil[37]{Department of Mechanical Engineering, University of Michigan, Ann Arbor, Michigan 48109, USA}
\affil[38]{Department of Physics and Astronomy, University College London, London, WC1E 6BT, UK}

\date{\today}

\maketitle

\begin{sciabstract}
Atomistic simulations of matter, especially those that leverage first-principles (\emph{ab initio}) electronic structure theory, provide a microscopic view of the world, underpinning much of our understanding of chemistry and materials science.   
Over the last decade or so, machine-learned force fields have transformed atomistic modeling by enabling simulations of {\em ab initio} quality over unprecedented time and length scales. 
However, early ML force fields have largely been limited by: (i) the substantial computational and human effort of developing and validating potentials for each particular system of interest; and (ii) a general lack of transferability from one chemical system to the next. 
Here we show that it is possible to create a general-purpose atomistic ML model, trained on a public dataset of moderate size, that is capable of running stable molecular dynamics for a wide range of molecules and materials. 
We demonstrate the power of the \MPz{} model — and its qualitative and at times quantitative accuracy — on a diverse set of problems in the physical sciences, including properties of solids, liquids, gases, chemical reactions, interfaces and even the dynamics of a small protein. 
The model can be applied out of the box as a starting or ``foundation'' model for any atomistic system of interest and, when desired, can be fine-tuned on just a handful of application-specific data points to reach {\em ab initio} accuracy. 
Establishing that a stable force-field model can cover almost all materials changes atomistic modeling in a fundamental way: experienced users get reliable results much faster, and beginners face a lower barrier to entry. 
Foundation models thus represent a step towards democratising the revolution in atomic-scale modeling that has been brought about by ML force fields.
\end{sciabstract}

\addtocontents{toc}{\protect\setcounter{tocdepth}{0}}

\section{Introduction}
An overarching goal in the field of atomistic modeling is to develop an interatomic potential (alternatively also called a ``force field'') that quickly and accurately predicts the total energy and atomic forces for an arbitrary chemical structure.
Existing methods are not capable of this feat: while \emph{ab initio} methods such as density functional theory (DFT) \cite{KohnDFT,qe2020,kuhne2020cp2k,VASP1, Hasnip2014Density,Jain2016Computational,Neugebauer2013Density} are widely applicable and accurate, their high computational cost prohibits their use in many important cases, including high-throughput workflows and those in which large ($\gg$1000 atom) systems need to be simulated over long timescales. 
Conversely, empirical force-field models that use simple functional forms are extremely cheap and so quick to use, but fail to accurately capture the important subtleties of the many-body interactions between collections of atoms induced by quantum mechanics\cite{finnis2003interatomic}. 
Finally, modern machine-learning based interatomic potentials (MLIPs) are capable of faithfully approximating \textit{ab initio} methods for orders of magnitude less cost, but typically require significant upfront investment and human effort when generating and labeling the training dataset \cite{behler2007, gap, THOMPSON2015316, schnet, deringer_machine_2019, drautz2019, lilienfeld2020, batzner2022, ko2023recent}.
Furthermore, these datasets and models typically need to be re-developed from scratch for each new system of interest \cite{deringer_chemrev2021}.
As a remedy to these issues, and as a step towards a truly universal MLIP, we present \MPz{}, a foundation model for materials chemistry that displays an impressive out-of-the-box ability to model a wide variety of chemical systems.
Crucially, we also demostrate that fine-tuning \MPz{} using just a handful of new configurations leads to quantitatively accurate models, dramatically reducing the cost and barrier to entry for the modeling of novel chemical systems.

\MPz{} uses the MACE architecture~\cite{batatia_mace_2023}, which unified the atomic cluster expansion (ACE)\cite{drautz2019, dusson2022atomic, lysogorskiy2021performant, witt2023acepotentials} and equivariant graph neural networks\cite{batatia2022design, batzner2022}. MACE was designed to keep only what appear to be essential components of the latter\cite{batatia2022design}: the element embedding~\cite{schnet, Darby2023Trace} and the equivariant messages constructed through the symmetric tensor product operation. 
MACE's unique innovations are that (i) it uses high body-order equivariant features in each layer (4-body in the present case), and consequently only two layers of message passing are sufficient;\cite{delle_rose_three_2023} (ii) it is only mildly nonlinear, as the only nonlinear activations are in the radial basis and the final readout layer, hence its classification as a graph tensor network, (iii) it uses tensor decomposition~\cite{Darby2023Trace} for efficient parameterization of high body-order features.
The MACE architecture allows \MPz{} to accurately model its training data while remaining competitively performant with other graph neural networks, presently allowing simulations of around a thousand atoms for
nanoseconds per day on a GPU.

Despite training \MPz{} on a dataset with a specific materials focus (MPtrj), the most striking finding is that our model shows remarkable out-of-distribution performance, and leads to stable molecular dynamics (MD) simulations for arbitrary systems over long timescales showing chemically sensible structures, reactions and transformations.
In the main body of this paper, we showcase the generality of \MPz{} by considering three disparate classes of chemical systems: solid and liquid water, heterogeneous catalysis, and metal-organic frameworks.
In the Supplementary Information, we further demonstrate \MPz{}'s capabilities on an unprecedented range of qualitative and quantitative examples drawn from computational chemistry and materials science, including running molecular dynamics simulations for a wide variety of chemistries, predicting phonon spectra, calculating activation energies for point defect and dislocation motion, simulating solvent mixtures, combusting hydrogen gas, modeling a complete rechargeable battery cell, and many more.

There are several versions of the \MPz{} model, all trained on the same data set, with minor variations in the model architecture. Unless otherwise stated, all results in this paper were obtained with the \MPzbt{} version, and this is emphasized in figures and captions, while we retrain the simpler ``\MPz{}'' name in the text for readability. All model versions are publicly available.

\clearpage
\section{Applications}

\subsection{Water and aqueous systems}\label{sec:water_results}

\begin{figure}[!htbp]
  \centering\includegraphics[width=0.9\textwidth,,keepaspectratio]{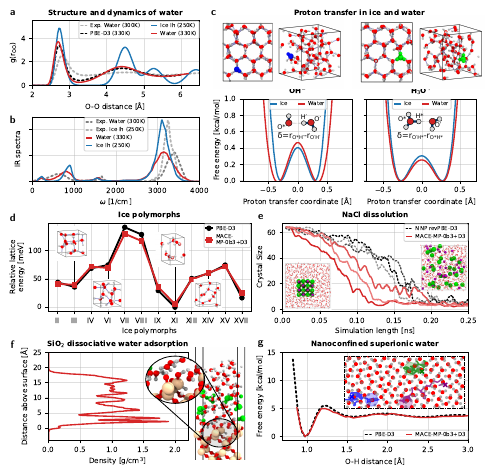}
  \caption{\textbf{Aqueous systems.}
    (a) Oxygen--oxygen radial distribution function for bulk water (experimental result from Ref.~\cite{Skinner2013/10.1063/1.4790861}) and ice Ih.
    (b) Experimental (Ref.~\cite{Bertie1996,Moberg2017}) and computed infrared spectra of bulk water and ice Ih.
    (c) Free energy profiles as a function of the proton transfer barrier for a hydroxide ion and excess proton in ice Ih at \SI{250}{\kelvin} and bulk water at \SI{330}{\kelvin}. Snapshots at the top show the simulation cells.
    (d) Performance of \MPzbt{} (red squares) on the relative lattice energies of the DMC-ICE13 dataset, compared to the reference method, PBE-D3\cite{grimme2010consistent} (black circles). 
    (e) Dissolution of a $4 \times 4 \times 4$ unit-cell \ch{NaCl} nanocrystal in water at \SI{400}{\kelvin}, monitoring the extent of dissolution over the simulation time via the crystal size. Performance of the \MPzbt{} (red lines) is compared to a neural network potential\cite{oneill2022crumbling} trained explicitly to capture \ch{NaCl} dissolution (black dashed lines).
    (f) \ch{SiO2}/water interface simulation showing density modulations and dissociative water adsorption, with an inset highlighting the deprotonation of water as indicated by a shoulder in the water density plot. \ch{H3O+} defects in the liquid are highlighted in green.
    (g) The free energy profile of the \ce{O-H} distance in the superionic phase of monolayer water in a confining potential. The inset shows a snapshot of the monolayer superionic phase with lines indicating the \SI{50}{\pico\second}-long trajectory of randomly chosen hydrogen atoms with ``$\times$'' indicating their initial positions.}
  \label{fig:water}
\end{figure}

Water is ubiquitous in nature and technology and has long been a major focus of computational work. Driven by the delicate balance between directional hydrogen bonding and primarily non-directional van der Waals interactions, aqueous systems remain a challenge for simulations~\cite{gillan2016perspective}.
For example, the study of proton transfer in water, a fundamental process characterized by the continuous breaking and forming of covalent bonds, has long required using {\em ab initio} molecular dynamics for detailed atomistic insight~\cite{Marx1999/10.1038/17579,Tuckerman2002/10.1038/nature00797,Agmon2016/10.1021/acs.chemrev.5b00736}. We demonstrate in this section how \MPz{} describes various aqueous systems.

We start by examining the structure of liquid water and hexagonal ice (ice Ih). The oxygen--oxygen radial distribution function, depicted in \cref{fig:water}a, shows reasonable agreement with reference simulations. The infra-red vibrational spectra of both phases, shown in panel \cref{fig:water}b, align well with experimental observations, albeit with a notable red shift in the stretching vibrations indicating a softer description of the \ch{O-H} bond as is well-known for PBE-D3~\cite{grimme2010consistent, gillan2016perspective}. In panel \cref{fig:water}d, the relative stabilities of 12 ice polymorphs with respect to ice Ih, used in a recent benchmark\cite{dmcice13}, show excellent agreement with respect to PBE-D3 with a MAE of around \SI{5}{meV}. Proton defects (\ch{OH^-} and \ch{H3O^+}) in ice Ih and liquid water were simulated, revealing robust descriptions of proton transfer, as shown in \cref{fig:water}c. The proton transfer barrier for hydroxide is higher than for hydronium in liquid water, consistent with experimental diffusion trends. 

Next, we evaluate \MPz{} for describing solid--liquid interfaces.
First, we focus on \ch{NaCl} in water in two cases: a \ch{NaCl}(001) interface in contact with water and a small nanocrystal surrounded by water.
Simulations were performed at \SI{400}{\kelvin} to promote dissolution, and compared to simulations with a custom-trained ML potential based on revPBE-D3 from Ref.~\cite{oneill2022crumbling}.
As expected, for the flat surface the model predicts no dissolution events on the timescale of the simulation (\SI{0.5}{\nano\second}).
Meanwhile, for the nanocrystal surrounded by water, \MPz{} captures a dissolution mechanism resembling that in Ref.~\cite{oneill2022crumbling} as shown in \cref{fig:water}e. The dissolution proceeds via a crumbling mechanism, where an initial steady loss of ions is followed by the rapid disintegration of the crystal. As ions dissolve from the crystal, they are hydrated by water. The dissolution process is stochastic, leading to an intrinsic variation between independent simulations, as shown by three examples. The final structure of the dissolved ions in water also displays the expected orientation of the water molecules with respect to the ions.

We then model the \ch{SiO2}/water interface, \cref{fig:water}f, revealing the expected density modulations in the first few contact layers. As before, the liquid phase is found to be overstructured, a common characteristic of the PBE functional \cite{gillan2016perspective} used by MPTraj and therefore by \MPz{}. \ch{SiO2} is known for its dissociative water adsorption, which we observe in our simulations. Deprotonation of water is evidenced by the shoulder in the water density plot and can also be seen in the inset of a snapshot of this system in \cref{fig:water}f. 

Finally, we investigate nanoconfined water in graphene-like nanocapillaries~\cite{Algara-Siller2015/10.1038/nature14295,Fumagalli2018/10.1126/science.aat4191}, which exhibits dramatically different properties from bulk water. \MPz{} proved robust in simulating nanoconfined water. Stable simulations were conducted at \SI{4}{\giga\pascal} and \SI{600}{\kelvin}, conditions under which a superionic phase with high ionic conductivity was previously predicted~\cite{kapil_first-principles_2023} using a custom-trained ML potential. The \MPz{} model accurately captured the dynamical characteristics of this phase, including extensive proton transfer on the ten pico-seconds timescale, as illustrated in the inset of \cref{fig:water}g. Comparing the free energy profile associated with the \ce{O-H} distance [Fig.~\ref{fig:water}g] against the PBE-D3 reference, \MPz{} shows near quantitative agreement and an overall very good description of nanoconfined water.
\subsection{Catalysis}\label{sec:catalysis}
\begin{figure}[!htbp]
  \centering\includegraphics[width=0.9\textwidth,keepaspectratio]{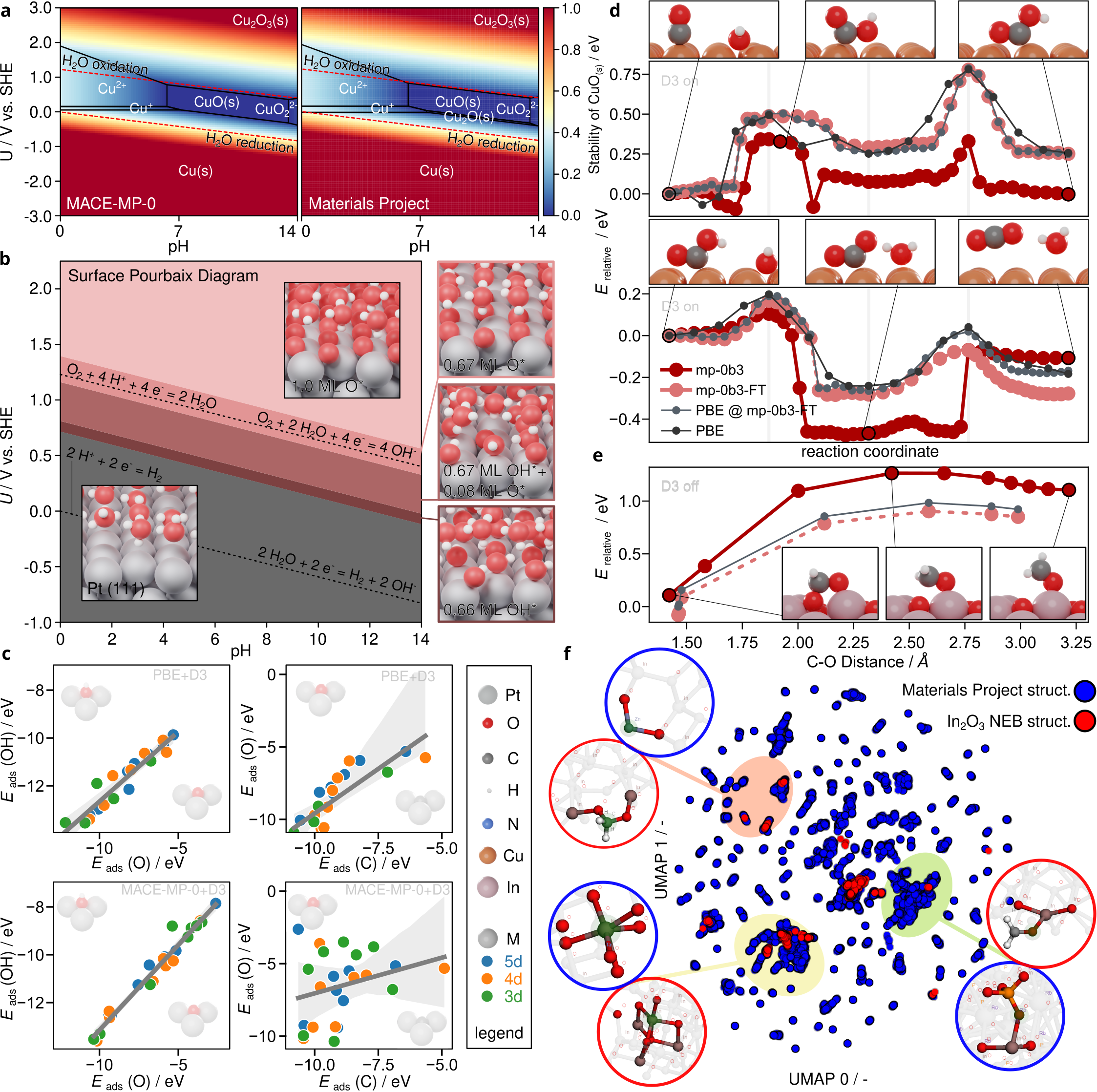}
  \caption{\textbf{Heterogeneous catalysis.}
    (a) Pourbaix diagrams of \ch{CuO} bulk systems constructed with \MPzbt{} (left) and Materials Project reference data (right).
    (b) {\MPzbt{}}+D3-calculated Pt(111) surface Pourbaix diagram, in overall good agreement with the literature\cite{Hansen2008-zz}.
    (c) The relative adsorption energy scaling relation between \ch{O} and \ch{OH} on transition metal surfaces is captured correctly by \MPzbt{}+D3, as is the lack of linear scaling between \ch{C} and \ch{O}~\cite{norskov2022nonlinear}. Metals are colored according to rows in the Periodic Table as 3d, 4d and 5d. 
    (d) Reaction profile of multistep electrochemical \ch{CO} oxidation on \ch{Cu}. \ce{CO-OH} coupling and dehydrogenation reactions are characterised in the upper and lower panel, respectively. Energy profiles from \MPzbt{}+D3 and PBE+D3 nudged elastic band (NEB) calculations show significant deviations although the qualitative features agree. Fine-tuning (FT) yields a model that is in excellent agreement with the reference. 
    (e) \MPzbt{} reaction profile for a key reaction step (\ch{CH2O2 -> CH2 + O}) in the \ch{CO2}-to-methanol conversion on \ch{In2O3}\cite{schaaf2023accurate} and the profile of an FT model.
    (f) Comparison of the atomic environments in the training data (blue) and in the \ch{In2O3} NEB images (red) in the form of a UMAP plot~\cite{mcinnes2018umap, rokasel_2023_10426282}. Insets show local environments with similar MACE features (inset frames in blue for training data and in red for NEB configurations), exemplifying which bulk training environments influence predictions for the out-of-domain catalytic test case.}
  \label{fig:cat_main}
\end{figure}

The study of heterogeneous  \cite{Nrskov2009Towards,Medford2015From,Bruix2019First-principles-based} and electrocatalysis \cite{Qin2023Cation-Coordinated, Man2011Universality, Auer2020Self-activation} is another major area where DFT excels. It provides atomistic insight into the underlying reaction mechanisms and enables the prediction of the properties of new catalytic materials,\cite{Wang2021Ternary} including reaction barriers and rates, which are in turn used to predict turnover frequencies\cite{Norskov2014}. The latter is essential for the computational discovery of new solid catalysts for overcoming the dependence on rare and toxic elements and improving the efficiency of critical processes for energy conversion. However, the computational cost of DFT is a serious impediment. Empirical interatomic potentials are typically inadequate for catalysis applications as they rarely describe chemical reactions accurately. Machine learning has already had strong impact in computational catalysis\cite{Margraf2023Exploring,schaaf2023accurate, Yang2023Neural}, \textit{e.g.}, enabling fast screening of materials spaces\cite{Tran2018Active,Foppa_Sandip_2022,Khatamirad2023}, and free energy calculations beyond the harmonic approximation\cite{schaaf2023accurate,Stocker2023Estimating,Tran2023Open}. However, developing such accurate potentials from scratch still requires significant human and computational effort. 
We now test the performance of \MPz{} for different catalysis applications and summarise the results in \cref{fig:cat_main}.

Potential--pH Pourbaix diagrams are central to understanding the aqueous stability of solid materials in an electrochemical environment\cite{pourbaix1966,pourbaix1973}, and thus allow predicting the active phase of an electrocatalyst under given conditions. Within the computational hydrogen electrode (CHE) framework \cite{Norskov2004-ow}, these diagrams can be computed without an explicit electrostatic model. \Cref{fig:cat_main}a--b show the  Pourbaix diagrams for bulk \ch{CuO} and a \ch{Pt}(111) surface calculated with \MPz{} using the D3 correction. The Pourbaix diagrams are constructed via the formalism described in \cite{Persson2012-kg,Singh2017-uu}, where only the energies of the relevant solids are calculated while corrected experimentally-derived energies are used for the aqueous ions. In both cases, the \MPz{} results show remarkably good agreement with DFT \cite{Hansen2008-zz}, predicting the correct sequence of stable phases (with the exception of a very narrow region of \ch{Cu2O} stability) and corresponding pH and potential ranges. While this accuracy may be expected for the bulk \ch{CuO} system that is represented in the training set, the electrosorption at the \ch{Pt}(111) surface is also well described despite being out of domain.

In \cref{fig:cat_main}c, adsorption energy scaling relations between atomic and hydrogenated adsorbates on transition-metal surfaces are shown for \MPz{} and PBE (see SI for more examples). Such scaling relations are central to understanding the activity of heterogeneous catalysts \cite{norskov2007linear, Lopez2019lsr}. \MPz{} captures these trends well, and the slopes of the linear fits are in reasonable agreement with DFT (\textit{e.g.} \num{0.71} for \ch{O} vs.\ \ch{OH}, compared to \num{0.64} for PBE). 
Importantly, the lack of correlation between \ch{O} and \ch{C} adsorption energies is also captured, indicating that the model is not merely sorting metals according to their general reactivity\cite{norskov2022nonlinear, GarciaMuelas2019redox}.  \Cref{fig:cat_main}d--e show reaction energy profiles for \ch{CO} oxidation on \ch{Cu}
\cite{Tiwari2020} and a key step in \ch{CO2} conversion to methanol on \ch{In2O3}\cite{dang2020rationally, schaaf2023accurate}, respectively. While these are not quantitatively accurate when compared to DFT, \MPzbt{} nevertheless captures the location and magnitude of the barriers surprisingly well. To obtain quantitative agreement, \MPzbt{} is fine-tuned with five single-point DFT calculations from each energy profile. NEB calculations with the fine-tuned (FT) model then yield excellent agreement with the DFT reference in almost all cases, with the exception of the energy of the final state of CO oxidation, which is slightly overestimated by the FT model. Here, describing the subtle non-covalent interactions between the surface and molecular \ch{CO2} and \ch{H2O} would require additional training. Nonetheless, this shows that fine-tuning with very small datasets is sufficient to obtain quantitatively accurate potentials for heterogeneous catalysis.

\Cref{fig:cat_main}f illustrates how \MPzbt{} generalizes to out-of-domain catalysis tasks from bulk training configurations. To this end, the high-dimensional MACE features are projected to 2D using a Uniform Manifold Approximation Projection (UMAP)\cite{mcinnes2018umap}, with local atomic environments in the training set shown in blue and those found in the \ch{In2O3} transition path shown in red. Representative environments with similar features are highlighted, indicating that the internal representation of the atomic environments in the NEB configurations is similar to the representation of under-coordinated environments and metal--organic systems in the training set. 

While \MPzbt{} is not always quantitatively accurate for the most challenging catalysis applications, its stability in MD and exploring reactive pathways is remarkable and provides a starting point for further optimisations. Relevant configurations or phase space regions thus identified may subsequently be validated either by first-principles calculations or serve to initiate active-learning for refining the model, as demonstrated for the NEB calculations. Even at its current foundation level, \MPzbt{} already allows a statistical sampling far beyond the present DFT-based state of the art which is still largely thermochemistry-centered, whereas the foundational MACE model will pave the way for true kinetic modeling by explicit evaluations of reaction profiles and the reactive flux along them.
\subsection{Metal--organic frameworks}\label{sec:mofs}
\begin{figure}[http!]
  \centering
  \includegraphics[width=0.87\textwidth,keepaspectratio]{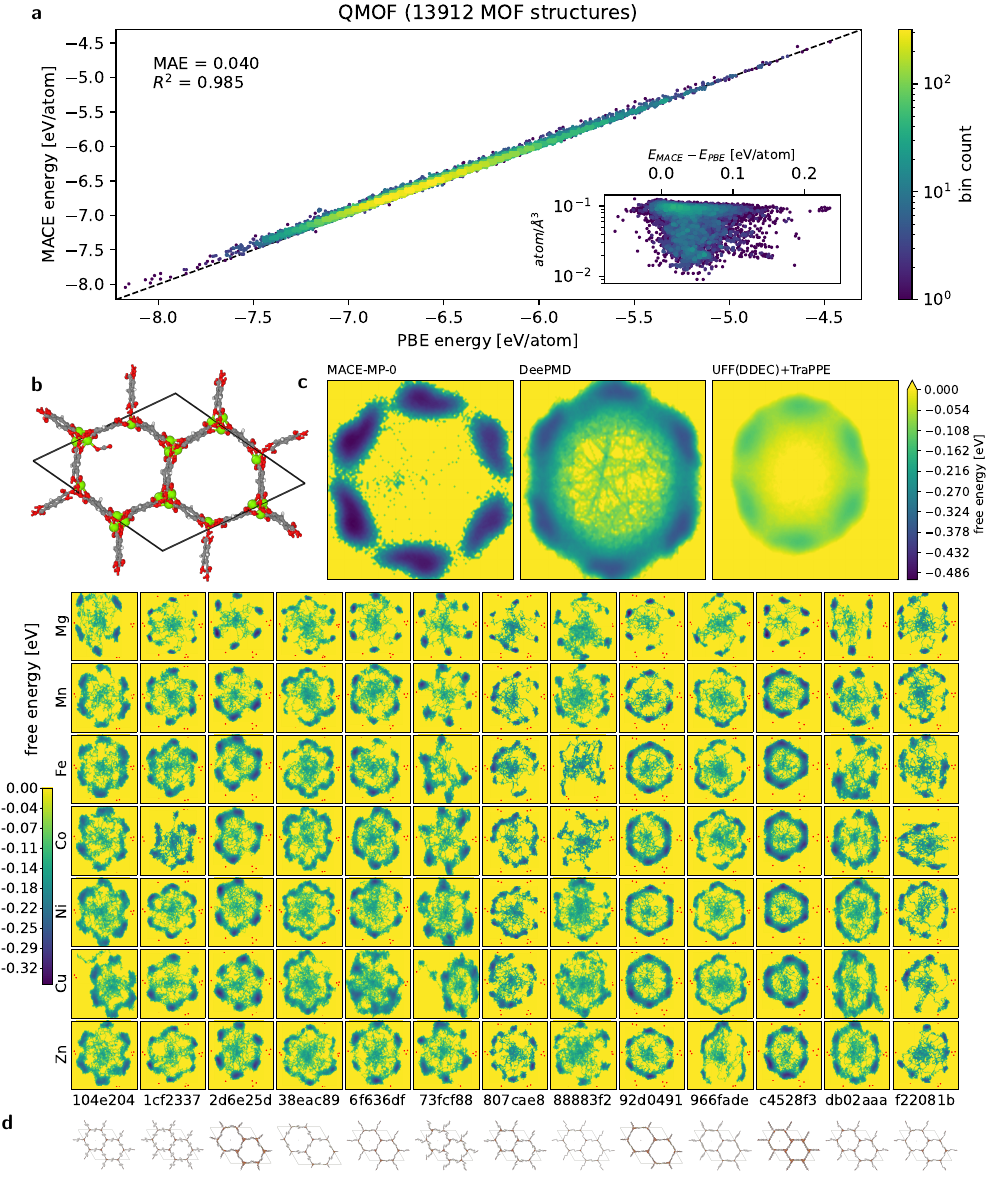}
  \caption{\textbf{Metal--organic frameworks.} (a) Comparison between \MPzbt{} and DFT (PBE) energies on 13,912 relaxed structures with compatible GGA calculations (i.e. without the Hubbard-U correction) and pseudopotentials in the QMOF database \cite{rosen2021machine,rosen2022high}. The inset presents the energy error distribution in relation with the atomic density (number of atoms per volume). The protocol for filtering incompatible calculations is provided in \Cref{sec:si_mofs}. (b) Mg-MOF-74 structure with chemisorbed \ch{CO2} optimized with \MPzbt{}. Color key: \ch{Mg} (orange), \ch{O} (red), \ch{C} (brown), \ch{H} (white). (c) Left: free energy landscape of \ch{CO2} in Mg-MOF-74. Middle: free energy landscape from Ref.~\cite{zeng2023deepmd} using a custom-trained DeePMD ML force field. Right: free energy landscape using the UFF classical force field \cite{rappe1992uff} with DDEC6 charges \cite{manz2016introducing} for the framework and TraPPE for \ch{CO2} \cite{potoff2001vapor}. (d) Free energy maps of 91 hypothetical MOF-74 analogues, with the QMOF ID of the parent \ch{Mg}-containing frameworks indicated at the bottom of each column and the transition metal to the left of each row.}
  \label{fig:mof}
\end{figure}

Metal--organic frameworks (MOFs) are a class of nanoporous materials comprised of metal cations or clusters connected by organic linkers arranged in a periodic lattice \cite{yaghi2019introduction}. Due to their large surface areas, tunable building blocks, and permanent porosity, MOFs hold substantial promise for various applications, including but not limited to catalysis, energy storage, gas adsorption and separations, and optoelectronic devices \cite{yaghi2019introduction}.
We tested our pre-trained model directly against version 14 of the Quantum MOF (QMOF) database, which contains DFT-computed properties at several levels of theory for 13,912 MOFs and structurally related coordination polymers \cite{rosen2021machine, rosen2022high}. \MPzbt{} was not trained on any data from the QMOF database, making this a challenging test of its transferability to largely unseen chemistries.

As shown in \cref{fig:mof}a, \MPzbt{} performs well out-of-box in predicting the PBE energies of MOFs, achieving an MAE of \SI{0.040}{eV/atom} (with the full range of energies spanning nearly 4~eV/atom, about $100$ times larger), despite the pronounced difference between the inorganic crystals of the MPtrj training set and the MOF structures that make up the QMOF database. This accuracy spans most of the periodic table, after exclusion of elements with incompatible pseudopotentials and calculation parameters (see \Cref{sec:si_mofs} and \Cref{fig:mof_si}).

To validate the use of \MPzbt{} for capturing dynamic processes, we investigate \ch{CO2} adsorption in a prototypical MOF known as Mg-MOF-74. The MOF-74 family, including the Mg-containing version, has been extensively studied for the selective adsorption of \ch{CO2} \cite{britt2009highly,queen2014comprehensive, choe2021mof}. Of particular note, the coordinatively unsaturated metal sites \cite{kokccam2020coordinatively} of Mg-MOF-74 enable chemical bonding interactions between the metal and \ch{CO2} adsorbate \cite{britt2009highly} that cannot be captured from classical force fields alone. We directly compare the adsorption dynamics against the results presented in Ref.\cite{zheng2023quantum}, which considered the same system using a custom-trained ML force field generated using DeePMD-Kit \cite{zeng2023deepmd} and PBE-D3 calculations in CP2K \cite{kuhne2020cp2k}.

\MPzbt{} accurately and efficiently captures the \ch{CO2} adsorption process in Mg-MOF-74. 
As shown in \cref{fig:mof}c, the \ch{CO2} adsorbate favorably binds to the Mg center in a tilted configuration that is in agreement with both experimental neutron diffraction data \cite{valenzano2010computational,queen2014comprehensive} and the previous custom-trained ML model \cite{zheng2023quantum}. The mean bond distance between the \ch{Mg} center and \ch{CO2} adsorbate is predicted to be \SI{2.27}{\Angstrom} from \MPzbt{} (\Cref{fig:mof_si_a}), in  agreement with the experimental value of \SI{2.27}{\Angstrom} \cite{queen2014comprehensive} and the value of \SI{2.23}{\Angstrom} from the custom ML model in Ref. \cite{zheng2023quantum}. The mean \ce{Mg-O-C} bond angle is predicted to be \ang{137.3} from \MPzbt{} (\cref{fig:mof_si_a}), substantially closer to the experimentally determined bond angle of 131$^\circ$ \cite{queen2014comprehensive} than the \ang{118.6} value from the ML model in Ref.~\cite{zheng2023quantum}. The projected density map for the \ch{CO2} adsorption site (\cref{fig:mof}b) is, again, in excellent agreement with prior work \cite{valenzano2010computational,zheng2023quantum} and shows how the adsorbed \ch{CO2} molecules are mobile but largely confined to the vicinity of the Mg binding site due to chemisorption.

To showcase an example of how one might use the foundation model in a high-throughput setting, we considered 91 hypothetical MOF-74 analogues derived from those in Ref.\ \cite{witman2016silico} based on 13 (out of 58) different frameworks and seven different metal cations (M) that have been used to synthesize M-MOF-74 \cite{queen2014comprehensive}. \Cref{fig:mof}e shows the resulting free energy maps, comprising over \SI{357}{\nano\second} of simulation altogether, displaying diverse and dynamic behaviour of the \ch{CO2} adsorbate across the range of hypothetical MOF-74 analogues.

Given the nature of our foundation model, we anticipate many additional application areas where \MPz{} (or one of its future variants) could be of value in the MOF field. Based on the \ch{CO2} adsorption example, we envision applications in capturing dynamic processes, particularly those that cannot be accurately modeled using classical force fields and are prohibitively expensive to carry out with {\em ab initio} MD given the large unit-cell size required to describe most MOFs. Foundation models are promising for modeling competitive multi-component physisorption and chemisorption processes, especially across many families of compositionally different MOFs and combinations of gas mixtures, for which training a system-specific, on-the-fly active learning model would be expensive or even prohibitive.
In addition to the compositional diversity relevant to high-throughput screening, not all MOFs can be described via a static picture and based on an ideal crystalline structure: in fact, there has been recent interest in liquid and amorphous MOFs \cite{bennett2018liquid,castel2022atomistic}, and the dynamic behavior of crystalline frameworks \cite{evans2020four} --- such as in the so-called ``flexible'' and ``breathing'' MOFs --- has been leveraged for highly selective separation processes \cite{taylor2018near}. This dynamic behavior cannot be completely captured from static DFT calculations alone, and accurate and easily accessible interatomic potentials are expected to accelerate the modeling of spatio-temporal processes in future studies \cite{van_speybroeck_towards_2021}.


\subsection{A wide range of applications and benchmarks}

In the Supplementary Information in 32 subsections, we provide a rather wide ranging set of application examples to support the claim that the \MPz{} is a robust modeling tool, and when fine-tuned can reach {\em ab initio} accuracy. We also give the results of a comprehensive set of benchmarks, including the performance on calculating phonon dispersions, bulk and shear moduli of crystals, atomisation energies and lattice constants of elemental solids, the cohesive energies of the S66 set\cite{S66} of molecular dimers and the X23 set\cite{x23_rt} of molecular crystals, the CRBH20 set\cite{crbh20} of reaction barrier heights, and the homonuclear diatomic binding curves. The full set of heteronuclear diatomic curves is provided in the Supplementary Materials. 

We also give more details on the training protocol, a graphical exploration of the data, including histograms of energies, forces, stresses, magnetic moments, and element and composition counts, and a discussion of the quantification of the uncertainty in the model predictions. 

\section{Fine-tuning}
\begin{figure}
    \centering
  \includegraphics[width=0.87\textwidth,keepaspectratio]{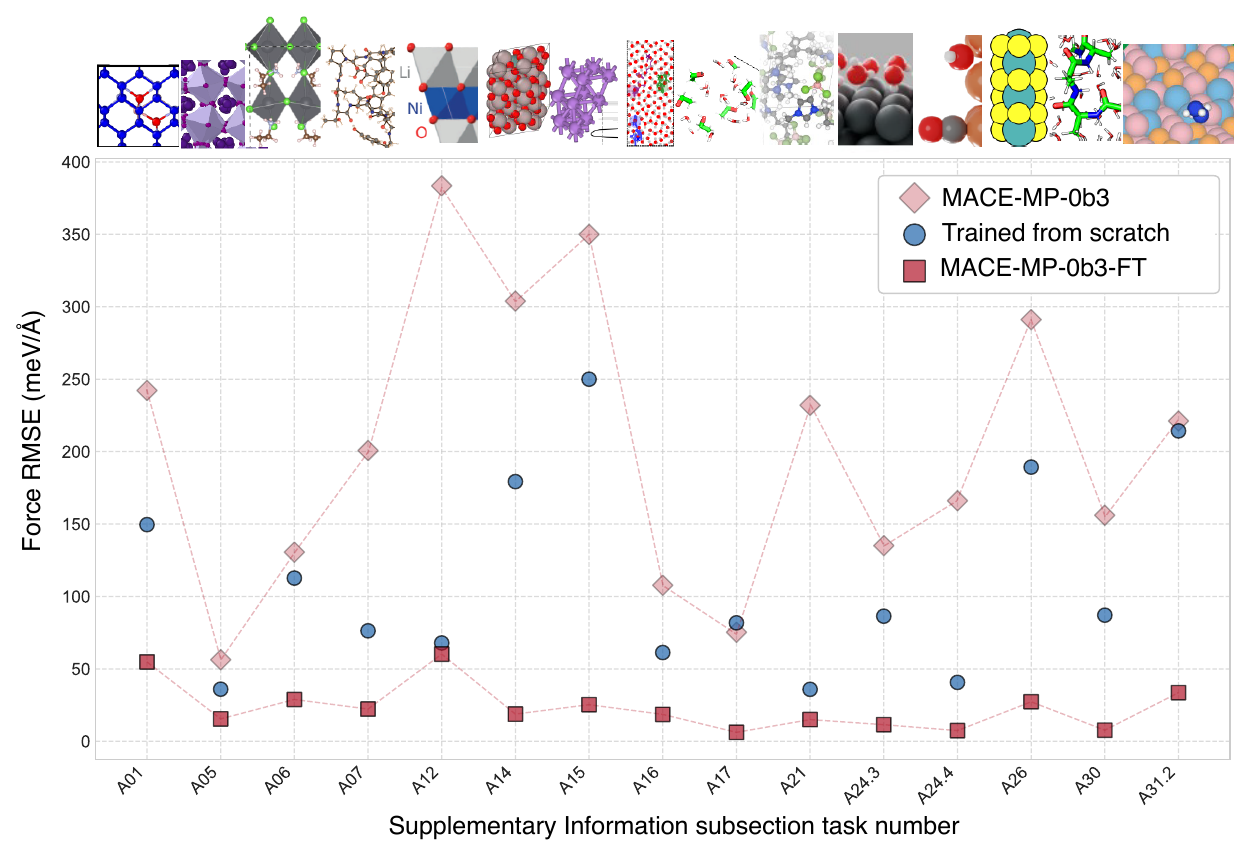}
  \caption{\textbf{Fine-tuning.} A comparison of force RMS error on selected applications in the SI for which fine-tuning was performed. The \MPzbt{} model is shown with pink diamonds and the fine-tuned model (\MPzbt-FT) for each application with red squares. For comparison, in each case we also show the results corresponding to a model trained ``from scratch'' only to the small amount of fine-tuning data (blue circles).}
    \label{fig:finetuning}
\end{figure}

Although the multitude of applications demonstrates that \MPz{} is a robust model, it is also clear that, in many cases, it is not accurate enough out of the box to rival or replace \textit{ab initio} calculations. For a selection of examples, we performed fine-tuning on configurations generated using \MPz{}, typically via molecular dynamics or other downstream tasks appropriate for the application. We used approximately 100 new configurations for each application during fine-tuning. 
To prevent catastrophic forgetting~\cite{McCloskey1989CatastrophicII} and retain the robustness of the foundation model, we introduce a new fine-tuning protocol: \emph{multi-head replay fine-tuning}. This approach includes a subset of the foundation model training data in the loss function while fine-tuning on the new data (see Appendix~\ref{sec:finetuningprotocol} for details). We train a separate model for each case using this multi-head fine-tuning protocol with replay. 
Figure~\ref{fig:finetuning} shows the resulting force errors, which decrease significantly in every case. For comparison, we also present the force errors of a MACE model trained just on the small fine-tuning dataset. In almost all cases, the force errors of the model trained from scratch are significantly worse than those of the fine-tuned model. In each corresponding subsection of the SI, we demonstrate the performance of the fine-tuned model on application-relevant observables, showing considerable improvement over the original model in every case.


\section{Related work: general purpose MLIPs}

The development of the \MPz{} models as a foundation models for atomistic materials simulation follows more than a decade of intense activity and progress in making MLIPs for specific materials. 
General purpose MLIPs -- {\it i.e.,} models that aim to target a wide range of chemical systems spanning many possible combination of elements -- are much more recent.
Here, we summarise the brief history of such general purpose models as well as the culmination of this trend into the creation of true ``foundation models''.
Within this commentary, we seek to highlight the particular merits of \MPz{} in comparison to existing alternatives. It is worth noting that the reason we call \MPz{} ``foundational''  is because of how it can be used, as already mentioned in the introduction:  
the model is suitable for many different applications as a tool for initial exploration, but it likely requires fine-tuning for specific simulation tasks to achieve quantatively accurate predictions.  

A key advance towards making general purpose MLIP models was made by MEGNet, introduced in 2019\cite{megnet2019}. This model, which provides property prediction for inorganic crystals, was trained on minimum energy configurations in the Materials Project (MP)~\cite{jain2013commentary} that includes most elements of the periodic table (89).%
Subsequently, models that predict atomic forces were also trained on MP-based datasets, including M3GNet~\cite{chen_universal_2022} and CHGNet~\cite{deng_chgnet_2023}, which were trained on snapshots of DFT relaxations of MP structures, with CHGNet using the MPtrj dataset introduced at the same time\cite{deng_chgnet_2023}. 
The ALIGNN-FF model\cite{choudhary2023unified} was trained on a database of inorganic crystals, JARVIS-DFT\cite{jarvis}, which covers 89 elements and uses the optB88vdW exchange-correlation functional\cite{optB88vdW}. 
The proprietary GNoME \cite{merchant_scaling_2023} (based on the NequIP architecture \cite{batzner2022}) model also starts from MP, but uses a complex active learning workflow to generate and train on a dataset of inorganic crystals nearly two orders of magnitude larger than MPtrj. 
The above models were created primarily for the purpose of ``materials discovery'',  i.e. predicting thermodynamic stability of hypothetical inorganic crystals. In addition, they were capable of molecular dynamics for such crystals, and indeed both CHGNet and GNoME were used to study alkali metal ion diffusion in battery materials. 
More recently, the DPA models (DPA-1\cite{dpa1} and DPA-2\cite{dpa2}) were trained to a wide variety of datasets (with 56 and 73 elements, respectively), a combination of some previously available and some released with the models (altogether 4M configurations). The second paper reports MD results for versions of the baseline model fine-tuned separately to specific systems (e.g. water, solid-state electrolytes, ferroelectric oxide).
To date, the most general and transferable force field for molecular dynamics is the PFP model \cite{takamoto2022towards} (TeaNet architecture \cite{takamoto2022teanet}), also proprietary (including its training set that originally covered 45 elements, recently updated to 72 elements\cite{Takamoto2023}, and is significantly larger than MP and also covers molecules and surfaces). PFP was demonstrated for running simulations on solid-state ionic conductors, and a molecular adsorption and a heterogeneous catalysis example---systems that formed part of its training data set.
There are also ML force fields specialized for organic molecules (with a much more limited number of elements)
such as the ANI (and later AimNET) series of models\cite{smith_ani-1_2017, Smith2019,aimnet2019} and the MACE-OFF models\cite{kovacs2023maceoff}, as well as for metal alloys\cite{lopanitsyna2023modeling}.
However, there has yet to be a comprehensive demonstration that a single ML potential can describe solid, liquid, and gaseous systems of materials and molecules across the periodic table and well beyond the distribution of the underlying training set.
\footnote{Since the first preprint version of this manuscript, a number of models have been fitted to the same MPtrj data set and also to larger extended datasets including the Alexandria\cite{Schmidt2024Improving} and OMat24.\cite{barroso-luque-2024-OpenMaterials}. Notable models (reported in preprint form) that showed high in-domain accuracy include SevenNet\cite{kim_sevennet_mf_2024} (based on the NEquIP architecture\cite{batzner2022}), GRACE\cite{Bochkarev2024Graph}, Orb\cite{Neumann2024Orb,orbv3}, EquiformerV2\cite{barroso-luque-2024-OpenMaterials}, MatterSim\cite{Yang2024MatterSim}and eSEN\cite{eSEN}. Of these, the MatterSim models, have been tested in the molecular dynamics for some materials including polymers and surfaces. The GRACE theoretical framework formally generalises MACE, but actual released GRACE models remain in close correspondence with the MACE design choices. To compare with these newer models we also include results for a new model termed MACE-MPA-0 in the SI, which has been trained on an extended dataset including MPtrj and Alexandria.}
%


\section{Outlook}

The stable MD propagation for a wide range of materials across the periodic table and the DFT-quality simulation (in some cases after  fine-tuning) that we have shown here are landmark achievements for a single machine-learned interatomic potential. 
In this sense, we expect that the present study will have implications for the wider development of the field, beyond any specific model parameterisation.
Yet there are a number of limitations of the current (``b3'' and ``MPA-0'') versions of the \MPz{} foundation model.
The exchange--correlation functional used in the MPtrj dataset is PBE~\cite{perdew1996generalized}, which must be augmented with Hubbard $U$ terms 
to improve electronic correlations for particular element combinations (introducing inconsistencies in the PES that must be 
compensated~\cite{Jain2016Computational}), and dispersion corrections, such as D3\cite{grimme2010consistent}. Recent developments in DFT are beginning to supersede conventional GGA functionals by achieving improved accuracy at comparable computational cost~\cite{Furness2020Accurate,Kingsbury2022Performance}, and methods beyond DFT such as hybrid functionals~\cite{henderson_accurate_2011} and the random phase approximation~\cite{harl_assessing_2010} improve 
upon this even further, but at much larger computational cost.
Refitting or fine-tuning the model to a more modern functional is expected to increase its predictive power, and will reduce the need for system-dependent corrections such as the use of Hubbard $U$ terms and dispersion.
(Note that the above mentioned inconsistency is not present in the more recent MATPES dataset,\cite{matpes} which removes the Hubbard U correction altogether.)

The MACE architecture that we used to fit the data presently does not contain explicit long-range interactions  (beyond the \SI{12}{\Angstrom} receptive field afforded by two steps of message passing), nor does it take into account magnetic or spin degrees of freedom. Despite the success in describing many different chemistries demonstrated herein, there will be observables, particularly in the context of dilute solutions and at interfaces, that cannot be calculated with a short-range model. There are several approaches to incorporating explicit electrostatic interactions into atomistic ML models in the literature\cite{Ghasemi2015Interatomic,Grisafi2019Incorporating,Ko2021fourth-generation,Vondrak2023q-pac}, as well as spin degrees of freedom\cite{mtpmagnetic2022,rinaldi2023noncollinear, deng_chgnet_2023}. In the future, foundation models could undoubtedly benefit from such an extension.

Considering the results for the diverse systems shown in the SI, a particular area where the model clearly needs improvement is describing intermolecular interactions. While the overarching goal of MD stability is achieved, for many systems there is room for improvement in a quantitative sense, for example in obtaining more accurate densities of molecular liquids, such as ethanol-water mixtures (section~\cref{sec:ethwat}).
The present version of the potential includes a repulsive pair potential\cite{byggmastar2019} that helps describe the repulsive interaction of atoms at close range, the accuracy of the model (e.g., in predicting the equation of state) at high pressures is limited due to the absence of data in this regime. This can easily be remedied either by active learning\cite{Yang2024MatterSim} or a more systematic approach, e.g. replicating part of the MP dataset at lower and higher densities. 

Although we have described an example of a model with wide generalisation, we expect that there will be considerable improvements possible both in the model architecture and in optimising the way in which data is assembled, and the model is fine-tuned.\cite{Gardner2023, dpa2,BenMahmoud-24-06}
It is an open question whether the biggest gains will be obtained by improving the underlying data (both the amount and the consistency) or by scaling the size and expressivity of the model. 
There is good evidence that reaching higher levels of electronic structure theory (such as improved XC functionals) from a DFT baseline and beyond requires significantly less data than fitting to DFT itself\cite{bartok_water_2013,Smith2019,dral2020}, and we show an example of this in the SI, where we fine-tune the model to data computed with the r2SCAN functional\cite{r2scan}.  

Finally, there is the tantalising possibility that with some improvements,  it will be possible to make an ML force field model that achieves quantitative agreement with explicit electronic structure theory across the full range of chemistry and structure. If this turns out to be true, future foundation models may truly provide a universal model for carrying out atomistic simulations at scale.

\clearpage
\section{Methods}
\label{sec:methods}

\paragraph{MACE} All models trained in the paper use the MACE~\cite{batatia_mace_2023} architecture implemented in PyTorch~\cite{Paszke2019PyTorchAI} and employing the {\em e3nn} library~\cite{geiger2022e3nn}. The MACE training and evaluation codes are distributed via GitHub under the MIT license, available at \url{https://github.com/ACEsuit/mace/}. The models used in this paper are available at \url{https://github.com/ACEsuit/mace-mp/}.
MACE is an equivariant message-passing graph tensor network where each layer encodes many-body information of atomic geometry.
At each layer, many-body messages are formed using a linear combination of a tensor product basis~\cite{batatia2022design, Darby2023Trace}.
This is constructed by taking tensor products of a sum of two-body permutation-invariant polynomials, expanded in a spherical basis.
The final output is the energy contribution of each atom to the total potential energy.
For a more detailed description of the architecture, see Refs.~\cite{batatia_mace_2023} and \cite{KovacsBenchmark2023}.

\paragraph{Model versions} Different model versions have been released based on this work, including the models used in the first version of the manuscript, now named MACE-MP-0a, and the model used in the present version, named \MPzbt. All previous models can be found at \url{https://github.com/ACEsuit/mace-mp/}. Unless otherwise stated in the text, all models used in this paper correspond to \MPzbt. We use the label ``\MPz'' to refer to this model series generally. 

\paragraph{Hyper-parameters} The model referred to in this work uses two MACE layers, a spherical expansion of up to $l_\text{max}=3$, and 4-body messages in each layer (correlation order 3).
The model uses a \num{128}-channel dimension for tensor decomposition.
We use a radial cutoff of \SI{6}{\Angstrom} and expand the interatomic distances into 10 Bessel functions multiplied by a smooth polynomial cutoff function to construct radial features, in turn fed into a fully-connected feed-forward neural network with three hidden layers of \num{64} hidden units and SiLU non-linearities.
We fit an $L=1$ model, corresponding to a ``medium sized'' model, as it represents a good compromise. More efficient models that only pass invariants during the message passing step ($L=0$) or those that pass higher order tensors ($L\geq2$) are straightforward to train, and can form part of the accuracy/efficiency tradeoff in selecting the optimal model in the future.  
The irreducible representations of the messages have alternating parity (in {\em e3nn} notation, $\texttt{128x0e + 128x1o}$). 

\paragraph{Distance transforms and pair repulsion}

Smooth behavior of the potential at close approach is essential for a broadly applicable model. We use a combination of Ziegler–Biersack–Littmark (ZBL)\cite{zbl} core potential to the short-range repulsive forces, and distance transformation to smoothly connect this behavior to equilibrium interactions. The ZBL energy is given by,
\begin{align}
    E_{\text{ZBL}} &= \sum_j \frac{14.3996 \cdot Z_i \cdot Z_j}{r_{ij}} \cdot \phi(r_{ij}/a) \cdot \text{Envelope}(r_{ij}, r_{\max}, p), \\
    \phi(r/a) &= c_0 e^{-3.2 (r/a)} + c_1 e^{-0.9423 (r/a)} + c_2 e^{-0.4028 (r/a)} + c_3 e^{-0.2016 (r/a)}.
\end{align}
where $Z_u$ and $Z_v$ are the atomic numbers of the interacting atoms, and $r_{ij}$ is the interatomic distance between atoms $i$ and $j$. The screening length $a$ is given by $a = 0.529 \cdot a_{\text{prefactor}} / (Z_i^{a_{\text{exp}}} + Z_j^{a_{\text{exp}}})$. The coefficients are $c = \{0.1818, 0.5099, 0.2802, 0.02817\}$. The maximum cutoff radius is defined as $r_{\max} = R_{\text{cov}}(Z_u) + R_{\text{cov}}(Z_v)$. The envelope function $\text{Envelope}(r, r_{\max}, p)$ is a polynomial cutoff function applied to smooth the potential. We use the same envelope as the radial basis. To smoothly transition from the the ZBL to the MACE energy, we use the Agnesi distance transform \cite{acejl2023},
\begin{equation}
    y_{ij} = \left(1 + \frac{a \cdot (r_{ij} / r_0)^q}{1 + (r_{ij} / r_0)^{q - p}} \right)^{-1}.
\end{equation}
where $y_{ij}$ is the transformed distance, and the parameters $a$, $q$, and $p$ control the shape of the transformation, $r_0 = \frac{1}{2} (R_{\text{cov}}(Z_u) + R_{\text{cov}}(Z_v))$. We then evaluate the radial basis in this transformed space instead of directly on the distances.

\paragraph{Normalization}
To ensure internal normalization of the weights and smooth extrapolation to high pressure systems, we divide the atomic basis in each layer by a learnable quantity called density normalization $e_{i}$,
\begin{equation}
 \label{eq:density-normalization}
   e_{i} = 1 +\sum_j \tanh \big( \text{SiLU} \big( \big [\sum_k W_{k} B_{k}(r_{ij}) \big ] \big)^{2} \big)
\end{equation}
where $B$ denotes a set of Bessel basis and $W$ are learnable weights. The predicted density normalization varies between 1 and the number of neighbors of atom $i$, depending on the local environment. This normalization corresponds to a smooth version of the node degree normalization in graph neural networks~\cite{kipf2017semisupervised}.
The node energy $\epsilon_{a}$ of atom $a$ is shifted by the isolated atoms energies.
Therefore, the prediction of the energy for the whole structure is constructed as
\begin{equation*}
\hat{E} = \sum_{a=1}^{N} \left[\sigma \left(\sum_{k=1}^K \epsilon_a^{(k)}\right) + \mu_{Z_a}\right]
\end{equation*}
where $K$ denotes the total number of message passing layers and $\epsilon_a^{(k)}$ is the energy of atom $a$ at layer $k$.
$\mu$ and $\sigma$ are the isolated atomic energies and the mean square of the atomic forces computed on the training set.
The predicted forces and stresses are computed as derivatives of the total energy with respect to the atomic positions and the strain tensor, respectively.

\paragraph{Training loss}
The models were trained using a weighted sum of Huber losses of energy, forces, and stress:
\begin{equation}
  \begin{aligned}
    \mathcal{L} & = \frac{\lambda_E}{N_b} \sum_{b=1}^{N_b}\mathcal{L}_\text{Huber}\biggl(\frac{\hat{E}_b}{N_a}, \frac{E_b}{N_a}, \delta_E\biggr) \\ 
    &\hphantom{=}+ \frac{\lambda_F}{3\sum_{b=1}^{N_b}N_a} \sum_{b=1}^{N_b} \sum_{a=1}^{N_a}\sum_{i=1}^{3}\mathcal{L}^\star_\text{Huber}\biggl(-\frac{\partial\hat{E}_b}{\partial {r}_{b,a,i}}, F_{b,a,i}, \delta_F\biggr) \\
                & \hphantom{=}+ \frac{\lambda_\sigma}{9N_b} \sum_{b=1}^{N_b}\sum_{i=1}^3\sum_{j=1}^3\mathcal{L}_\text{Huber}\biggl(\frac{1}{V_b}\frac{\partial\hat{E}_b}{\partial {\varepsilon}_{b,ij}}, \sigma_{b,ij}, \delta_\sigma\biggr),
  \end{aligned}
  \label{eq:loss}
\end{equation}
where $\lambda_E, \lambda_F, \lambda_\sigma$ are predetermined weights of energy ($E$), forces ($F$), and stress ($\sigma$) losses, the symbols under a hat correspond to predicted values, and $N_b$ and $N_a$ are the batch size and the number of atoms in each structure. In the last term involving the stress, $\varepsilon_b$ and $\sigma_b$ correspond to the strain and stress tensors, respectively. 
We used $(\lambda_E, \lambda_F, \lambda_\sigma) = (1, 10, 10)$ and Huber deltas of $\delta_E = 0.01, \delta_F = 0.01,  \delta_\sigma = 0.01$.
We use a conditional Huber loss $\mathcal{L}^\star_\text{Huber}$ for forces, where the Huber delta $\delta_F$ is adaptive to the force magnitude on each atom.
The Huber delta $\delta_F$ decreases step-wise by a factor from \num{1.0} to \num{0.1} as the atomic force increases from \num{0} to \SI{300}{eV\per\Angstrom}.
For more details, see the section~\ref{sec:trainingprotocol} in the SI.

\paragraph{Optimization} The models are trained with the AMSGrad~\cite{Reddi2019} variant of Adam~\cite{Kingma2014} with default parameters $\beta_1 = 0.9$, $\beta_2 = 0.999$, and $\epsilon=10^{-8}$.
We use a learning rate of 0.001 and a exponential moving average (EMA) learning scheduler with decaying factor of 0.99999.
We employ a gradient clipping of 100. The training curves for the medium model is presented in \cref{fig:metrics} in the SI.
Model is trained for 100 epochs on \numrange{40}{80} NVIDIA H100 GPUs across \numrange{10}{20} nodes.
Training the medium-sized model took approx.\ 2,600 GPU hours. We find that \MPz{} achieves an energy MAE of \SI{18}{\milli\eV\per atom} and a force MAE of \SI{39}{\milli\eV\per\Angstrom} for the medium model. After fine-tuning with higher weights for energies for an additional 50 epochs, the small model is able to achieve an energy MAE of \SI{13}{\milli\eV\per atom} (see SI \ref{sec:trainingprotocol}).

\paragraph{Performance}
The speed of evaluation of the \MPz{} model depends on the atomic density, hardware, floating point precision, size of model, \textit{etc.} (see section SI~\ref{sec:timings} for details), but a rough guide is that on a single NVIDIA A100 GPU with 80GB of RAM, it can do several nanoseconds per day for 1000 atoms. When run in parallel using domain decomposition, weak scaling at \SI{0.1}{\nano\second\per day} is perfect up to 32,000 atoms and 64 GPUs for a dense metallic alloy.  

\paragraph{Training data}

The \MPzbt{} model was trained on the \texttt{MPtrj} dataset which was compiled originally for CHGNet~\cite{deng_chgnet_2023}. This dataset consists of a large number of static calculations and structural optimization trajectories from the Materials Project (MP)~\cite{jain2013commentary}.
These include approx.\ $1.5$M configurations (roughly ten times the approx. $150$k unique MP structures), mainly small periodic unit cells (90\% under \num{70} atoms) describing inorganic crystals with some molecular components.  The DFT calculations use the PBE exchange-correlation functional with Hubbard $U$ terms applied to some transition metal oxide systems, but no additional dispersion correction~\cite{MP_calc_details}.

Since the potential we fit calculates the energy based only on structural information, ideally we would like to use consistent electronic calculation parameters and the lowest energy electronic state for each configuration. One significant source of inconsistency is the application of Hubbard $U$, which is used in MP calculations only when \ch{O} or \ch{F} are present together with any of 8 transition metals (\ch{Co}, \ch{Cr}, \ch{Fe}, \ch{Mn}, \ch{Mo}, \ch{Ni}, \ch{V}, \ch{W})~\cite{MP_Hubbard_U}.
The application of $U$ leads to a shift in energy correlated with the value of $U$, \textit{i.e.} a few \unit{eV}, not explicitly accounted for in our fit. Thus, energies from calculations using those \num{8} elements with and without \ch{O} or \ch{F} are inconsistent (in the sense that the energy along a continuous deformation path that removes the \ch{O} or \ch{F} atoms from around these metals would be discontinuous).
The pre-trained CHGNet fit to MPtrj used energies corrected to account for the presence or absence of $U$~\cite{Jain_PRB_2011}.
In our fit, this shift only occurs between structures with different compositions and for any given composition the energies should be consistent.
As a result, we expect configurations that include local regions of these metals with very different \ch{O} or \ch{F} content, \textit{e.g.} an interface between a metal and an oxide, may be poorly described.

In addition, the current fitting database includes a variety of magnetic orders generated as part of a systematic search for the magnetic ground state~\cite{Horton_npjCM_2019}, chosen from the full database only based on calculation type (``GGA Static'' and ``GGA Structure Optimization'') and energy-difference criteria~\cite{deng_chgnet_2023}.
To quantify the effect of this additional and unaccounted-for degree of freedom, we classify the magnetic order associated with each calculation task into one of four categories: 1) no atomic magnetic moment listed, 2) moment converged to zero on all atoms, 3) converged to ferromagnetic order, and 4) converged to another magnetic order.
Of the approx.\ $150$k MP-IDs present, about $48$k have more than one magnetic order present in the fitting database.
In the vast majority of cases, this includes a calculation where the moments are {\em unknown} (\textit{i.e.} not recorded) and a single other magnetic order, and we can hope that they are actually consistent.
However, for \num{5186} MP-IDs we find multiple non-trivial magnetic orders.
To quantify the effect on the fitting quantities, we calculate the minimum energies of each magnetic order for each material, and
analyze the range of minima values seen for each material (distribution is plotted in SI \cref{fig:E_magmom_range}).
While the vast majority of materials have negligible variation, there are hundreds with variation \SI{>100}{meV/atom} (\textit{i.e.} an order of magnitude larger than the energy error on the validation set), and a few that vary by \SI{<0.5}{eV/atom}.

\paragraph{Long-range dispersion corrections}

Dispersion interactions, sometimes called van der Waals interactions, are crucial for describing the weak, long-range interactions between electrons.
Common approximations in DFT, such as PBE \cite{perdew1996generalized}, cannot capture such long-ranged interactions, motivating the use of additive non-local corrections, such as DFT-D3 \cite{grimme2010consistent} or rVV10 \cite{sabatini2013rVV10}.
Inclusion of a dispersion correction to DFT is necessary to describe the dynamics of liquid water \cite{lin2012water}, the geometries and binding energies of layered solids \cite{terentjev2018layered}, and stability of metal--organic frameworks \cite{formalik2018MOFvdW}, among many other examples.

Additive dispersion corrections typically employ a physical model for dispersion interactions with empirical parameters optimized to cut off the correction at interatomic distances where approximate DFT is reliable.
DFT-D3 is an interatomic potential which uses tabulated values of atomic polarizabilities to describe two-body and, optionally, three-body Axilrod--Teller \cite{axilrod1943threebody} dispersion interactions.
As \MPzbt{} is trained to PBE energies, forces, and stresses, it inherits PBE's lack of long-range dispersion interactions.
An optional, additive DFT-D3 dispersion correction can be applied to \MPzbt{}.
The PyTorch implementation of DFT-D3 used in this work is described in Ref. \cite{takamoto2022towards}.
The same parameters used in PBE-D3(BJ), i.e., DFT-D3 with a Becke-Johnson damping function \cite{grimme2011effect}, are used in the D3 correction to \MPzbt{}.

\section*{Author contributions}

\textbf{Model training:} YC, PB, IB, CL; \textbf{Data/Model analysis:} PB, YC, JR, NB, RE, MCK, ES, IB; \textbf{MACE code:} IB, YC, SWN, DPK, PB, WCW, MA, SV, ES, CL; \textbf{Application examples:} WJB (\ch{CsPbI3}, \cref{sec:pero_ino}); LLS (catalysis: \ch{In2O3}, \cref{sec:catalysis,sec;in2o3}); IB (a-C quenches, \cref{sec:C-mq}); ZEM (a-C graphitisation, \cref{sec:C-gra}); NK (Si interstitials, \cref{sec:si_si}); EVU, XRA, NON (aqueous interfaces, \cref{sec:water_results,sec:aqueous_interfaces}); YC (molten salts, \cref{sec:molten_salts}); CSc, VK, FDP, XRA (water and ice, \cref{sec:water_results,sec:water-bulk-nano}); SPN and AGS (\ch{LiNiO2}, \cref{sec:LNO}); SWN (lithiated graphite, \cref{sec:LiC}); AME (zeolites \cref{sec:zeo}); JJ (transition metal dichalcogenides, \cref{sec:tmd}); JHM (ethanol/water, \cref{sec:ethwat}, trialanine, \cref{sec:ala}); GL, LAMR (a-Si \cref{sec:si_amo}); DK (carborane, \cref{sec:carborane}, ammonia-borane, \cref{sec:amonia_borane}); VC (S polymerisation, \cref{sec:S}); JR, JaG, JoG and AAN (phonons, \cref{sec:phonons}); JR, REAG (materials discovery: formation energy, \cref{sec:formation}); KSJ (materials discovery: stoichiometric substitutions, \cref{sec:element-substitution}); ADK (materials discovery: highly-coordinated structures, \cref{sec:coordination}); ASR, YC and AME (MOFs, \cref{sec:mofs,sec:si_mofs}); MV (solvent mixtures, \cref{sec:mixtures}); DPK, ES, SMB (hydrogen combustion, \cref{sec:h_comb}); NKa, CSu (HOIPs, \cref{sec:pero_organic}); FF, JT (protein dynamics and stability \cref{sec:protein_folding}); PG, YW, TDS, JRK and CO (point and extended defects in BCC metals, \cref{sec:bcc_defects}); BXS, FB (molecule-surface interactions, \cref{sec:molecule_surf_interactions}) ;WCW (HEA, \cref{sec:timings}); EF, SD (catalysis: linear scaling relationships, \cref{sec:catalysis,sec:lsr}); HJ, HHH (catalysis: \ch{CO} oxidation on \ch{Cu}, \cref{sec:catalysis,sec:oxi}); SH, SD (catalysis: Pourbaix diagrams, \cref{sec:catalysis,sec:pour}); MCK (benchmarks: bulk and shear moduli, \cref{sec:bulk_moduli}); FDP (benchmarks: cohesive energies and lattice constants of solids, \cref{sec:lattice-energies}, atomization energies \cref{sec:atomization-energies}, and reaction barrier heights, \cref{sec:reaction-barrier}); TKS (\ch{Al2O3}, \cref{sec:alumina}, diatomics, \cref{sec:dia}); JPD (Arsenic random structure search, \cref{sec:rss}); IBM (high-pressure hydrogen, \cref{sec:h2}); IBM, CvdO (electrode-electrolyte interface / battery system, \cref{sec:battery}); JK, VS and KH (\ch{CeO2}, \cref{sec:ceria}); FZ (ionic liquids, \cref{sec:ionic_liquids}); 
\textbf{Fine-tuning}: IB, NB, TW
\textbf{Uncertainty quantification:} FBi, MC, SC, CHH, CO and YW (appendix \ref{sec:uq});
\textbf{Supervision of research:} AB, ACF, AM, ASR, CH, CO, CPG, CSu, GC, HHH, JaG, JK, JTM, KAP, KH, KR, MA, MC, SD, SMB, TV, VLD, WH; \textbf{Drafted manuscript:} IB, NB, YC, GC, SD, HHH, MCK, JR, ASR, CSc, JTM; \textbf{Edited manuscript:} IB, NB, YC, GC, VLD, JaG, REAG, JR, MCK, KAP, ASR, LLS, JTM, AM, CO, AME, WCW, JLAG; \textbf{Supervised manuscript writing:} NB, GC, VLD, VK, JTM, CSc.

\section*{Acknowledgments}

Model training made use of resources of the National Energy Research Scientific Computing Center, a DOE Office of Science User Facility supported by the Office of Science of the U.S. Department of Energy under contract no. DE-AC02-05CH11231 using awards BES-ERCAP0023528 and BES-ERCAP0022838.
Part of this work was performed using the Cambridge Service for Data-Driven Discovery (CSD3), part of which is operated by the University of Cambridge Research Computing on behalf of the STFC DiRAC HPC Facility (www.dirac.ac.uk). The DiRAC component of CSD3 was funded by BEIS capital funding via STFC capital grants ST/P002307/1 and ST/R002452/1 and STFC operations grant ST/R00689X/1. DiRAC is part of the National e-Infrastructure.
The work of YC, JR, ADK, MCK, MA and KAP was supported by the US Department of Energy, Office of Science, Office of Basic Energy Sciences, Materials Sciences and Engineering Division, under contract no. DE-AC02-05-CH11231 (Materials Project program KC23MP).
We could not have done this work without the DFT relaxation trajectories freely provided by the Materials Project and carefully curated into the MPtrj training set by Bowen Deng \cite{deng_chgnet_2023}.
JR acknowledges support from the German Academic Scholarship Foundation (Studienstiftung).
YC acknowledges financial support from UC Berkeley and Taiwan-UC Berkeley Fellowship from the Ministry of Education in Taiwan.
MCK acknowledges support by the National Science Foundation Graduate Research Fellowship Program under Grant No. DGE-2146752. Any opinions, findings, and conclusions or recommendations expressed in this work are those of the author(s) and do not necessarily reflect the views of the National Science Foundation.
NB was supported by fundamental-research base-program funding from the U.S. Naval Research Laboratory.
ASR acknowledges support via a Miller Research Fellowship from the Miller Institute for Basic Research in Science, University of California, Berkeley.
LLS acknowledges support from the EPSRC Syntech CDT with grant reference EP/S024220/1.
AM and XRA acknowledge support from the European Union under the ``n-AQUA" European Research Council project (Grant no. 101071937). SWN, AB, TV acknowledge support from the European Union's Horizon 2020 research and innovation program under the Marie Skłodowska-Curie Actions (Grant Agreement 945357) as part of the DESTINY PhD program. GC, CPG, TV, AB and SWN acknowledge support from the European Union's Horizon 2020 research and innovation program under Grant Agreement 957189 (BIG-MAP). VK acknowledges support from the Ernest Oppenheimer Early Career Fellowship and the Sydney Harvey Junior Research Fellowship, Churchill College, University of Cambridge. V.K. acknowledges computational support from the Swiss National Supercomputing Centre under project s1209.
ZEM acknowledges support from the EPSRC Centre for Doctoral Training in Theory and Modeling in Chemical Sciences (TMCS), under grant EP/L015722/1.
VLD acknowledges support from UK Research and Innovation [grant number EP/X016188/1] and the John Fell OUP Research Fund.
CH and FZ acknowledge support by the Deutsche Forschungsgemeinschaft (DFG, German Research Foundation) in the framework of the priority program SPP 2363, “Utilization and Development of Machine Learning for Molecular Applications - Molecular Machine Learning” Project No. 497249646 as well as further funding though the DFG under Germany's Excellence Strategy - EXC 2075 - 390740016 and the Stuttgart Center for Simulation Science (SimTech).
AME's work used the DiRAC Extreme Scaling service (Tursa) at the University of Edinburgh, which is part of the STFC DiRAC HPC Facility (www.dirac.ac.uk) and scarf cluster (www.scarf.rl.ac.uk/) maintained by Scientific Computing Department STFC. AME's access to DiRAC resources was granted through a Director’s Discretionary Time allocation in 2023/24, under the auspices of the UKRI-funded DiRAC Federation Project. AME's work was also supported by Ada Lovelace centre at STFC (https://adalovelacecentre.ac.uk/), Physical Sciences Databases Infrastructure (https://psdi.ac.uk) and EPSRC under grants EP/W026775/1 and EP/V028537/1.
IB, RE and NK were supported by the Harding Distinguished Postgraduate Scholarship.
HJ gratefully acknowledges support from the Alexander-von-Humboldt (AvH) Foundation. HHH, JTM and KR acknowledge support from the German Research Foundation (DFG) through DFG CoE e-conversion EXC 2089/1. 
FB acknowledges the Alexander von Humboldt Foundation for a Feodor Lynen Research Fellowship and the Isaac Newton Trust for an Early Career Fellowship. BXS acknowledges support from the EPSRC Doctoral Training Partnership (EP/T517847/1).
IB, DPK, XRA, WJB, FDP, RE, VK, DK, GL, NON, LLS, CSc, TKS, CvdO, EVU, WCW acknowledge access to CSD3 GPU resources through a University of Cambridge EPSRC Core Equipment Award (EP/X034712/1).
We acknowledge project/application support by the Max Planck Computing and Data Facility.
KH, JK and VS acknowledge the Swedish Research Council (Vetenskapsrådet,
project number 2021-06757) and the National Strategic
e-Science program eSSENCE for funding, as well as the Swedish National Infrastructure for Computing
(SNIC/NAISS) for providing computer resources used in this project. 
SW, AB and TV acknowledge the Pioneer Center for Accelerating P2X Materials discovery (CAPeX), DNRF Grant number P3. 
YW acknowledges support from the Shanghai Jiao Tong University.
WCW acknowledges support from the EPSRC (Grant EP/V062654/1).
CO and CHH acknowledges support from NSERC (Discovery Grant GR019381) and NFRF (Exploration Grant GR022937). 
JaG, JoG and AN would like to acknowledge the Gauss Centre for Supercomputing e.V. (https://www.gauss-centre.eu) for funding workflow-related developments by providing generous computing time on the GCS Supercomputer SuperMUC-NG at Leibniz Supercomputing Centre (www.lrz.de) (Project pn73da). JaG was supported by ERC Grant MultiBonds (grant agreement Nº 101161771; Funded by the European Union. Views and opinions expressed are however those of the author(s) only and do not necessarily reflect those of the European Union or the European Research Council Executive Agency. Neither the European Union nor the granting authority can be held responsible for them.)
SH and JJ acknowledge funding from EPSRC (EP/T001038/1, EP/S022953/1 ).
ADK acknowledges the Savio computational cluster resource provided by the Berkeley Research Computing program at the University of California, Berkeley (supported by the UC Berkeley Chancellor, Vice Chancellor for Research, and Chief Information Officer). ACF acknowledges funding from EU Graphene Flagship, ERC grants Hetero2D, GIPT, EU grants Graph-X, CHARM, EPSRC grants EP/K01711X/1, EP/K017144/1, EP/N010345/1, EP/L016087/1, EP/V000055/1, EP/X015742/1.
WB, CSu, and CG thank the US AFRL for partial funding of this project through grant FA8655-21-1-7010. JPD, JRK and GC acknowledge funding from the NOMAD Centre of Excellence (European Commission grant agreement ID 951786). PG and TDS acknowledge the support from the Cross-Disciplinary Program on Numerical Simulation of CEA, the French Alternative Energies and Atomic Energy Commission. PG and TDS used access to the HPC resources of IDRIS under the allocation A0120913455 attributed by GENCI.
SC and MC acknowledge the support by the Swiss National Science Foundation (Project 200020\_214879). FB and MC acknowledge support from NCCR--MARVEL, funded by the Swiss National Science Foundation (grant no. 182892). MC acknowledges funding from the European Research Council (ERC) under the European Union’s Horizon 2020 research and innovation programme (grant no. 101001890--FIAMMA).
We acknowledge the Jean Zay cluster of access to compute as part of the Grand Challenge: GC010815458 (Grand Challenge Jean Zay H100).
GC is grateful to \'Agnes Borsz\'eki for help with graphics.

\bibliographystyle{ieeetr}
\bibliography{references}

\clearpage
\appendix
\setcounter{figure}{0}  
\setcounter{table}{0}   
\renewcommand{\thefigure}{S\arabic{figure}}
\renewcommand{\thetable}{S\arabic{table}}
\renewcommand{\theequation}{S\arabic{equation}}

\section*{Supplementary Information}
\label{sec:si}

The following sections contain a diverse set of examples where the \MPzbt\ foundation model is applied
to a variety of material and chemical systems with each subsection containing one application with one or more related examples.

\subsubsection*{Similarity statement}

Each subsection also contains a statement (both qualitative and quantitative) about the extent to
which the training data contains configurations similar to those relevant to the application in that section.
This should inform the reader about the degree of extrapolation inherent in the particular example. In order
to facilitate further scrutiny, we provide a data file for most applications that can be used in conjunction with
the \texttt{chemiscope} tool (at \href{https://chemiscope.org}{\url{chemiscope.org}}) to explore the chemical environments in the training data and
the application example and their relation to one another.

\subsubsection*{Performance summary}

Each subsection also contains a concise statement summarising the performance of the \MPzbt{} model in the application. 

\tableofcontents

\addtocontents{toc}{\protect\setcounter{tocdepth}{2}}



\clearpage

\section{Further Applications}
\label{sec:applications}

\subsection{Self-interstitials in silicon}\label{sec:si_si}

\begin{figure}[!ht]
  \centering
  \includegraphics[width=\textwidth,keepaspectratio]{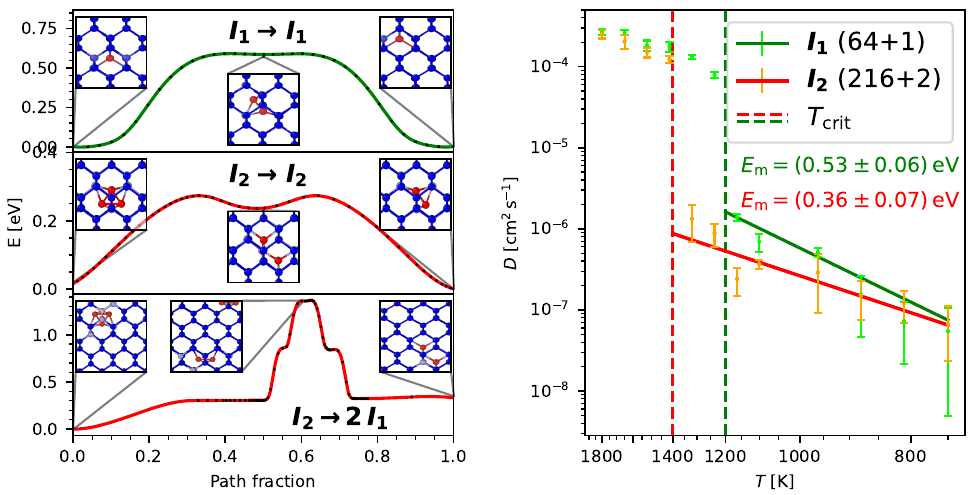}
  \caption{Single- ($I_1$) and di-interstitial ($I_2$) defects in silicon. Left: Nudged elastic band paths between two metastable sites with a subset of images shown (black dots). Right: Diffusion coefficient against inverse temperature and Arrhenius laws with migration energy~$E_\mathrm{m}$ and phonon instability temperature~$T_\mathrm{crit}$. 
  }
  \label{fig:diinterstitial}
\end{figure}

Di-interstitial silicon ($I_2$) constitutes a test case for the transferability of \MPz{} to point defects in a periodic lattice. In the following, self-diffusion coefficients $D$ and interstitial migration energies $E_\mathrm{m}$ are obtained from MD simulations without a D3 dispersion correction.
Consistency tests with a $(64+2)$-atom silicon structure from \cite{bartok2018machine} relaxed with the PW91 functional are performed. Relaxing with \MPz{} showed no change in energy at a force tolerance of \SI{0.05}{eV\per\Angstrom}.

The $I_2$ structure was generated by relaxing a $(216+2)$-atom
diamond structure with lattice constant \SI{5.4}{\Angstrom}.
After running NVT MD, the distance to the closest lattice site for each atom was plotted.
The interstitials propagate consistent with the accepted mechanism~\cite{yoshida2015defects}.
Isolating the trajectories of the point defect shows a characteristic
jump length corresponding to expected jumps between stable sites.

At temperatures below $\SI{700}{\kelvin}$,
the interstitials almost exclusively remain bound in a state corresponding to the ground state with $C_{1h}$ symmetry~\cite{richie2004complexity} and transition between symmetry-equivalent ground states. At higher temperatures, higher-energy states are sampled in which the interstitials separate into tetrahedral interstitial states, lying at a higher energy $\SI{0.3}{\eV}$ (\cref{fig:diinterstitial}).
We calculate $D$ from the mean-square-displacement (MSD) of all atoms in the unit cell~\cite{sahli2005ab}.
The fit to the Arrhenius law (\cref{fig:diinterstitial}) gives 
$E_\mathrm{m}=\SI[separate-uncertainty = true]{0.36(7)}{\eV}$, in agreement with~\cite{posselt2005migration} and~\cite{du2005fast}.

The above calculations were repeated for a single interstitial ($I_1$) in a 64-atom cell.
The interstitial predominantly occupies the tetrahedral state and transitions between symmetry-equivalent states via the split $\langle 110\rangle$ state.
While these states are expected, the occupancy of each state shows larger deviations from those reported in~\cite{sahli2005ab}.
\Cref{fig:diinterstitial} shows a nudged elastic band (NEB) between two tetrahedral interstitial sites via a $\langle 110 \rangle$ split interstitial site, from which an energy barrier of $\SI{0.6}{\eV}$ is calculated.
Repeating the MD simulations
at several temperatures (\cref{fig:diinterstitial}) gives 
$E_\mathrm{m}=\SI[separate-uncertainty = true]{0.53(6)}{\eV}$, in agreement with LDA~\cite{sahli2005ab}.
While the prefactor is strongly system-dependent~\cite{pichler2012intrinsic}, the migration energy is an intrinsic property of the energy landscape and may be compared with other calculations.
However, the finite system size and the interaction between interstitials results in the difference between the single- and di-interstitial migration energies.

\MPzbt{} displays the correct qualitative dependence of the phonon instability temperature $T_\mathrm{crit}$ on interstitial density (fig. \ref{fig:diinterstitial}). Specifically, $T_\mathrm{crit}=\SI{1400}{\kelvin}$ of the $(216+2)$-system, which has an interstitial density of approximately $0.9\%$, is lower than the melting temperature of $\SI[separate-uncertainty = true]{1449(10)}{\kelvin}$ calculated for cubic diamond silicon with PBE~\cite{dorner2018melting}. A further decrease to $T_\mathrm{crit}=\SI{1200}{\kelvin}$ is observed for the $(64+1)$-system, which has a larger interstitial density of $1.5\%$. However, the phonon instability temperature for the $(64+1)$-system, calculated with LDA, is $\SI{1473}{\kelvin}$~\cite{sahli2005ab}, which \MPzbt{} consequently underestimates.  The melting points and phonon instability temperatures are expected to have a dependence on the exchange-correlation functional.

\subsubsection*{Similarity statement}

The MP dataset contains 41 pure silicon structures, including the diamond structure but no self-interstitial defects.

\subsubsection*{Performance summary}

Silicon interstitials display the correct set of local minima, while their relative occupancy at finite temperature is not correct. The predicted activation energies for self-diffusion of the single and di-interstitials are consistent with previous force field and DFT calculations. The dependence on interstitial density is at least qualitatively correct, whereas the phonon instability temperature of the single interstitial system is too low by about \SI{20}{\%}. 

\subsubsection*{Fine-tuning}

\begin{figure}[!ht]
    \centering
    \includegraphics[width=\linewidth]{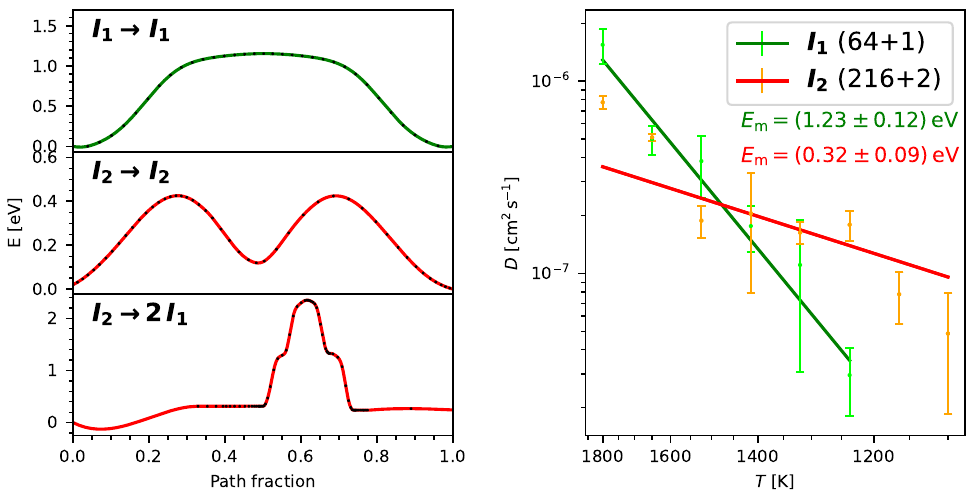}
    \caption{Results for the finetuned model. Plots are the same as in Fig. \ref{fig:diinterstitial}.}
    \label{fig:diinterstitial-2}
\end{figure}

Fine-tuning was performed on $100$ configurations sampled from an MD trajectory at $1800\mathrm{K}$ from the $I_2$
system. 
Fig. \ref{fig:diinterstitial-2} shows the NEBs and diffusion coefficients, recomputed with the finetuned model.
The barrier heights of all NEBs have increased.
The migration energy for $I_1$ has increased to $E_\mathrm{m}=\SI[separate-uncertainty = true]{1.23(12)}{\eV}$, in agreement with the increase in barrier height. However, this is significantly larger than the LDA value in \cite{sahli2005ab}.
For $I_2$, the migration energy has not changed, within the error, and remains consistent with the barrier height.
Finally, we note that the phonon instability temperatures for $I_1$ and $I_2$
are now above the melting temperature of cubic diamond silicon calculated with PBE\cite{dorner2018melting}.


\clearpage
\subsection{Amorphous silicon from melt--quench simulations}\label{sec:si_amo}

\begin{figure}[htbp!]
  \centering
  \includegraphics[width=\textwidth,keepaspectratio]{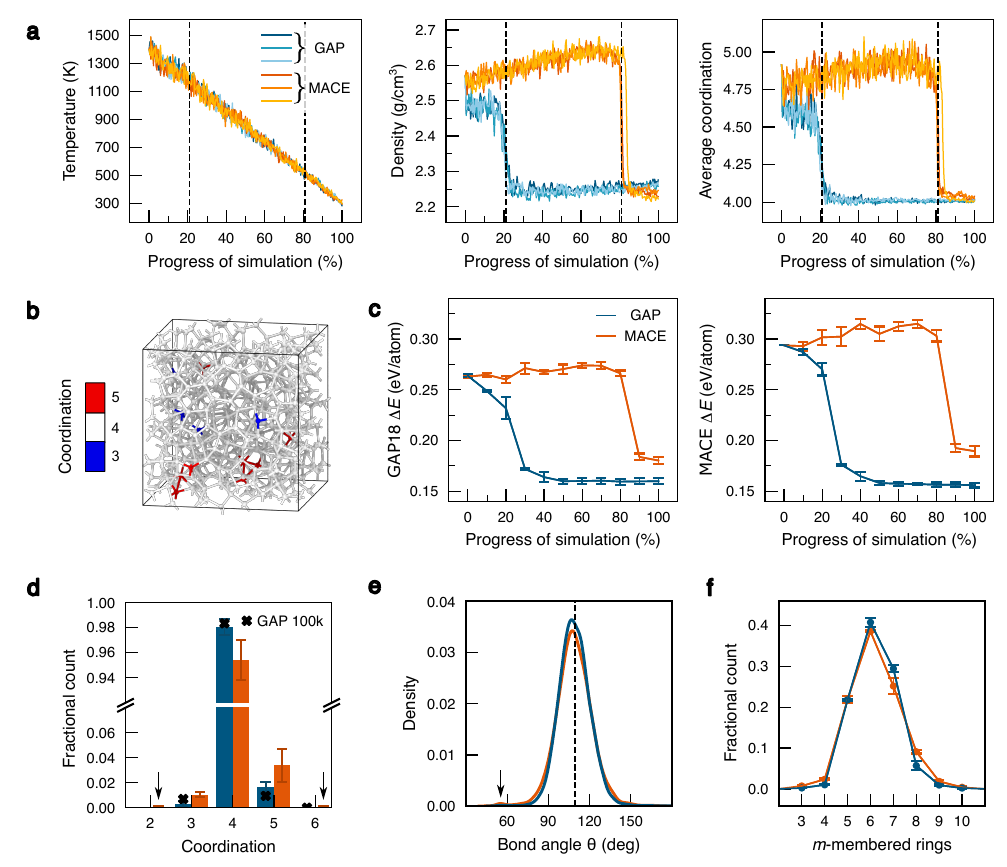}
  \caption{\textbf{Amorphous silicon.} We characterise structural models of amorphous silicon simulated by melt-quenching using \MPzbt{} and benchmark against the well-established Si-GAP-18 model \cite{bartok2018machine}. (a) Evolution of selected properties during the quench simulations: temperature (left), density (center), and average coordination number, defined using a bond-length cutoff of \SI{2.85}{\Angstrom} (right). (b) Amorphous silicon structure generated using the \MPzbt{} potential, where atoms are color-coded by coordination number. (c) Excess enthalpies (\textit{$\Delta$E}) relative to diamond-type silicon, computed using Si-GAP-18 (left) and \MPzbt{} (right), for relaxed snapshots from simulations with either potential. (d) Coordination count, where the black crosses represent a study of a 100,000-atom system driven by Si-GAP-18 \cite{deringer2021origins}. (e) Bond-angle distribution, where the dashed line is the equilibrium \SI{109.5}{\degree} angle for tetrahedral environments. (f) Ring-size distribution.}
  \label{fig:silicon-mp}
\end{figure}

Amorphous silicon (a-Si) is a prototypical disordered material and has served as an example of both the physical and chemical insight afforded by machine-learned potentials \cite{bernstein_quantifying_2019,deringer2021origins} and as a challenging benchmark for the development of new methodology and potential models \cite{bartok2018machine,lysogorskiy2021performant,morrow_indirect_2022}.

Here, we assess the performance of the MACE-MP-0b3 potential on standard melt-quench simulations, and compare it to the well-established Si-GAP-18 potential, a general purpose GAP model for Si \cite{bartok2018machine}, that has been extensively validated in previous literature \cite{deringer2018realistic, bernstein_quantifying_2019, deringer2021origins}. We perform melt-quench simulations of 512-atom systems quenching from the liquid state at \SI{1400}{\kelvin} to the amorphous state at \SI{300}{\kelvin} at a rate of \SI{1e12} {\kelvin\per\second}, for a total of \SI{1100}{\pico\second}. The simulations were run in LAMMPS \cite{plimpton1995fast}, in the NpT ensemble using a Nosé-Hoover thermostat and barostat with a time step of \SI{1}{\femto\second}. A well-equilibrated liquid starting structure was generated by annealing a random hard-sphere model at \SI{1400}{\kelvin} for \SI{10}{\pico\second} using Si-GAP-18. The protocol was repeated three times for each model to assess the variability in predictions.

\cref{fig:silicon-mp} (a) presents the temperature, density and average coordination within a cutoff of \SI{2.85}{\Angstrom} as a function of the progress of the quench simulation.
The density of the liquid structure stays constant for the Si-GAP-18 model, but increases slowly to around \SI{2.65}{\gram\per\cubic\centi\meter} for the \MPzbt{} model, above the experimental density of liquid Si at \SI{2.57}{\gram\per\cubic\centi\meter}\cite{Dharma-wardana-20-08}.
When quenched from the liquid state, Si undergoes a vitrification transition that is accompanied by a sudden decrease in density and concomitantly by a decrease in the average atomic coordination number, as the highly-coordinated metallic liquid transforms into the mainly four-fold coordinated semiconducting a-Si. The \MPzbt{} model predicts a glass transition temperature around \SI{600}{\kelvin}, compared to around \SI{1050}{\kelvin} for Si-GAP-18, and experimental observations of around \SI{1000}{\kelvin} \cite{Hedler-04-11, McMillan-04-11}. Hence the \MPzbt{} model appears to overstabilise the liquid phase below the experimental glass transition temperature. 
After the transition, the resulting bulk structures are mostly fourfold-connected, with average densities of \SI{2.266}{\gram\per\cubic\centi\meter} for Si-GAP-18 and \SI{2.233}{\gram\per\cubic\centi\meter} for \MPzbt{}, close to the experimental density of \SI{2.285}{\gram\per\cubic\centi\meter}\cite{Custer-94-01}.

The final structure from one of the three quench simulations driven by \MPzbt{} is presented in \cref{fig:silicon-mp} (b), where the Si atoms are color-coded by coordination numbers. This confirms the insight from \cref{fig:silicon-mp} (a), which is that the majority of atoms are 4-fold coordinated, with some 3- and 5-fold coordination defects.

We further compare the excess energies ($\Delta E$), calculated relative to crystalline diamond-type silicon (\textbf{dia}), as shown in \cref{fig:silicon-mp} (c). These values were computed by taking structural snapshots throughout the quench and relaxing them either with Si-GAP-18 (left) or \MPzbt{} (right), similar to Ref.~\cite{deringer2018realistic}. Both Si-GAP-18 and \MPzbt{}  show small energy fluctuations between individual runs for the Si-GAP-18 simulations, while \MPzbt{} had some minor variability in each run. \MPzbt{} predicts the same overall trend in energies as Si-GAP-18, but the energy predictions have an offset of around +\SI{0.05}{\electronvolt\per\atom} for the liquid structures, while the predictions from both models on the amorphous structures are very close. The final structures quenched by \MPzbt{} have a higher energy (\SI{0.18}{\electronvolt\per\atom}) than those quenched by Si-GAP-18 (\SI{0.16}{\electronvolt\per\atom}). This compares to a previous result of \SI{0.14}{\electronvolt\per\atom} for a 4,096-atom system quenched at a rate of \SI{1.0e11}{\kelvin\per\second} driven by Si-GAP-18 ~\cite{deringer2018realistic} and the experimental excess enthalpy of a-Si after deposition and annealing of \SI{0.14}{\electronvolt\per\atom}~\cite{roorda1991structural}.
The present quench rate is one order of magnitude faster, hence higher average enthalpies and variability can be expected. Nonetheless, the excess enthalpies predicted by \MPzbt{} are qualitatively in the correct ballpark, viz.\ slightly above the corresponding crystalline phase. 

Probing the short range order of the structure, \cref{fig:silicon-mp} (d) presents the distribution of Si coordination, averaged over all three repeats for each model. While the Si-GAP-18 structures present fewer than 2\% of 3- and 5- fold defects, the \MPz{} structures features around 4\% of coordination defects, mostly 5-fold coordinated atoms, but also 2-, 3- and 6-fold defects in smaller concentrations. The 100k-atom simulations in Ref.~\cite{deringer2021origins} predicted well-relaxed a-Si to contain on the order of \numrange{1.5}{2}\unit{\%} of defects, consistent with the 4,096-atom simulations of Ref.~\cite{deringer2018realistic}. Faster quenching leads to higher defect counts, as seen from the Si-GAP-18 results in Fig.~\ref{fig:silicon-mp}, whereas the overall defect count predicted by \MPz{} is still notably higher than that of Si-GAP-18 at the same quench rate. Another measure of the quality of the short-range structure is the bond angle distribution, shown in \cref{fig:silicon-mp} (e), where both distributions appear very similar, but for a small feature at \SI{60}{\degree} for the \MPz{} structures [arrow in \cref{fig:silicon-mp} (e)].

Finally, we assess the quality of the medium-range order of the network using shortest-path ring statistics with the \texttt{matscipy} package \cite{grigorev2023matscipy}, depicted in \cref{fig:silicon-mp} (f). The distribution of $m$-membered rings is very similar for both models, but the \MPz{} structures have an increased count of small ($m<5$) and large ($m>7$) rings, with the presence of $m=3$ rings that supports the feature in \cref{fig:silicon-mp} (e) at \SI{60}{\degree} and the cluster of 5-fold coordinated atoms in \cref{fig:silicon-mp} (b).

\subsubsection*{Similarity statement}

The MP dataset includes 41 different silicon-only structures, however, many of them are very high-density (high coordination number) or crystalline (all 4-fold). There were no cases of wide coordination number distribution, as seen in liquid Si -- however, we found 5 unique a-Si structures (with 100 atoms each) with a mix of slightly higher and lower coordination numbers, providing information about a-Si. Based on a UMAP analysis, the closest structures in the training set are mp-1244971, mp-1245242 and mp-1245041. To help with visualization, we provide \texttt{amorphous-silicon.json} on \url{chemiscope.org}.

\subsubsection*{Performance summary}
The model performs reasonably well for the description of the melt--quench process, leading to good-quality a-Si structures, albeit markedly underestimating the vitrification temperature and the excess enthalpy and overestimating the number of 5-fold coordinated atoms compared to a domain-specific ML potential.

\clearpage
\subsection{Amorphous carbon}\label{sec:c_amo}
\begin{figure}[htbp!]
  \centering
  \includegraphics[width=\textwidth,keepaspectratio]{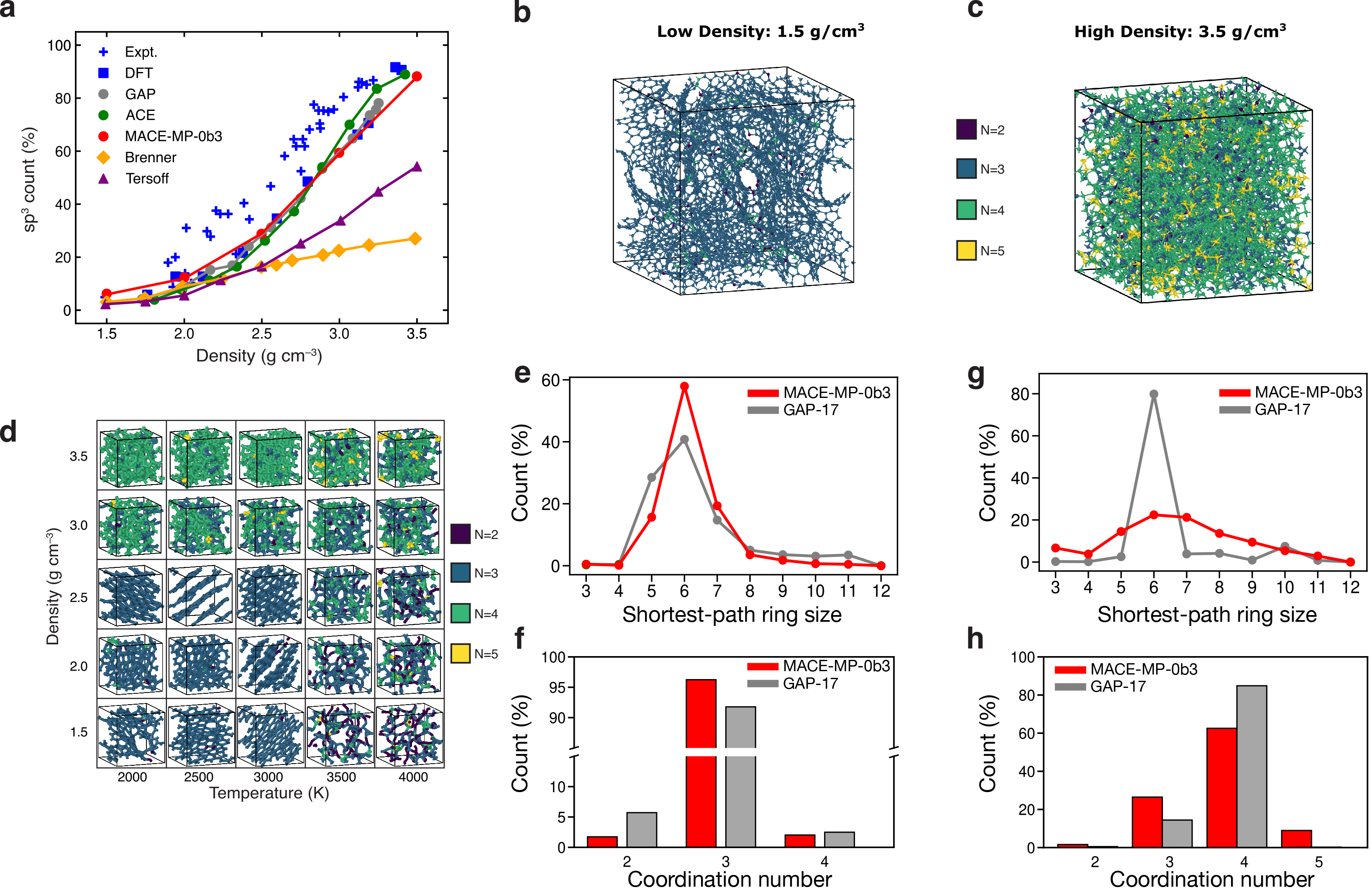}
  \caption{\textbf{Amorphous carbon.} (a) Count of sp$^{3}$ (fourfold coordinated) carbon atoms in melt-quenched carbon structures as a function of density. The results obtained with the \MPzbt{} model are compared to computational and experimental data compiled in Ref.~\cite{DeringerC2017} and references therein, as well as Refs.~\cite{Jana2019} and \cite{Qamar2023}. 
  (b--c) 4,096-atom structures generated using the \MPzbt{} potential. 
  (d) Results from 25 $\times$ 200 atom graphitisation simulations spanning relevant temperature and density ranges, similar to Ref.~\cite{Stenczel2023}. The structures in panels (b--d) are colour-coded according to coordination numbers as indicated in the legends. 
  (e) Shortest-path ring size count for 4,096-atom low-density structures as determined using \texttt{matscipy}\cite{grigorev2023matscipy}. 
  (f) Coordination number count for 4,096-atom low-density structures. 
  (g--h) Same as panels (e) and (f) but for high-density structures.
  The results in panels (e--h) are shown for a C-GAP-17-driven simulation (gray) and for a comparable simulation driven by the \MPzbt{} potential (red).}
  \label{fig:carbon-mp}
\end{figure}

\subsubsection{Melt--quench simulations}\label{sec:C-mq}
Carbon forms many different crystalline and amorphous modifications. The structural diversity of amorphous carbon (a-C), characterised by the simultaneous presence of three-fold coordinated (sp$^2$) and four-fold coordinated carbon atoms (sp$^3$), makes it a challenging system for both classical and ML force fields~\cite{deTomas2019,marchant_exploring_2023}. The correct description of its growth mechanism has been among the early successes of ML-driven materials modeling~\cite{Caro2018}.

We assess the accuracy of the \MPz{} model in reproducing the structural complexity of amorphous phases by plotting the concentration of four-fold coordinated atoms (sp$^3$) as a function of density in \cref{fig:carbon-mp}. To generate amorphous structures with a given density with the \MPz{} model, we perform melt-quench simulations. We start by melting diamond structures at a given density by running NVT simulations at \SI{8000}{\kelvin} for \SI{3}{\pico\second}. We then perform a fast quench, reducing the temperature from \SI{8000}{\kelvin} to \SI{300}{\kelvin} at a cooling rate of \SI{1000}{\kelvin\per\pico\second}. Finally, we optimize the geometry with LFBGS to obtain the final structure and determine the count of sp$^{3}$ atoms using a bond-length cutoff of \SI{1.85}{\Angstrom}. We observe in \cref{fig:carbon-mp} that the \MPz{} model predictions reproduce the trend observed in both the DFT~\cite{Jana2019} and the experimental data extracted from~\cite{DeringerC2017}. We also see good agreement with results of quenches using the carbon ACE reported by Qamar et al.\ in Ref.~\cite{Qamar2023}, and using C-GAP-17 reported in Ref.~\cite{Jana2019}, noting that both potentials had been specifically trained on large carbon datasets.

\subsubsection{Graphitisation}\label{sec:C-gra}

\MPz{} was used for two annealing runs for simulation cells containing 4096 atoms at low density (\SI{1.5}{\gram\per\cubic\centi\meter}) and high density (\SI{3.5}{\gram\per\cubic\centi\meter}), respectively. The low-density simulation was run at 2000 K and the high-density simulation was run at 4000 K. Additionally, $25 \times 200$ atom annealing runs spanning from 2000 to 4000~K and \num{1.5} to \SI{3.5}{\gram\per\cubic\centi\meter} were conducted to explore a finer grid of parameters. 
Both 4,096-atom structures were compared to structures generated using C-GAP-17 using the same protocol, which was also used recently for an ``on-the-fly'' generated GAP potential~\cite{Stenczel2023}. The protocol has two stages. The goal of stage I is to prepare the starting configuration for the annealing in stage II, and it begins with a random structure with a hard-sphere constraint of $r_{min}$ \SI{\ge 1}{\Angstrom} and equilibrating it at\SI{9000}{\kelvin} for \SI{40}{\pico\second}, followed by cooling to \SI{5500}{\kelvin} over \SI{40}{\pico\second} and subsequent quenching to \SI{300}{\kelvin} over \SI{10}{\pico\second}. The structures are then held at \SI{300}{\kelvin} over \SI{50}{\pico\second} before being rapidly heated up to the annealing temperature over \SI{10}{\pico\second}. This concludes stage I. In stage II, the structure is annealed at \SI{2000}{\kelvin} or \SI{4000}{\kelvin} for \SI{350}{\pico\second} using a time step of \SI{1}{\femto\second}. We used C-GAP-17 to perform stage I because \MPz{} was found to be unstable at \SI{9000}{K}.

For the low-density case, the structures generated by \MPz{} in annealing simulations agree with the predictions of the established C-GAP-17 model in terms of overall trends (Figs.~\ref{fig:carbon-mp}b and \ref{fig:carbon-mp}d). For more detailed insight, Fig.~\ref{fig:carbon-mp}e shows the shortest-path ring distribution for the low-density structure, indicating that \MPz{} predicts a greater number of 6-membered rings and fewer large rings for the low-density structure compared to C-GAP-17 (thus suggesting a higher degree of crystallinity in the \MPz{} prediction) -- this might be correlated with the higher relative count of graphite versus amorphous-like structures in the training dataset, although further analysis is required. 

Figure \ref{fig:carbon-mp}g shows poor agreement between \MPz{} and C-GAP-17 in terms of ring count for the high-density structure.
Whereas annealing with C-GAP-17 led to a partly crystallised structure under the conditions chosen (indicated by the large abundance of 6-membered rings, as are found in the diamond structure), the \MPzbt{} model gave rise to a highly disordered structure -- this is evident from a visual inspection of Fig.~\ref{fig:carbon-mp}c, and reflected in the ring-count plot in Fig.~\ref{fig:carbon-mp}g showing a notable number of (presumably strained) 3-membered rings as well as larger ring sizes of $>7$. In other words, the \MPzbt{} appears to fail to describe the crystallisation in this case (we note that a partially ordered structure was obtained in a 200-atom simulation at 3000 K but not at 4000 K at the same density; Fig.~\ref{fig:carbon-mp}d).
Finally, Figs.~\ref{fig:carbon-mp}f and \ref{fig:carbon-mp}h show the coordination number for both 4,096 atom structures. \MPzbt{} leads to more sp$^{2}$ environments and fewer sp and sp$^{3}$ environments compared to C-GAP-17 (again likely consistent with a higher degree of crystallinity, as suggested by the ring count in Fig.~\ref{fig:carbon-mp}e).
In the high-density 4,096-atom structure obtained with \MPzbt{}, a notable amount of 5-fold coordinated atoms are observed, on the order of about 10\% -- a behaviour that would not be expected in a simulation of carbon at diamond-like density. 

We note that these issues are not expected to be a fundamental shortcoming of the MACE architecture: we show that training on a wider-ranging dataset with more crystal structures like the Alexandria dataset (which notably has no amorphous structures) results in a model, MACE-MPA-0, that describes the structure of amorphous carbon in better agreement with C-GAP-17 at high and low density, as shown in the section dedicated to the MACE-MPA-0 model (\ref{sec:alexandria}).

\subsubsection*{Similarity statement}

The MP dataset contains 89 different all-carbon structures, most of which correspond to diamond and related stacking polytypes (lonsdaleite-like and more complex ones, all purely sp$^3$-bonded), as well as graphite in various forms. These structures include a range of mixed configurations with sp$^{2}$/sp$^{3}$ coexistence -- the latter is expected to be critical for a correct description of a-C. The dataset also contains a number of compressed and defective fullerene configurations, with one of those cells containing sp-, sp$^{2}$-, and sp$^{3}$-like environments, and a few hypothetical allotropes (notably ``T-carbon'' and a cubane-motif-based form, representing 3- and 4-membered shortest-path rings, respectively). In essence, the dataset does contain relevant carbon environments but does not contain a significant share of highly disordered carbon configurations. Based on UMAP analysis, we find that the closest structures in the training set are mp-568028 and mp-568806. We provide \verb|amorphous_carbon.json| to help visualize the interactive UMAP on \url{chemiscope.org}.

\subsubsection*{Performance summary}
The \MPz{} model correctly captures the sp$^{3}$ content as a function of density in melt--quench simulations. Detailed analysis of long-annealing simulations shows qualitative agreement with a purpose-trained ML force field for low-density a-C graphitization and poor agreement for high-density annealing.

\clearpage
\subsection{Ceria nanoparticles}\label{sec:ceria}

\begin{figure*}[htbp!]
  \centering
  \includegraphics[width=\linewidth,keepaspectratio]{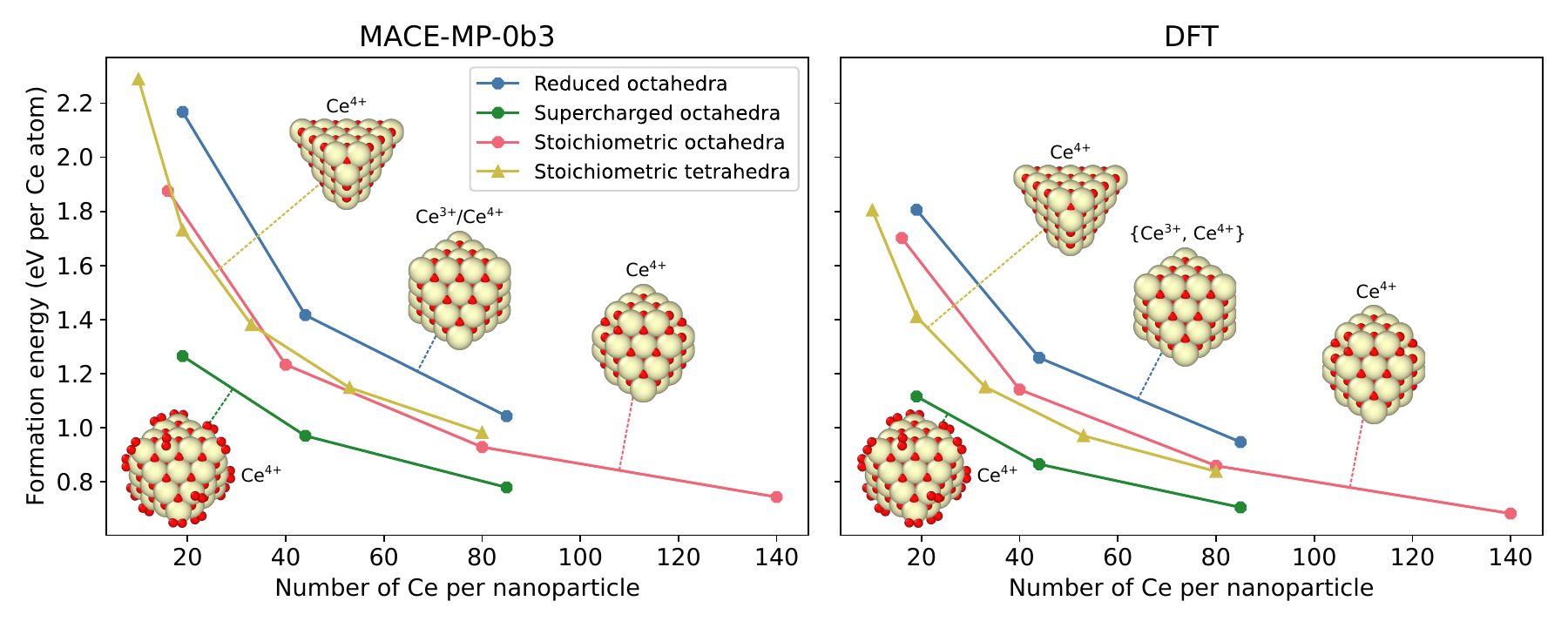}
  \caption{Size-dependent formation energies for different shapes and stoichiometries of ceria nanoparticles (NPs). Left panel: Results calculated with the \MPz{} model, which was not trained on any data from ceria surfaces or NPs.  Right panel: Independent validation data from PBE+U calculations. The NP images are just illustrations; the optimized structures from MACE-MP and DFT are compared in more detail in \cref{fig:ceriacloseup}. The formation energy is calculated with respect to stoichiometric bulk \ch{CeO2} (and gas-phase \ch{O2} molecules as needed).
  }
  \label{fig:UU_ceO2}
\end{figure*}
Cerium oxide (ceria, \ch{CeO2}) is a reducible metal oxide with intriguing chemical and physical properties, and important technological applications especially for nanostructured ceria. Experiments in the literature have for example shown that the oxygen storage capacity (OSC) of ceria at the nanoscale is strongly shape- and size-dependent.

Behind the versatile usage of ceria lies one overriding feature, namely, its exceptional reduction-oxidation (redox) properties enabled by the duality of the cerium ion (\ce{Ce^4+ <-> Ce^3+}). It is generally a formidable task to try to mimic interactions, structure and energetics simultaneously for a compound like ceria without having access to explicit electrons. In earlier work\cite{broqvist2015reaxff}, we constructed a reactive interaction model using the ReaxFF modelling framework \cite{Senftle2016Reaxff} with the aim of handling stoichiometric and partially reduced ceria bulk, surfaces, and nanoparticles (NPs). The model was based on a training set of DFT calculations for a large number of ceria systems in various forms and configurations (bulk, clusters, surface systems; stoichiometric as well as reduced systems). With some exceptions that model performed very well. In the present study, instead, our forcefield is \MPz{}, where the content of ceria in the training set is only bulk structures, namely exactly 18 bulk polymorphs (stoichiometric or partially reduced).

Here we assess the ability of the \MPz{} model to describe small ceria nanoparticles of different shapes, sizes and reduction degrees without the training ever including any ceria NPs or surfaces. The optimized NP structures/shapes as well as their formation energies (with respect to stoichiometric bulk \ch{CeO2} and gas-phase \ch{O2} as needed)
will be assessed. 

The left panel in \cref{fig:UU_ceO2} shows the \MPz{} results for optimized particles up to 140 formula units: stoichiometric tetrahedra, stoichiometric truncated octahedra, and perfect octahedra (which are partially reduced by virtue of their shapes). The bottommost curve pertains to ``supercharged'' NPs, i.e. perfect octahedra that are decorated with oxygen molecules. The right panel shows the corresponding results from independent reference calculations at the DFT (PBE+U) level, taken from Refs. \cite{kullgren2013supercharged,broqvist2015reaxff} 

\begin{figure*}
    \centering
    \includegraphics[width=0.9\linewidth,keepaspectratio]{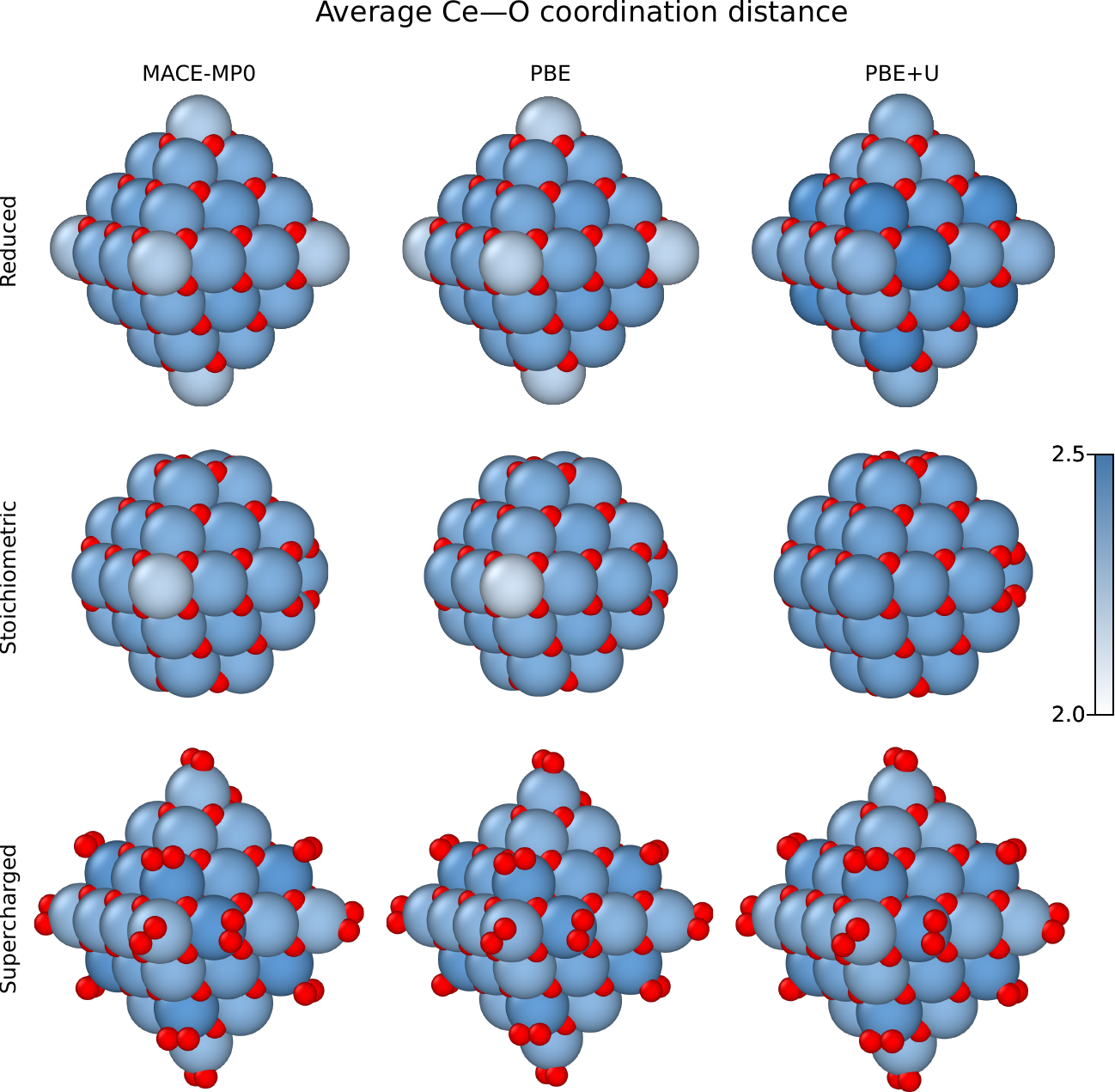}
    \caption{Optimized structures of the perfect \ch{Ce44O80} octahedron with three different methods: \MPz{}, PBE, and PBE+U. The purpose of the figure is to highlight the ``pattern of distances'' rather than quantitative values. The large spheres are the Ce ions (regardless of charge), and the small red spheres are oxygens. The colour scheme indicates the optimized interatomic distances in the following way: for each Ce atom, the distances to  its O neighbours in the coordination figure of nearest-neighbours 
    is measured and the average value is  reflected in the colour of the sphere. Light blue indicates a short average Ce--O distance, dark blue indicates a long average Ce--O distance. The distance scale is shown in the colour bar to the right which covers the range from \SI{2.0}{\Angstrom} to \SI{2.5}{\Angstrom}.
    }
    \label{fig:ceriacloseup}
\end{figure*}

The agreement between the energetics in the two panels in \cref{fig:UU_ceO2} is good overall, which is satisfying. However, we note that as far as structures are concerned, the  \MPz{} model (leftmost panel of \cref{fig:ceriacloseup}) 
is unable to distinguish between \ch{Ce^3+} and \ch{Ce^4+} ions, both of which should in fact be present in a partially reduced perfectly octahedral ceria nanoparticle. This deficiency of the \MPz{} structure is evident from a comparison with the independent electronic PBE+U calculations in the rightmost panel of \cref{fig:UU_ceO2}, labelled ``DFT+U''. Such calculations involve a Hubbard correction which enforces a stronger and more adequate localization of electrons at \ch{Ce^3+} sites than what is achieved by standard PBE without U, which is the DFT method used in the Materials Project for ceria. The presence of both \ch{Ce^3+} and \ch{Ce^4+} ions in the PBE+U results is seen to lead to local relaxations of the nearest-neighbour oxygen ions around the Ce ions, resulting in symmetry breaking of the NP; see 
for example the lack of symmetry with respect to the NP edges in the rightmost panel.
    
Neither the proper local relaxation nor the symmetry breaking, both seen in the PBE+U results, is captured by the \MPz{} model. On the other hand, the middle panel of \cref{fig:ceriacloseup} shows our PBE-optimized results for the same NP. 
The structural similarity between the \MPz{} result 
and the PBE-optimized nanoparticle is evident.

\subsubsection*{Similarity statement:} There are altogether 18  \ch{CeO2} and \ch{CeO_{2-x}} bulk structures present in the MP dataset. No examples of stoichiometric or reduced ceria surface structures or ceria nanoparticles are present in MP.

\subsubsection*{Performance summary}
Broadly correct prediction of the energy of nanoparticles as a function of size, including overoxidised particles, with respect to reference DFT results.

\clearpage
\subsection{Inorganic halide perovskite}\label{sec:pero_ino}

\begin{figure}[htbp!]
  \centering
  \includegraphics[width=0.6\textwidth,keepaspectratio]{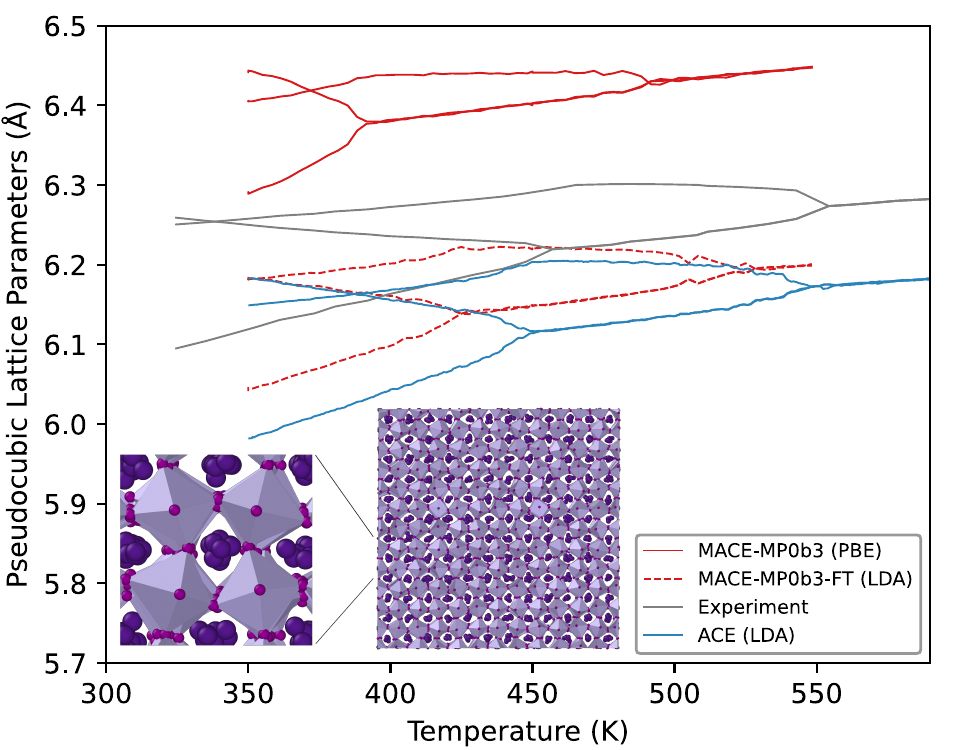}
  \caption{Variation of pseudo-cubic lattice parameters with temperature for \ch{CsPbI3}. The \MPz{} model is compared to experimental data reported by Even and co-workers \cite{Even2018} and an atomic cluster expansion (ACE) model trained for this material\cite{ACESmall2023}. Inset shows an illustration of the system used to calculate the lattice constant dependence on temperature.}
  \label{fig:perovskite_simulation_cell}
\end{figure}

Halide perovskites have been shown to exhibit subtle phase transitions and long-range structural correlations. The \MPz{} model has been applied to these systems by predicting phase transitions in the inorganic perovskite \ch{CsPbI3}. This material shows two solid-solid phase transitions between room temperature and \SI{600}{\kelvin}, both of which involve small rotations of the octahedral units and accompanying changes in pseudo-cubic lattice parameters\cite{Even2018}.

To analyze these transitions, we ran constant pressure simulations of a 14000 atom supercell with a slowly varying temperature. \Cref{fig:perovskite_simulation_cell} shows the variation in pseudo-cubic lattice parameters with temperature, compared to experimental data. These data were obtained from a \SI{3}{ns} simulation during which the temperature was raised from \SI{350}{K} to \SI{550}{K}. The \MPz{} model correctly predicts the qualitative nature of both phase transitions. There is a shift in both the transition temperatures and the average lattice constant, which has also been observed in other studies of these materials with DFT\cite{ACESmall2023}. It is also known that the choice of exchange-correlation functional has a large effect on transition temperatures for these materials\cite{fransson2023, Jinnouchi2019}.

\subsubsection*{Similarity statement}

There are 57 structures in the MP dataset containing some combination of \ch{Cs}, \ch{Pb} and \ch{I}, without other elements. Of these, 5 structures contain all three of these elements in different compositions spanning several phases of this material. Based on UMAP analysis, these 5 structures are close to the training dataset. In particular, the cubic and orthorhombic phases which are studied in this example are present. Several similar structures with \ch{Br} replacing \ch{I} are also in the training set.

\subsubsection*{Performance summary}
Both structural phase transitions and their transition temperatures well captured, and the 10\% discrepancy in the latter with respect to experimental values is likely due to the PBE functional.

\subsection*{Fine-tuning}

Fine-tuning was performed using 100 configurations sampled randomly from the database used in reference \cite{ACESmall2023}, wherein an ACE model was constructed specifically for studying phase transitions and nanostructural features of \ch{CsPbI3} and the LDA exchange correlation functional was used. This database was obtained by running NPT and NVT MD of the material at a range of temperatures.   Figure \ref{fig:perovskite_simulation_cell} compares the predictions of the lattice constant variation from the \MPz{} foundation model to that of the finetuned model, and the ACE model trained in reference \cite{ACESmall2023}. One can see that the model prediction is shifted substantially towards that of the ACE model. 

\clearpage
\subsection{Hybrid Organic-Inorganic Perovskites (HOIPs)}\label{sec:pero_organic}
\begin{figure*}[htbp!]
  \centering
  \includegraphics[width=0.9\textwidth,keepaspectratio]{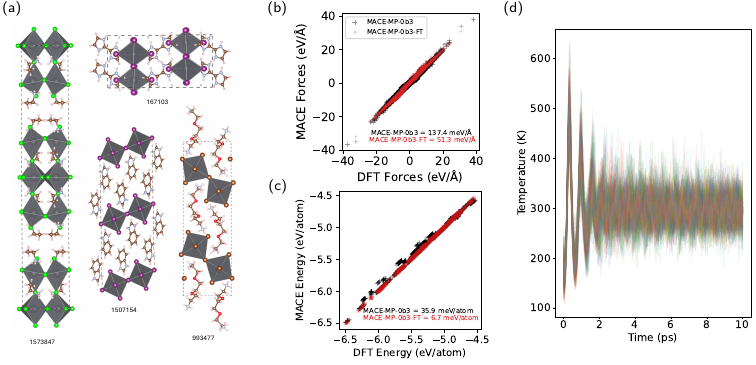}
  \caption{(a) Four selected HOIP structures \cite{HOIP_ex1,HOIP_ex2,HOIP_ex3,HOIP_ex4} with CCDC deposition numbers given below each structure. (b) Force parity and (c) energy parity plots, with samples taken from MD trajectories using both the \MPz{} and a version fine-tuned on 159 configurations, and compared directly with the corresponding DFT values. (d) The MD trajectories for 159 compositions for 10 ps in NPT ensemble at $T=300$ K. }
  \label{fig:hoip}
\end{figure*}
Hybrid organic-inorganic perovskites (HOIPs) \cite{hoip_original} are a promising class of perovskites that have been studied extensively due to their strong chiral response, optical absorption, high carrier mobility. However, the most-studied three-dimensional \ce{MAPbI3} suffers considerably from low stability \cite{hoip_intro}. Two-dimensional HOIPs have the advantages of enhanced stability and structural tunability, which makes them candidates for promising applications in photoluminescence (PL), solar cells and light emitting diodes (LEDs) \cite{hoip}. However, 2D HOIPs are computationally expensive to examine using DFT because of their complicated unit cells formed when the organic cations separate the inorganic layers in (100) direction, giving the modified general formula \ch{{A'}_{\textit{m}}A_{\textit{n}-1}B_{\textit{n}}X_{\textit{3n}+1}}, corresponding to $n$ layers of the 3D-parent \ch{ABX3} structure, separated by a layer of A' organic cations that carry either a single charge ($m$ = 2) or two charges ($m$ = 1); the diversity of these systems can be seen in \cref{fig:hoip}a.

Using \MPz{}, we investigated a set of 159 experimentally synthesized 2D HOIPs from the  Cambridge Structural Database (none are in the MP) with B = \ch{Pb} and X =  \ch{Cl}, \ch{I}, and \ch{Br}. The organic cations (A' and A) are comprised of only the elements \ch{C}, \ch{H}, \ch{N} and \ch{O}, with either a +1 or +2 charge. MD simulations were performed within the NPT ensemble using \MPz{} with the D3 correction at a temperature of 300 K and pressure of 1 atmosphere for 10 ps. For all the 159 MD the trajectories remain stable and reach equilibirum in about 3 ps (\cref{fig:hoip}a. No bond-breaking or surface cleavage between the organic/inorganic layers occurred. To test the accuracy of the model, we applied it to a set of 3,007 configurations drawn from a recent study of HOIPs \cite{doi:10.1021/jacs.4c06549}, also sampled from multiple MD trajectories, with the errors in forces (RMSE = \SI{137}{meV/\Angstrom}) and energies (RMSE = \SI{35.9}{meV/atom}) (\cref{fig:hoip}b and c).

To fine-tune the model, a total of 159 configurations (one per composition) was randomly sampled from the MD trajectories. The errors in forces and energies are significantly reduced to \SI{51}{meV/\Angstrom} and \SI{6.7}{meV/atom}, and are  similar to the errors reported in \cite{doi:10.1021/jacs.4c06549}. However, in that study, the training procedure required 2457 configurations from MD trajectories, about 15$\times$ more than what is used in the fine-tuning in this study.

\subsubsection*{Similarity statement}
In the MP training set, there are in total 398 structures with \ch{PbX} (X: \ch{Cl}, \ch{Br}, \ch{I}), and \num{1627} structures with organics made of the elements \ch{C}, \ch{H}, \ch{N} and \ch{O}. Based on UMAP analysis, we observe that some have similarities in environments to the MP training set, but only 22 structures have a mixture of \ch{PbX} and CHNO. From these, there are 14 3D HOIPs in which 12 of them have methylammuniumm (MA) as the organic cation, 3 0D HOIPs, and 4 cases of non-perovskite systems. In the MP, there is one 2D HOIP, which is the most similar to our dataset (mp-1194995), but this structure was not in the our dataset.

\subsubsection*{Performance summary}
Stable NPT MD at ambient conditions for all 159 2D hybrid perovskite materials.

\clearpage
\subsection{Protein Dynamics and Stability}\label{sec:protein_folding}
\begin{figure*}[htbp!]
  \centering
  \includegraphics[width=0.9\textwidth,keepaspectratio]{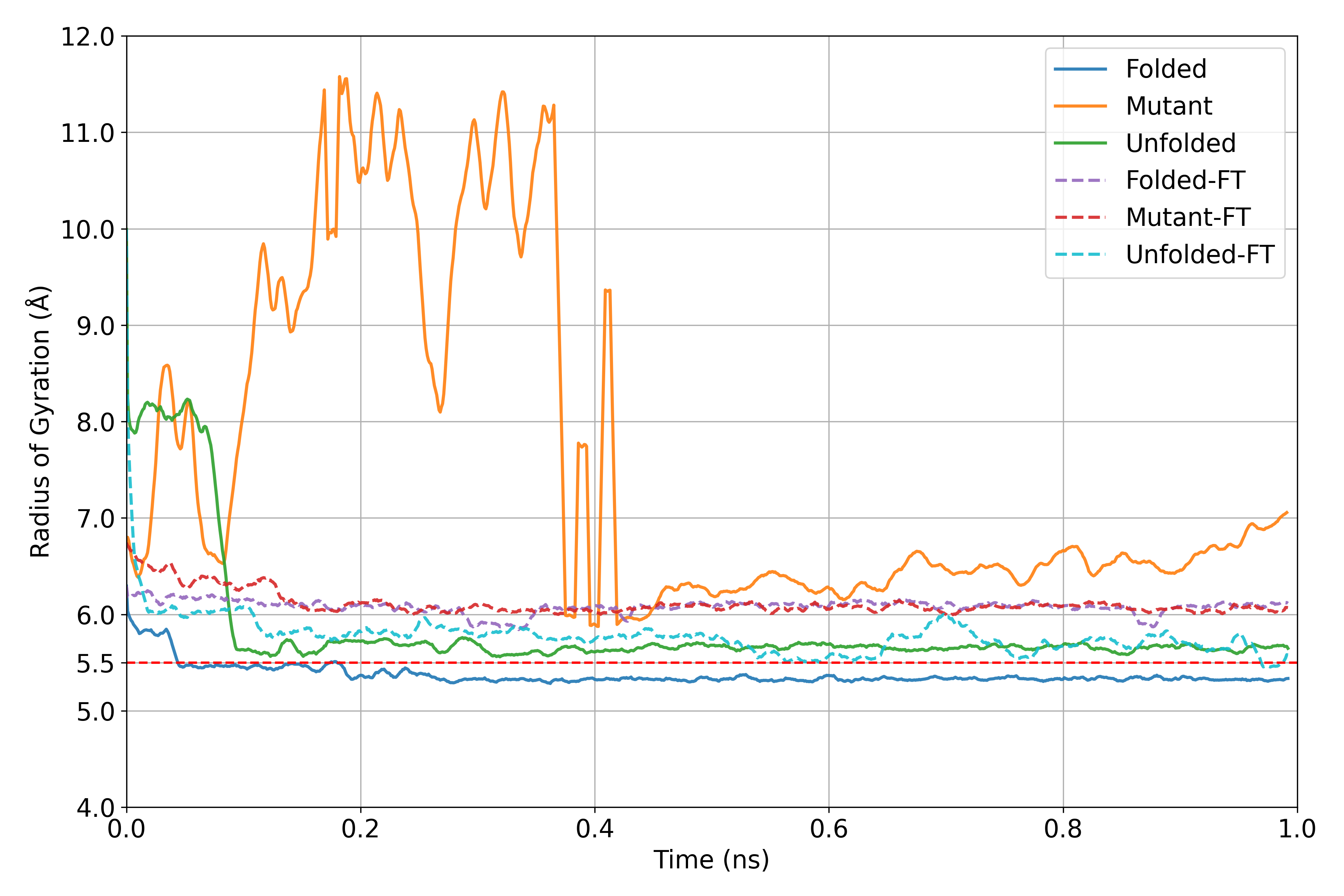}
  \caption{Plot of the radius of gyration (\textup{\AA}) versus the time of simulation (ns) for the three simulations performed on Chignolin (PDB: 1UAO) and Chignolin mutant (PDB: 5AWL) with both \MPzbt{} and the fine-tuned (FT) version. The radius of gyration curves are smoothed with a running mean over 10 steps to better visualise their overlap. The "Unfolded" simulation trajectories start from an unfolded structure of Chignolin, the "Mutant" lines correspond to the simulations for the Chignolin mutant and "Folded" corresponds to the simulations that start from a partially folded structure of Chignolin. The red line corresponds to the computed radius of gyration for the PDB 1UAO structure.}
  \label{fig:protein_stability_rg}
\end{figure*}
Understanding the dynamics of proteins is crucial for deciphering their biological function, and remains a core challenge of computational chemistry. Machine learning potentials have the capability of modelling non-covalent interactions, which are key components of secondary and tertiary structures of proteins, due to the quantum mechanical data on which they are trained on. 
In this section, we perform simulations on the well-known engineered Chignolin peptide (PDB: 1UAO), an artificial prototype for the protein folding phenomenon. This study diverges significantly from the majority of studies discussed in this manuscript as we are employing a machine-learning potential trained on inorganic crystals for a biological purpose. 

We perform three separate simulations for each trained or fine-tuned model:
\begin{itemize}
    \item A simulation starting from a partially folded structure of Chignolin.
    \item A simulation starting from an unfolded structure of Chignolin.
    \item A simulation of a mutant of Chignolin (PDB: 5AWL) starting from its PDB structure which contains 12 crystallographic waters.
\end{itemize}

Representative starting structures for the folded and unfolded Chignolin were obtained from Ref \cite{Wang2023}, specifically the first (folded) and last (unfolded) geometries from the 9543 conformations sampled by replica exchange molecular dynamics. 
We protonate negatively charged residues (1UAO: ASP3, GLU5, GLY10 | 5AWL: ASP3, GLU5, TYR10) and remove the proton on GLY1 for 1UAO and on TYR1 for 5AWL. We are aware that the real stable state of Chignolin is non-neutral and solvated but this test is performed in a neutral environment as \MPz{} was not trained with a charge-based loss and testing stability of the simulation is our primary goal.
Simulations were performed at 300K for 1 ns at 1 fs time-step and a sampling frequency of 1 ps in the NVT ensemble using the Atomic Simulation Environment (ASE) package. The Langevin thermostat was employed with a friction coefficient of 10 ps$^{-1}$. \MPzbt{} with the D3 dispersion correction was used for all simulations. Most simulations completed successfully in less then a day (elapsed-time) while one simulation ("Mutant") with the pre-trained \MPzbt{} model took about three days.

We first analyse the results obtained from the pre-trained-only version of \MPz{}.
From Figure \ref{fig:protein_stability_rg} we can see that from its unfolded state (green line), Chignolin is quickly compacting at the start of the simulation and then maintains a stable radius of gyration, while the folding state (blue line) is maintained when starting from a partially folded state.   
Both simulations of Chignolin try to converge around the radius of gyration computed for the crystal structures 1UAO (red line), with the "Folded" simulation under-shooting and the "Unfolded" simulation slightly over-shooting it.
On the other hand, the Chignolin mutant (yellow line) shows a perturbed radius of gyration due to some of the crystallographic waters losing contact to the protein early on in the simulation. Note that the radius of gyration was computed considering all the system's atoms. This was done to highlight any lack of stability in the simulation.

\begin{figure*}[htbp!]
    \centering
    \includegraphics[width=\linewidth]{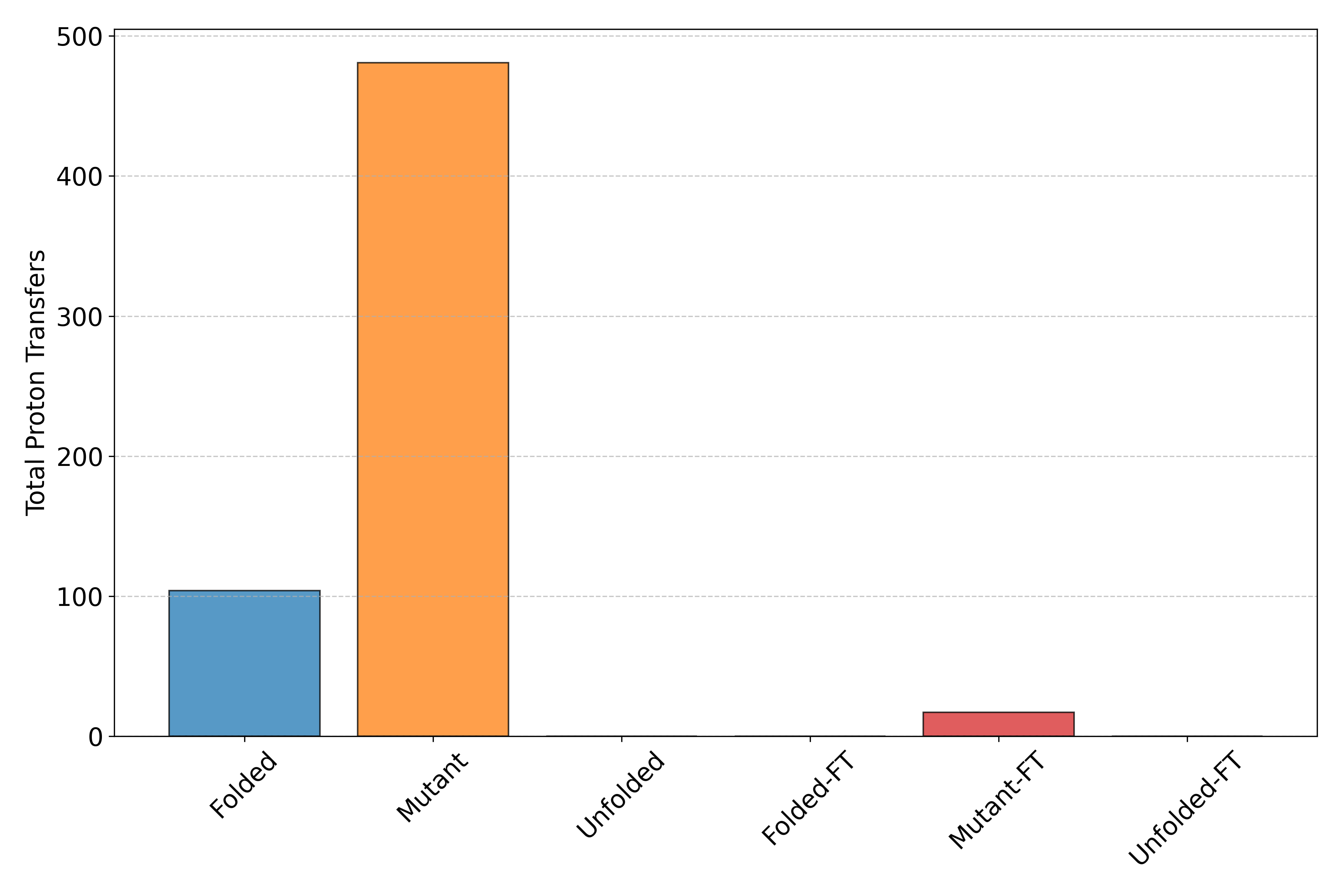}
    \caption{Histogram of protons transfers occurring during the 1 ns simulations. Total count of proton transfers throughout the simulation for each system.}
    \label{fig:protein_folding_proton_transfer}
\end{figure*}

The simulations pursued with the fine-tuned (FT) model show improvements in terms of stability for the Chignolin Mutant (red line) as the crystallographic waters interact with the protein at any given time during the simulation. The unfolded Chignolin simulation (light-blue line) has an even faster formation of a compact protein structure compared to the pre-trained-only model, while the folded Chignolin simulation (purple line) is stable but has a shift towards a higher-average radius of gyration. To investigate this further, we performed a proton transfer analysis, shown in Figure \ref{fig:protein_folding_proton_transfer}.

On the left, we estimate the amount of proton transfers happening across time steps of each simulation. We use a simple distance-based detection to define the transfer of a proton. At every time step, we keep track of the closest heavy-atom neighbour of each proton. If in a time-step the proton has a new heavy atom neighbour with respect to the previous time-step and its distance to it is less or equal to 1.2 \AA{} (resembling the fact that a bond has likely formed), then we increase the total transfer count by one. We observe a large number of proton transfers for the folded Chignolin and Chignolin mutant in the pre-trained \MPz{}, while none in the unfolded. The number of proton transfers drops drastically in the fine-tuned version of the model. In theory, we expect the transfers to be low as the system is neutralised and there are little to no water solvation shell with which the system could interact. Thus, it is clear that the fine-tuned model is closer to our expectations. This however might also mean that the fine-tuned model is biased towards weaker covalent and non-covalent interactions, which could also explain the higher radius of gyration observed in Figure \ref{fig:protein_stability_rg} for all of the fine-tuned simulations. 

We want to point out that some proton transfers are still observed in the Chignolin mutant even after fine-tuning. The proton transfer encountered happens between the neutralised COOH and NH$_{2}$ functional groups of GLY1 and GLY10 of Chignolin. This is a typical case of acid-base chemistry where we would expect the proton to transfer from the more acidic carboxylic acid group to the basic amine group, forming a zwitterion in physiological conditions (i.e. neutral pH).
Overall this small case-study shows the capability of MACE in modelling reactivity in molecular dynamics simulations. We point out that these observed processes of proton transfer cannot be observed with classical force fields as they are generally not parametrised for such effects. Moreover, we conclude that fine-tuning the pre-trained foundation model is overall necessary to find a better agreement with the experiment and realistic physicochemical behaviour. However, more real-world tests, which are out of the scope of this section, are needed to confirm that the model behaves correctly. Such tests would include a fully solvated system in physiological pH conditions and use of a better-tailored fine-tuning dataset.

\subsubsection*{Similarity statement}
\begin{figure*}[htbp!]
    \centering
    \includegraphics[width=0.9\textwidth,keepaspectratio]{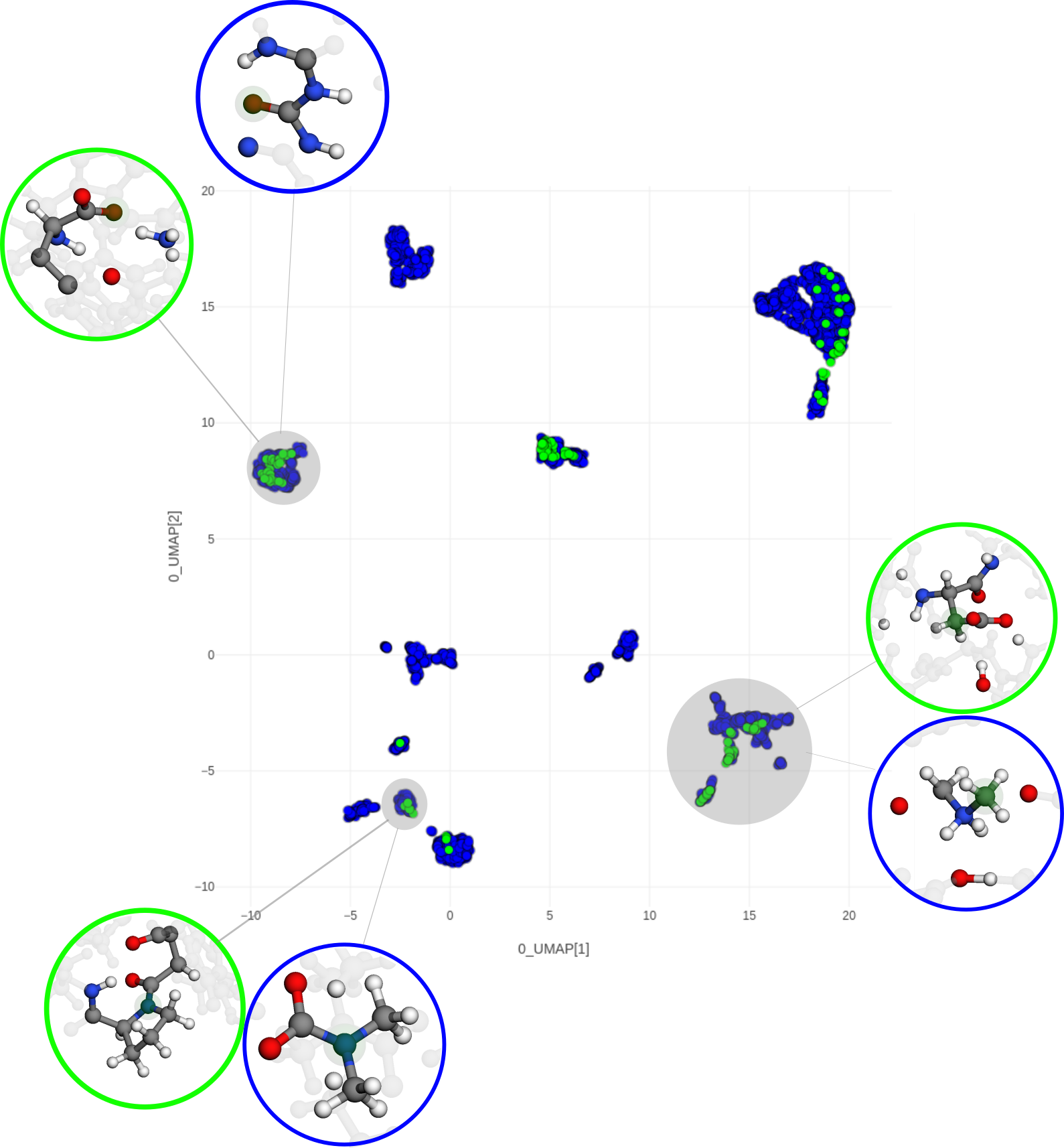}
    \caption{Comparison of the atomic environments in the 99 filtered structure of the training data (blue dots) and in the Chignolin mutant structure (green dots) in the form of a UMAP plot. Insets of some local environments are shown for a few key functional groups of the protein (green circles) with corresponding similar clustered environments from the training set (blue circles)}
    \label{fig:umap_protein_folding}
\end{figure*}

The MP dataset encompasses only 99 structures exclusively composed of the elements hydrogen (H), carbon (C), oxygen (O), and nitrogen (N). Based on the UMAP analysis, the atomic environments of Chignolin and Chignolin mutant are clustered similarly to those of the filtered 99 structures (see Figure \ref{fig:umap_protein_folding}), which could mean that the environments of the training set are fairly similar to those of the test structures . However, after manual inspection, we observe that the filtered structures only resemble the protein under study by some functional groups such as carboxylic acid (-COOH), amino groups (-NH\textsubscript{2}), aromatic and amide groups. Some of the most similar structures are \textit{e.g.} mp-998880, mp-1203308, mp-556151, mp-707289 and mp-1203544.

\subsubsection*{Fine-tuning}
Fine-tuning of the \MPzbt{} model was performed with 100 configurations obtained from the simulations done with the model itself. Specifically, 50 samples of the "Unfolded" Chignolin simulation with 25 in the first third of the simulation where the system is less compact and 25 in the rest of the simulation, 25 of the "Folded" Chignolin simulation and 25 of the "Mutant" Chignolin. For the latter, we only kept into account the water molecules that were interacting with the system and removed those that went towards very large distances from the protein center of mass (>20 \AA{}).  

\subsubsection*{Performance summary}
Most simulations performed showed no specific un-physical phenomena (e.g. no "explosions" are observed), however many proton transfers that can be deemed to be unchemical are observed for the pre-trained model. For a few systems, the radius of gyration is maintained or converges to the value computed from the PDB structures. However, we note that residues that were charged in the PDB structures were protonated, thus the experimental folding (i.e. formation of secondary structure) cannot be fully recovered in this context.

\clearpage
\subsection{Hydrogen combustion}\label{sec:h_comb}
\begin{figure}[htbp!]
  \centering
  \includegraphics[width=0.8\textwidth,keepaspectratio]{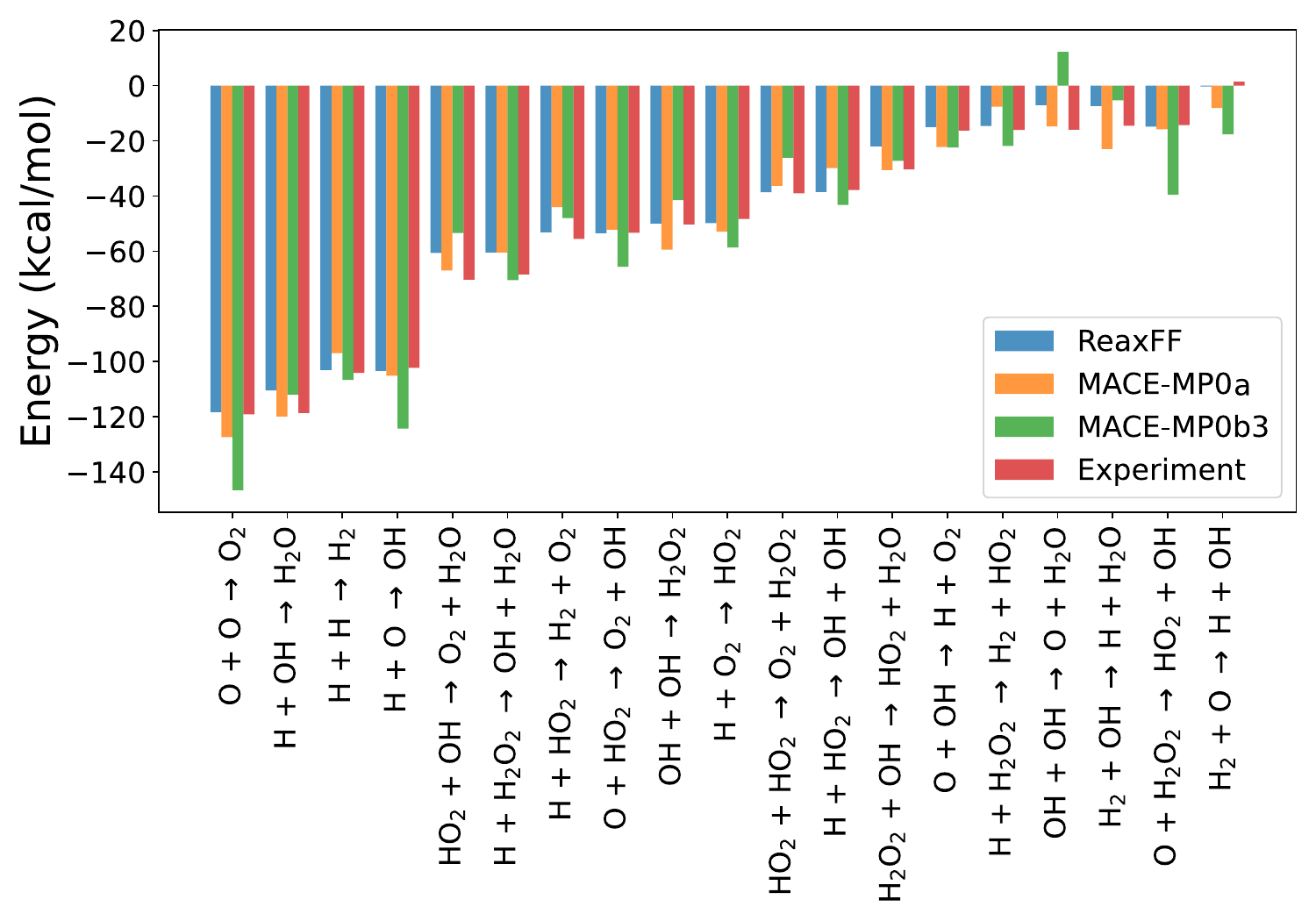}
  \caption{Comparison of Heats of Reaction of Key Hydrogen Combustion Reactions with ReaxFF and Literature}
  \label{fig:hydrogen_combustion_heats}
\end{figure}

Describing the complex reactivity in hydrogen combustion systems is a challenging task, often approached through thermodynamics via quantum mechanical (QM) calculations. However, accurately capturing kinetics in molecular dynamics simulations is hindered by the lack of a suitable transferable empirical force field, a difficulty compounded by the challenge of collecting experimental data in high-pressure, extreme explosion conditions. In this context, we compare our machine learning force field, \MPz{}, against the reactive classical empirical force field ReaxFF\cite{agrawalla_reaxff}. ReaxFF model was created by fitting to an extensive QM dataset. The creation involved the identification of key reactions and components, and the collection of data on formation heats, reaction heats, and energy barriers, as well as computing bond stretching energies and valence angle distortion energies for all combinations of hydrogen and oxygen. In our study, the performance of \MPz{} in describing these reactions is evaluated and compared to both ReaxFF and experimental values. It is important to note that \MPz{} has not been specifically trained for hydrogen combustion reactions as the training data primarily comprises periodic representations of strongly bonded inorganic materials and a smaller fraction of molecular crystals; see \cref{sec:hydrogen_combusion_similarity}.

As seen in \cref{fig:hydrogen_combustion_heats}, the ReaxFF model agrees well with experimental values \cite{baulch_2005} for heats of reaction, with an RMSE of \numlist{4.82}~\unit{kcal\per mol}. The \MPzbt{} model (green in the figure) that is used in almost all other examples, while qualitatively showing the right trend, is significantly less accurate (with an RMSE of 14.7~\unit{kcal\per mol}). It is interesting to note that this example is one of the very few cases where the original version of the foundation model, \MPz{}a (orange in the figure) does much better, with an RMSE of 6.6~\unit{kcal\per mol} (and its ``large'' version, with more free parameters, reaches the accuracy of ReaxFF). The latest version of the foundation model is significantly more stable in general than the original versions, due to better architectural choices. Nevertheless, this example shows that there is still more understanding to be gained, and in the future it should be possible to improve the out-of-the-box accuracy of the foundation model without sacrificing its stability. Furthermore, a detailed investigation of the accuracy of different foundation models would need to necessarily include a study of the accuracy of DFT and the exchange-correlation functional that is used to generate the training data, rather than assuming that it is close to the experimental values. For example, the PBE functional has a mean absolute deviation of about 9~\unit{kcal\per mol} for a set of reactions relevant to hydrocarbon combustion\cite{gga_reaction}.

\begin{figure}
  \centering
  \includegraphics[width=\linewidth,keepaspectratio]{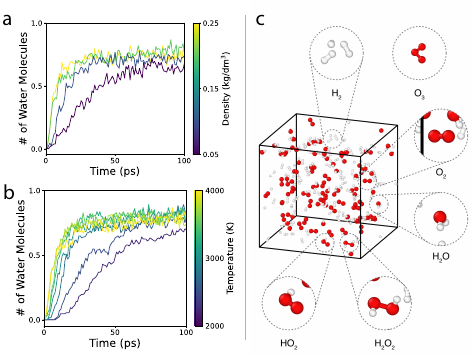}
  \caption{Analysis and visualization of hydrogen combustion in MD simulations using the \MPzbt{} potential. Water formation as a function of elapsed time for a range of (a) densities and (b) temperatures. (c) Representative snapshot during MD simulations, with key species highlighted. Note, although \ce{O3} is not present in the particular simulation frame shown, it is occasionally observed in other simulations.}
  \label{fig:hydrogen_combustion}
\end{figure}

We used Packmol\cite{martinez2009packmol}  to randomly arrange a 1:1 fuel mixture comprising 128 \ch{H2} and 64 \ch{O2} molecules within cubic cells (side length, a = \numrange{25}{42}~\unit{\Angstrom}), yielding densities ranging from \numrange{0.05}{0.25}~\unit{\kilogram\per\cubic\deci\meter}. Employing NVT simulations for \SI{100}{\pico\second} with the \MPzbt{} potential, we tracked the evolution of \ce{H2}/\ce{O2} mixtures. Reactivity analysis focused on water formation, identified via pairwise cutoffs derived from the first minima of the radial distribution function. The relationship between temperature/density variations in the fuel mixture and the water formation rate is depicted in \cref{fig:hydrogen_combustion}a and \cref{fig:hydrogen_combustion}c. We find qualitative agreement with the water formation curves of Ref.~\cite{agrawalla_reaxff}, with a max conversion of approx. \SI{80}{\%}.

\subsubsection*{Similarity statement} \label{sec:hydrogen_combusion_similarity}
We analyze the MPtrj training dataset for the key species in hydrogen combustion (\ce{O2}, \ce{H2}, \ce{H2O}, \ce{H2O2}, \ce{HO2}, and \ce{OH}). These species are present as minority units in other structures, appearing in 2277, 1310, 1342, 232, 21, and 0 structures, respectively. We find 21 molecular crystals composed exclusively of \ce{O2}, 17 for \ce{H2}, 11 for \ce{H2O}, 2 for \ce{H2O2}. There are only eight structures made up of multiple key reaction species. UMAP analysis reveals only 2 MPtrj structures (mp-684678 and mp-1181087) with high similarity to frames within MD simulations.

\subsubsection*{Performance summary}
Heats of reaction show the right trend for 19 reactions. Chemically correct species produced during combustion, with final yield also consistent with reference methods. 

\clearpage
\subsection{Sulfur polymerisation}\label{sec:S}

\begin{figure}[htbp!]
  \centering
  \includegraphics[width=0.65\textwidth,keepaspectratio]{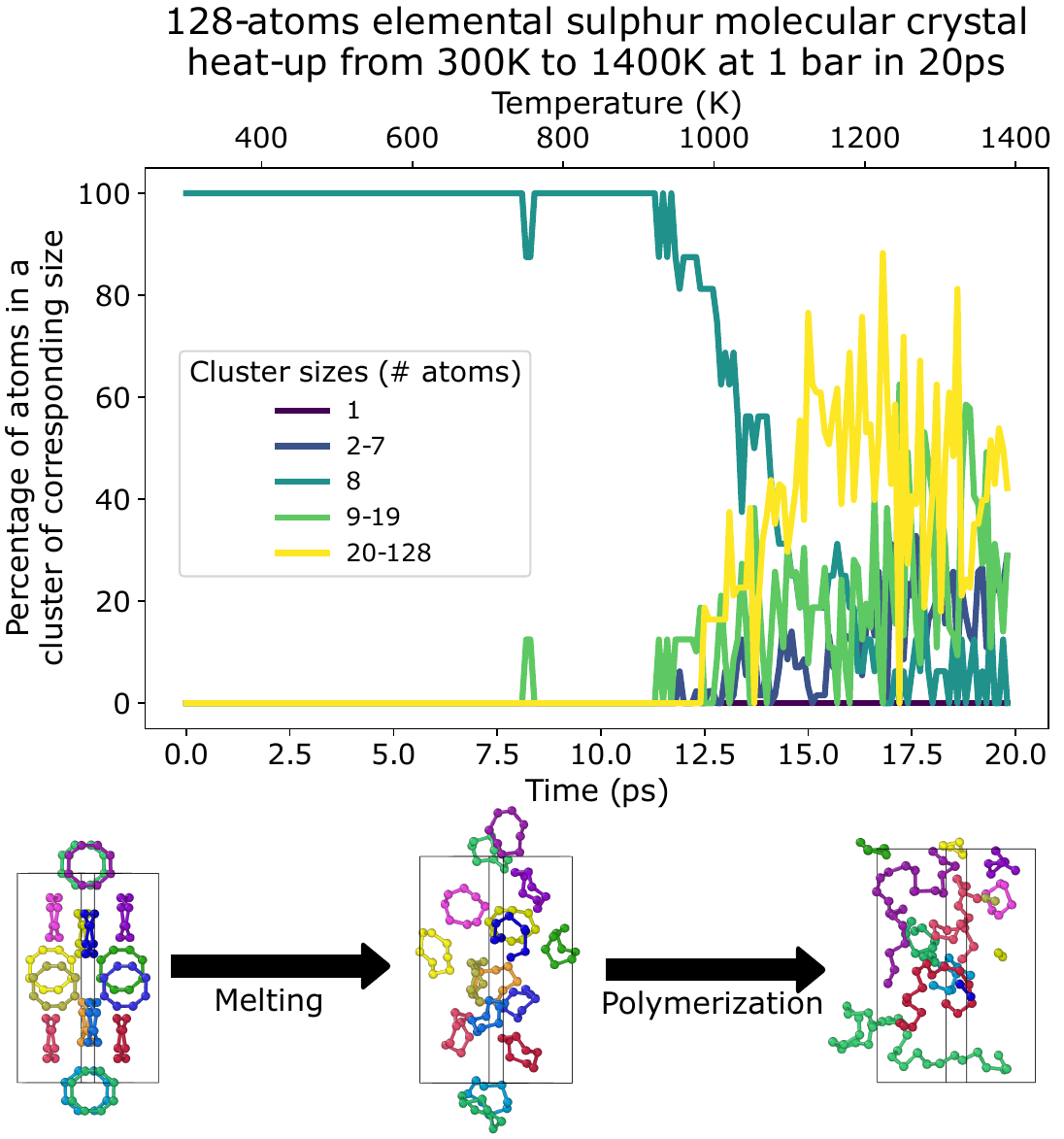}
  \caption{{\bf Elemental sulfur.} Evolution of the cluster size as a function of temperature in a \SI{20}{\pico\second} heat-up simulation with a linear increase in temperature applied over time. The pictures below are representative snapshots from the simulation, visualised in OVITO \cite{stukowski2009visualization}.}
  \label{fig:sulfur}
\end{figure}

We ran a \SI{20}{\pico\second} NPT heat-up of a 128-atom structural model of elemental sulfur from \num{300} to \SI{1400}{\kelvin} at \SI{1}{bar}. At ambient pressure, experiments show that the molecular crystal formed of \ch{S8} rings melts at \SI{392}{\kelvin} and starts polymerizing at \SI{432}{\kelvin}~\cite{steudel2003elemental}, forming large chains that result in a 4-fold increase in viscosity of the liquid. We can qualitatively reproduce this melting and chain formation with \MPzbt{}+D3 correction (\cref{fig:sulfur}). The simulated melting temperature does not exactly match the experiment, as expected for a very fast run with only 128 atoms -- however, the simulation does qualitatively reproduce the onset of de-polymerisation with increasing temperature, where large chains break down into smaller ones. For these reasons, we can say that \MPzbt{}+D3 is at least qualitatively applicable to simulate the polymerisation of elemental sulfur starting from $\alpha$-rhombohedral crystalline \ch{S8} (with further work being required to test the quantitative agreement).

\subsubsection*{Similarity statement}

The MP dataset contains 31 structures that only contain the element sulfur. Based on UMAP analysis, we see that a large part of the atomic environments in the example system are similar to environments in the training data. The database contains geometry optimizations of sulfur crystals formed of rings with various sizes: \footnotesize6 (mp-7), 7 (mp-557559), 8 (mp-77), 9 (mp-556269), 10 (mp-557031), 11 (mp-561370), 12 (mp-558014), 13 (mp-583072), 14 (mp-561513), 10x6 \normalsize (there exists a sulfur crystal form comprised of \ch{S_{10}} and \ch{S6} rings, \footnotesize 557031), 18 (mp-555915) and 20 (mp-558964)\normalsize. It also contains crystals with planar strands of sulfur \footnotesize(mp-1179643)\normalsize, trigonal polymeric sulfur \footnotesize(mp-555760)\normalsize, and so-called fibrous sulfur (quenched polymeric liquid sulfur, \footnotesize mp-1196831)\normalsize, as well as isolated dimers \footnotesize(mp-1179639)\normalsize, trimers \footnotesize(mp-655141)\normalsize \ and single atoms \footnotesize(mp-1063988)\normalsize. It does not contain melt or polymeric liquid structures. 
Based on the UMAP analysis, the closest (most relevant) structures in the training set are: \footnotesize mp-556269, mp-555915, mp-83, mp-557031, mp-557559, mp-666931\normalsize.
We provide \verb|sulfur.json| to help visualize the interactive UMAP on \url{chemiscope.org}.

\subsubsection*{Performance summary}
Qualitatively correct polymerisation, with large clusters forming then subsequently breaking down to smaller size with an increase in temperature.

\clearpage
\subsection{Zeolites}\label{sec:zeo}

\begin{figure*}[htbp!]
  \centering\includegraphics[width=\linewidth,keepaspectratio]{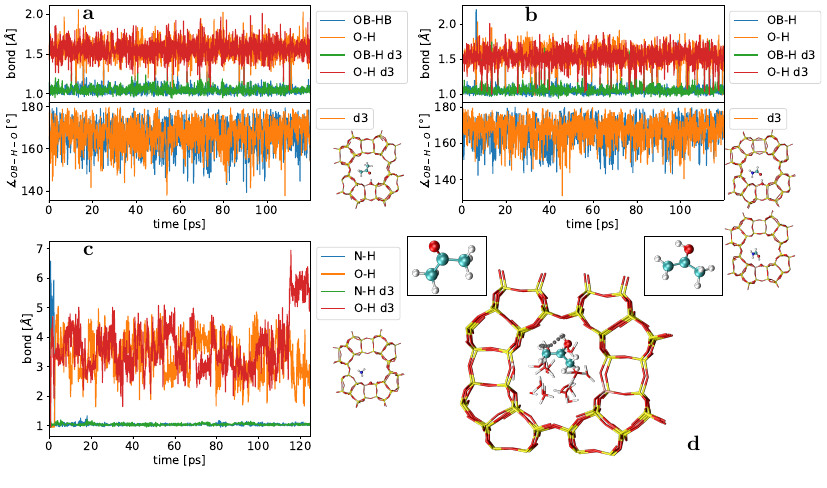}
      \caption{a) MOR-\ch{Al} with acetone, instantaneous bond lengths between \ch{O}@BAS, \ch{H}@BAS and \ch{O}@acetone and \ch{H}@BAS upper panel and angles lower panel - for both \MPzbt{} and D3 corrected a hydrogen bond is formed. b) same as a) but for formamide, one can see deprotonation events when \ce{O-H} bond value approaches  \SI{1}{\Angstrom}, c) \ch{NH3} case with bonds same as a) \ch{NH3} quickly deprotonates the BAS} to form \ch{NH4^+} d) NEB~\cite{Lindgren2019,Makri2019} path for keto--enol tautomerism of acetone, the path is indicated by the \ch{H} atom traces, only final position of the rest of the atoms is shown
  \label{fig:zeo}
\end{figure*}

Zeolites are mesoporous materials with an important role as heterogeneous catalysts in several industrial processes. In this section, we assess the suitability of \MPz{} to model these materials. We chose two zeolites, Modernite and Zeolite Socony Mobil–5 - ZSM-5 or MOR and MFI by their International Zeolite Association names. We have investigated the dynamic stability of the zeolite frameworks themselves, MOR with \ch{NH3}, acetone and formamide inside one channel and MFI with water, cyclohexane and a mixture of \ch{N2} in one channel and \ch{CO2} in the other. Each zeolite was modified by adding a Br{\o}nstead acide site - BAS, \ch{Al}, and the compensating \ch{H} on the adjacent oxygen, see \cref{fig:zeo}. Another set of simulations was carried out on MOR-\ch{Al}, where in addition to acetone, we introduced 20 water molecules and 32 water molecules, and similarly for MFI-\ch{Al} with cyclohexene instead of acetone.
\MPz{} correctly identified the adsorption sites, for ammonia, acetone and formamide and the structural motifs in agreement with DFT calculations from, \cite{Trachta2023} in NPT ensemble~\cite{melchionna1993,melchionna2000} simulations carried at \SI{300}{K} and \SI{400}{K} for \SI{125}{ps} each using ASE. Furthermore, in the case of ammonia \MPz{} correctly predicts the formation of the \ch{NH4^+} and its stabilization around the BAS by the creation of hydrogen bonds with adjacent oxygen atoms, see \cref{fig:zeo} panel c. \MPz{} also correctly reproduces the DFT findings that acetone does not deprotonate the BAS but forms hydrogen bonds, while formamide predominantly forms hydrogen bonds but deprotonates the BAS occasionally.
Additionally, for the system MOR-\ch{Al} with acetone and 20 water molecules, we have computed the barrier of the keto-acetone to enol-acetone conversion. \MPz{} gave a barrier of \SI{2.11}{eV} and with D3 correction, \SI{2.20}{eV}, numbers are in good agreement with PBE calculations reported in \cite{Cucinotta2006}. The code used to generate the trajectories is available in the repo \cite{elena2023}.

\subsubsection*{Similarity statement}

For the system MOR-\ch{Al} the training set contains \num{145} structures that have \ch{Si}, \ch{O}, \ch{Al}, and \ch{H} elements on their own or along with other elements. Based on UMAP analysis, we see almost all atomic environments in the example system are similar to environments in the training data. Similar findings hold true for MFI-\ch{Al} with \num{145} structures matches.
The structures with adsorbates inside have very low structural matches, for examples, acetone in MOR-\ch{Al}, matches only three structures, and formamide only one, and none very close to the studied zeolites. Adsorbants on their own match 1029 structures for formamide, \num{1892} for acetone and \num{3139} for \ch{NH3}. If we consider only \ch{Si}, \ch{O}, \ch{Al}, and \ch{H} elements we have only \num{11} similar structures for both zeolites considered and none is an exact match but they offer good representability of the local environments. 

The closest (most relevant) structures in the training set are \ch{CO2} (mp-556034, mp-20066, mp-995224, mp-11725, mp-644607, mp-1102227, mp-1190685, mp-995198, mp-1190699, mp-1077906, mp-1077316, mp-729728). \ch{CO2} alone matches \num{4896} structures with \ch{C}, \ch{O} and alongside other elements.

We provide 
\begin{itemize}
\item \verb|MOR-Al_FilterType.exclusive_SiOAlH_chemiscope_input.json|
\item \verb|MFI-Al_FilterType.exclusive_SiOAlH_chemiscope_input.json|
\item \verb|MFI-Al-H2O_FilterType.exclusive_SiAlOH_chemiscope_input.json|
\item \verb|MFI-Al-H2O-cyclohexene_FilterType.exclusive_SiAlOCH_chemiscope_input.json| 
\item \verb|MFI-Al-H2O_FilterType.exclusive_SiAlOH_chemiscope_input.json| 
\item \verb|MFI-Al-cyclohexene_FilterType.exclusive_SiAlOCH_chemiscope_input.json|
\item \verb|MOR-Al_FilterType.inclusive_SiOAlH_chemiscope_input.json| 
\item \verb|MFI-Al_FilterType.inclusive_SiOAlH_chemiscope_input.json|
\end{itemize}
to help visualize the interactive UMAP on \url{chemiscope.org}.

\subsubsection*{Performance summary}
Correct prediction of binding sites, and qualitatively correct reaction behaviour for a range of structures and ligands, including good agreement of predicted reaction barrier with DFT.

\clearpage
\subsection{Open-circuit voltage of lithiated graphite}\label{sec:LiC}
\begin{figure*}[htbp!]
  \includegraphics[width=\linewidth,keepaspectratio]{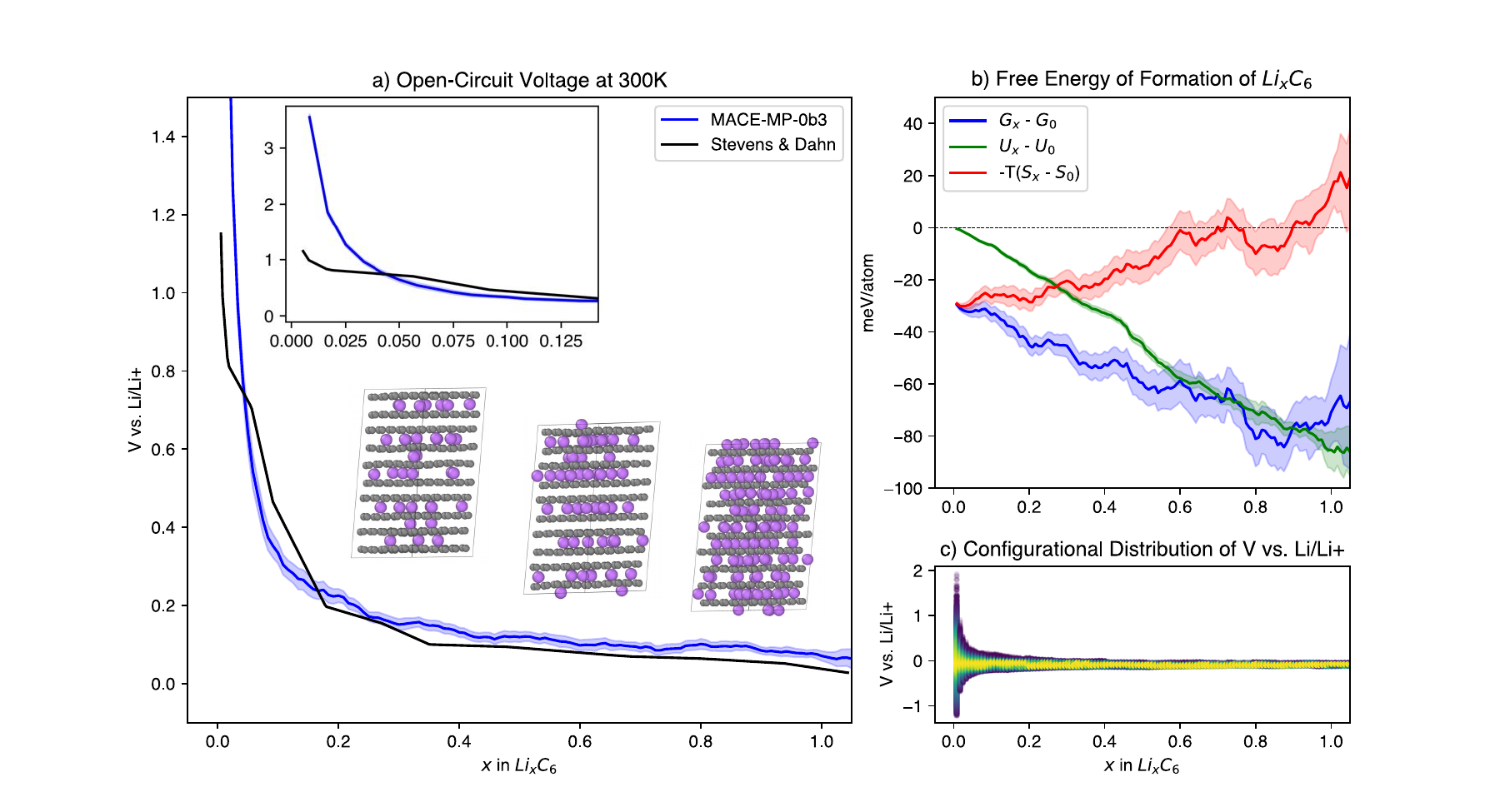}
  \caption{(a) The open-circuit voltage profile of lithium in graphite versus lithium metal, computed with \MPzbt{} (blue) using a hybrid Grand Canonical Monte Carlo(GCMC)/Molecular Dynamics protocol, contrasted with an experimental reference (black) \cite{stevensdahn2001}. The variance of the simulated voltage is estimated over 100 bootstrapped samples of GCMC/MD trajectories. Representative lithium-graphite configurations are shown at $x = 0.3, 0.5, 1.0$. (b) The contributions to the free energy of formation of sampled lithium-graphite phases. (c) The density plot of individual Li/graphite structures sampled during a GCMC/MD simulation, showing the distribution of potential values over the configurational ensemble.}
  \label{fig:OCV}
\end{figure*}

The open-circuit voltage profile of an electrode material is an example of a technologically relevant macroscopic observable that can be accessed through atomistic simulation. We apply \MPzbt{} to model the thermodynamics of lithium-ion intercalation in graphite using hybrid grand canonical Monte Carlo/molecular dynamics (GCMC/MD).

Beginning from a \num{720}-atom cell of pristine graphite containing 10 graphene layers, we generate 40 parallel simulation trajectories of 30,000 steps each at a system temperature of \SI{300}{\kelvin}. In our GCMC/MD protocol, at every simulation step, we update the ionic positions according to Verlet dynamics. Every 5 steps, we generate a Monte Carlo proposal on the system volume, followed by a proposal on the system’s composition. For the volume proposal, we sample a perturbation of the unit cell: this is a set of 3 Euclidean vectors sampled component-wise from a normal distribution with a mean of zero and variance of \SI{0.01}{\Angstrom}. We add these random vectors to the existing lattice vectors and rescale the atomic positions to generate the proposed unit cell update. For the composition proposal, we randomly make one of three modifications to the population of lithium ions: insertion, deletion, or swapping. We choose one of these types of modification at random and then generate 5 candidate structures, each with either a single new lithium atom placed in a void in the host lattice (insertion), an existing lithium atom displaced into a void (swapping), or an existing lithium atom removed (deletion). As the lattice in our simulations is \textit{not} fixed, but evolves under molecular dynamics, we use Voronoi triangulation to identify void sites in the atomic lattice, excluding all sites subject to steric overlap according to the atomic radii. Once the set of composition candidates is generated, their energy is evaluated, and the lowest-energy candidate is used as the composition proposal. If a composition proposal is accepted, before proceeding to the next simulation step, we relax the ionic positions and the unit cell for up to ten ionic steps with a force tolerance of \SI{0.05}{eV\per\Angstrom} using the FIRE algorithm.

After sampling configurations with this protocol, we compute the open-circuit voltage as a function of lithium concentration over the sampled ensemble. Following previous work \cite{huang2019anode}, the open-circuit voltage is estimated as the negative of the free energy of formation per atom of the phase with composition \ch{Li_{x}C6} from reference states of graphite (\ch{C6}) and metallic BCC lithium, divided by the lithium concentration: $V(x) = - \Delta G_{f,Li_{x}C_{6}} / x$. The free energy per atom of the metallic lithium reference state is taken as the potential energy predicted by \MPz{} of BCC lithium after structural optimization, neglecting entropy. To determine the free energies per atom of the \ch{C6} and \ch{Li_{x}C6} phases, we compute the internal energy \( U \) and Gibbs entropy \( S \) as Boltzmann averages over the sample distribution at concentration $x$: $G(x) = U(x) - TS(x)$, $U(x) = \sum_{j}(E_j \cdot p_j)$, $S(x) = - k_B \cdot \sum_{j}(p_j \cdot \ln(p_j))$, with probabilities 
$p_j = e^{-\frac{E_j}{k_B T}} / {\sum_{j} e^{-\frac{E_j}{k_B T}}}$, where $E_j$ is the potential energy per atom predicted by \MPz{} and $j$ indexes the set of all simulation frames with composition \ch{Li_{x}C6}. 

\MPz{} reproduces the experimentally known voltage profile of \ch{Li}/graphite with good quantitative accuracy (\cref{fig:OCV}a). In the regime of $x$ \num{ > 0.04}, the error is \SI{< 0.1}{V}, which reflects the combined error of the model as well as the limitations of the GCMC/MD protocol. This may be compared favorably with a recent purpose-developed model for lithium-graphite energetics \cite{babar2021graphite}, which reported \SI{< 0.1}{V} error for $x$\num{ > 0.0833} in an open-circuit voltage profile produced through GCMC; that model was trained on more than 8,000 system-specific DFT calculations, while \MPz{} obtains comparable accuracy zero-shot. We note that at very low concentrations, our predicted voltage is higher than the experimental voltage by as much as a factor of 3, indicating overstabilization of dilute lithium. Since the lithium fraction appears in the denominator of the expression for the open-circuit voltage, very slight energetic deviations are magnified in this range; moreover, the free energy at low concentrations is dominated by the entropic contribution (\cref{fig:OCV}b), for which the limited sample size introduces uncertainty. Beyond this lowest-concentration regime, \MPz{} provides good agreement with experiment.

\subsubsection*{Similarity statement}

There is a skew towards battery materials in MP. Given this, there are several \ce{Li-C} structures that are relevant to this application: mp-1210743 (\ch{Li2C}), mp-976060 (\ch{Li3C}), mp-1223102 (\ch{Li7C120}), mp-1378 (\ch{LiC}), mp-1021323 (\ch{LiC12}), mp-1232339 (\ch{LiC12}), mp-1001581 (\ch{LiC6}). There also exist 62 pure carbon structures including graphite (mp-48).

\subsubsection*{Performance summary}
Correct prediction of voltage as a function of Li concentration with reference to experimental curve. 

\clearpage
\subsection{Jahn-Teller Distortions in \ch{LiNiO2}}\label{sec:LNO}

\begin{figure*}[htbp!]
  \centering
  \includegraphics[width=\textwidth,keepaspectratio]{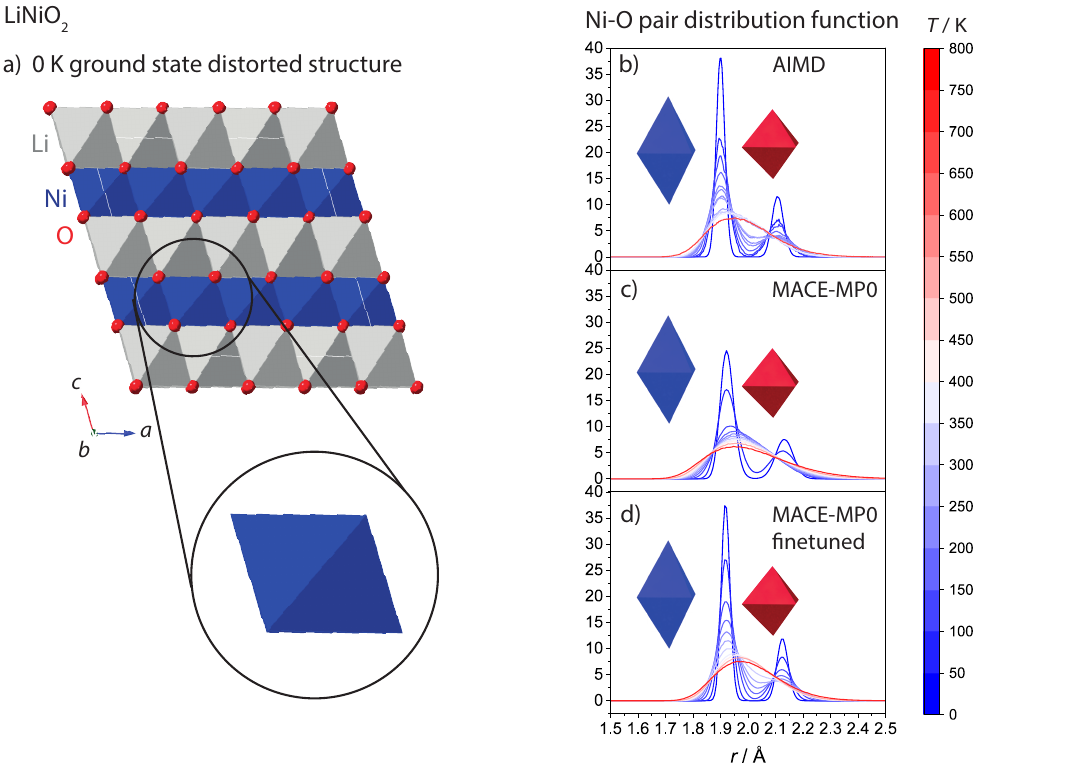}
  \caption{a) Ground state structure of LNO. Magnified insert shows a \ch{NiO6} octahedron with Jahn-Teller distortion producing long and short axes.
  b) \ch{Ni-O} pair distribution functions (PDF) as a function of ion separation computed by AIMD.\cite{GenreithSchrieverChemrxiv} Line colour corresponds to trajectory temperature as shown in the colorbar. c) PDFs computed by \MPzbt{}. Inserts indicate shapes of \ch{NiO6} octahedra (blue: distorted, red: undistorted). d) PDFs computed by fine-tuned MACE model as described in text. The same temperature values are used in panels c) and d), slightly fewer values are used in AIMD but spanning a similar range of temperature.
  }
  \label{fig:LNO_PDFs}
\end{figure*}

\ch{LiNiO2} (LNO) is an important material in lithium-ion battery research, serving as a model for future high-performance cathode materials with reduced or no cobalt content. Structural and chemical degradation of these materials is a key challenge, so understanding their thermodynamic and dynamic properties is critical. In particular, the presence of ordered regions of Jahn-Teller distorted \ch{NiO6} octahedra may influence electron and Li-ion transport. Recent {\it ab initio} molecular dynamics (AIMD) and variable-temperature X-ray diffraction (VT-XRD) measurements have suggested a phase transition from a low-temperature phase with Jahn-Teller distortions to a dynamic phase without permanent Jahn-Teller distortions at high temperature\cite{GenreithSchrieverChemrxiv}. Above this displacive phase transition the material does not exhibit Jahn-Teller distortions either in the time-averaged bulk structure or in instantaneous snapshots of the local structure. A similar displacive transition was also observed in the related sodium-ion battery cathode material \ch{NaNiO2}.\cite{NagleCocco24}

We have investigated the ability of \MPz{} to reproduce the temperature dependence of the structural distortions in LNO.
We used a 256-atom supercell, initialized from the established DFT ground state structure which has distorted \ch{NiO6} octahedra with zigzag long-ranged ordering of their long axes (see fig.~\ref{fig:LNO_PDFs}a). We performed NPT-ensemble dynamics using \MPz{} without a D3 correction, in line with the earlier AIMD simulations at the PBE+{\it U} level of theory. 

Trajectories were propagated for approximately \SI{250}{ps} at 13 temperature values from \SI{25}{\kelvin} to \SI{1000}{\kelvin}. These trajectories were used to analyse the phase transition behaviour (fig.~\ref{fig:LNO_PDFs}c), and a sample of 104 configurations was used to fine-tune a new model following the protocol outlined in sec.~\ref{sec:finetuningprotocol}. Specifically, we extracted configurations from each trajectory at \SI{20}{ps} intervals, after discarding the first \SI{100}{\ps} for equilibration. The performance of the fine-tuned model is compared with that of the original \MPz{} in the next section.

\subsubsection*{Performance}
The \MPz{} model demonstrates stable dynamics during a heating trajectory from \SI{0}{\kelvin} to \SI{1000}{\kelvin} with temperature increments of \SI{10}{\kelvin\per\pico\second}, time step \SI{1}{\femto\second}. Long simulations (at least \SI{500}{\pico\second}) could be run at \SI{1000}{\kelvin} without significant energy drift or noticeably unphysical behavior.

The Ni-O pair distribution functions predicted by both AIMD and \MPz{} (Fig.~\ref{fig:LNO_PDFs}b and c) show two peaks at low temperature corresponding to the short and long Ni-O bonds in Jahn-Teller distorted octahedra. At higher temperatures one broad asymmetric peak appears, which is characteristic of octahedra in a displacive high-temperature phase.\cite{GenreithSchrieverChemrxiv} \MPz{} thus  correctly predicts a phase transition from Jahn-Teller-distorted to undistorted octahedra on heating. However, it significantly underestimates the temperature at which the transition occurs. We diagnose the onset of the transition at the lowest temperature where the PDF has only one maximum and the second peak (at longer bond lengths) becomes a shoulder on the short-bond peak. Above this temperature the trajectories show dynamic Jahn-Teller distortions, i.e.~reorientations of the long O-Ni-O axes over time. This onset occurs between 50 and \SI{100}{\kelvin} in the \MPz{} simulations, compared to ca.~\SI{250}{\kelvin} with AIMD.\cite{GenreithSchrieverChemrxiv}  The transition is complete when the Ni-O PDF exhibits only one broad asymmetric peak without a shoulder. \MPz{} predicts that this completion temperature is below \SI{150}{\kelvin}, compared with \SI{350}{\kelvin} in AIMD.

By contrast, the fine-tuned model predicts an onset temperature between 250 and \SI{300}{\kelvin}, and the transition to complete by \SI{350}{\kelvin}, both in excellent agreement with the AIMD results.
This improvement suggests that the deficiency of the original model results from undersampling of important configurations in the training data.

\subsubsection*{Similarity statement}

Battery materials are well represented in the MP database. 1393 structures in the training set contain \ch{Li}, \ch{Ni}, or \ch{O} atoms, 143 contain all three elements, and 23 have the exact formula \ch{LiNiO2} representing different crystal structures. However, many of these structures are obtained from geometry optimization and hence neglect the temperature dependence of the equilibrium geometry.
The substantial improvement of \MPz{} performance for this system after fine-tuning on variable-temperature trajectories suggests that incorporating these temperature-dependent configurations is necessary to describe structural properties of the material.

\subsubsection*{Performance summary}
\MPz{} correctly captures the loss of Jahn-Teller distortions on heating, with an underestimation of the transition temperature compared with AIMD that is largely corrected with a small amount of fine-tuning.

\clearpage
\subsection{Point and extended defects in BCC metals}\label{sec:bcc_defects}

This test explores bulk and extended defect properties of three prototypical BCC metals: W, Mo and Nb. An accurate description of these properties is essential to enable predictive modelling of mechanical responses to applied loads such as dislocation glide\cite{rodney2017ab}, dislocation climb through interaction with point defects, grain boundary motion, and the competition between cleavage and dislocation emission that underpins the brittle to ductile transition in fracture.
Across-the-board accuracy for bulk and defect properties in these systems is challenging even for bespoke machine learning potentials fit to carefully curated datasets~\cite{Goryaeva2021}.

\begin{table}[h!]
    \centering
    \begin{tabular}{lrrr}
    \hline
    &     DFT &  \MPzbt &  \MPzbt-FT \\
    \hline
    W       &  3.185 &   3.203 & 3.191 \\ 
    Mo      &  3.163 &   3.182 & -- \\
    Nb      &  3.322 &   3.321 & -- \\
    \hline
    \end{tabular}
    \caption{Lattice constants (in units of \AA{}) for W, Mo and Nb computed by variable-cell minimisation. DFT reference data is from Ref.~\cite{Hiremath2022} for W, Ref.~\cite{Smirnova2020} for Mo and Ref.~\cite{Yang2019_Nb} for Nb.}
    \label{tab:BCC-lattice}
\end{table}

\begin{figure}[h!]
  \centering
  \includegraphics[width=0.95\textwidth,keepaspectratio]{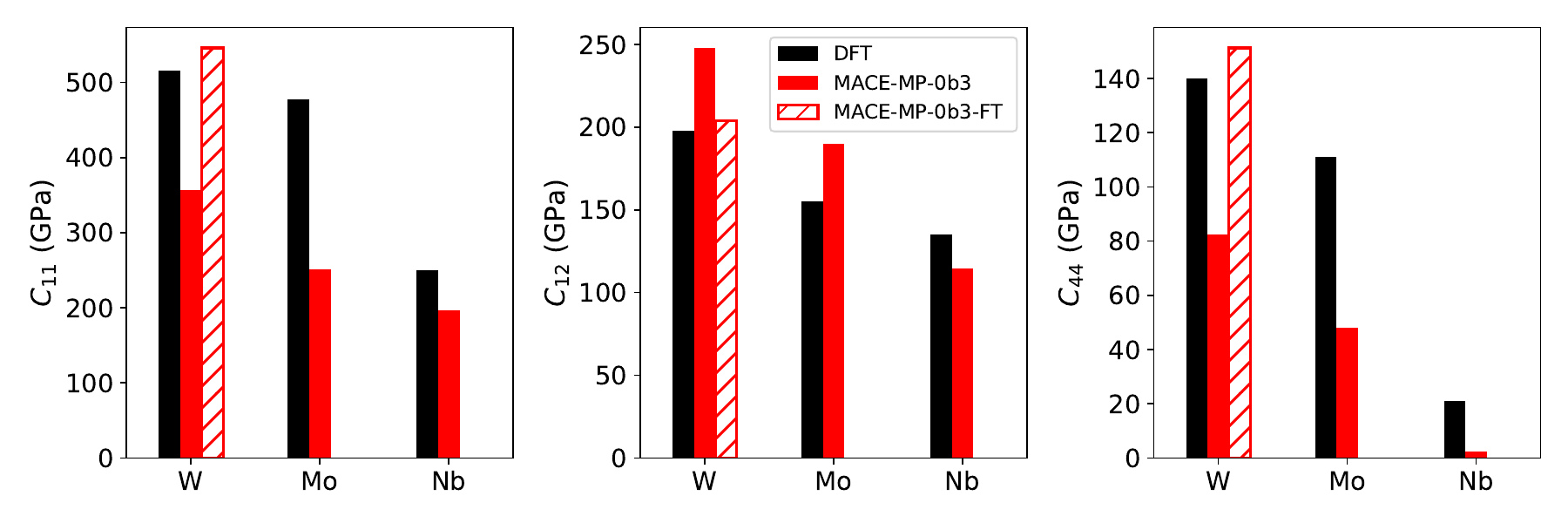}
  \caption{Cubic elastic constants for W, Mo and Nb computed with linear stress-strain fits for DFT (black; from Ref.~\cite{Hiremath2022} for W, Ref.~\cite{Smirnova2020} for Mo and Ref.~\cite{Yang2019_Nb} for Nb), the \MPzbt{} (solid red). \MPzbt-FT model finetuned on W data is shown with hatched red bars.}
  \label{fig:BCC-elastic}
\end{figure}

\begin{figure}[h!]
  \centering
  \includegraphics[width=0.95\textwidth,keepaspectratio]{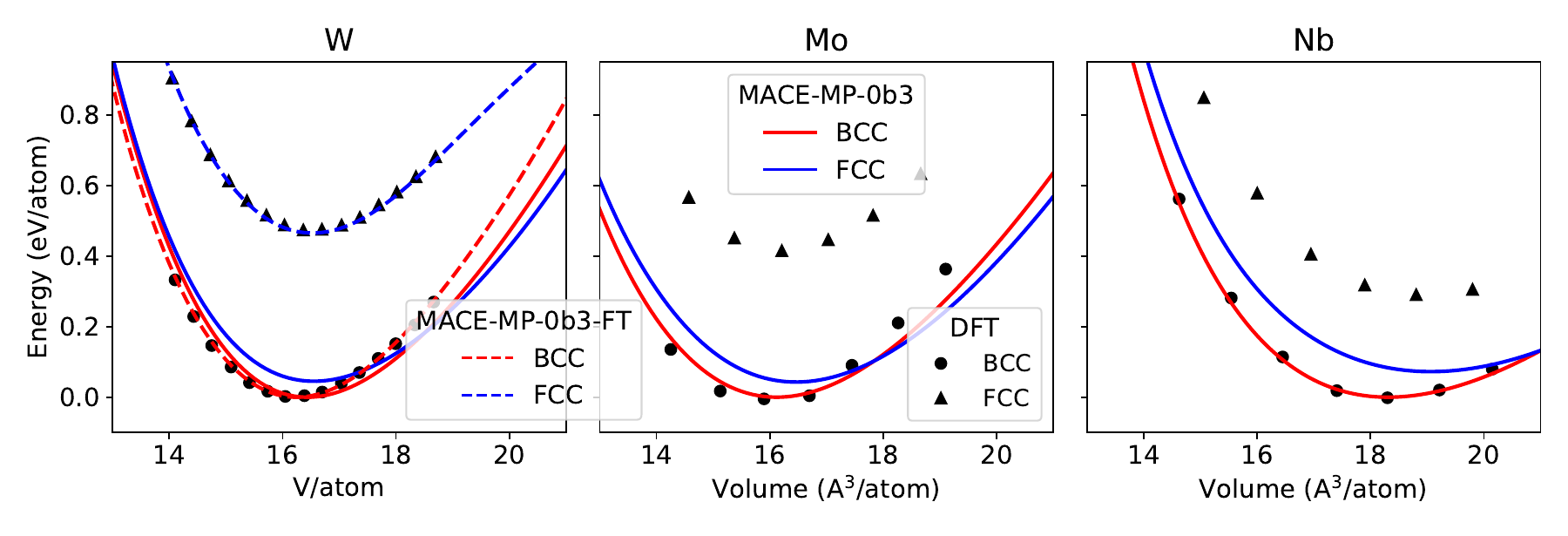}
  \caption{Energy vs volume curves for W, Mo and Nb computed with the medium \MPzbt{} model (solid lines) FCC (red) and BCC (blue) crystals. \MPzbt{}-FT model finetuned on W data is shown with dashed lines. The DFT reference data shown with the black circular (BCC) and triangular (FCC) points is from Ref.~\cite{Cak2014}.}
  \label{fig:BCC-eos}
\end{figure}

Lattice and elastic constants are shown in Table~\ref{tab:BCC-lattice} and Fig.~\ref{fig:BCC-elastic}, respectively. Lattice constants are generally well reproduced, but there is a general softening of the elastic response.
In Fig.~\ref{fig:BCC-eos} we compare the energy-volume (E-V) curves predicted by the \MPzbt{} model for BCC and FCC phases of the three metals with reference DFT data from Ref.~\cite{Cak2014}. The BCC cases show generally good agreement. There is room for improvement in the curvature of the E-V curves (critical for the elastic properties) for W and Mo, while Nb is well described.  FCC energies are underestimated while the curvature is approximately correct giving reasonable predictions of the elastic response.

\begin{figure}[h!]
  \centering
  \includegraphics[width=0.9\textwidth,keepaspectratio]{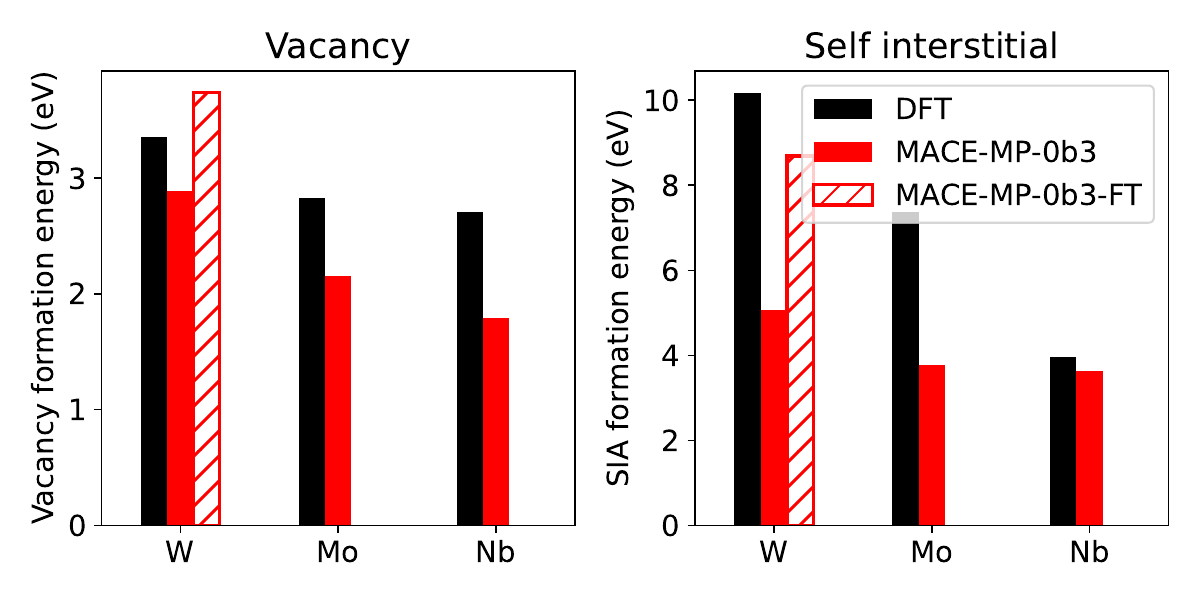}
  \caption{Vacancy and SIA formation energies for W, Mo and Nb computed with the \MPzbt{} model (solid red), W finetuned \MPzbt{}-FT model (hatched red)  and DFT reference data from Ref.~\cite{Ma2019} (black).}
  \label{fig:BCC-point-defects}
\end{figure}

We next investigated point defect formation energies, including vacancies and self-interstitial atoms (SIAs). Calculations were performed in a $5\times5\times5$ supercell and were relaxed to a force tolerance of \SI{1.e-5}{eV\per\Angstrom}. For the SIAs, a short MD run was performed to escape the initial metastable configuration. The results, illustrated in Fig.~\ref{fig:BCC-point-defects} show good agreement with reference DFT data is from~\cite{Ma2019}: vacancy energies are predicted within ca. 20\% of the DFT value and SIA energies within 50\%. For all three elements the \MPzbt{} predicts that the $\langle111\rangle$ dumbbell is the most stable SIA configuration, in agreement with DFT.

\begin{figure}[h!]
  \centering
  \includegraphics[width=1.0\textwidth,keepaspectratio]{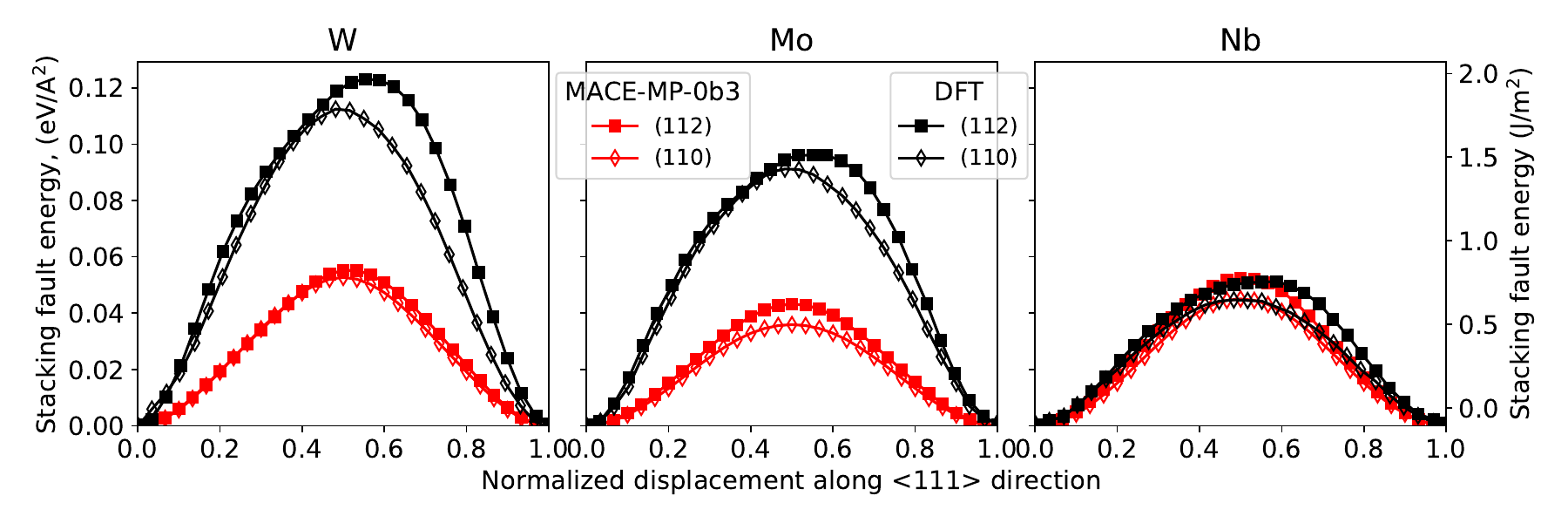}
  \caption{Generalised stacking fault profiles for $(112)[111]$ and $(110)[111]$ $\Gamma$-surfaces predicted by the \MPzbt{} model, shown in red. The DFT reference data, shown in black, is from Ref.~\cite{Starikov2024}.}
  \label{fig:BCC-gamma}
\end{figure}

We next looked at generalised stacking fault energy profiles for the $(112)$ and $(110)$ $\Gamma$-surfaces along the $[111]$ direction as shown in Fig.~\ref{fig:BCC-gamma}. These results were obtained with constrained minimisation where atoms were allowed to move only in the direction perpendicular to the cut surface and with a force tolerance of \SI{1.e-3}{eV\per\Angstrom}. The details of the method are explained in Ref.~\cite{Ventelon2010}. Nb is very well described by the \MPzbt{} model, while there is an underestimate in the stacking fault energies in W and Mo by around a factor of two that explains the underestimates in the screw dislocation glide barriers discussed below.

Dislocations in BCC materials lie predominantly in the 
$\langle111\rangle\{110\}$ and $\langle100\rangle\{010\}$ slip systems.
We investigate the characteristics of $\langle111\rangle$ screw and $\langle100\rangle$ edge dislocations by calculating the transition pathways and Peierls barriers using the nudged elastic band (NEB) method and the \MPzbt{} potential
comparing against DFT~\cite{Dezerald2014} and hybrid
QM/MM calculations~\cite{grigorev2023calculation,Swinburne2017} where the data is available.

The cells contained $\approx$1400 atoms for the [111] screw dislocation and $\approx$2200 for the [100] edge dislocation. Geometry optimisations to obtain starting configurations used the FIRE algorithm with a force tolerance of \SI{1e-6}{eV\per\Angstrom}. To create and analyze atomistic dislocation configurations we employed the \texttt{matscipy.dislocation} module~\cite{grigorev2023matscipy}.
The transition path calculation is performed with an adaptive ODE solver~\cite{Makri2019} following the approach of Refs.~\cite{grigorev2020hybrid, grigorev2023calculation}, using fifteen intermediate images with stopping force tolerance of \SI{0.025}{eV\per\Angstrom}.
Starting positions for the NEB relaxation were obtained by linear interpolation between initial and final configurations.

\begin{figure}[h!]
  \centering
  \includegraphics[width=1.0\textwidth,keepaspectratio]{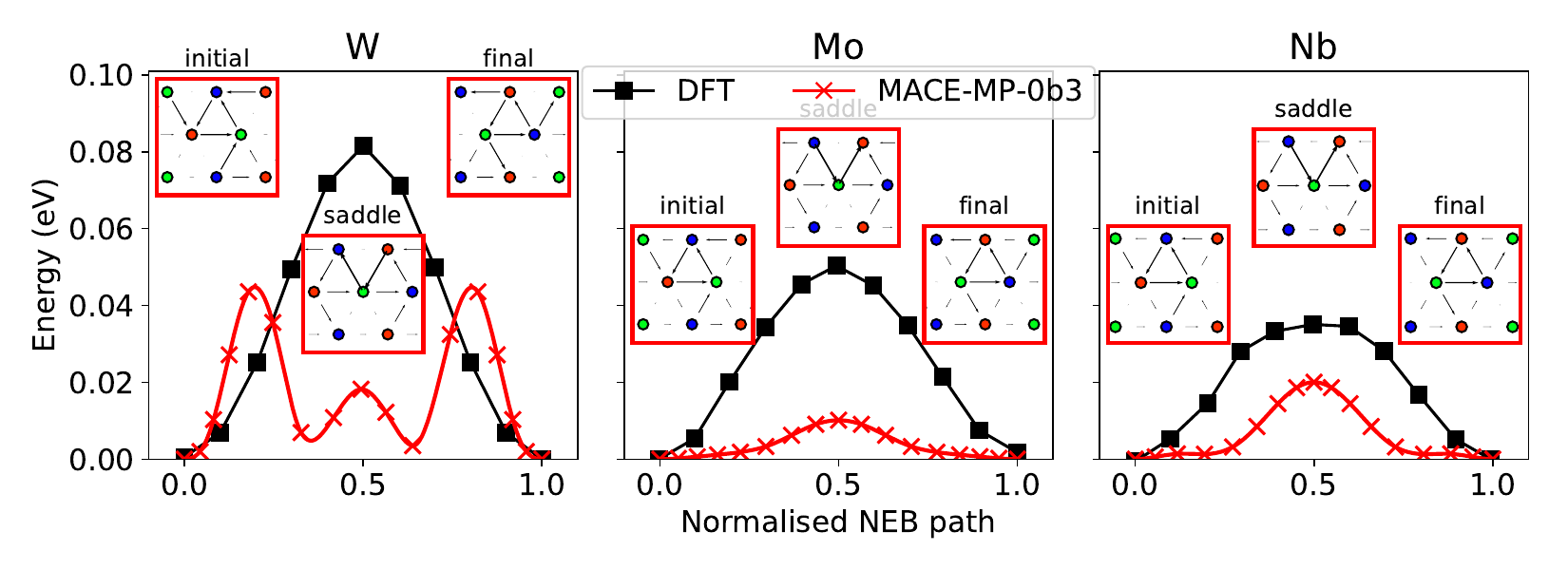}
  \caption{Screw dislocation glide barriers for W, Mo and Nb. DFT data from~\cite{Dezerald2014}}
  \label{fig:BCC-screw-glide}
\end{figure}

\begin{figure}[h!]
  \centering
  \includegraphics[width=1.0\textwidth,keepaspectratio]{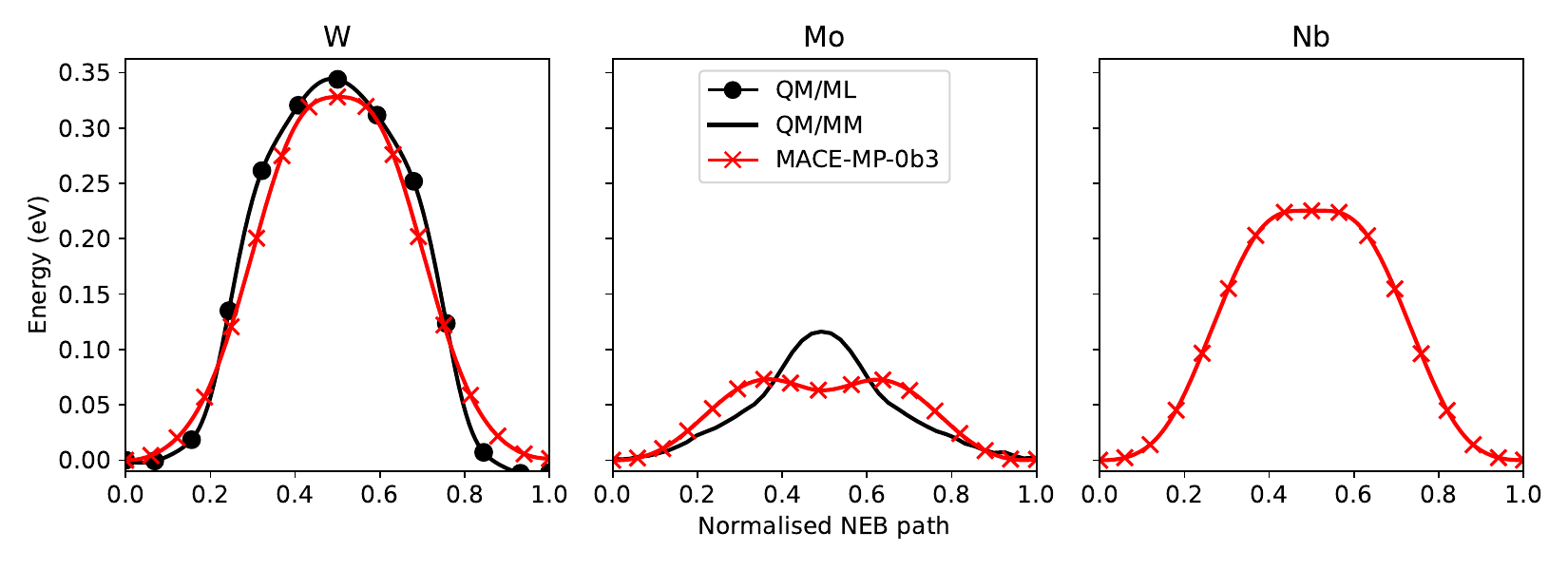}
  \caption{Edge $\langle100\rangle$ dislocation glide barriers for W, Mo and Nb. QM/ML data for W from Ref.~\cite{grigorev2023calculation}, QM/MM data for Mo from Ref.~\cite{Swinburne2017} }
  \label{fig:BCC-edge-glide}
\end{figure}

Figs.~\ref{fig:BCC-screw-glide} and \ref{fig:BCC-edge-glide} illustrate the NEB minimum energy path depicting the Peierls barriers for $\langle111\rangle$ screw and $\langle100\rangle$ edge dislocations in W, Mo and Nb. We compare to DFT results where they are available.
Insets within Fig.~\ref{fig:BCC-screw-glide} illustrate the dislocation core structures at the initial, intermediate, and final positions along the \MPzbt{} minimum energy path.
Screw dislocations are known to be a sensitive probe of potentials, since the accuracy required is on the meV/atom level.
DFT predicts dislocations to move between highly symmetric easy core configuration via split core configuration at the saddle point~\cite{Dezerald2014}. For all three metals the degenerate core is incorrectly predicted to be the most stable configuration by \MPzbt{} model. At the same time, split core configuration at the saddle point is correctly predicted for Mo and Nb, while for W the single hump shape of the glide barrier is not reproduced with some intermediate configurations having energy, close to the degenerate core.
The barrier height is underestimated for all three metals compared to reference DFT results~\cite{Dezerald2014}. For the edge dislocation, where energy differences are larger, we find that the barrier height aligns well with QM/ML results in W~\cite{grigorev2023calculation} and Mo~\cite{Swinburne2017}. However, the presence of a minimum along the transition path for Mo results in a spurious stable dislocation configuration. 

We anticipate the performance of \MPzbt{} for all properties considered here would be substantially improved by enhanced accuracy in stress and a more precise agreement on the elastic constants, followed by fine-tuning on defect configurations where necessary.

\subsubsection*{Fine-tuning}

Fine-tuning was performed on 142 W configurations containing isolated atom, deformed perfect BCC and FCC structures, vacancy in BBC and Molecular Dynamics snapshots for BCC. This is sufficient to correct softening of BCC and the over stabilty of FCC as seen in the updated elastic constants and equation of state plots in Fig.~\ref{fig:BCC-elastic} and Fig.~\ref{fig:BCC-eos}, and also improves the agreement of the lattice parameter and the vacancy and SIA formation energies as seen in Table~\ref{tab:BCC-lattice} and Fig.~\ref{fig:BCC-point-defects}. The improvement in the SIA formation energy demonstrates transferability as there are no interstitial configurations included in the fine-tuning dataset.

\subsubsection*{Similarity statement}
The MP dataset includes 7 elemental tungsten, 7 elemental molybdenum, and 4 elemental niobium structures. They are all crystalline without any defects. Based on UMAP analysis, we find that the closest structures in the training set are mp-8641 for tungsten, mp-8637 for molybdenum and mp-8636 for niobium. We provide \verb|W_input.json|, \verb|Mo_input.json| and \verb|Nb_input.json| to help visualize the interactive UMAP on \url{chemiscope.org}.

\subsubsection*{Performance summary}

Energy-volume curves for BCC are well reproduced, while for FCC structures the \MPzbt{} model shows a ca.~0.5 eV/atom shift in energy; this is corrected by fine-tuning. Stacking fault profile energies for Nb are well reproduced, while for W and Mo they are underestimated by a factor of around two with respect to DFT. Relaxed point defect structures are reasonable in all cases, with formation energies within 50\% of reference DFT values (again, improved by fine tuning). Peierls energy barrier profiles for dislocation glide are qualitatively correct for the edge dislocations and underestimated for the screw dislocation together with incorrect screw dislocation core stability. There is a small spurious local minimum near the top of the barrier for the edge dislocation in Mo.

\clearpage
\subsection{Alumina defects and bulk diffusion}\label{sec:alumina}

\begin{figure*}[htbp!]
  \centering
  \includegraphics[width=\linewidth,keepaspectratio]{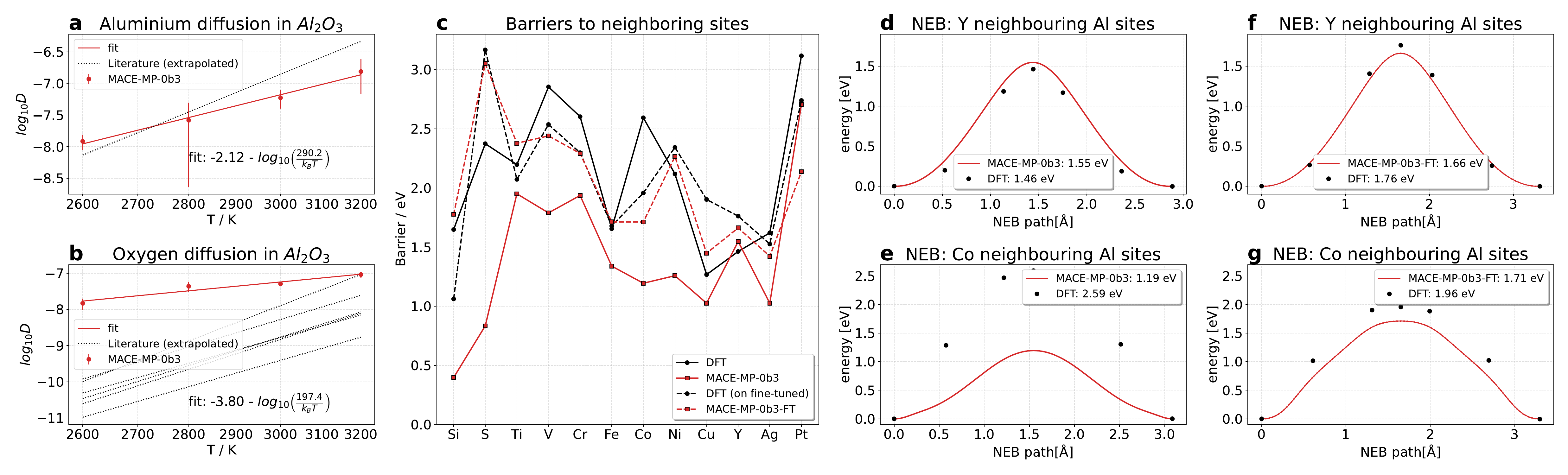}
  \caption{
    (\textbf{a}--\textbf{b}) Arrhenius plot of elemental diffusion in \ch{Al2O3} compared with experimental results from \cite{diffusion_in_alumina_2006};
    (\textbf{c}) Comparison of \MPzbt{} NEB barrier paths (red) and PBE single point evaluation of the \MPzbt{} transition state (black) for elements in \ch{Al} or \ch{O} sites in \ch{Al2O3} moving to neighboring sites (connecting lines are guides to the eye), same protocol repeated with \MPzbt{}-FT as well shown with dashed lines;
    (\textbf{d--e}) lowest energy NEB path for \ch{Y} (where \MPzbt{} is accurate) and \ch{Co} (where there is a substantial discrepancy), with single-point PBE evaluations on selected images of the obtained NEB  path, indicated by black points;
    (\textbf{f--g}) \ch{Y} and \ch{Co} NEBs with the fine-tuned \MPzbt{}-FT model and the corresponding single point PBE evaluations.
  }
  \label{fig:al2o3}
\end{figure*}

\subsubsection{Bulk diffusion}
One \ch{Al} and one \ch{O} vacancy were introduced into a 270-atom alumina supercell, and over \SI{2}{ns} (at \SI{2.5}{fs}) MD was used at temperatures between \numrange{2600}{3200}~\unit{\kelvin} to measure diffusivities of the two elements.
Diffusivities agree within one order of magnitude for \ch{Al}, and activation energies are underestimated for both compared to extrapolated experimental values~\cite{diffusion_in_alumina_2006}.
This demonstrates the long-timescale stability of the model, even at high temperatures and for long MD trajectories, but highlighting a shortcoming of the model for quantitative prediction of macroscopic observables.

\subsubsection{Elemental defects}
\label{sec:alumina-elemental}

Elemental defects in \ch{Al2O3} were investigated by substituting \ch{Si}, \ch{S}, \ch{Ti}, \ch{V}, \ch{Cr}, \ch{Fe}, \ch{Co}, \ch{Ni}, \ch{Cu}, \ch{Y}, \ch{Ag}, and \ch{Pt} into lattice sites in a $2 \times 2 \times 1$ supercell (120 atom) and minimal energy paths to neighbouring sites were obtained using NEB\cite{neb_aseneb, neb_jk_optimiser} starting from a linear interpolation. Paths were converged (max 100 steps, \SI{0.3}{eV\per\Angstrom} tolerance on projected forces) and the lowest energy one was tested with PBE single point evaluations using CASTEP \cite{clark2005first}.
Comparing \MPzbt{} and PBE on \cref{fig:al2o3}c--e there are large discrepancies, with a total force component RMSE of \SI{0.35}{eV\per\Angstrom} across 152 structures evaluated.

\subsubsection*{Finetuning}

Fine-tuning was performed on 72 configurations, which included substitutions of each element and snapshots from paths found with \MPzbt{}.
The barrier energies computed with the finetuned model are shown on \cref{fig:al2o3}c, showing improved agreement with the PBE reference in barrier energies and a lower \SI{0.21}{eV\per\Angstrom} force component RMSE on 149 PBE evaluations.
Notably, the \MPzbt{} model found a lower energy path for \ch{S}, \ch{Ni}, \ch{Cu}, and \ch{Y} than the finetuned model.

\subsubsection*{Similarity statement}

There are 109 structures in the MP dataset containing exclusively \ch{Al} \& \ch{O}, pure \ch{Al2O3} appears as \verb#mp-1143# (used to generate supercells).
There is a total of 243 structures in the training set with \ch{Al} \& \ch{O} and exactly one of \ch{Si}, \ch{S}, \ch{Ti}, \ch{V}, \ch{Cr}, \ch{Fe}, \ch{Co}, \ch{Ni}, \ch{Cu}, \ch{Y}, \ch{Ag} or \ch{Pt}.

\subsubsection*{Performance summary}
Activation energies for self-diffusivity are underestimated compared to experimental values (extrapolated from lower temperatures). Dopant atom migration minimum energy paths are all stable, and are sometimes accurate (e.g., \ch{Y}) and sometimes only qualitative (e.g., \ch{Co}) with respect to DFT single-point reevaluations. Fine-tuning substantially improves agreement with DFT.

\clearpage
\subsection{Random structure search: Arsenic}\label{sec:rss}
\begin{figure*}[ht!]
  \centering
  \includegraphics[width=0.99\textwidth,keepaspectratio]{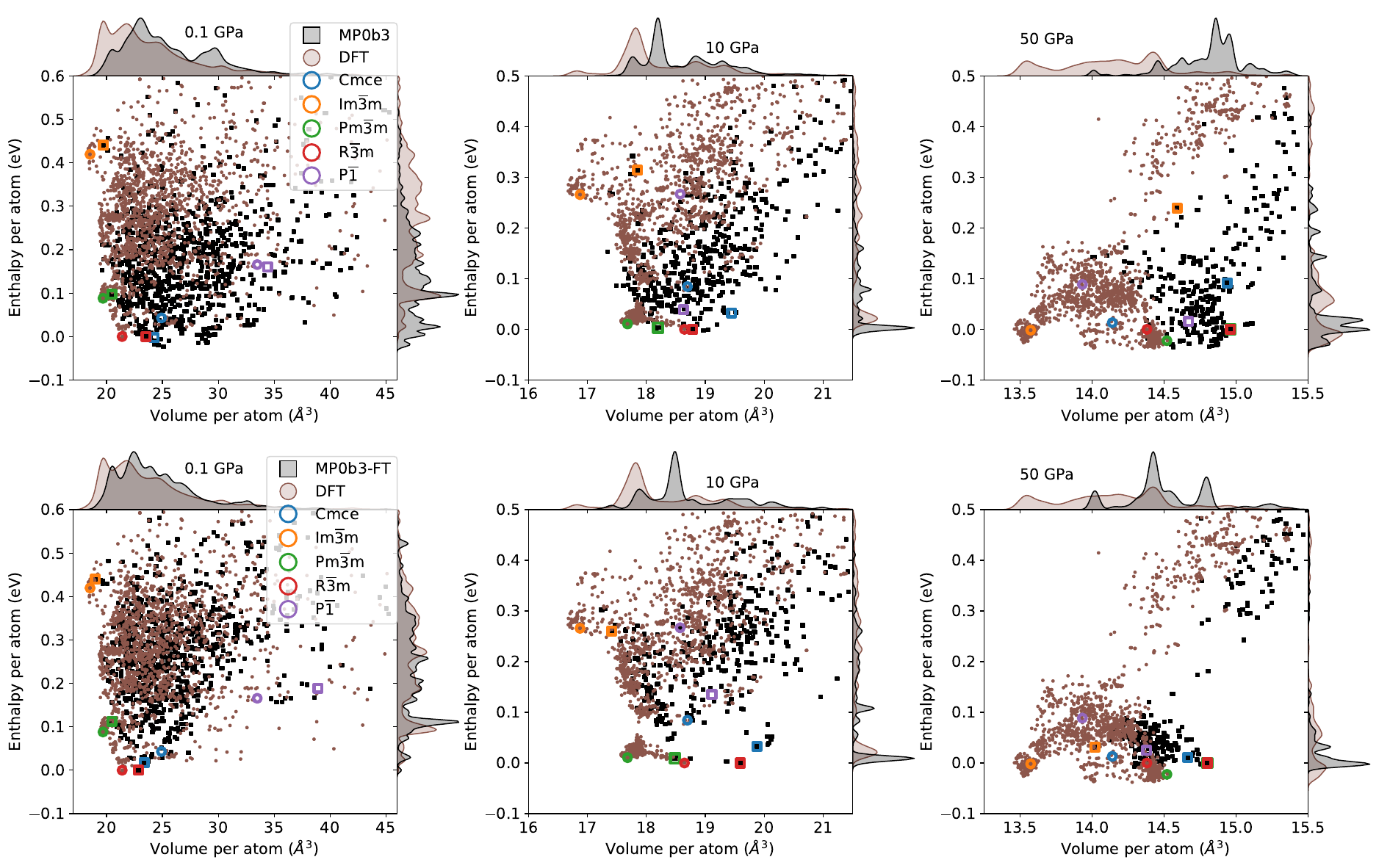}
  \caption{Densities of states of random structure search (RSS) minima for As at \num{0.1}, \num{10} and \SI{50}{\giga\pascal} obtained using DFT (brown, circles) and \MPzbt{} (black, squares) before  (top row) and after fine-tuning (bottom row). Known As structures are highlighted in various colors (see legend). }
  \label{fig:As_RSS}
\end{figure*}

\emph{Ab initio} Random Structure Searching (AIRSS) \cite {pickard2011ab} is a simple, yet highly successful, approach for discovering new materials computationally. Multiple candidate structures are generated randomly, subject to physically motivated constraints, and then relaxed to local enthalpy minima using ab-initio methods such as DFT. There is great interest in accelerating structure prediction by using surrogate models \cite{bernstein2019novo, podryabinkin2019accelerating, pickard2022ephemeral, merchant_scaling_2023}, such as ML potentials, to perform the initial structural relaxations. Here we test the suitability of \MPz{} (without the D3 correction) for this task by searching for structures of Arsenic at \num{0.1}, \num{10} and \SI{50}{\giga\pascal}. The exceptional structural variety encountered during RSS probes the robustness of the model in an extremely extrapolative regime; there are only six \ch{As} structures and no high-pressure data in the training set.

At each pressure, 2000 ($\times 100\, n$) random structures were generated using $n=2-6$ atoms per primitive unit cell, 2-4 randomly chosen symmetry operations, minimum distance constraints of \SI{2}{\Angstrom} and a volume per atom of \num{15}--\SI{40}{\cubic\Angstrom}. The structures were then relaxed with \MPz{} (ASE, force tolerance of \SI{1e-3}{eV\per\Angstrom}) and CASTEP \cite{clark2005first}; PBE exchange-correlation functional \cite{perdew1996generalized}, \SI{400}{eV} cutoff energy, k-point spacing of $2\pi\times$ \SI{0.05}{\per\Angstrom} and Vanderbilt ultrasoft pseudopotentials \cite{vanderbilt1990soft} with a force tolerance of \SI{0.05}{eV\per\Angstrom} and stress tolerance of \SI{0.1}{\giga\pascal}. With these settings 96\% and 99\% (at \num{0.1} and \SI{10}{\giga\pascal}) of structural relaxations were successful with DFT and \MPz{} respectively. The distributions of relaxed structures are depicted in \cref{fig:As_RSS} and the known structures listed in \cref{tab:As_RSS} are highlighted with colored symbols. 
The energy and volume distributions are visually similar with the relative energy differences between the highlighted structures, particularly the low-energy ones, generally being small compared to the overall range. Fine-tuning improves the agreement further and reduces the systematic shift in volume distribution.  Inspection of the structures at \SI{0.1}{\giga\pascal} reveals that similar 3-fold coordinated 3D, layered, and 1D structures are found with both \MPz{} and DFT. Furthermore, all known structures listed in \cref{tab:As_RSS} were found using \MPz{}, including a simplified packing of the \ch{As4} tetrahedra found in yellow As.

\begin{figure*}[ht!]
  \centering
  \includegraphics[width=0.99\textwidth,keepaspectratio]{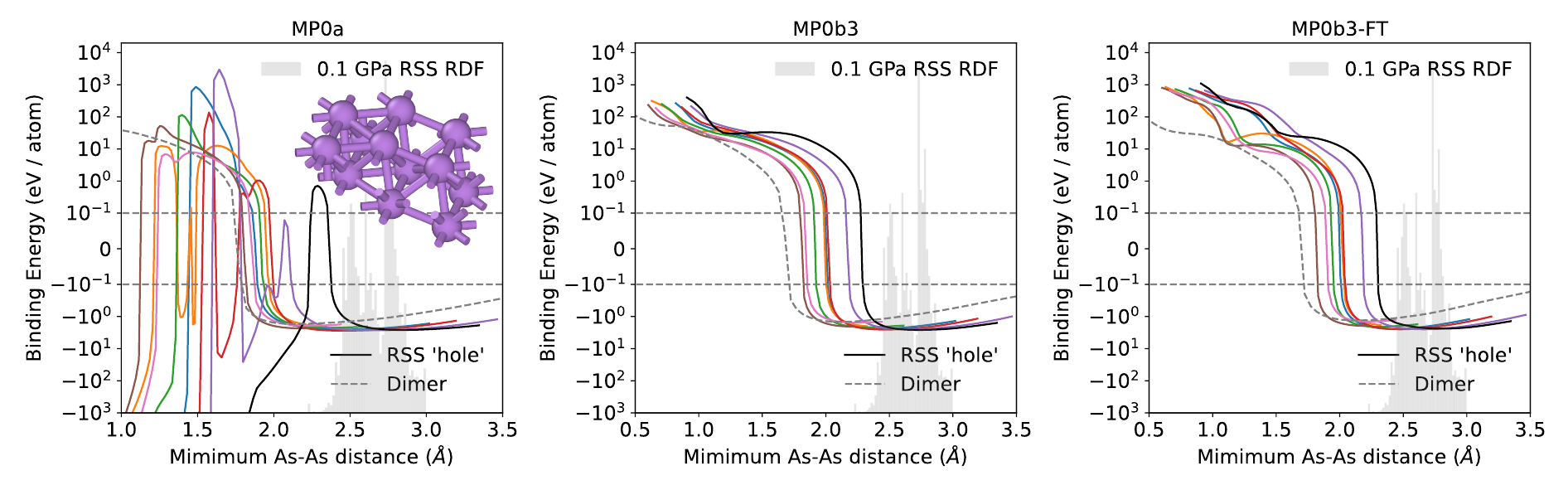}
  \caption{  The binding energy of 7 (unrelaxed) randomly generated structures (colored lines), and one pathological structure found during RSS with \MPz{}a, the original version of our model, (black line and inset), are shown as the structures are compressed uniformly with fixed fractional atomic coordinates. The \ce{As-As} dimer energy and the RDF for the \SI{0.1}{\giga Pa} \ch{As} RSS structures are shown for comparison. Left panel corresponds to the original \MPz{}a model, the middle panel to \MPzbt{}, while the right panel to the fine-tuned \MPzbt{}-FT model.}
  \label{fig:As_vscan}
\end{figure*}

The original \MPz{}a foundation model suffers from ``holes'' in the potential energy surface, where exceptionally dense, highly coordinated structures are predicted to be overly stable, as shown in \cref{fig:As_vscan}. The repulsion seen in the As--As dimer curve indicates that these holes are caused by higher body-order terms in an extrapolative regime - holes occurs at shorter \ce{As-As} distances than occur in the radial distribution function of the \SI{0.1}{\giga\pascal} RSS results. These holes are typically not an issue during ambient pressure MD, due to the large energy barriers seen in \cref{fig:As_vscan}. The updated model does not exhibit these holes due to the use of the ZBL repulsive pair potential and density normalisation (see Methods). Bar a systematic shift in the volume distribution, there is reasonable agreement with DFT even at 50 GPa.

\begin{table}[htbp!]
  \centering
  \caption{Summary of known Arsenic structures. The white \ch{P} structure type is used as a proxy for yellow \ch{As} as the structure is unknown \cite{hart2018one}.}
  \resizebox{\textwidth}{!}{
    \begin{tabular}{cccccc}
      \\ \hline
      \hline
      \textbf{}     Structure               & Pressure       & In training set? & Space Group  & Z  & Found with \SI{0.1}{\giga\pascal} \MPzbt{}? \\
      \hline
      A7, grey As \cite{schiferl1969crystal} & ambient        & yes              & $R\bar{3}m$  & 2  & yes                         \\
      black P \cite{smith1975structures}     & ambient        & yes              & $Cmce$         & 4  & yes                         \\
      white P                                & ambient        & no               & $P\bar{1}$   & 24 &  \ch{As4} tetrahedra found      \\
      simple cubic  \cite{silas2008density}  & \num{27}--\SI{57}{\giga\pascal}      & yes              & $Pm\bar{3}m$ & 1  & yes                         \\
      bcc \cite{silas2008density}            & \SI{\geq 110}{\giga\pascal} & no               & $Im\bar{3}m$ & 1  & yes                         \\ \hline
    \end{tabular}}
  \label{tab:As_RSS}
\end{table}

\subsubsection*{Fine-tuning}
Fine-tuning was performed on a total of 200 configurations. The 5 known structures were relaxed at each pressure contributing 15 configurations to the fine-tuning. A further 185 configurations were selected from the relaxed structures from the original RSS using furthest-point sampling on averaged (across atomic sites) MACE descriptors. This procedure yielded a roughly even split between configurations at each pressure with 77, 51 and 72 configurations selected at 0.1, 10 and 50 GPa respectively.

\subsubsection*{Similarity statement}

The MP dataset contains six pure As structures. Grey arsenic ($R\bar{3}m$) \cite{schiferl1969crystal} and the orthorhombic allotrope ($Cmce$, isostructural with black phosphorus) \cite{smith1975structures} have been observed at ambient conditions whilst the simple cubic structure ($Pm\bar{3}m$) \cite{silas2008density} is stable at moderate pressure between \num{27}-\SI{57}{\giga\pascal}. The remaining three structures are \SI{> 0.4}{eV/atom} above grey arsenic. There are an additional 3857 unique structures that contain As and other elements. Within these structures there are a total of 22047 As environments of which 1606, 1537, 534  and 12 are 1, 2, 3 and 4-fold coordinated by neighbouring As atoms respectively (\SI{2.7}{\Angstrom} cutoff). Many of the 3-fold As environments are found in \ch{AsX} compounds where X is a group I or II element and the As atoms are arranged in local clusters. There is one As atom which is bonded to 4 neighboring As atoms (\ch{Cs7(InAs2)3}, mp-1203378), one structure containing isolated \ch{As4} tetrahedra (\ch{AsO3}, mp-1215144) and two structures containing connected \ch{As4} tetrahedra (\ch{Re4As5S4} mp-1209063 and \ch{Re4As6S3} mp-1219545).

\subsubsection*{Performance summary}
All expected low enthalpy stable structures found.

\clearpage
\subsection{Properties of bulk and nanoconfined water}\label{sec:water-bulk-nano}

See main text \cref{sec:water_results} for results and discussion.

\subsubsection*{Similarity statement}

The MP dataset contains 21 structures composed of \ch{O} and \ch{H} elements and 7769 structures that have \ch{O} and \ch{H} elements alone or together with other elements. Based on UMAP analysis, we see that some atomic environments in the example system are similar to environments in the training data. For instance, bulk water and ice comprise typical molecular environments (\textit{e.g.}, the environment of atom 20 in structure 13 of \verb|water_exclusive_OH_chemiscope_input.csv|) but also environments of hydrogen peroxide (\textit{e.g.}, the environment of atom 1 in structure 14 of \verb|ice_exclusive_OH_chemiscope_input.csv|). Despite being two dimensional, the superionic phase also comprises distinct environments mimicking those of water molecules (\textit{e.g.}, the environment of atom 3 in structure 20 of\\ \verb|superionic_exclusive_OH_chemiscope_input.csv|) and dissociated environments mimicking those of hydrogen peroxide (the environment of atom 12 structure 12 of\\ \verb|superionic_exclusive_OH_chemiscope_input.csv|). The environments farthest from the MP dataset are the monolayer oxygen environments surrounded by a (flat) hexagon of 6 other oxygen atoms.

\subsubsection*{Performance summary}
The \MPz{} model demonstrates stability and reliable performance in conducting simulations across diverse conditions for both bulk and confined water. It maintains stability in NVT simulations at experimental densities and temperatures for bulk water, ice Ih, and reactive proton defects (\ch{OH^-} and \ch{H3O^+}). The model describes extensive proton transfer in nanoconfined water at \SI{4}{\giga\pascal} and \SI{600}{K}, in good agreement with reference methods.

\clearpage
\subsection{Ethanol-water density-composition curves}\label{sec:ethwat}

\begin{figure}[htbp!]
  \centering
  \includegraphics[width=0.6\linewidth,keepaspectratio]{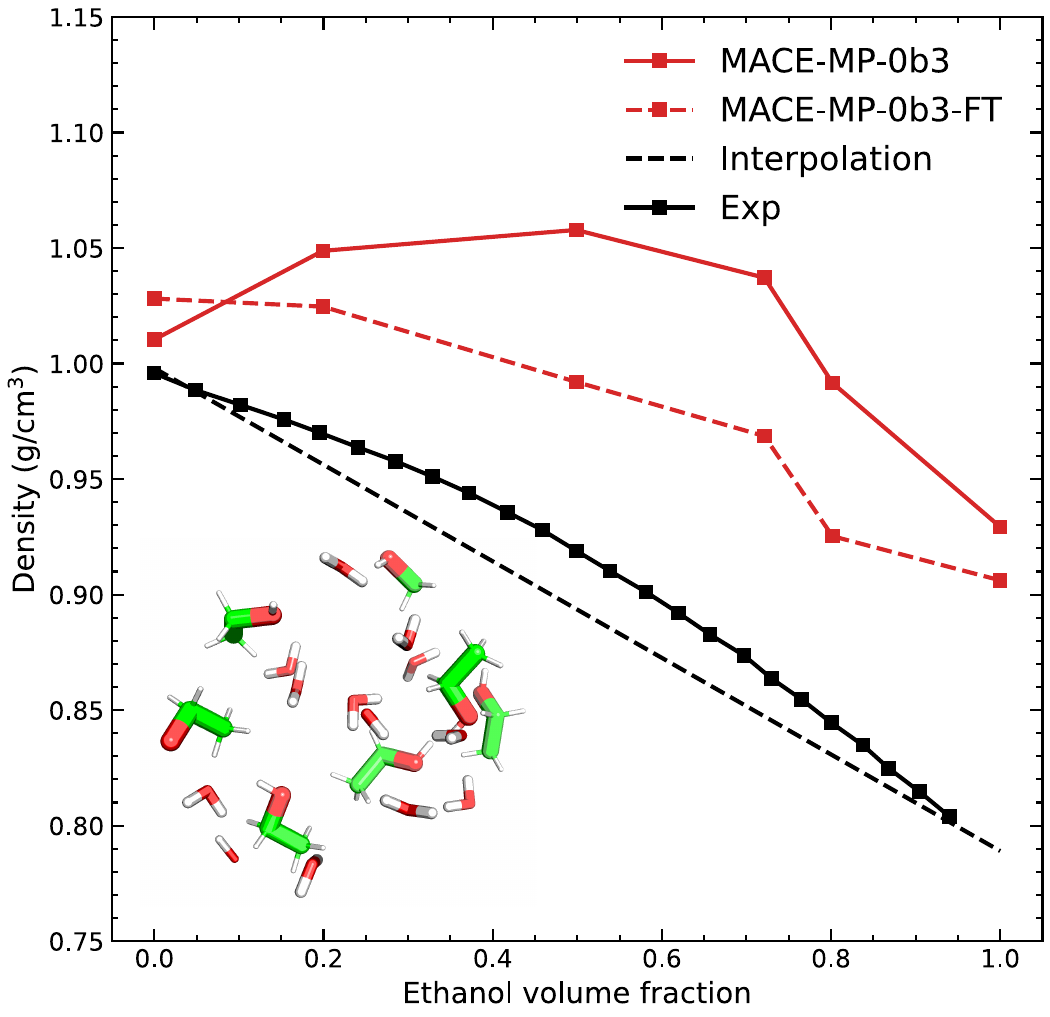}
  \caption{Ethanol--water density curves obtained by NPT MD using \MPzbt{} in single precision, compared to experimental data taken from Ref.~\cite{southard2018perry}. }
  \label{fig:ethanol_water_density}
\end{figure}

In this section, we investigate the ability of the \MPzbt{}{} model to describe mixtures of molecular liquids. In particular, we study the density-composition curve for a range of volume fractions of ethanol in water. Initial configurations were generated with Packmol~\cite{martinez2009packmol}, with 120 molecules per box, and the initial box vectors were set to be slightly below the experimental density for each composition.  Initial structures were minimised to a tolerance of 0.01 eV/\AA{} with the L-BFGS algorithm. Trajectories were generated in the NPT ensemble using ASE, including a D3 dispersion correction with the Becke-Johnson damping function. Final densities were computed as the averaged of the final 1000 snapshots from the simulation, once the density had converged.

\subsubsection*{Similarity statement}

The MP dataset contains 37 structures that contain only the elements \ch{C}, \ch{H}, and \ch{O}. Based on UMAP analysis, we observe that almost all atomic environments in the example system are similar to environments in the training set. On closer inspection, we find that the most similar environments to the majority of the example configurations are clusters primarily containing water, hydroxide and atomic hydrogen and oxygen, with a few examples containing small hydrocarbon-type fragments.

\subsubsection*{Performance summary}
For low volume fractions of ethanol, \MPzbt{} predicts a spurious density maximum at 50\% volume fraction, whilst correctly predicting the lower density for high ethanol volume fractions. \MPzbt{}-FT densities more closely follow the shape of the experimental curve, successfully capturing the deflection from linear behaviour. \MPzbt{}-FT overpredicts the absolute densities by approximately 4\% with respect to experiments, which is not uncommon for a GGA DFT functional on molecular liquids.


\clearpage
\subsection{Solvent mixtures}\label{sec:mixtures}

\begin{figure}[htbp!]
  \centering
  \includegraphics[width=\textwidth,keepaspectratio]{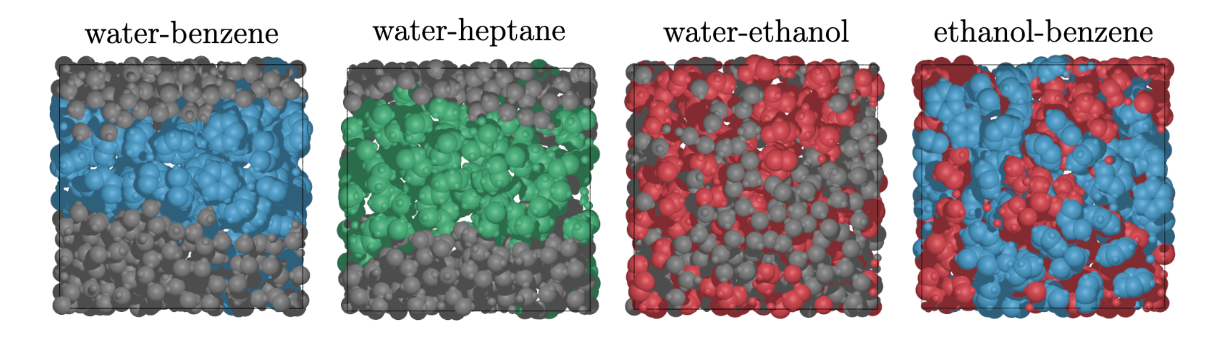}
  \caption{Snapshots of solvent mixtures after \SI{1}{\nano\second} of NVT MD with \MPzbt{}. The solvents shown are benzene (blue), heptane (green), ethanol (red), and water (grey). Axis orientations in the figures were chosen to highlight the phase separation in these systems.}
  \label{fig:mixtures}
\end{figure}

Modeling solvent mixtures requires an accurate description of intermolecular forces within highly disordered systems. To investigate the performance of \MPz{} in this setting, MD simulations were performed for four mixtures of solvents of varying polarity. The investigated systems are water-benzene, water-heptane, water-ethanol, and benzene-ethanol. Simulations were performed at \SI{300}{\kelvin} in the NVT ensemble via the ASE interface. A time step of \SI{1}{\femto\second} and a friction constant of \SI{0.001}{\per\femto\second} were used. In the case of immiscible solvents, a mixture of equal volumes of both solvents with their corresponding densities was assumed. In the case of miscible solvents, the experimental density of the mixtures was used. All systems were initialized with a uniform random mixture of both solvents using the \texttt{packmol} code. \cite{martinez2009packmol}

\Cref{fig:mixtures} shows the states of all systems after \SI{1}{\nano\second}. Notably, mixtures of water with apolar solvents (heptane and benzene) quickly form separate phases, whereas the ethanol-water and ethanol-benzene systems remain mixed on the timescale of the simulation. This is in good agreement with experiment.

\subsubsection*{Similarity statement}

The MP dataset contains 37 structures composed exclusively of \ch{C}, \ch{H}, and \ch{O}, and 1902 structures that contain \ch{C}, \ch{H}, and \ch{O} along with other elements. Regarding the specific molecules, several ice structures but none of the other molecules are included as pure compounds. The closest to benzene (with the ratio of \ch{C}:\ch{H} 1:1) is mp-995197 containing chains of dimethylbenzenes with methyl-methyl bridges. The UMAP analysis shows that many atomic environments from our structures have similar environments in the training data. However, no liquid configurations are included in the MP. We provide two files to visualize the interactive UMAP on \url{chemiscope.org}. \verb|solvents_mixtures_CHO.json| contains structures exclusively containing \ch{C},\ch{H}, and \ch{O}. \verb|solvents_mixtures_CHOplus.json| includes structures containing \ch{C}, \ch{H}, and \ch{O} along with other elements.

\subsubsection*{Performance summary}
Miscibility of all four mixtures correctly predicted, at least on the limited timescale of the MD simulations.  

\clearpage
\subsection{Aqueous interfaces}\label{sec:aqueous_interfaces}
\begin{figure}[ht!]
  \centering
  \includegraphics[width=\textwidth,keepaspectratio]{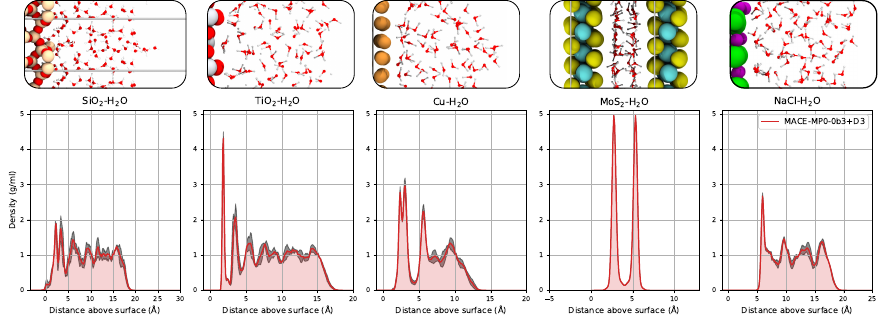}
  \caption{\textbf{Water structure and density at various interfaces}. Representative snapshots of investigated systems are shown with corresponding water density profiles. The red curve represents the average water density profile obtained from three independent MD simulations, the grey shading indicates the standard deviation. The systems depicted are water on silicon dioxide (\ch{SiO2}), titanium dioxide (\ch{TiO2}), copper (\ch{Cu}), between two layers of molybdenum disulfide (\ch{MoS2}), and on sodium chloride (\ch{NaCl}).}
  \label{fig: aqueous_density}
\end{figure}
Simulating complex systems, such as solid-liquid interfaces, is a difficult endeavor, as the potential must simultaneously describe the two materials and their interface. We tested the effectiveness of \MPz{} on a wide range of aqueous interfaces, from oxides and metals to confinement.

NVT MD simulations were performed on a variety of surfaces at a temperature of \SI{330}{K}. The average density of water above the surface is shown in \cref{fig: aqueous_density}.

\ch{SiO2} and \ch{TiO2} were two notable oxide systems in which dissociative and molecular adsorption was observed, respectively. Deprotonation of water was expected on the surface of silicon dioxide, which is evidenced by the shoulder in the water density plot. These figures show that the interfacial water property is accurately reproduced; however, the liquid phase is overstructured, which is a common characteristic of the PBE functional \cite{gillan2016perspective} used in the Materials Project.

Water in confinement was also investigated within \ch{MoS2} slit pores. The simulation captures the pronounced stratification characteristic of the aqueous phase perpendicular to the two-dimensional layers. This was also observed for water confined between graphene sheets and boron nitride nanotube. In particular, in \cref{fig: aqueous_density} we show sharply defined interfacial water layers between the \ch{MoS2} sheets. Upon the addition of interlayer spacing (not shown), we also capture additionally smoother intermediate layers, noting that with more layers, we lose the sharp peaks at the surfaces.

Finally, the \ch{NaCl} (001) surface in contact with water was simulated. The system comprised a ($3\times4$) \ch{NaCl}(001) supercell containing \num{24} atoms, with 3 \ch{NaCl} layers and a unit cell lattice constant of \SI{5.72}{\Angstrom} on top of which were \num{89} water molecules. A subsequent \SI{25}{\Angstrom} of vacuum was added between the adsorbed water layer and the lower layer of the surface.
The layered structure of the water as previously observed in \textit{ab initio} PBE simulations in Ref. \cite{liu2008density} is captured by the \MPz{} model, with the positions of the density minima and maxima qualitatively agreeing with the PBE simulations.

\begin{figure}[ht!]
  \centering
  \includegraphics[width=\textwidth,keepaspectratio]{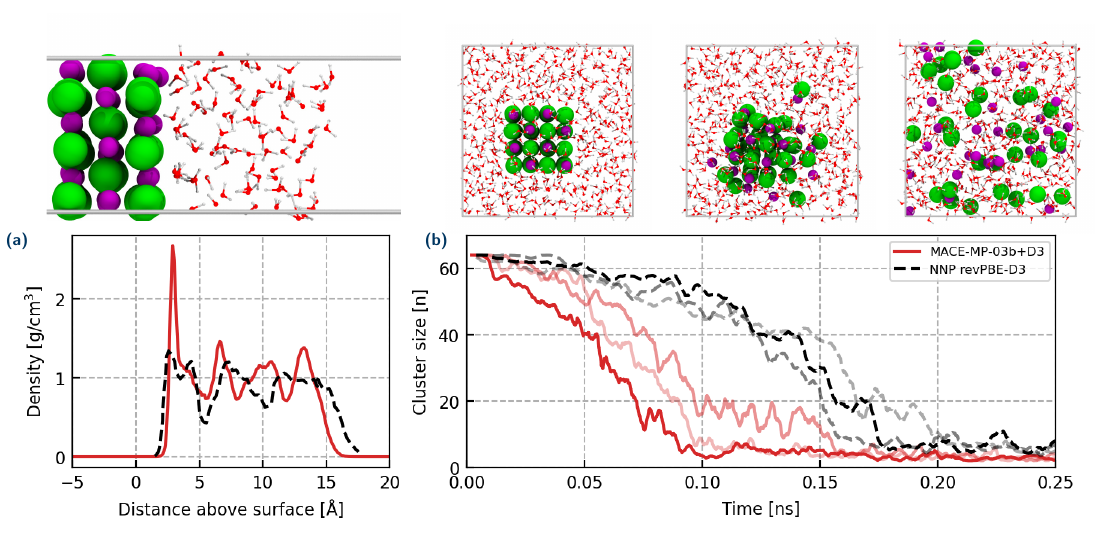}
  \caption{\textbf{Dissolution of NaCl in water}. (a) Density profile of water in contact with \ch{NaCl}(001) surface, with a representative snapshot from the simulation showing no dissolution events from the pristine surface. (b) Evolution of crystal size of $4\times4\times4$ \ch{NaCl} nanocrystal in water over time, comparing the \MPz{} (red line) with an ML model explicitly trained to capture \ch{NaCl} dissolution (black dashes) ~\cite{oneill2022crumbling}. Representative snapshots showing the dissolution progress of the crystal are shown above the plot.}
  \label{fig: NaCl_dissolution}
\end{figure}

Simulation of dissolution processes is another challenge for the \MPz{}  model. It must be able to describe the very different chemical environments of the bulk crystal surrounded by water going through the stages of ions detaching from the crystal to fully solvated ions in solution. In \cref{fig: NaCl_dissolution}, we compare the \MPz{} model in NVT simulations of the pristine \ch{NaCl} (001) interface in contact with water and a \ch{NaCl} nanocrystal surrounded by water at \SI{400}{\kelvin}.
The nanocrystal system simulated comprised a $4\times4\times4$ \ch{NaCl} nanocrystal comprising 32 ions, with lattice constant \SI{5.72}{\Angstrom} surrounded by 625 water molecules, giving a final concentration when dissolved of \SI{2.84}{mol/kg}.

As expected, for a pristine \ch{NaCl} surface, the model predicts no dissolution events on the time scale of the simulation.
Meanwhile, for the nanocrystal surrounded by water, the model captures a dissolution mechanism similar to that reported by Ref. ~\cite{oneill2022crumbling} with an ML model trained specifically to capture \ch{NaCl} dissolution at revPBE-D3 level of theory. The dissolution proceeds via a crumbling mechanism, where an initial steady loss of ions precedes rapid disintegration of the crystal. Moreover, the dissolution process is stochastic, leading to an intrinsic variation between independent simulations, as shown
by three examples. The resulting solution of ions in water also displays correct expected orientation of the water molecules with respect to the ions.

\subsubsection*{Similarity statement}
The MP dataset contains 460, 100, 112, 13 and 29 structures composed exclusively of [H, O, Si], [H, O, Ti], [H, O, Cu], [H, O, Mo, S] and [H, O, Na, Cl], respectively. The corresponding number of structures inclusive of the given atoms along with other elements is 477, 215, 435, 260 and 190. 
Based on UMAP analysis, the closest atomic environments for each of these systems are mp-626085, mp-626550, mp-697660, mp-990086 and mp-504600.
Two files are provided for each of the systems for visualising using \verb|chemiscope|, one for inclusive and one for exclusive matches in the training set.

\subsubsection*{Performance summary}
All interface structures correctly predicted, including dissociative adsorption on \ch{SiO2} and molecular adsorption on \ch{TiO2}. At the salt/water interface correctly predicted dissolution from nanocrystal and no dissolution from flat surface on nanosecond time scale. 

\clearpage
\subsection{Molten salts}\label{sec:molten_salts}
\begin{figure*}[htbp!]
  \centering
  \begin{subfigure}[b]{0.6\linewidth}
    \includegraphics[width=\linewidth,keepaspectratio]{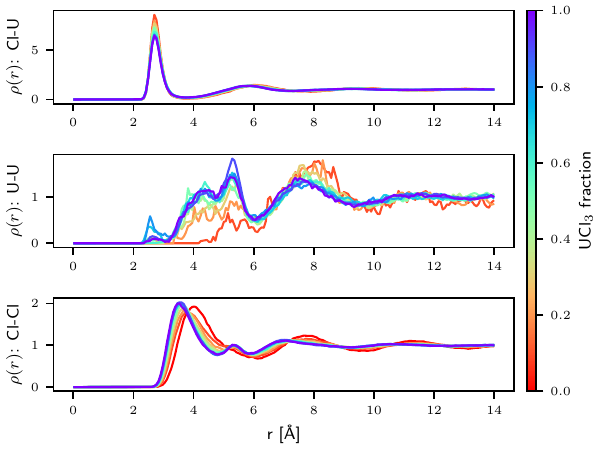}
    \caption{MACE-MP-0b3+D3}
  \end{subfigure}
  \hfill
  \begin{subfigure}[b]{0.39\linewidth}
    \includegraphics[width=\linewidth,keepaspectratio]{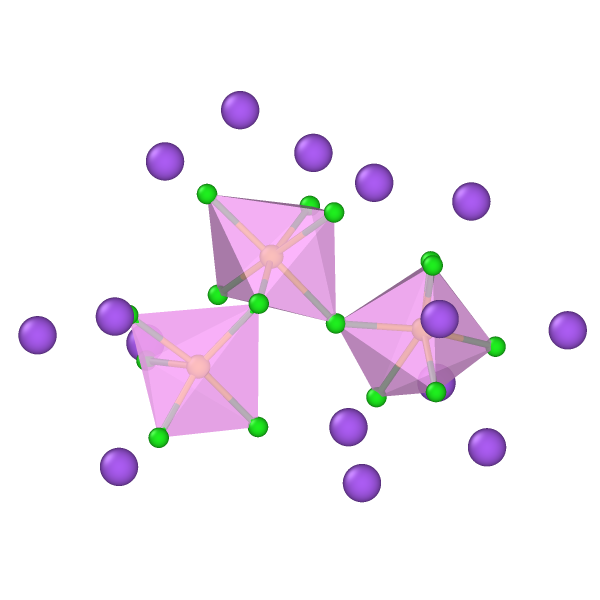}
    \caption{}
  \end{subfigure}
  \caption{\ch{NaCl-UCl3} molten salt mixtures at 1100K. (a) Pair correlation function of \ch{NaCl-UCl3} mixtures at \SI{1100}{\kelvin}. (b) Example \ch{U-Cl} oligomers forming vertex sharing coordination polyhedra (\ch{U}: yellow, \ch{Cl}: green, \ch{Na}: purple).}
  \label{fig:NaCl-UCl3}
\end{figure*}

With increasing interests in molten-salt energy technologies, we have simulated binary \ch{NaCl-UCl_3} salt mixtures \ch{(NaCl)_{(1-x)}(UCl_3)_x} at different compositions using \MPz{}. The initial structures were randomized using Packmol in a cubic cell at the density estimated by the linear interpolation of the densities of constituent solid-state salt at \SI{0}{\kelvin}. \cite{martinez2009packmol,Chiang_muse_2023}. We then implemented geometric optimization using Lennard-Jones potential and further relaxed structures using \MPz{}+D3 with a two-step process: NVT relaxation through annealing at $1.2\times$ target temperature for \SI{5}{ps}, and NPT relaxation at \SI{1100}{\kelvin} and zero pressure for \SI{10}{ps}. \Cref{fig:NaCl-UCl3}a presents the pair correlation functions between \ce{Cl-U}, \ce{U-U}, \ce{Cl-Cl} in salt mixtures. The characteristic peaks and transitions are consistent with previous polarizable ion models \cite{van2021coupled} and AIMD simulations \cite{andersson2022ab}, except for a noticeable shift of U-U peak from \num{4.5} to \SI{4}{\Angstrom} and a U-U peak formation at \SI{5.5}{\Angstrom} at high \ch{UCl3} concentration. The shift could be explained by the lack of Hubbard $U$ correction for rare earth elements in MP, leading to unrealistic ionic radii and solvation shell in the mixture. We also note that there is a small U-U peak around \SI{2.5}{\Angstrom}. This peak is absent in previous molten salt studies at high temperature \cite{andersson2022ab}, but as it is close to the equilibrium distance between \ch{U} as demonstrated by the homonuclear diatomic curve (\cref{fig:homonuclear-2b}), its appearance indicates the formation of a few \ce{U-U} bonds at a high fraction of molten \ch{UCl3} salt.

\subsubsection*{Similarity statement}

The MP dataset contains 573 structures composed of at least one Na, U, or Cl atom, 14 elemental Na crystals, 14 elemental U crystals, and 3 \ch{Cl2} molecular crystal structures. Based on UMAP analysis, we see that all atomic environments in the example system are similar to environments in the training data. We found that \ch{Cl2} molecular crystals are close to the molten salts but most of the pure \ch{U} metals are found separated from the molten salt in terms of MACE descriptors. We provide 
\begin{itemize}
    \item \verb|T_1100-P_0-seed_3-npt-5_chemiscope_input.json|
\end{itemize}
to help visualize the interactive UMAP of molten \ch{Cl64Na28U12} on \url{chemiscope.org}.

\subsubsection*{Performance summary}
Correct pair distribution peaks and variation of peak positions as a function of concentration, with a notable shift in the first U-U peak position, due to absence of Hubbard-$U$ correction. 

\clearpage
\subsection{Room temperature ionic liquids}\label{sec:ionic_liquids}

\begin{figure}[htbp!]
  \centering
  \includegraphics[width=\linewidth]{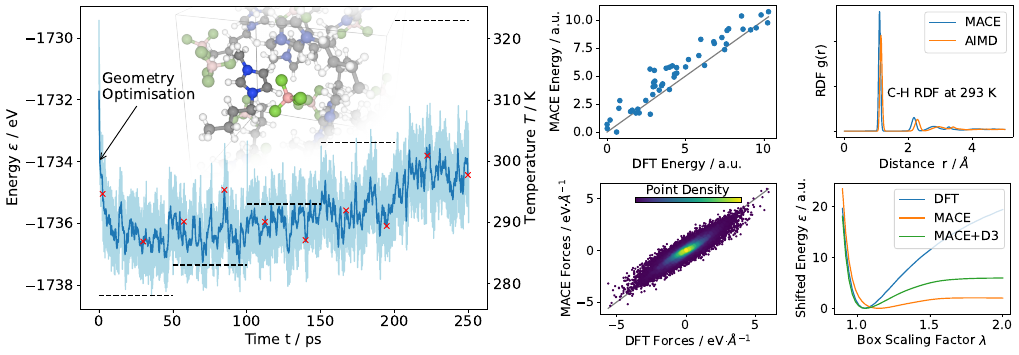}
  \caption{MD simulation of the BMIM \ch{BF4} room temperature ionic liquid. The left panel shows energy
  as a function of the trajectory, starting from an energy minimization, followed by MD simulation in an NVT ensemble with a step-wise increase in temperature (dashed lines, right axis). The middle column of panels
  shows energy and force parity plots for configurations from the trajectory (red markers). The right top panel compares RDF to AIMD, and the right bottom panel shows a rigid-molecule volume-scan test \cite{magduau2023machine}.
  }
  
  \label{fig:rtil_bmim_bf4}
\end{figure}

Room temperature ionic liquids provide a class of organic solvents with desirable properties such as low melting and high boiling points, chemical inertness, and good ionic conductivity, making them applicable to different chemical and physical applications. Furthermore, these properties can be tuned by changing substituents on the anion or cations. The vast availability of substituents makes simulations at quantum mechanical accuracy to optimize these \textit{in silico} a very interesting approach.

For the given example, the class of imidazolium-based ionic liquids was chosen.
Simulations using the \ch{BF4-} anion with the 1-butyl-3-methylimidazolium (BMIM) cation were conducted.
A single MD simulation was performed starting at the experimental density~\cite{rebeloDetailedThermodynamicAnalysis2004} of BMIM \ch{BF4} at \SI{273}{\kelvin}.
The temperature was stepwise increased from \SI{273}{\kelvin} to \SI{323}{\kelvin}.
Between each increase, the cell was adjusted to the new density and equilibrated over \SI{500}{\femto\second}.
All simulations are conducted using the \MPzbt{} with additional D3(BJ) corrections.

At each temperature, an NVT simulation using a Langevin thermostat was conducted for \SI{50}{\pico\second} with a time step of \SI{0.5}{\femto\second}.
From the final trajectory spanning \SI{250}{\pico\second} over 5 different temperatures, data points were uniformly selected, and energies and forces were compared to DFT~\cite{perltFindingBestDensity2018}.
Additionally, the radial distribution function was compared to an AIMD simulation, indicating a shift in the hydrogen positions compared to DFT.
Furthermore, interatomic interactions are probed using a volume scan, demonstrating the importance of the additional D3(BJ) correction to stabilize the correct volume (see Figure~\ref{fig:rtil_bmim_bf4}).
Finally, a MD simulation in an NPT ensemble at \SI{1}{atm} and \SI{300}{\kelvin} showed that the model reproduces the experimental density within 5\%.

All models struggle to run simulations using a \ch{Cl-} anion instead of \ch{BF4-}, as this results in \ch{Cl} atoms bonding to the aromatic ring.  
To address this, fine-tuning was performed on 100 configurations of 16 ion pairs of BMIM Cl.  
A workflow using IPSuite\cite{zillsCollaborationMachineLearnedPotentials2024a} was used to set up the simulation box from \verb|SMILES|\cite{weiningerSMILESChemicalLanguage1988}, generate GROMACS\cite{abrahamGROMACSHighPerformance2015} input files and sample random configurations.  
The CL\&P force field\cite{canongialopesModelingIonicLiquids2004} was used in the 50~ns NVT sampling simulation.  
After fine-tuning, the \MPzbt{}-FT model was able to run stable simulations for BMIM Cl.

\subsubsection*{Similarity statement}
There are 52 structures in the MP dataset that explicitly include the \ch{BF4} anion.
Although there are organic nitrogen-containing molecules as well as heterocyclic systems, there are no alkyl-substituted imidazolium derivatives like BMIM in the training dataset.
We provide \verb|BMIM_BF4.json| for a comparison of snapshots from the MD trajectory to the training dataset on \url{chemiscope.org}.

\subsubsection*{Performance summary}
Stable MD for \ch{BF4} anion, but \ch{Cl} anion bonded to imidazolium.
Bond formation is prevented by applying fine-tuned models.
Intermolecular distance distribution shows small peak shift, and intermolecular attraction underestimated.

\clearpage
\subsection{High-pressure hydrogen}\label{sec:h2}

\begin{figure}[htbp!]
  \centering
  \includegraphics[width=\linewidth,keepaspectratio]{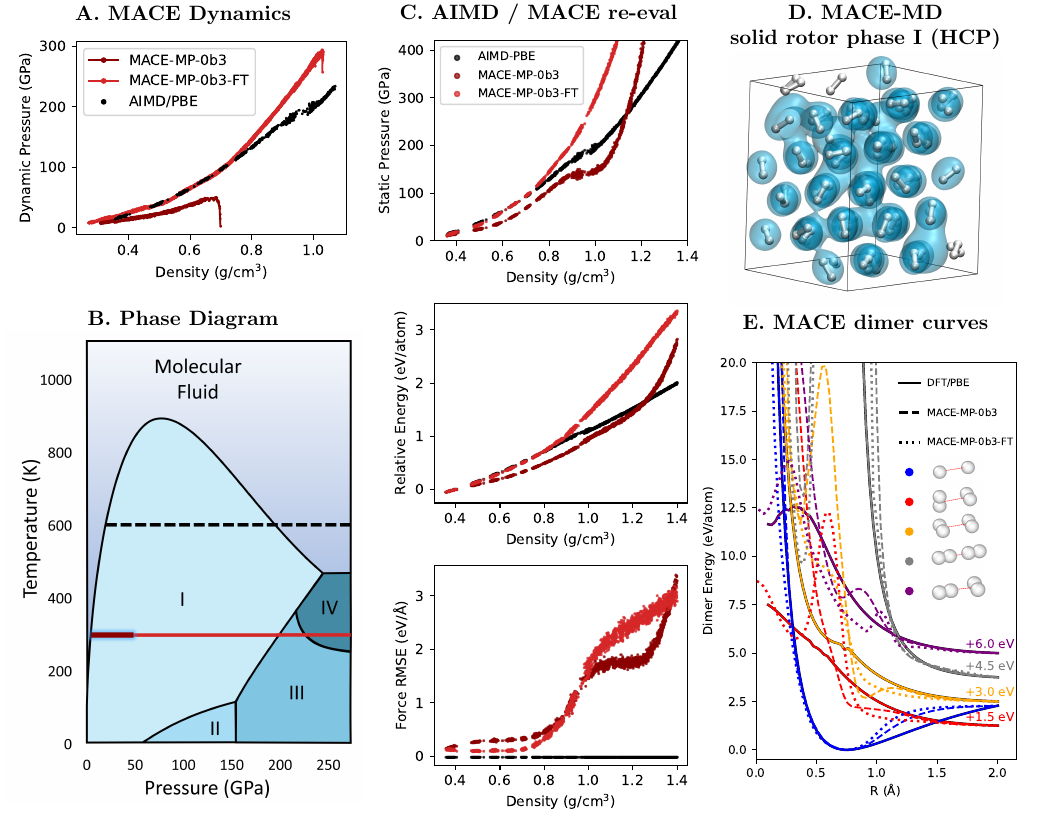}
  \caption{(A) Pressure/density dependence in MACE-MD compared to AIMD. (B) Illustrative phase diagram of high-pressure hydrogen \cite{gregoryanz2020everything, magduau2017simple}. Horizontal lines show the NPT-MD simulations: \MPzbt{} (dark red), \MPzbt -FT (red) (see text) and AIMD/PBE (black)~\cite{magduau2017theory}. (C) MACE properties (pressure, energy, forces) evaluated on the AIMD trajectory and compared to the original PBE result. (D) MACE reproduces solid hydrogen phase I as a hexagonal-close-packing (HCP) lattice of free rotors. (E) MACE energy curves computed for different orientations of the \ch{H2}-\ch{H2} dimers compared to DFT/PBE results. The distance is measured between the centers of mass of the two molecules. The curves were offset along the y-axis to improve readability.}
  \label{fig:HPH}
\end{figure}

Condensed-phase hydrogen is an exotic state of matter that forms at extreme conditions in the core of larger planets and in specially-designed  laboratory diamond anvil cells. Despite the simplicity of the hydrogen atom and molecule, the condensed phase exhibits fascinating phenomena such as entropy-driven phase transitions \cite{pickard2012density, magduau2013identification}, phonon localization \cite{howie2014phonon, magduau2017infrared}, quantum rotor solid phases \cite{cooke2020raman}, and an insulator-to-metal transition \cite{cheng2020evidence, zong2020understanding}. AIMD has been used extensively to study the molecular mechanisms underlying these phenomena, however, simulations are often affected by finite-size effects. Bespoke ML potentials \cite{bischoff2024hydrogen}, fitted to reproduce the interaction of hydrogen molecules with ab initio accuracy, have demonstrated simulations at an unprecedented level of detail, unlocking new scientific observations \cite{cheng2020evidence}. Being so different from other materials in the MP database, solid hydrogen is a uniquely challenging test for \MPzbt{}.

The stability of the potential on this system was tested by running MD simulations at high pressure and by investigating the H-H and \ch{H2}-\ch{H2} dimer curves for molecules in different orientations. The MD simulations started from an AIMD thermalized Pc-48 crystal structure \cite{pickard2012density, magduau2013identification, ackland2020structures} and pressure was slowly ramped up at constant temperature. The \MPzbt{} potential is stable up to a density of \SI{0.7}{\gram\per\centi\meter\cubed} and pressure of \SI{50}{\giga\pascal}, however it does not accurately reproduce the AIMD equation of state (EoS). At low pressures, the dynamics does reproduce the expected behaviour of hydrogen in phase I -- HCP crystal of freely rotating hydrogen molecules. A fine-tuned variant of the potential, \MPzbt{}-FT, was obtained by sampling 50 configurations with densities ranging from \num{0.5} -- \SI{0.9} {\gram\per\centi\meter\cubed} from a MD trajectory generated with an earlier MP potential. These configurations were evaluated with PBE and the \MPzbt{} model was fine-tuned on this new dataset. The resulting potential is more robust and remains stable up to a remarkable \SI{300}{\giga\pascal} and a density of \SI{1.0} {\gram\per\centi\meter\cubed}. Additionally, the \MPzbt{}-FT model accurately reproduces the AIMD EoS up to \SI{130}{\giga\pascal}.

The accuracy of both potentials was also quantified by re-evaluating the pressure, energy and forces on an existing AIMD trajectory \cite{magduau2017theory}. \MPzbt{} follows the overall trend of the PBE EoS up to \SI{0.9} {\gram\per\centi\meter\cubed}, but shows a constant erroneous shift to lower pressures. \MPzbt-FT significantly improves the result and accurately captures the EoS up to \SI{0.8} {\gram\per\centi\meter\cubed}. A similar result is found for energy and forces, where \MPzbt-FT significantly improves the agreement with PBE in the density range \num{0.5} -- \SI{0.9} {\gram\per\centi\meter\cubed}, where new data was added. Finally, the energy dissociation curves for \ch{H2}-\ch{H2} dimers are smooth at separation distances above \SI{1.4}{\Angstrom}, which corresponds to distances observed in equilibrium MD. Similar smooth behaviour is observed for the H-H energy dissociation curve, where both models accurately reproduce PBE around the \SI{0.75}{\Angstrom} equilibrium bond length of the H$_2$ molecule.

\subsubsection*{Similarity statement}

The MP dataset contains 17 structures crystal structure composed exclusively of H. From these, only 2 structures are above the \SI{0.2}{\gram\per\cubic\centi\meter} density value: mp-1096977 (hexagonal P4/mmm \SI{0.24}{\gram\per\cubic\centi\meter}) and mp-754417 (hexagonal P6/mmm \SI{0.24}{\gram\per\cubic\centi\meter}), yet \MPzbt{} was found to be stable up to around \SI{0.7}{\gram\per\cubic\centi\meter}. We provide \verb|hydrogen_exclusive.json| for visualization on \url{chemiscope.org}.

\subsubsection*{Performance summary}
Correct solid hydrogen structure reproduced at low to moderate pressures by \MPzbt{} with inaccurate EoS. At pressures above \SI{50}{\giga\pascal}, the potential failed and resulted in unphysical structures. A \MPzbt{}-FT potential fine-tuned on 50 configurations successfully reproduced the EoS  up to \SI{130}{\giga\pascal} and remained stable up to \SI{300}{\giga\pascal}.


\clearpage
\subsection{Ammonia and borane thermal decomposition}\label{sec:amonia_borane}
\begin{figure}[htbp!]
  \centering
  \includegraphics[width=\textwidth,keepaspectratio]{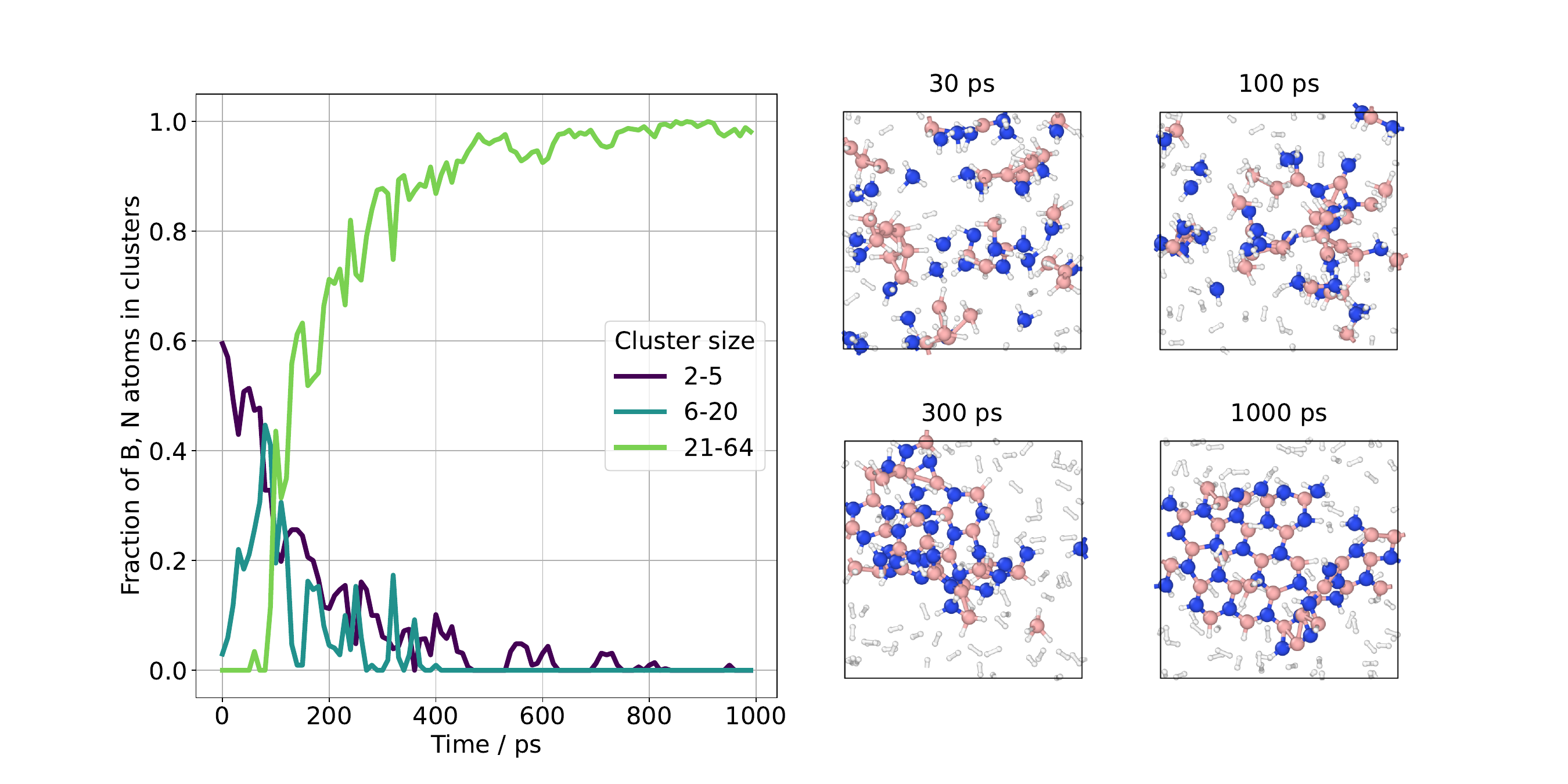}
  \caption{Decomposition of ammonia and borane at \SI{1600}{K}. The plot shows the time evolution of cluster size in terms of the fraction of heavier atoms (\ch{B} and \ch{N}) found in each cluster size group. Snapshots of the system at different times show the growth of \ch{BN} clusters and evolution of \ch{H2} molecules.}
  \label{fig:borane-ammonia}
\end{figure}

Ammonia and borane form an adduct \ch{NH3BH3}. At high temperatures, these molecules lose hydrogen gas to give increasingly heavier \ch{B}- and \ch{N}-containing molecules, ultimately resulting in the growth of hexagonal boron nitride (hBN) \cite{Frueh2011}. We simulated this process with \MPzbt{} by running NVT molecular dynamics simulation of 32 ammonia and 32 borane molecules in a cubic box of length \SI{16}{\Angstrom}, at a temperature of \SI{1600}{\kelvin} for \SI{1}{\nano\second} with \SI{0.5}{\femto\second} time step. We analysed the time evolution of the system in terms of the sizes of heavier atom clusters (excluding hydrogen), illustrated in \cref{fig:borane-ammonia}. Initially, the formation of ammonia borane adduct and small borane clusters is seen, while in 100 ps timescales \ch{B} and \ch{N} atoms are increasingly more clustered, with preference for \ch{B}-\ch{N} bonds over homonuclear bonds. Ultimately, an hBN-like fragment is formed.

\subsubsection*{Similarity statement}

The training set contains 67 structures composed of \ch{H}, \ch{B}, \ch{N} elements.
The training set contains various structures encountered during the simulation including ammonia, borane, and \ch{HBN} compounds of various stoichiometries, for example borazine (\ch{B3N3H6}) and \ch{(BNH2)_n} chains. We performed UMAP analysis for 100 frames taken from a 1 nanosecond \MPz{} MD simulation against training data containing at least one of the \ch{HBC} elements and any other elements. Based on the UMAP values, most of the simulation atomic environments are clustered near the training data, with exceptions being species with unusual valency (e.g. \ch{BH2}). The closest structures in the training set are mp-1197795, mp-1203334 (both containing B and N, among other elements) and mp-1214811 (\ch{B6N6H10} bicyclic aromatic compound). We provide \verb|ammonia-borane.json| for visualization on \url{chemiscope.org}.

\subsubsection*{Performance summary}
Model correctly predicts hydrogen production and \ch{BN} cluster formation from thermal decomposition of \ch{NH3BH3}. 

\clearpage

\subsection{Heterogeneous Catalysis}\label{sec:hetcat}

Computational heterogeneous catalysis evolves around the exploration of \textit{operando} catalyst stability and catalytic reaction mechanisms to provide information about the nature of the active site that defines a catalyst's performance. This information provides a basis for screening applications to find efficient and ideally non-precious and non-toxic catalysts. To this end, a variety of atom-scale properties are investigated, including bulk and surface energies to evaluate catalyst stability in (surface) phase- or Pourbaix diagrams, as well as adsorption energies, reaction thermodynamics, and reaction barriers that are key to elucidating mechanisms and catalytic activity \cite{Norskov2014}. Local geometry optimizations and transition state searches via \textit{e.g.} NEB calculations \cite{neb_aseneb, neb_jk_optimiser} that yield target properties are usually conducted on slab models that exemplify the catalyst surface. By expanding the usual surface science approaches via thermodynamic referencing of protons and electrons to pH and applied potential on basis of the computational hydrogen electrode (CHE) \cite{Norskov2004-ow}, concepts in thermal catalysis can be extended to electrocatalysis. This approach provides fairly robust results even though the simulation of the electrolyte, charged species or an applied potential and thus the direct influence of the electrified solid/liquid interface is omitted. The computationally involved methodology is fully transferable to \MPz{} and we apply it in full to the examples presented in the main text and below.

\subsubsection{Pourbaix diagrams}\label{sec:pour}
\begin{figure}[!ht]
  \centering
  \includegraphics[width=0.8\textwidth,keepaspectratio]{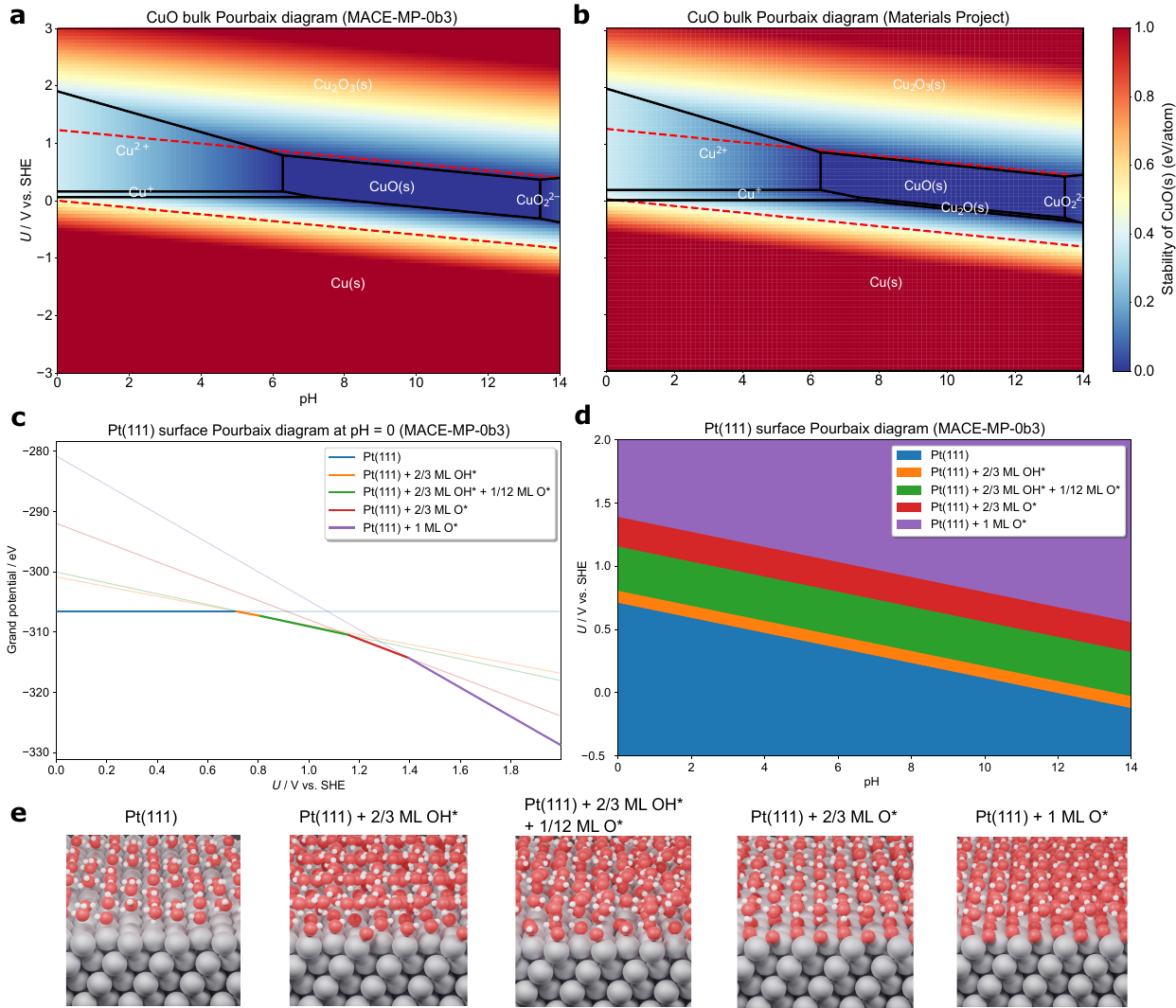}
  \caption{Pourbaix diagrams of \ch{CuO} bulk systems with energies of relevant solid compounds taken from (a) the {\MPz{}} calculations and (b) the MP reference. (c) and (d) shows the {\MPz{}}-calculated Pt(111) surface Pourbaix diagrams at pH = 0 and at various pH, respectively, which are in good agreement with \cite{Hansen2008-zz}. Different stable surface structures are represented in different colors. (e) shows the stable Pt(111) surface structures from low to high applied potentials. The red dashed lines indicate the stable window of water ranging from $U_{\mathrm{RHE}}$ = \SI{0}{V} to $U_{\mathrm{RHE}}$ = \SI{1.23}{V} ($U_{\mathrm{RHE}} \approx U_{\mathrm{SHE}} + 0.059 \cdot \mathrm{pH}$).}
  \label{fig:pourbaix}
\end{figure}

In \cref{fig:pourbaix} we show the Pourbaix diagrams, calculated by {\MPz{}} with D3 corrections, which illustrate the aqueous stability for a CuO bulk and a Pt(111) surface in dependence of applied potential and pH as referenced by the CHE. Structures for bulk CuO and all other related oxide and peroxide compounds are taken from MP and are subsequently optimized (both atomic positions and cell parameters) using \MPz{}. The energy corrections for oxides and peroxides, as well as the free energies for aqueous ions, are consistent with the values used in MP. As shown in \cref{fig:pourbaix}a and b, the overall trend of the CuO stability predicted by {\MPz{}} is well-aligned with the result given by MP, except for the narrow region of the \ch{Cu2O} phase that is not reproduced by {\MPz{}}.  As depicted in \cref{fig:pourbaix}c–e, {\MPz{}} predicts that the Pt(111) surface starts to oxidize at $U_\mathrm{SHE}$ = 0.72 V (pH=0), followed by a step-wise increasing OH*/O* surface coverage with more positive electrode potential. This is generally in good agreement with the Pt(111) surface Pourbaix diagram reported previously (58)\cite{Hansen2008-zz}, despite the predicted starting oxidation potential being 0.06 V too low.

\subsubsection{Linear Scaling Relationships (LSR)}\label{sec:lsr}
\begin{figure}[ht!]
  \centering\includegraphics[width=0.8\textwidth,keepaspectratio]{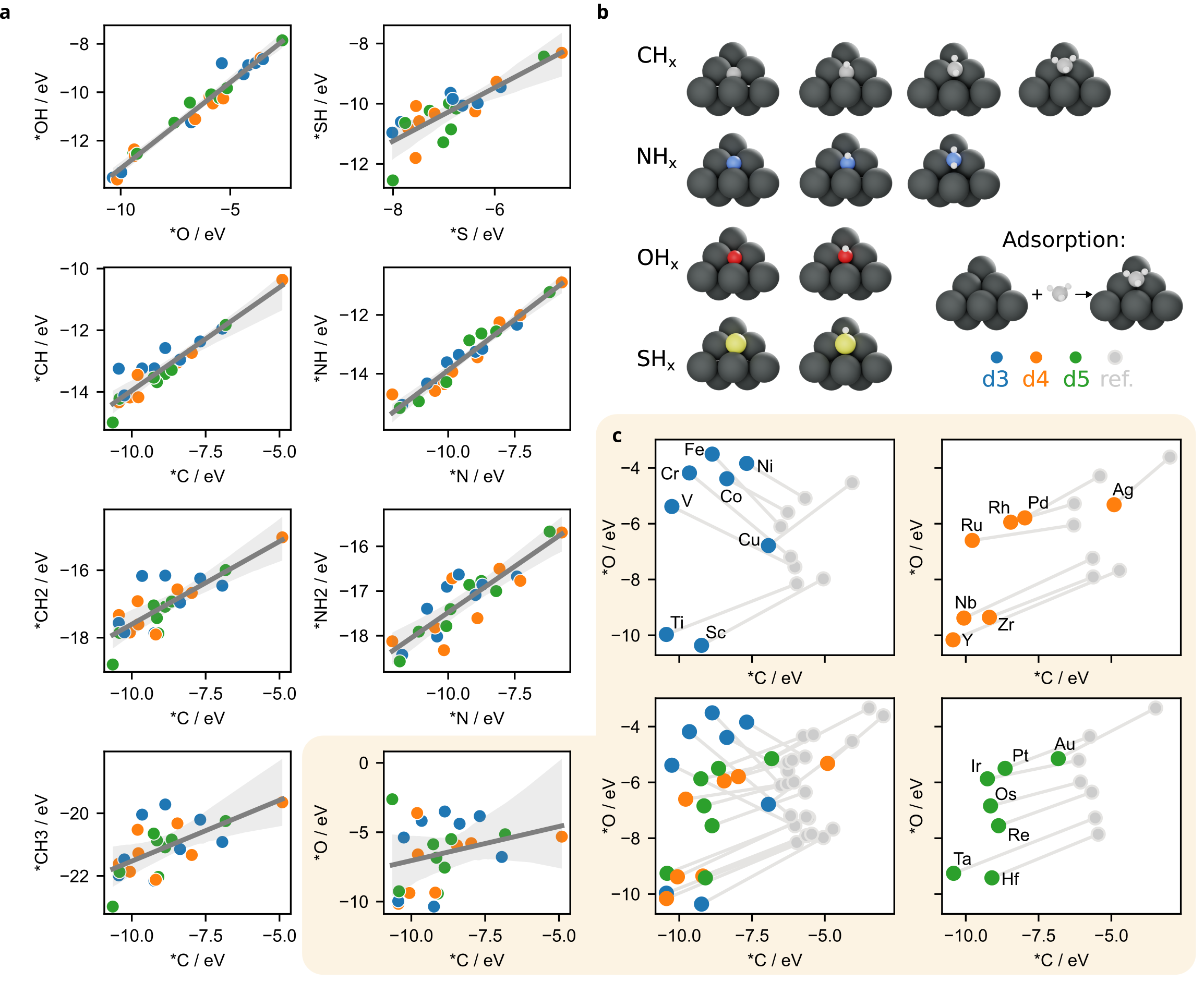}
  \caption{Correlation plots (a) between the adsorption energies of two intermediates at the same (hollow site) of tightly packed metal surfaces (b). The correlation between *O and *C (c) is not linear (in agreement with the literature).}
  \label{fig:catalysis_lsr}
\end{figure}

Adsorption energies of molecules and intermediates are indicative of catalyst reactivity and often used as descriptors in screening studies for catalyst materials. The adsorption energies are governed by electronic and geometric factors. Provided a consistent geometric environment (\textit{e.g.} a hollow site of a tightly packed metallic lattice) across different metal surfaces, the trend in adsorbate binding energies resulting primarily due to electronic effects can be observed. A well-known property of catalytic surfaces (\textit{e.g.} transition metals) is that binding energies of individual intermediates are not independent of each other, as a consequence of the varying degree of occupation of the metallic d-band~\cite{norskov2007linear,norskov2022nonlinear}. Linear scaling relationships (LSR) were found for a range of metallic surfaces and molecules that bind to this surface through the same atom (\textit{e.g.} E scales with EH$_\mathrm{x}$, where E = C, O, N, S and x = 1,2,3), however, this scaling is not linear when comparing adsorbates that bind through different atoms (\textit{e.g.} C versus O).

In \cref{fig:catalysis_lsr} we show the correlation between the adsorption energies of EH$_\mathrm{x}$, where E = C, O, N, S, and x = 0, 1, 2, 3. The structures (\cref{fig:catalysis_lsr}b) were relaxed with the {\MPz{}} model with D3 correction (cutoff = 4 nm), and the adsorption energy was computed as $\Delta E_{\textrm{ads}} = E(a*) - E(*) - E(a)$, with $a$ as the adsorbate and * as the empty surface site. The observed correlations are linear in all cases except for the correlation between \ch{O} and \ch{C}. In \cref{fig:catalysis_lsr}, the {\MPz{}} computed adsorption energies (blue/green/orange circles) are compared to the corresponding DFT values (connected with faint gray lines to faint gray circles) as reported by Norsk{\o}v~\cite{norskov2022nonlinear}. In this plot, although the absolute error of the obtained adsorption energies in comparison to the DFT values is high (which is not surprising as the model is extrapolating in this example), the trend of grouping metals into passive (noble, \textit{e.g.} Au) catalytic (so-called Pt group) and non-reducible (\textit{e.g.} Zr) is correctly captured and the essence of the LSR relationships was reproduced with the {\MPz{}}+D3 model.

\subsubsection{\ch{CO} (electro-)oxidation on \ch{Cu}}\label{sec:oxi}
\begin{figure}[ht!]
  \centering
  \includegraphics[width=0.9\textwidth,keepaspectratio]{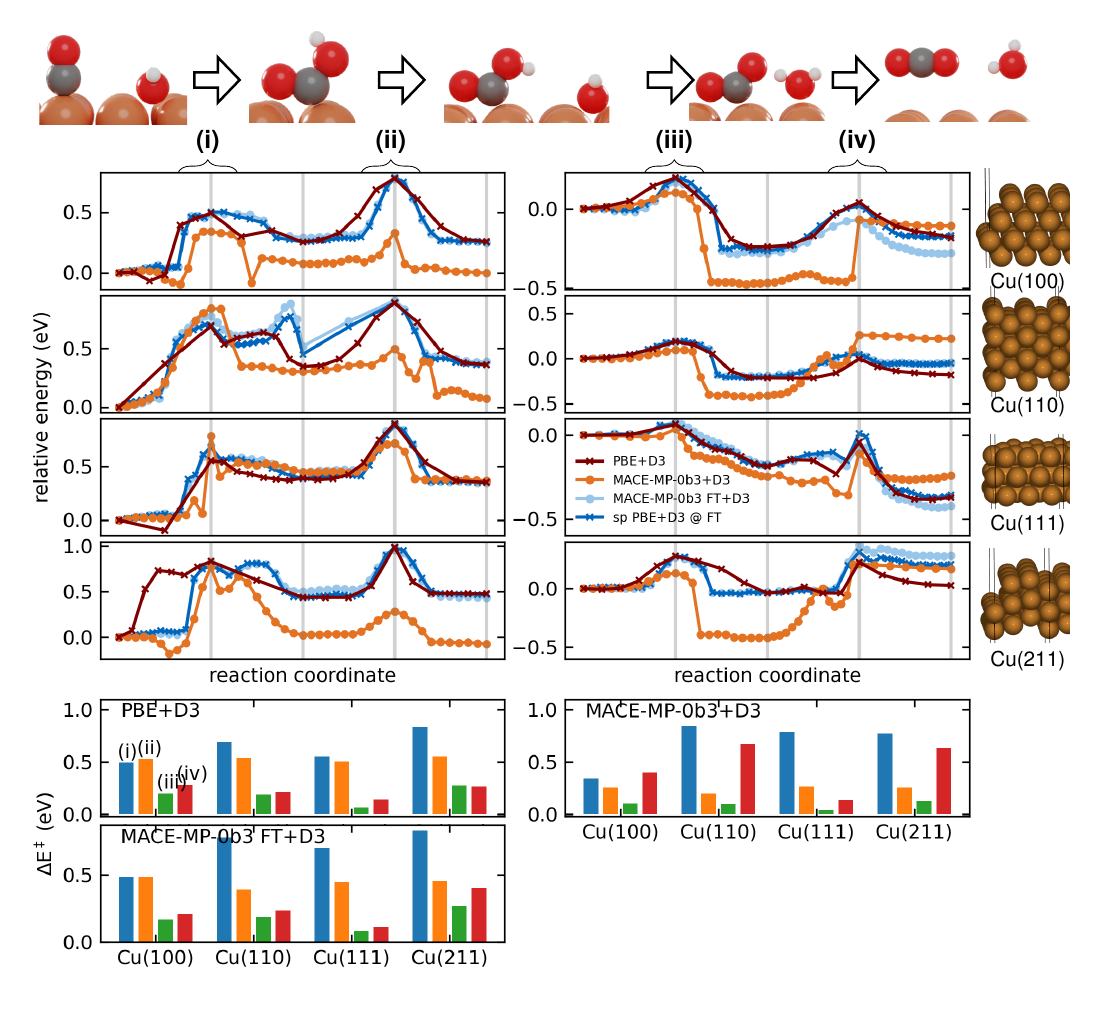}
  \caption{NEB profiles (top) and extracted barriers (bottom) of reactions (i-iv) computed for the potential independent steps in the multistep reaction mechanism of CO electro-oxidation on Cu for the low index facets (100), (110), (111), and (211). The reaction mechanism without implicit solvent is adapted from \cite{Tiwari2020}. The NEB calculations were carried out with \MPzbt{} (orange), a finetuned model (lightblue) and DFT (darkred), each with the D3 dispersion correction. Additionally, single point PBE+D3 calculations (deepblue) were performed for the finetuned NEB-path. Faint vertical lines indicate minima and saddle points along the reaction coordinate for reaction steps i-iv.}
  \label{fig:catalysis_eCOoxCu}
\end{figure}

We test the ability of \MPz{} to predict the catalytic reaction mechanisms for the oxidation of CO on different facets of Cu, as previously explored via DFT \cite{Tiwari2020}. Specifically, we evaluate two complex potential-independent reaction steps of the CO oxidation, OH* + CO* $\rightleftharpoons$ COOH* and OH* + COOH* $\rightleftharpoons$ CO$_2$(g) + H$_2$O(g) as shown in Fig.~\cref{fig:catalysis_eCOoxCu}. Each reaction contains two reaction barriers, labeled i, ii and iii, iv, respectively. Other reaction steps of the CO oxidation which we do not include in our evaluation, are the adsorption of CO* and  OH*, where the latter is an electrochemical process that can be described via the CHE \cite{Norskov2004-ow}. We recompute the reaction pathways for two Cu terraces (111) and (100) and two step-sites (110) and (211) via geometry optimizations of the initial and final states and subsequent NEB calculations (one for each reaction barrier i-iv) in lattice-parameter-adjusted simulation cells. 

The reaction profiles and reaction barriers from the converged NEB calculations are shown in Fig.~\cref{fig:catalysis_eCOoxCu} for PBE+D3, \MPz{}+D3, and fine-tuned model (+D3). Fine-tuning was performed via a training set of 70 structures which was collected by drawing every fifth \MPz{} NEB trajectory image and recomputing energies and forces with the computational settings of the \texttt{MPtrj} dataset. The foundation model \MPz{}+D3 shows in most cases qualitative agreement to the PBE+D3 reference, capturing the trends in barrier height differences for reaction steps i-iv. Quantitatively, the \MPz{} barriers are underestimated, in line with systematic softening behavior noted in \cite{deng2025systematic}. Except to this are the cases of CO$_2$ and H$_2$O desorption (vi) for the (110) and (211) facet. Here, the barriers are overestimated due to an underestimated final binding energy and (relatedly) a different final state geometry. In contrast, the fine-tuned model shows near-quantitative agreement in terms of the barrier heights, deviating on average by only 0.056 eV). This close agreement does not become immediately apparent when comparing the NEB profiles which show some deviations. This can be attributed to a different number of images in the climbing image NEB calculations (43 for MACE and 19 for PBE+D3). To better demonstrate the accuracy of the fine-tuned MACE, we additionally performed PBE+D3 single point calculations for images of the fine-tuned NEB path, illustrating excellent agreement with the PBE-D3 energies.

\subsubsection{\ch{In2O3}}\label{sec;in2o3}

As a final test system we investigate a key step (\ch{CH2O2 -> CH2 + O}) in carbon dioxide hydrogenation to methanol over indium oxide via an NEB transition state search. This reaction has been extensively studied with \textit{ab initio} methods due to indium oxide's promising selectivity compared to conventional modified copper catalysts~\cite{dang2020rationally, schaaf2023accurate}. First, we perform a global geometry optimization of the reactant near an oxygen vacancy. \MPz{} correctly identifies the three-oxygen-coordinated indium as the active site~\cite{schaaf2023accurate}. Following a NEB calculation \MPz{} predicts the reaction barrier within 15\% of that investigated with DFT (\SI{1.16}{eV} vs.\ \SI{0.98}{eV}), as visible in \cref{fig:cat_main}. Fine-tuning with just five single-point DFT calculation recovers the barrier with quantitative accuracy.%

\subsubsection*{Similarity statement} 
With the exception of the \ch{CuO} bulk Pourbaix diagram, which is based on structures from the Materials Project, all presented examples treat surface slab models (with and without reacting adsorbates) and represent a significant extrapolation. The dataset does not include any such slab models but only related bulk structures. These bulk structures include 6 bulk structures that contain Pt, O, H and 15 bulk structures that contain Pt and O for the Pt Pourbaix diagram and the corresponding LSR example (similar number for other metals in the LSR), 8 different Cu-bulk phases and 111 structures composed of Cu, O, Cu, and H along with other elements for the example of CO oxidation on different Cu facets, and 9 structures that contain In, O, H and C and a 825 bulk structures that contain indium and oxygen for the \ch{In2O3} example. The most similar configurations for the CO oxidation example are \ch{Cu2H4C4N3}O with the Materials Project ID mp-686268 and for the \ch{In2O3} example \ch{(NH4)In(OH)PO4} with the ID mp-764968. We provide \verb|Pt_LSR.json| and \verb|COoxCu_closest_training_points.csv| for the LSR and CO oxidation example to help visualize the interactive UMAP on \url{chemiscope.org}.

\subsubsection*{Performance summary}
Solid energies accurately predicted leading to correct Pourbaix diagrams. Adsortion energies overestimated, but linear scaling relationships between different surface/adsorbate pairs preserved. Minimum energy paths for reactive steps qualitatively correct, in some cases small (\SI{0.2}{eV}) in other larger (\SI{0.5}{eV}) energy errors. 

\clearpage
\subsection{Carborane rearrangement}\label{sec:carborane}
\begin{figure}[!ht]
  \centering
  \includegraphics[width=\textwidth,keepaspectratio]{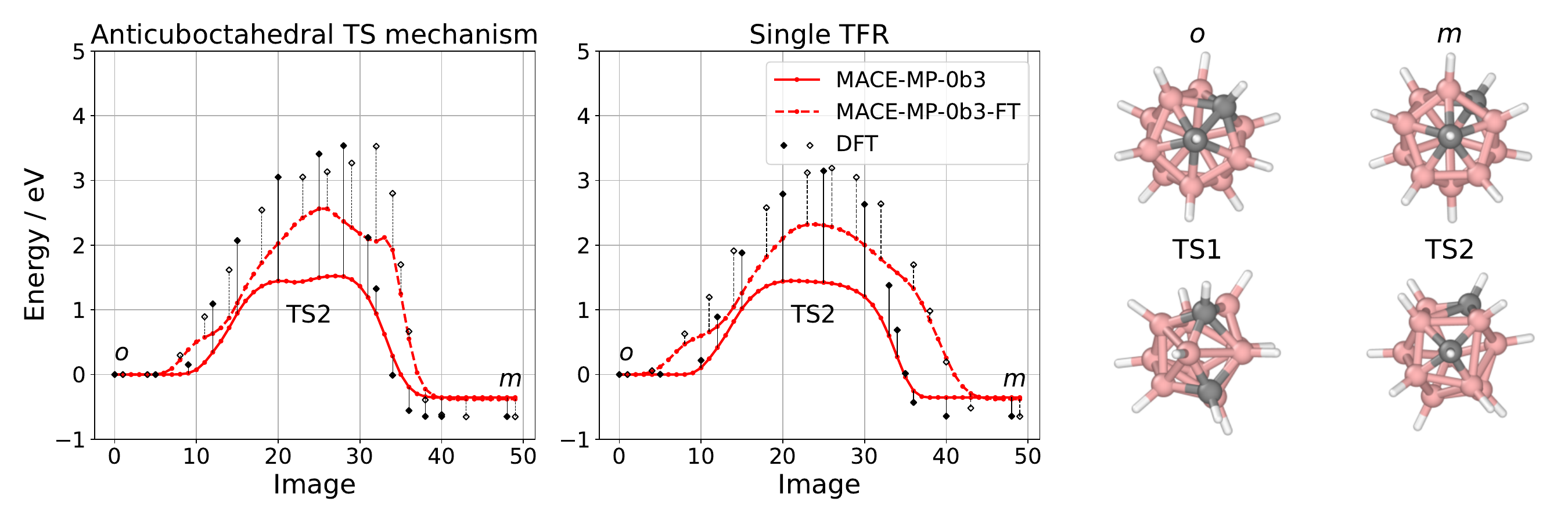}
  \caption{\MPzbt{} and \MPzbt{}-FT NEB minimum energy paths for the isomerization reaction of \textit{ortho}-carborane to \textit{meta}-carborane. DFT energies, evaluated on selected structures, are shown. All energies are shown relative to the \textit{ortho} isomer energy. On the right, end point and transition state structures are illustrated.}
  \label{fig:carborane}
\end{figure}

Carborane (\ch{C2H12B10}) is an organoboron compound with uses in drug discovery \cite{Marfavi2022} and organometallic chemistry \cite{Sivaev2000}. It adopts icosahedral cluster structures, with three isomers based on different relative positions of carbon atoms: \textit{ortho} (\textit{o}), \textit{meta} (\textit{m}), and \textit{para} (\textit{p}). The thermally activated rearrangements between these isomers have been thoroughly studied \cite{Brown2006}, with several mechanisms proposed involving triangular face rotation (TFR). We used \MPzbt{} to study two pathways from the \textit{ortho} isomer to the \textit{meta} isomer: one involving an anticuboctahedral transition state (by mutual rotation of two opposite faces), the other involving the rotation of a single triangular face.

We used \MPzbt{} to obtain rearrangement pathways for each mechanism. The \textit{o} and \textit{m} isomer structures were relaxed with a force tolerance of \SI{0.01}{eV\per\Angstrom} using \MPzbt{}. Pathways were obtained using nudged elastic band with 50 images, relaxed with a force tolerance of \SI{0.05}{eV\per\Angstrom}, first with spring constants of \SI{0.5} and then \SI{0.05}{eV\per\square\Angstrom}. We also evaluated energies of the images at the PBE/def2-TZVPPD level of theory using ORCA 5.0.3 \cite{Neese2022}.

\subsubsection*{Fine-tuning}

We finetuned \MPzbt{} using the \textit{ortho}, \textit{meta}, \textit{para} structures and 7 high-energy structures from various pathways connecting these isomers. \MPzbt{} and \MPzbt{}-FT pathways, along with some reference DFT energies, are shown in \Cref{fig:carborane}. \MPzbt{}-FT gave larger energy barriers than \MPzbt{}: 2.56 eV vs 1.52 eV for the anticuboctahedral mechanism and 2.32 vs 1.45 eV for the single TFR mechanism. While fine-tuning improved energy errors in the high-energy regions, the reaction energy errors were not improved significantly. The reference DFT reaction energy is -0.647 eV while \MPzbt{} and \MPzbt{}-FT predicted -0.355 and -0.380 eV, respectively.

\subsubsection*{Similarity statement}

The training set contains 5 structures composed of H, B, C elements, and 21837 structures that have H or B or C along with any other elements. The dataset contains 2 structures or containing icosahedral \ch{C2H11B10} clusters linked by a C-C bond to form a dimer, 1 icosahedral borane \ch{B12H12} cluster and 18 other borane clusters containing trigonal \ch{B3} faces. Along these, the dataset contains several hundred structures involving derivatives of borane and carborane clusters such as salts, metal complexes, and halogenated species. The closest structures in the training set are mp-1194548 (\ch{C4H22B20} containing two icosahedral carborane clusters joined with a C-C bond), and metal complexes containing carborane ligands: mp-759303, 705569, 1199795, 1198024.

\subsubsection*{Performance summary}
Carborane rearrangement pathways found with \MPzbt{} are qualitatively correct but do not lie on the reference DFT minimum energy path, as shown by the \textgreater 1 eV energy errors relative to DFT in the high energy regions of the pathway. 

\clearpage
\subsection{Transition Metal Dichalcogenides}\label{sec:tmd}

\begin{figure}[!ht]
  \centering
  \includegraphics[width=\linewidth,keepaspectratio]{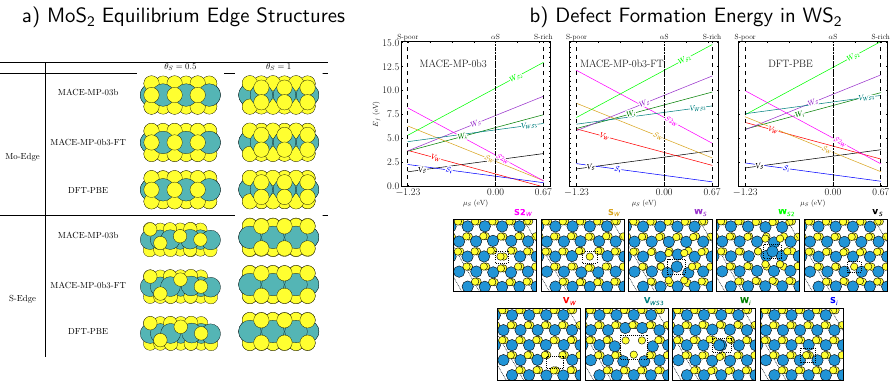}
  \caption{Panel (a) shows four \ch{MoS2} edge models of different type and sulfur coverage ($\theta_S$). DFT-predicted structures are obtained from \cite{rosen2018comprehensive}. Geometry optimizations using \MPzbt{} and \MPzbt{}-FT starting from DFT-optimized structures are shown. Panel (b) compares \ch{WS2} defect formation energies predicted by \MPzbt{}, \MPzbt{}-FT, and DFT-PBE results from \cite{kieczka2023defects}. Geometries of defects considered here as relaxed using \MPzbt{} are also shown in panel (b).}
  \label{fig:TMDs}
\end{figure}

Edges in \ch{MoS2} and TMDs more broadly are known to be sites of high reactivity with much relevance to TMD-catalyzed reactions and material aging studies. Studies have recently addressed the structure of \ch{MoS2} edges under various conditions using DFT \cite{rosen2018comprehensive}. Here, we examine stable Mo-edge and S-edge configurations with sulfur coverage $\theta_S = 0.5$ and $\theta_S = 1$ from \cite{rosen2018comprehensive}. The edge models are multi-layer models, i.e stacked infinite stripes. \Cref{fig:TMDs}a shows the \ch{MoS2} edge structures considered. Geometry optimization  performed on the DFT-optimized structures using \MPzbt{} and \MPzbt{}-FT generally preserved the configurations. One notable deviations was observed for the \MPzbt{} model prediction. The $\theta_S=0.5$ S-edge configuration slightly deviated from the DFT-predicted zigzag structure which was corrected following fine-tuning. To assess the extent of deviation, the $\theta_S=0.5$ S-edge was further geometry-optimized using DFT, quickly converging in 9 steps to the DFT minimum which was \SI{1.6}{eV} lower in energy.
We further assessed the MD stability of \MPzbt{} + D3 dispersion on these structures by running \SI{100}{ps} of MD (NVT ensemble) at \SI{300}{K}. This was also repeated using the fine-tuned model \MPzbt{}-FT. The MD simulations were found to be stable for the edge configurations above, using both models. Select MD trajectories can be found in supplementary files. 

Defects have a significant impact on the optical and electronic properties of 2D TMDs and come in various types including vacancies. We assess the ability of \MPz{} to describe defect formation energies ($E_f$) of various defects in \ch{WS2} as compared to PBE results from \cite{kieczka2023defects}. The formation energy is calculated using the formula:
\begin{equation}
  E_{\text{f}} = E_{\text{defect}} - E_{\text{pristine}} - \sum \Delta n_i \mu_i
\end{equation}
where \(E_{\text{defect}}\) and \(E_{\text{pristine}}\) are energies of the \ch{WS2} with and without defects, \(n_i\) and \(\mu_i\) are the number of atoms and chemical potential of element \(i\). The chemical potential of \ch{S} obeys the equilibrium condition $\mu_{\ch{WS2}} = \mu_{\ch{W}} + 2 \mu_{\ch{S}}$, and is bounded by predefined S-poor and S-rich conditions: $\mu_S^{\text{bulk}_{\ch{W}}} \leq \mu_S \leq \frac{1}{2}\mu_S^{\ch{S2}}$. It is calculated with respect to the \(\alpha\)-S as the reference state \cite{kieczka2023defects}. \Cref{fig:TMDs}b shows $E_F$ values predicted by \MPzbt{}, \MPzbt{}-FT and DFT are in qualitative agreement with regards to trend. \MPzbt{}-FT showed significant improvements in quantitative agreement  with DFT-PBE particularly for high-energy defects such as $V_{WS_3}$ and $V_W$.

\subsubsection*{Fine-tuning}

We finetuned \MPzbt{} on a collection of data comprising both \ch{MoS2} and \ch{WS2} structures. For \ch{MoS2}, 60 configurations were added in total. These comprise 15 snapshots from \MPzbt{} MD runs of each of the four edge models. For \ch{WS2}, 45 configurations in total were added which include three rattled copies of each of the defects and reference state configurations: pristine \ch{WS2}, $\alpha$-sulfur, tetragonal \ch{WO3}, body-centered cubic tungsten, \ch{S2} and \ch{SO2}. 

\subsubsection*{Similarity statement}

The majority of the \ce{Mo-S} or \ce{W-S} containing structures in the database include primitive units of layers of \ch{MoS2} or \ch{WS2}. However, clusters of \ce{Mo-S} and \ce{W-S} were found as well. Edge models for \ch{MoS2} are rare, with three structures identified (mp-990083, mp-989179 and mp-990086). These include a variant of the $\theta_S = 1$ S-edge and the $\theta_S = 0$ Mo-edge. No defect models of \ch{WS2} were found in the dataset.

\subsubsection*{Performance summary}
Geometric reconstruction of nanoribbon edges mostly correct, apart from small deviations. Ordering of defect formation energies qualitatively correct with an overall tendency to underestimate precise values. 

\clearpage
\subsection{Electrode-electrolyte interface / Battery system}\label{sec:battery}

\begin{figure}[ht]
    \centering
    \includegraphics[width=\textwidth,keepaspectratio]{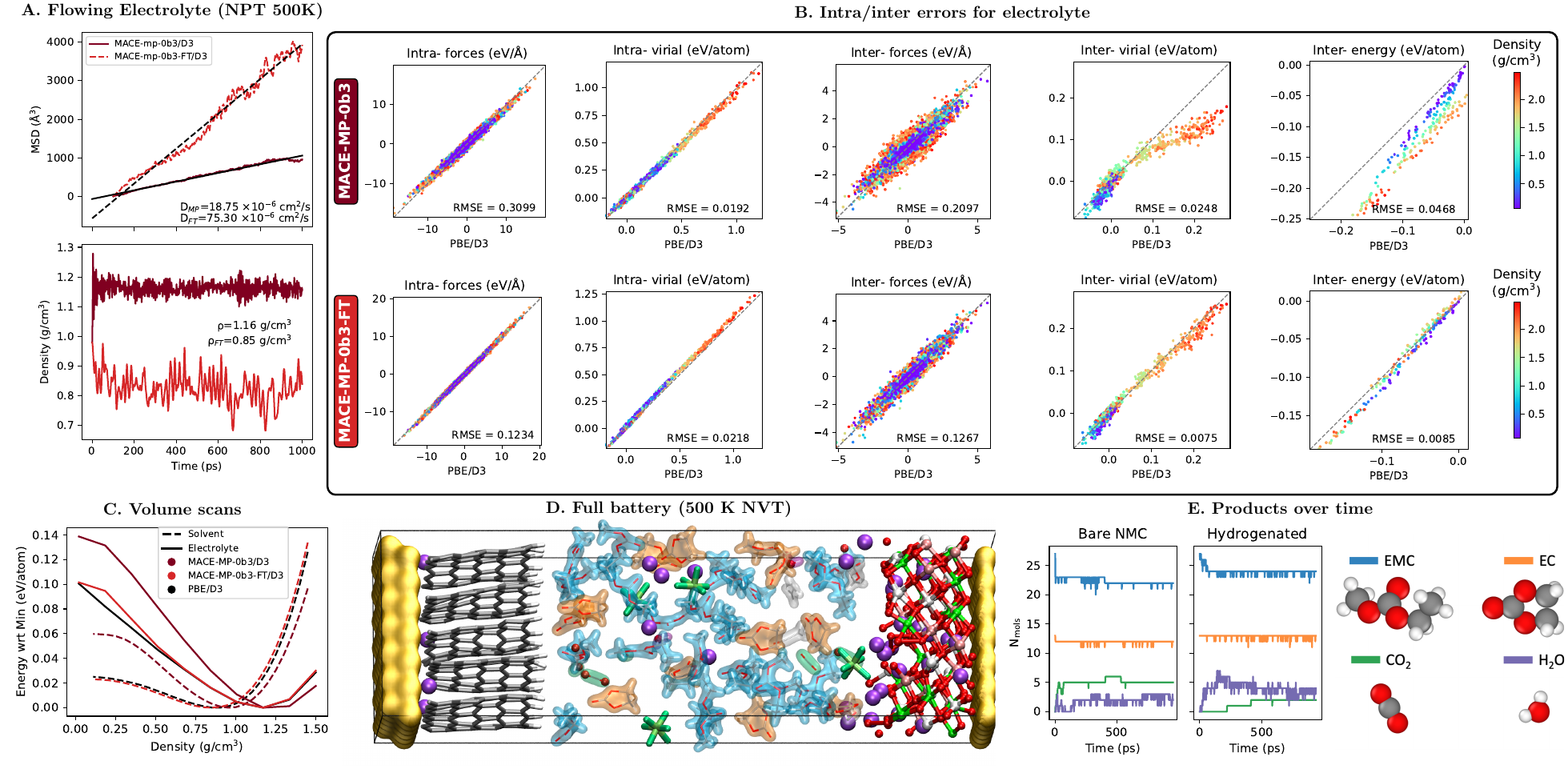}
    \caption{(A) Flowing EC/EMC \ch{LiPF6} liquid electrolyte with stable density and intact molecules at \SI{500}{\kelvin} NPT MD. (B) \MPzbt{} and \MPzbt{}-FT Intra-/Inter-molecular energy / forces / virial evaluated on independent PBE test set spanning all possible compositions of EC/EMC LiPF$_6$ electrolyte at densities between \num{0.1}--\SI{2.5} {\gram\per\centi\meter\cubed}. Note that the 10 FT configurations were randomly selected from this set of 200 configurations. (C) Rigid-molecule volume-scan test \cite{magduau2023machine} for both neat EC/EMC solvent and full EC/EMC \ch{LiPF6} electrolyte, compared to PBE. (D) Full battery simulation (Cu | H-capped graphite+Li | EC/EMC+\ch{LiPF6} | NMC+Li \cite{fang2021formation}), final snapshot of \SI{500}{\kelvin} \MPzbt{} NVT-MD, showing degraded solvent (grey iso-surface molecules), new \ch{CO2} (green iso-surface) and \ch{H2O} (purple iso-surface) molecules, oxygen atoms originating from the cathode floating in the electrolyte (solid red spheres). (E) Time progression of the predominant molecular species in the electrolyte for the two NVT simulation settings (1. neat NMC, 2. H-capped NMC).}

\end{figure}

Atomic-level interactions between the electrodes and electrolyte play a crucial role in determining the performance of electrochemical devices, including batteries, fuel cells, and electrocatalysts. Understanding these interactions is essential for optimizing the energy storage, conversion, and catalytic properties of these devices and to this end molecular modelling plays a crucial role. The remaining challenge is that processes underpinning transport and degradation in these devices take place on a long time scale, inaccessible to ab initio simulations. MLIPs are ideally suited to bridge this gap, bearing in mind that these complex materials and heterogeneous interfaces cover an extensive chemical space which poses a big challenge to ML models. Here we test the performance of \MPzbt{} on two separate systems -- pure (EC/EMC \ch{LiPF6}) electrolyte and the complete battery including the copper interface, anode, electrolyte and NMC cathode (totalling 9 chemical elements). Additionally, we test a fine-tuned model, \MPzbt{}-FT, for the pure electrolyte. We performed MD simulations at 500 K using the models plus the D3 correction to stress-test the qualitative robustness of the potentials. Further, we quantitatively assessed the potentials on a separate 200-config PBE test set in order to establish the accuracy of describing intra- and intermolecular interactions.

Previous work~\cite{magduau2023machine,niblett2024transferability,ju2025application} has shown that modelling even the neat solvent is a challenge to MLIPs owing to the weak, but crucially important, inter-molecular interactions. Here we find that \MPzbt{} is stable in the NPT ensemble at \SI{500}{\kelvin}, the density is preserved while the electrolyte (solvent+salt) remains liquid and all molecules remain intact for the entire duration of the simulation (\SI{1}{\nano\second}). A second NPT simulation was performed with \MPzbt{}-FT -- a model fine-tuned on 8 electrolyte and 2 pure solvent configurations with densities between \num{0.1}--\SI{2.3} {\gram\per\centi\meter\cubed} that were selected from the 200-config PBE data set. The dynamics was found to be similarly robust with stable density and intact molecules, however the densities and diffusivities found with the fine-tuned model are more physically realistic, meaning the electrolytes is less dense and and significantly more diffusive, as expected at \SI{500}{\kelvin}.  The accuracy of both models was independently tested on the 200-config PBE test set of electrolytes spanning all physical compositions and densities (will be published elsewhere). The \MPzbt{}-FT model yields lower RMSEs on both intra- and inter-molecular properties, demonstrating that fine-tuning can significantly specialize the potential with minimal number of new training data points (10 configs). That \MPzbt{}-FT reproduces inter-molecular interactions better than \MPzbt{} is also clearly illustrated on the rigid-molecule Volume Scans \cite{magduau2023machine} test. \MPzbt{}-FT accurately describes the PES as compared to PBE for both the neat solvent and the full electrolyte. Note that the potential correctly captures the $1/r$ Coulomb scaling of the energy up to large volume (low density), despite its short-range nature.

The full battery was simulated by \MPzbt{} in the NVT ensemble, since the volume of the entire system was not stable in NPT simulations (possibly the result of large compressibility differences along the x-direction normal to the liquid electrolyte layer vs the y/z-directions along the solid electrode slabs). All NVT simulations were stable at \SI{500}{\kelvin} for the entire simulation time (\SI{1}{\nano\second}). The electrolyte \ch{Li}-ions were found to deintercalate from both the graphite anode and NMC cathode and the electrolyte was mobile. The H-capped graphite was found to be inert, whereas the cathode-electrolyte interface exhibited pronounced reactivity. Evident from the start of the simulation was the extensive proton transfer from the carbonate solvent (EMC in particular) to the oxygen atoms in NMC. This in turn led to continuous breakdown of solvent molecule (which became a radical) and chemisorption onto the cathode surface, possibly demonstrating the initial steps of SEI formation. Notably, substantial amounts of \ch{CO2} and \ch{H2O} were generated in the process. Furthermore, oxygen atoms were easily extracted from the cathode leaving behind binding sites for the oxygen-rich carbonate molecules. A separate simulation setting was tested where the exposed oxygen atoms of the cathode were hydrogenated before the simulation. Similar reactivity was observed albeit with different outcomes, notably more water molecules and less carbon dioxide was generated in the process. These early simulations demonstrate \MPzbt{} is robust for battery interfaces and showcase the initial steps in modelling the SEI formation with ab initio accuracy -- which has been a long-held dream of the scientific community.

\subsubsection*{Similarity statement}

To perform the similarity analysis, 100 representative (decorrelated) structures were taken from the previously described simulations. While the MP dataset does not contain liquids, for all three simulations the UMAP analysis showed that all atomic environments were well represented in the MP dataset. More specifically, the pure electrolyte was found to contain environments close to mp-995234 and mp-995218 which correspond to \ch{HCO} and \ch{H_4C_5O_2}, respectively. These configurations were also found to be similar for the snapshots obtained by running the electrolyte-anode interface, and also notably included mp-707412 corresponding to \ch{H22C10O3}. For the entire battery system configurations such as mp-1194779 and mp-698267 were found to be similar which correspond to \ch{CuH_3C_3O_4} and \ch{CoHCO_3} respectively. We provide:
\begin{itemize}
\item \verb|interface_chemiscope_input.json|
\item \verb|battery_chemiscope_input.json|
\item \verb|electrolyte_chemiscope_input.json|
\end{itemize}
to help visualize the atomic environments against the MP dataset at \url{chemiscope.org}.

\subsubsection*{Performance summary}
Fine-tuning considerably improved solvent and electrolyte property (density, diffusivity) prediction. Full battery simulations were stable at fixed volume but unstable with variable volume, and show electrolyte reactions at unpassivated electrode.

\clearpage
\subsection{Metal--organic frameworks}\label{sec:si_mofs}

\begin{figure}[http!]
 \centering
 \begin{subfigure}{0.45\textwidth}
 \includegraphics[width=\textwidth,keepaspectratio]{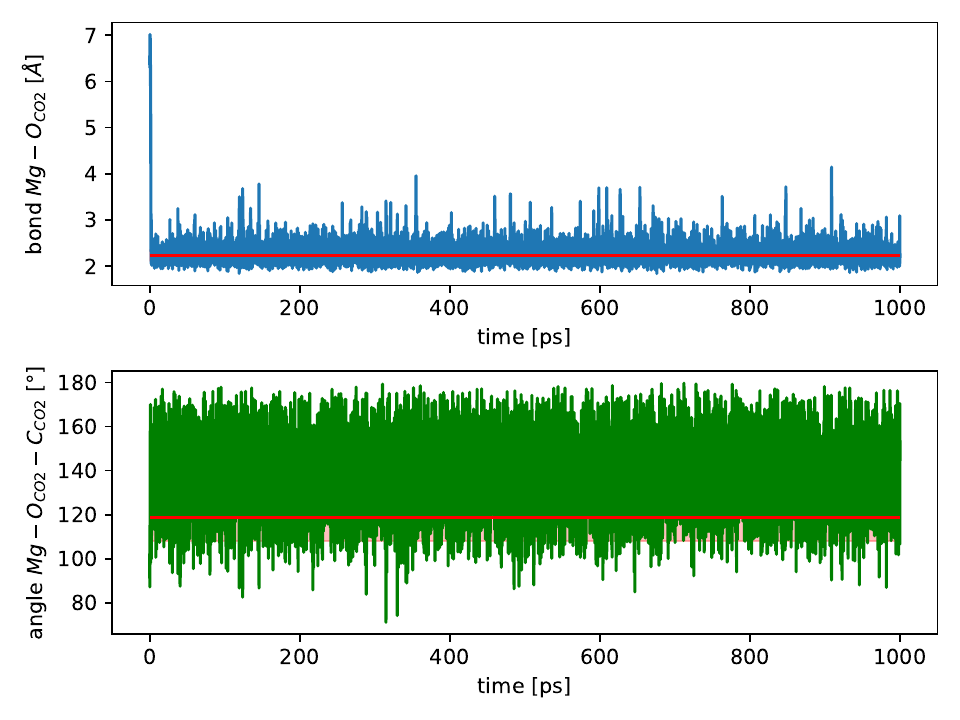}
 \caption{}
 \label{fig:mof_si_a}
 \end{subfigure}
 \begin{subfigure}{0.5\textwidth}
 \includegraphics[width=\textwidth,keepaspectratio]{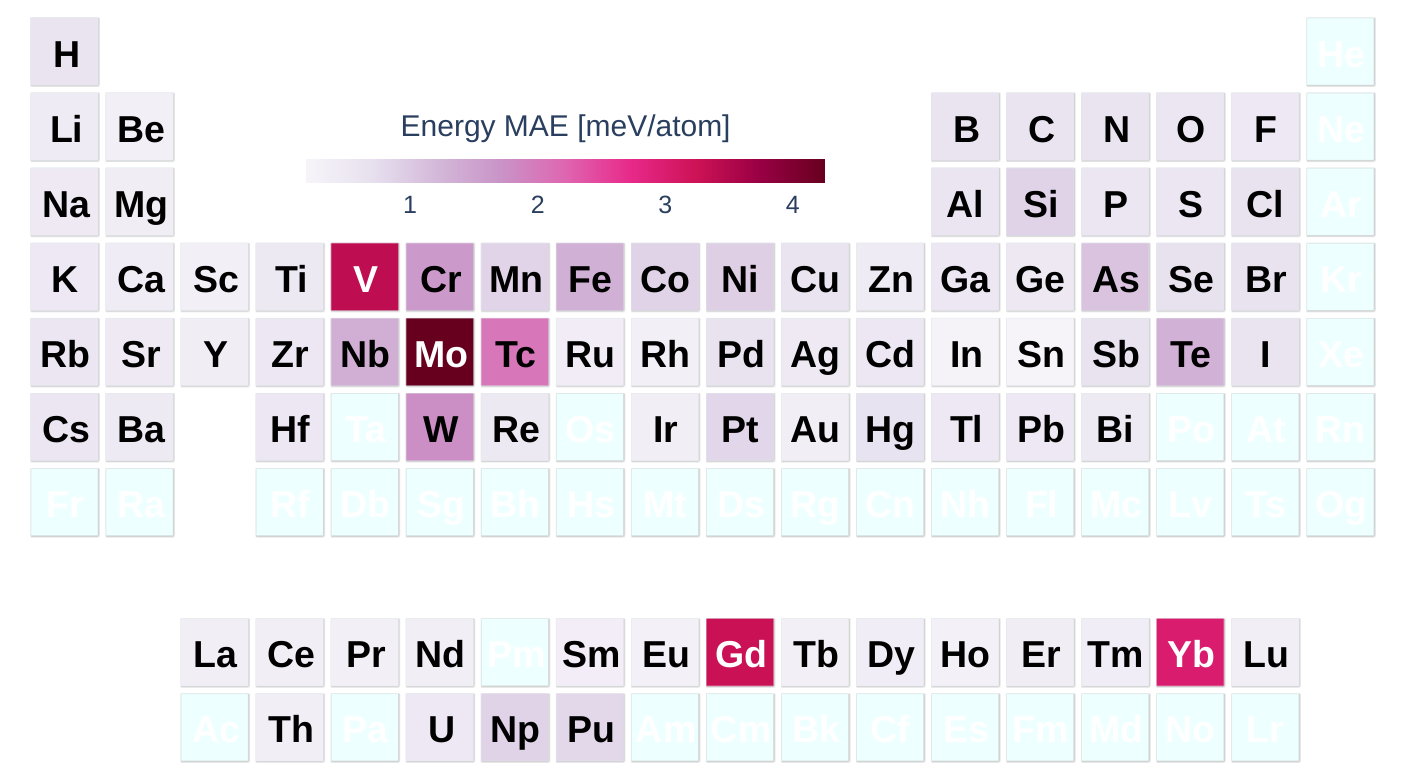}
 \caption{}
 \label{fig:mof_si_b}
 \end{subfigure}
 \begin{subfigure}{0.45\textwidth}
     \includegraphics[width=\textwidth,keepaspectratio]{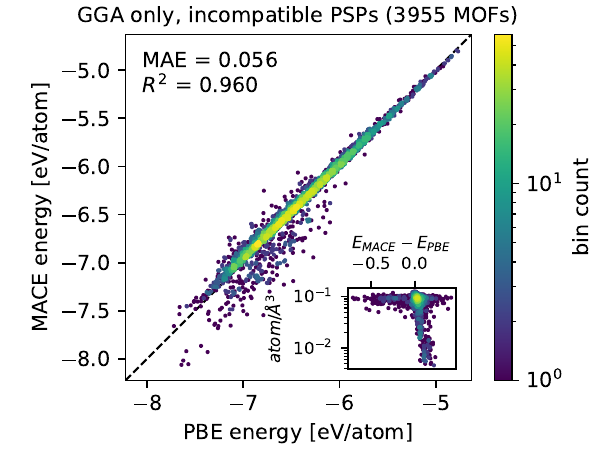}
     \caption{}
     \label{fig:mof_si_incompt_psps}
 \end{subfigure}
 \begin{subfigure}{0.45\textwidth}
     \includegraphics[width=\textwidth,keepaspectratio]{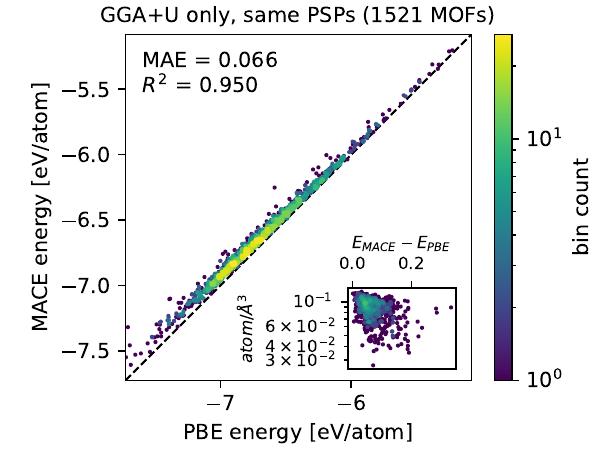}
     \caption{}
     \label{fig:mof_si_ggapu}
 \end{subfigure}
 
  \caption{(a) Top panel: instantaneous distance between the Mg center and the closest oxygen from \ch{CO2} in Mg-MOF-74, the red line shows the predicted average value from previous work. \cite{zeng2023deepmd} Bottom panel: instantaneous angle between Mg center, closest oxygen and carbon in the \ch{CO2}, red line indicates an average value from previous work from Ref.~\cite{zeng2023deepmd} (b) Elemental dependence of energy error distribution over the periodic table on complete QMOF dataset (20,375 relaxed structures) \cite{rosen2021machine,rosen2022high}, correlating with incompatible pseudopotentials used between MPtrj training and QMOF test sets (\Cref{tab:mp-qmof-potcars}). The absolute error of total energy was equally distributed to all atoms in the single MOF structure, and the average of each absolute error across different MOF structures were averaged for each element. (c) \MPzbt{} energy predictions compared to QMOF GGA-PBE calculations using pseudopotentials (PSPs) incompatible with the MPTrj. Structures that would be calculated at GGA-PBE+U level by MP standard, \textit{i.e.} structures containing oxygen (or fluorine) and any element of V, Cr, Mn, Fe, Co, Ni, W, or Mo, are excluded. (d) \MPzbt{} energy predictions compared to QMOF GGA-PBE calculations on \textit{only} the structures that would be calculated at GGA-PBE+U level by MP standard. QMOF structures calculated with incompatible PSPs are excluded.}
  \label{fig:mof_si}
\end{figure}

\subsubsection{QMOF}

Given that \MPz{} is pretrained against PBE, whereas QMOF was constructed from PBE-D3(BJ) calculations to account for dispersion corrections \cite{grimme2010consistent,grimme2011effect}, in the comparison with QMOF, we compare \MPz{} predicted energies with QMOF PBE energies by subtracting the dispersion correction from total QMOF energies.

We noted that most of the high-energy error MOF structures have high atomic density (\Cref{fig:mof_si_incompt_psps} inset), and that the errors cannot be canceled by adding dispersion correction. In \Cref{fig:mof_si_b}, we analyze the element-wise energy error per atom by distributing the absolute energy error per atom to the constituent elements by the corresponding composition in each MOF structure. As presented in \Cref{fig:mof_si_b}, there is a strong elemental dependence of energy error per atom. Most of high error elements can be attributed to the difference of chosen pseudopotentials used by MP and QMOF databases (in particular the choices of which electrons are treated as valence states), see \Cref{tab:mp-qmof-potcars}.

\Cref{fig:mof_si_incompt_psps} and \Cref{fig:mof_si_ggapu} present the incompatible QMOF entries excluded from the analysis in \Cref{fig:mof}a. The incompatibility stems from (i) different, incompatible pseudopotentials (PSPs), or (ii) mismatched calculation parameters (with or without Hubbard U correction) used in QMOF and MP database, or the combinations of both. \Cref{fig:mof_si_incompt_psps} compares MACE energies with QMOF PBE energies calculated with incompatible PSPs (case i). \Cref{fig:mof_si_ggapu} compares MACE energies with QMOF PBE calculations on structures that would be calculated at PBE+U level by MP standard at the curation time of MPTrj.

\begin{table}[htp!]
  \centering
  \begin{adjustbox}{width=\textwidth}
    \renewcommand{\tabcolsep}{3pt}
    \begin{tabular}{c|cccccccccccccccccc}
      \textbf{MPtrj} & \verb|Be|   & \verb|Bi|   & \verb|Cu_pv| & \verb|Eu|   & \verb|Fe_pv| & \verb|Gd|   & \verb|Li_sv| & \verb|Mg_pv| & \verb|Mo_pv| & \verb|Nb_pv| & \verb|Ni_pv| & \verb|Os_pv| & \verb|Re_pv| & \verb|Ti_pv| & \verb|V_pv| & \verb|W_pv| & \verb|Yb_2| \\
      \hline
      \textbf{QMOF}  & \verb|Be_v| & \verb|Bi_d| & \verb|Cu|    & \verb|Eu_3| & \verb|Fe|    & \verb|Gd_3| & \verb|Li|    & \verb|Mg|    & \verb|Mo_sv| & \verb|Nb_sv| & \verb|Ni|    & \verb|Os|    & \verb|Re|    & \verb|Ti_sv| & \verb|V_sv| & \verb|W_sv| & \verb|Yb_3| \\
    \end{tabular}%
  \end{adjustbox}
  \caption{Difference in VASP POTCARs used by MPtrj and QMOF}
  \label{tab:mp-qmof-potcars}
\end{table}

\subsubsection{\ch{CO2} adsorption}

All the calculations for the \ch{CO2} dynamics with MOFs were performed with \MPz{} by adding the D3 dispersion correction\cite{grimme2010consistent} to the \MPz{} potential. The simulations were carried out with ASE \cite{larsen2017atomic} on a cell containing 165 atoms with one \ch{CO2} molecule, initialised at the centre of the pore, using NVT Langevin dynamics~\cite{VANDENEIJNDEN2006} with a friction factor of \SI{5e-3}{\per\femto\second}. The temperature was set to \SI{600}{K} with a time step of \SI{1}{\femto\second}. Twenty-four \SI{1}{\nano\second} trajectories were generated using different initial velocities, and all quantities presented were averaged over all of them, discarding the first \SI{2}{\pico\second} from each to account for equilibration.  All structures had their cells and positions optimised at start using {\tt FrechetCellFilter} from ASE. The code used to generate the trajectories is available in the repo~\cite{elena2023} and ~\cite{elliott_2024_14001356}.

\subsubsection*{Similarity statement}

The MP dataset does not contain MOFs. There are 8 structures containing all \ch{MgOCH} elements, and 62 structures that have \ch{MgOHC} elements on their own or along with other elements. Based on UMAP analysis, we see that most atomic environments, both \ch{MgO} and linkers, in the example system, are similar to environments in the training data but none is Mg-MOF-74 specific.

The closest (most relevant) structures in the training set are \ch{CO2} (mp-556034 mp-20066 mp-995224 mp-11725 mp-644607 mp-1102227 mp-1190685 mp-995198 mp-1190699 mp-1077906 mp-1077316 mp-729728). \ch{CO2} alone matches 4896 structures with \ch{C}, \ch{O} and alongside other elements.

We provide 
\begin{itemize}
    \item \verb|co2_FilterType.exclusive_OC_chemiscope_input.json|
    \item \verb|mg-mof-74-co2_FilterType.exclusive_MgOCH_chemiscope_input.json| 
    \item\verb|mg-mof-74-co2_FilterType.exclusive_MgOCH_chemiscope_input.json| 
\end{itemize}
        that contain exact matches of \ch{Mg}, \ch{O}, \ch{C} and \ch{H}  and the inclusive versions 
\begin{itemize}
    \item \verb|mg-mof-74-co2_FilterType.exclusive_MgOCH_chemiscope_input.json|
    \item \verb|mg-mof-74-co2_FilterType.inclusive_MgOCH_chemiscope_input.json|  
    \item \verb|mg-mof-74_FilterType.inclusive_MgOCH_chemiscope_input.json|
\end{itemize}
to help visualize the interactive UMAP on \url{chemiscope.org}.

\subsubsection*{Performance summary}
Excellent energy prediction for large database of MOFs. Correct prediction of binding structure and free energy of \ch{CO2} in Mg-MOF-74. 


\clearpage

\subsection{Combinatorial Materials Discovery}\label{sec:materials discovery}

\subsubsection{Formation energy of hypothetical materials}\label{sec:formation}

\begin{figure}[htbp!]
  \centering
  \includegraphics[width=0.49\linewidth]{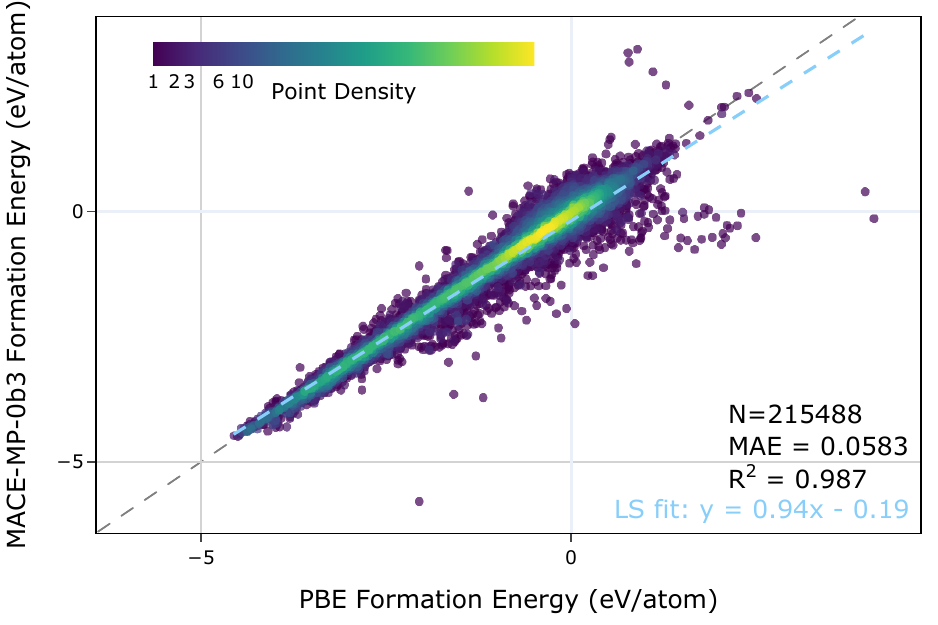}
  \label{fig:e-form-parity-mace-mp-0b3}
  \hfil
  \includegraphics[width=0.49\linewidth]{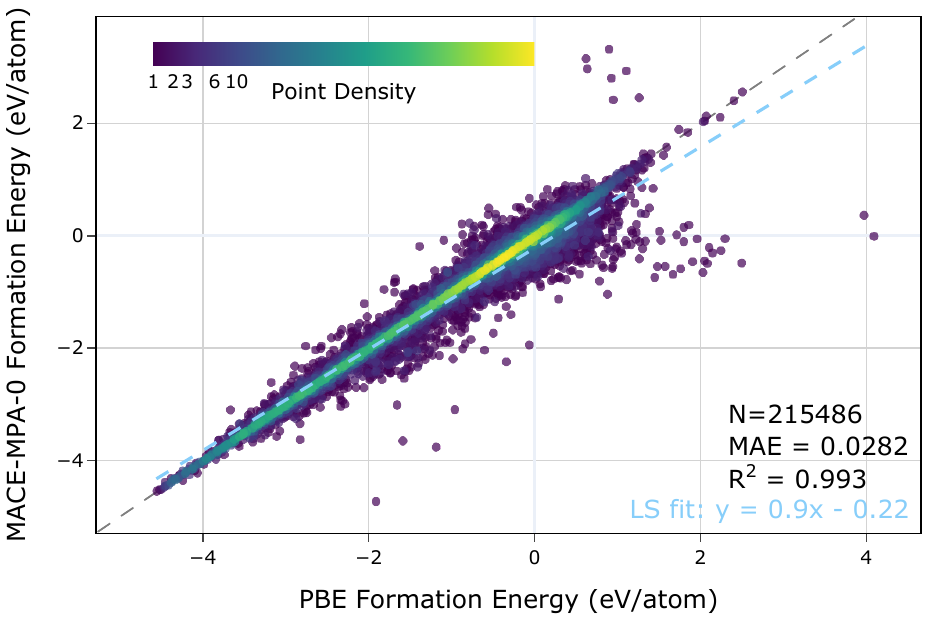}
  \label{fig:e-form-parity-mace-mpa-0}
  \caption{Formation energy parity plot showing the difference between the DFT-relaxed energy and the \MPzbt{}-relaxed energy (left) and \MPAz-relaxed energy starting (right) from WBM initial structures.}
  \label{fig:parity-e-form-mace-IS2RE}
\end{figure}

\MPzbt{}\ trained on the MPtrj dataset and \MPAz{} trained on the combined MPtrj and \href{https://huggingface.co/datasets/fairchem/OMAT24#salex-dataset}{sAlex} \cite{barroso-luque-2024-OpenMaterials} datasets generalize to out-of-distribution (OOD) chemistries as shown by their performance on the WBM dataset \cite{wang2021predicting} which was generated using elemental substitutions drawn according to a data-mined chemical similarity measure \cite{glawe_optimal_2016}.
The initial set of 9,524 structures for substitution were taken from the Materials Project convex hull.
The substituted structures were relaxed using the \texttt{MPRelaxSet} PBE DFT workflow.
After this, the convex hull was recalculated using the new structures and then subsequent rounds of substitution were carried out on the new structures that ended up on the combined MP plus growing WBM convex hull.
In total, 5 rounds of substitutions were carried out yielding a dataset of 257,487 inorganic crystals that are OOD with respect to MP and therefore well suited to benchmarking.
For this investigation, we use the cleaned version of the WBM dataset released in Matbench Discovery (MBD)\cite{riebesell_matbench_2023}, which first discards 524 crystals with unphysical or missing labels. Subsequently, all structures in WBM with composition+prototype matching a structure in MP are removed.
Within WBM composition+prototype, duplicates are dropped leaving only the lowest energy structure.
The final test set consists of \num{215 488} materials, and of these we obtained \MPzbt{} relaxed structures for \num{215 488} and \num{215 486} for \MPAz{}.

Following the MBD protocol, we use \MPzbt{}\ to relax the initial substituted structures and compare these predictions against the ground truth formation energy calculated with DFT.
The predictions on the \MPzbt{}\ self-relaxed structures result in an MAE of \SI{58}{meV/atom} as shown in \cref{fig:parity-e-form-mace-IS2RE}.
\MPAz{} cuts this error more than in half, achieving a formation energy MAE of \SI{28}{meV/atom}.
Similarly, \MPAz's $R^2$ improves on \MPzbt's $R^2$, going from 0.987 to 0.993.
However, both model's energy error exhibit a systematic tendency to underpredict the PBE formation energy as seen in the least-square fits of $y = 0.94x - 0.19$ for \MPzbt{} and $y = 0.9x - 0.22$ for \MPAz{}.
Interestingly, despite the lower error and higher $R^2$, \MPAz{}'s error is more systematic since slope and intercept are further from 1 and 0, respectively.
Thus a simple affine fit to correct for the systematic component in the energy error would reduce \MPAz{}'s already lower error more than \MPzbt{}'s.
While Alexandria is overall a highly valuable training set, this result is consistent with the observation that it is even more biased towards low-energy structures than MPtrj.

The OOD nature of the WBM dataset can be seen in the increase in MAE between WBM batches from \SI{48}{meV/atom} for the first batch to \SI{78}{meV/atom} for the final batch as the structures become increasingly dissimilar to MP due to accumulated substitutions.
When these predictions are used to attempt to classify whether the structure lies above or below the MP convex hull training set, \MPzbt{} achieves an F1 score of 0.66 and a discovery acceleration factor (DAF) of 3.95 compared to F1 of 0.85 and DAF of 5.58 for \MPAz.
The DAF is the ratio of the precision (TP/PP) to the prevalence (P/N) of the test set. Here, TP = True Positives, PP = Predicted Positives, P = Total Positives and N is the test set size.
These results show that \MPzbt{} and especially \MPAz{} can extrapolate to novel chemistries and are well-suited to high-throughput materials discovery.

\subsubsection*{Similarity statement}
By construction there is no overlap in terms of both composition and prototype together between MP and the WBM test set studied here. However there is overlap for both compositions and prototypes separately. MP contains \num{105 583} unique reduced formulae, whilst the WBM test set contains \num{160 055}. Of these \num{15 782} overlap MP albeit with all instances being examples of different prototype structures. MP contains \num{32 933} different isopointal prototypes, whilst the WBM test set contains just \num{2 816}. Of the prototypes found in WBM \num{1813} are also found in MP. Of the isopointal prototypes not seen in MP the 5 most common are (occurrences in parentheses): {ABC2\_oI8\_71\_a\_b\_f} (323), {ABC2\_hP12\_181\_c\_d\_i} (215), {AB2\_hP9\_189\_f\_adg} (156), {ABC2\_oI8\_44\_a\_b\_c} (117), {AB4C6\_mC22\_8\_a\_2ab\_2a2b} (106). These arise due to changes in symmetry during the relaxation of the substituted structures.

\clearpage
\subsubsection{Stoichiometric substitutions}\label{sec:element-substitution}

\begin{figure*}[!htbp]
  \centering
  \includegraphics[width=\linewidth,keepaspectratio]{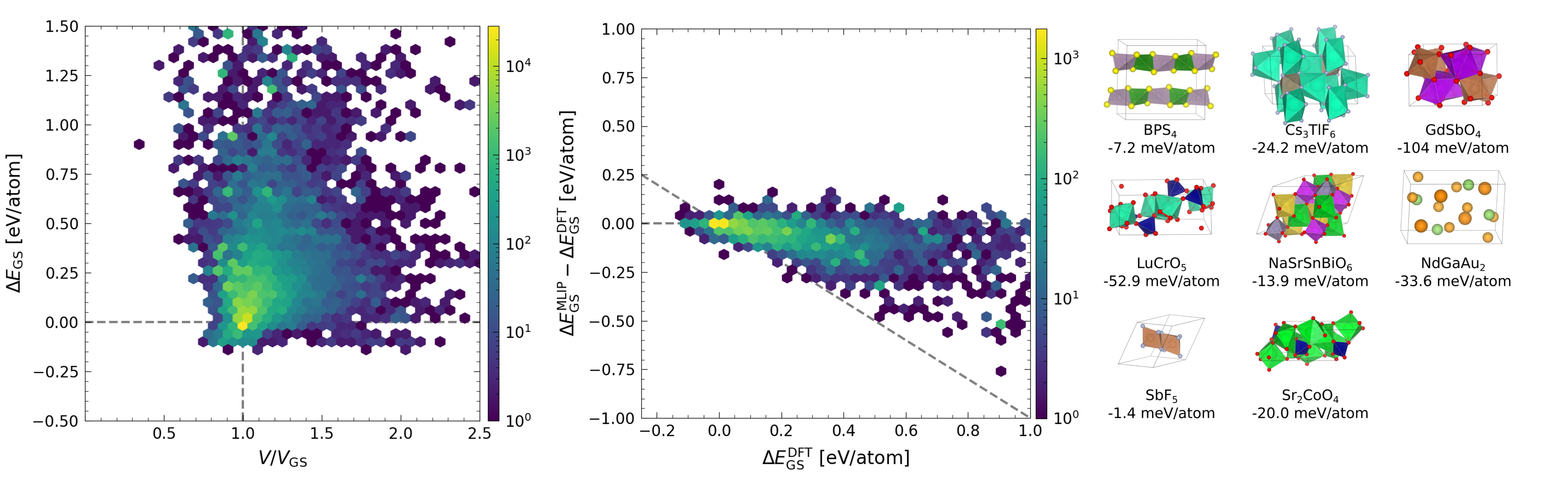}
  \caption{Left: Relative energies vs. volumes of ca. 150k relaxed structures generated via exhaustive element substitutions for 100 MP compositions. Center: Error of relative energies calculated with PBE and \MPzbt{} for 6909 randomly drawn structures from the left plot, as a function of the PBE relative energy. Right: DFT validated examples of newly discovered stable phases, along with their energies relative to the MP convex hull.}
  \label{fig:element_substitution}
\end{figure*}

To test the interpolative and extrapolative capabilities of the model within MP chemistry, exhaustive element substitutions for 100 randomly selected compositions were performed. Specifically, MP was first filtered to remove all compositions with more than 16 atoms in the reduced formula, to ensure that a sufficient number of possible substitutions. This set was randomly split into host and target compositions with a 60:40 ratio. 100 compositions were randomly drawn from the target split and substituted into each stoichiometrically matching host, yielding 154,685 substituted structures. These were optimized with \MPz{} using full unit cell relaxations (without the D3 correction). In \cref{fig:element_substitution}, the results are shown in terms of energies and volumes relative to the most stable structure of that composition within the MP. The distribution is sharply peaked at $\frac{V}{V_\mathrm{ref}}=1$ and ${E}-{E_\mathrm{ref}}=0$, indicating that the substituted cells often relax back to the known ground state structure from MP. This makes \MPz{} suitable for predicting the crystal structures of unknown materials. For a random sample of 16126  structures, the \MPz{} relative energies were validated with MP-compatible PBE DFT calculations, yielding an RMSE of 0.097 eV/atom.

Beyond recovering the MP ground state, the wide range of substitutions tested also yields a large number of alternative structures for each composition. 3522 of these have relative energies of zero or lower and are thus predicted to be more stable than the MP reference. To validate these predictions, DFT relaxations were performed for all structures with negative relative energies. The results confirms that 2724 of them are indeed more stable than the corresponding MP reference. Importantly, not each of these is a new stable phase, as some relaxations converged to the same minimum and multiple lower energy structures are found for some compositions. Nonetheless, for eight of the 100 compositions considered here, new structures with energies below the current MP hull were discovered (for \ch{BPS4}, \ch{Cs3TlF6}, \ch{GdSbO4}, \ch{LuCrO5}, \ch{NaSrSnBiO6}, \ch{NdGaAu2}, \ch{SbF5}, and \ch{Sr2CoO4}). These constitute genuine predictions of new, thermodynamically stable phases relative to the MP convex hull. These phases were also not reported in the recent GNoME effort (which focused on compositions not included in MP). However, the WBM dataset does report slightly more stable structures for \ch{Cs3TlF6} and \ch{SbF5}. It also includes structures for \ch{NdGaAu2}, which are less stable than the one reported herein.

\subsubsection*{Similarity statement}

The MP contains at least one (\textit{e.g.} \ch{NaNiIO6}, \ch{NaSrSnBiO6}) and at most 423 (\ch{Sr2CoO4}) structures that contain all elements in the discovered stable phases. These matches are similar to the discovered structures, as host and reference structures are present in the MP. Yet, the discovered phases are unique and by definition not part of the training set. For each discovery, the closest structures in the training set are: mp-1106139 (\ch{Bi3Pb}), mp-559695 (\ch{Cs3BiF6}), mp-561827 (\ch{Cs3TlF6}), mp-1211553 (\ch{LuCrO5}), mp-545399 (\ch{NaNiIO6}), mp-1522253 (\ch{NaSrSnBiO6}), mp-1220399 (\ch{NdGaAu2}), mp-1100068 (\ch{Sr2CoO4}), mp-775149 (\ch{TiS2O8}). We provide json files to visualize interactive UMAPs on \url{chemiscope.org}.


\clearpage
\subsubsection{Analysis of highly-coordinated theoretical structures}\label{sec:coordination}

\begin{figure}[h]
  \centering
  \includegraphics[width=4in,keepaspectratio]{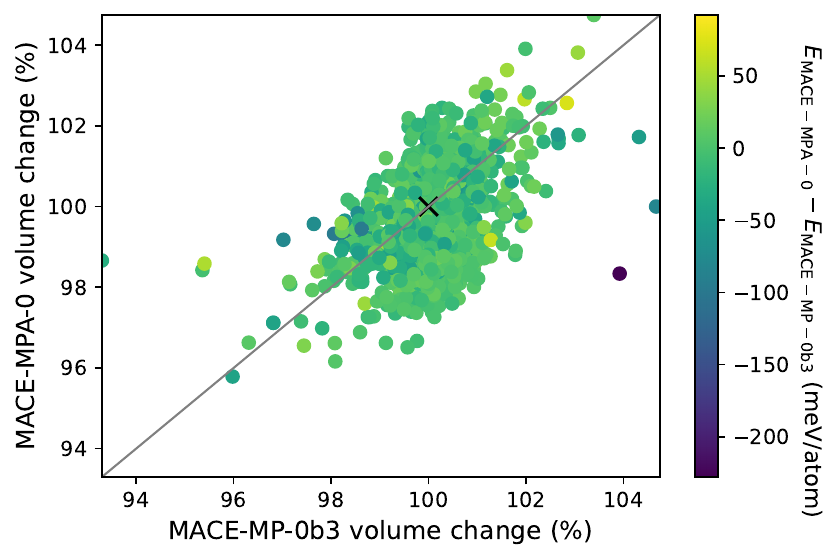}
  \caption{
    Comparison of the percentage volume change under relaxation relative to the initial structure volume as taken from the GNoME dataset.
    \MPzbt{} is plotted on the horizontal axis, and \MPAz{} on the vertical.
    The line indicates perfect agreement between both MACE models, and the ``x'' indicates perfect agreement with the GNoME volume.
    We observe a slight tendency of \MPzbt{} to predict larger volume changes, and thus larger volumes, than \MPAz.
    The points are colored by the difference in energy between \MPAz{} and \MPzbt, thus darker colors indicate \MPzbt{} predicts a lower total energy than \MPAz, and brighter colors indicate that \MPzbt{} predicts a higher total energy than \MPAz.
  }
  \label{fig:mace_mace_GNoME_comp}
\end{figure}

The recent GNoME database \cite{merchant_scaling_2023} published approximately 381,000 stable materials which could not be matched to any experimental structure in the Inorganic Crystal Structure Database (ICSD) \cite{zagorac_JAC_2019}, nor any public repository of theoretical structures such as the Open Quantum Materials Database (OQMD) or Materials Project.
An analysis of the local coordination environment of these structures using the \texttt{CrystalNN} algorithm \cite{zimmerman2020crystalnn} reveals 21,300 structures with maximum coordination number greater than 16.

To evaluate the performance of \MPzbt{} and \MPAz{} against the GNoME potential which generated these structures, we have re-relaxed 1,199 structures with a predicted maximum coordination number greater than 20.
Additionally, to compute formation energies, we re-relaxed all elemental structures in the Materials Project identified to be the lowest energy configuration from PBE$/+U$ calculations.
The {\tt FrechetCellFilter} class in the Atomic Simulation Environment (ASE) \cite{larsen2017atomic} was used to relax structures until the maximum (absolute) Cartesian component of any inter-atomic force was less than \SI{2e-2}{eV\per\AA{}}.
This force convergence setting is consistent with the most stringent force relaxation criterion used by the Materials Project.
No dispersion correction was used to augment the MACE models, consistent with the lack of a dispersion correction in GNoME.
All relaxations ran successfully on CPU resources.

The MACE-relaxed structures are highly similar to the GNoME ones.
\ref{fig:mace_mace_GNoME_comp} plots the percentage change in volume,
\begin{equation}
    \delta V = (100\%) \times \frac{V_\mathrm{MACE}}{V_\mathrm{GNoME}}
\end{equation}
for both \MPzbt{} and \MPAz{}.
\MPAz{} tends to predict smaller volumes than \MPzbt, with a mean absolute deviation (MAD) from the GNoME structure volumes of 0.117 \AA{}$^3$/atom, compared to the 0.085 \AA{}$^3$/atom MAD of \MPzbt.
However, their energies (indicated by the colors of the points) are nearly the same in all cases.

\begin{figure}
    \centering
    \includegraphics[width=\columnwidth,keepaspectratio]{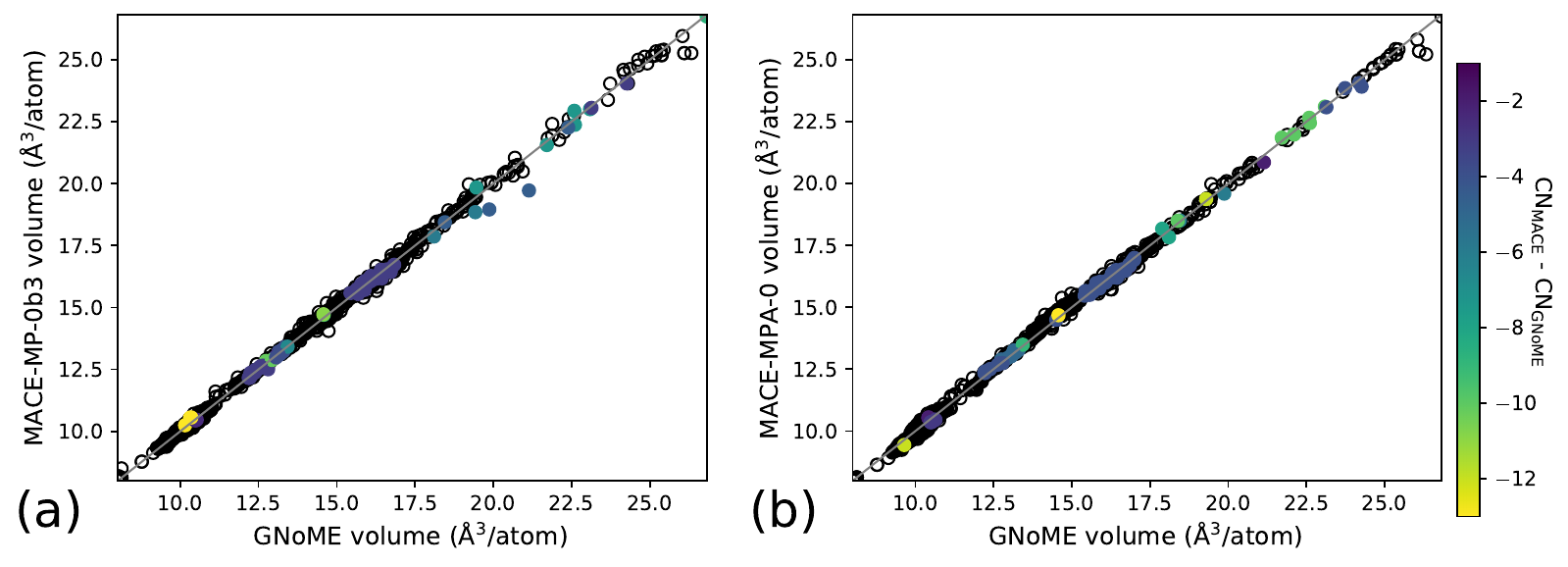}
    \caption{
        The final volume of the highly-coordinated GNoME materials before (horizontal) and after relaxation with a MACE model (vertical).
        Panel (a) plots the MACE-MP-0B3 model results, and (b) the MACE-MPA-0 results.
        The points are colored by their change in maximum observed coordination number as measured by \texttt{CrystalNN} \cite{zimmerman2020crystalnn}: a hollow circle indicates no change.
        Nonzero changes in the maximum coordination number, suggesting a decrease in dense bond patterns, are indicated by increasingly light colors for larger magnitude changes.
        In all cases, the maximum coordination number was observed to decrease or stay the same.
        \label{fig:mace_gnome_GNoME_comp}
    }
\end{figure}

In virtually all cases, the MACE models relax structures to an indistinguishable configuration from their GNoME starting point.
Only 14 (1) structures undergo an energy lowering of greater than \SI{10}{meV\per atom} during relaxation with \MPzbt{} (\MPAz).
15 (14) structures underwent an increase of symmetry under relaxation with \MPzbt{} (\MPAz), as measured by their space group number.
Only one structure decreased in symmetry when relaxed with \MPzbt. 
The maximum coordination number decreased for 82 (87) structures under relaxation with \MPzbt{} (\MPAz), and stayed the same otherwise.
Figure \ref{fig:mace_gnome_GNoME_comp} presents these observations for both MACE models.

\begin{figure}
    \centering
    \includegraphics[width=4in,keepaspectratio]{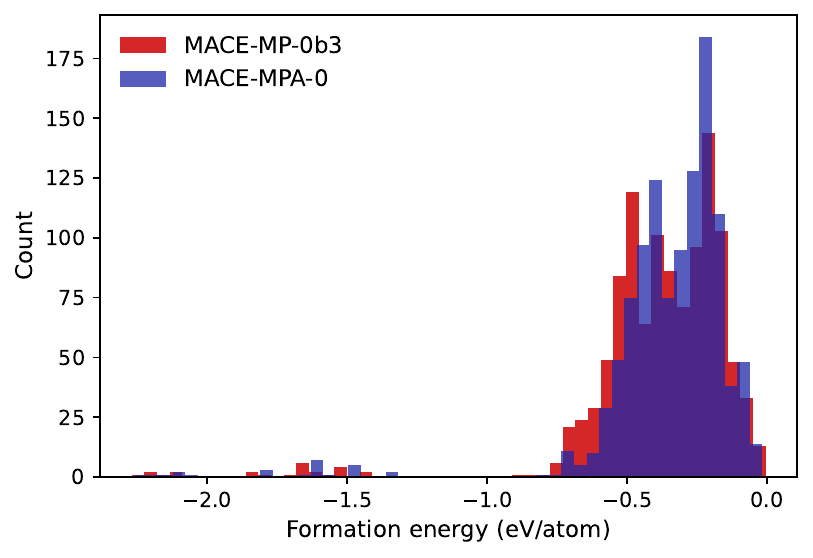}
    \caption{
        The distribution of formation energies (zero-pressure limit of the formation enthalpies at zero temperature) of the highly-coordinated GNoME structures after relaxation with \MPzbt (red) and \MPAz (transparent purple).
        Note the dual peaks in formation energies: between -1 and 0 eV/atom, most structures are intermetallics or covalently bound; below -1 eV/atom, most structures are oxides.
        \label{fig:eform_gnome}
    }
\end{figure}

Last, we consider the formation enthalpies, defined as
\begin{equation}
    \delta H_f = \frac{1}{\sum_i c_i} \left[H_\mathrm{solid} - \sum_i c_i \mu_i \right],
    \label{eq:enthalpy_form_gnome}
\end{equation}
the total enthalpy of a solid $H_\mathrm{solid}$ minus the chemical potentials of its elemental constituents $\mu_i$ weighted by their stoichiometry $c_i$ in the structure.
Note that $\sum_i c_i$ is the total number of atoms in the computational cell of the structure.
After relaxation, the enthalpies and chemical potentials in Eq. \ref{eq:enthalpy_form_gnome} should be well-approximated by energies.
Figure \ref{fig:eform_gnome} presents the distribution of formation energies of the GNoME structures.
Two distinct peaks can be observed, one for the intermetallics and covalently bound structures between -1 and 0 eV/atom, and one for the oxides below -1 eV/atom.
Also note that no materials were found to be unstable as indicated by a positive formation energy.

Cumulatively, these observations suggest that both MACE models perform comparably to the GNoME UIP for these highly-coordinated structures, and yield equilibrated structures of equal or higher symmetry than the GNoME UIP.

\subsubsection*{Similarity statement}

By construction, the materials in the GNoME set do not exist in the MP dataset.
The GNoME set was constructed in part by substituting elements on structures that originated in the Materials Project. 
However, some of the GNoME materials exist in the same chemical space as those in MP.
For the 1,199 GNoME materials considered here, 131 (10.9\%) contain exactly the same elements as structures in MP.
The chemical space of maximum overlap, Cu-La-Zn, contains four GNoME (7ee54b7a37, 3422b9acb0, f36439bb45 and 68a5e7535f) and two MPtrj (mp-1223296 and mp-1093834) materials.
We provide two JSON-format dictionaries, \verb|MPtraj_chem_env.json.gz|, and \verb|GNoME_chem_env_w_ovlp.json.gz|, which tabulate the chemical environments spanned by MP and the GNoME subset, respectively.
The GNoME subset is further categorized by overlap with MP.

\subsubsection*{Performance summary}
Excellent energy prediction of WBM and GNoME hypothetical materials, and able to newly discover stable materials as validated by DFT.  

\clearpage
\subsection{Alanine Tripeptide free energy surface}\label{sec:ala}

\begin{figure}[htbp!]
  \centering
    \includegraphics[width=\linewidth,keepaspectratio]{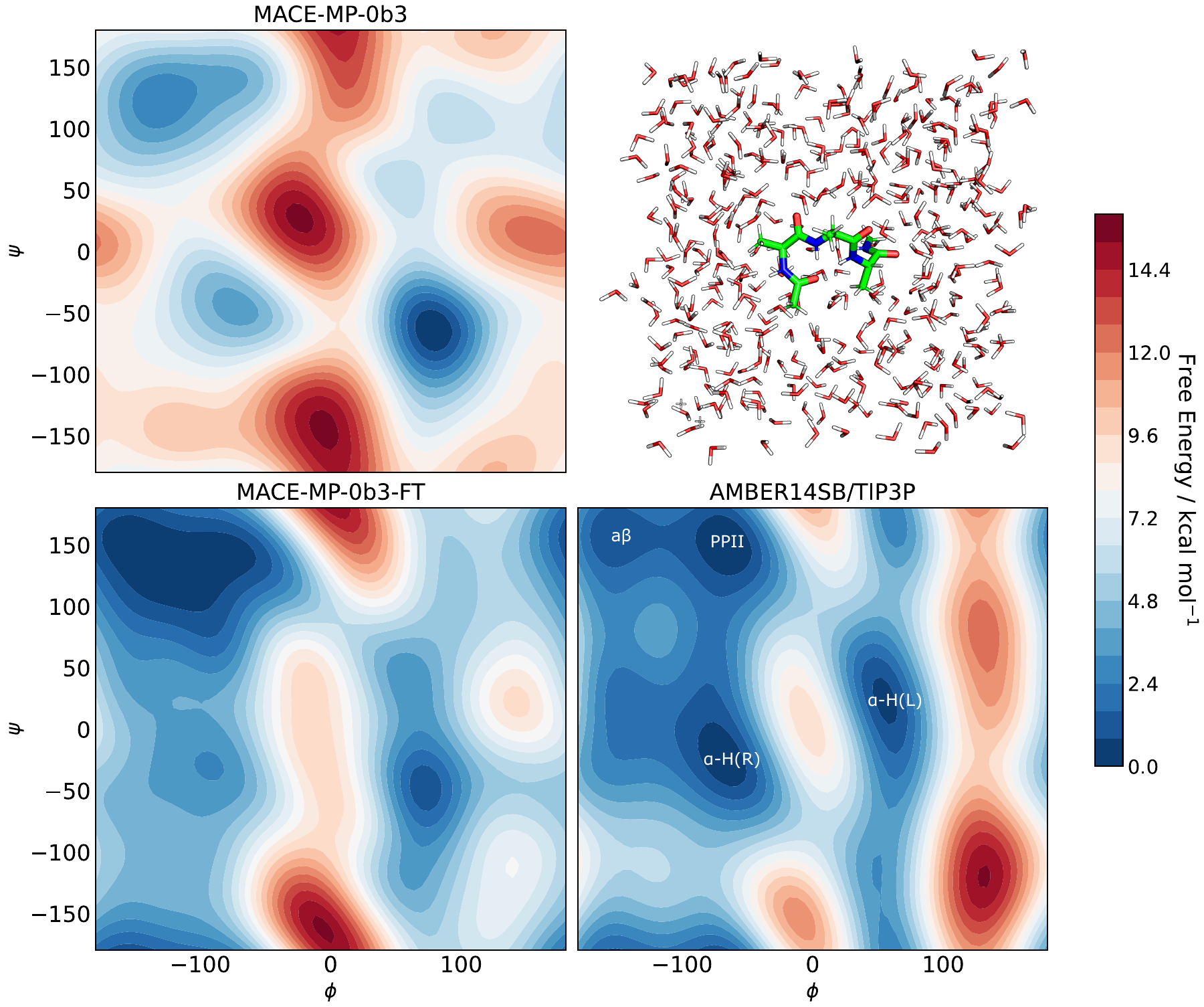}
  \label{fig:ala_tripeptide_fes}

  \caption{Free energy surfaces constructed from \SI{1}{\nano\second} sampling with well-tempered metadynamics. Enhanced sampling was performed on the central $\psi$ and $\phi$ backbone angles.}
\end{figure}

In this section, we investigate the ability of \MPzbt{} to construct the free energy surface of a simple peptide. We simulate a periodic box containing an alanine tripeptide solvated with explicit water for \SI{1}{\nano\second} without D3 correction. Sampling of the central backbone torsions was accelerated by well-tempered metadynamics, as implemented in OpenMM, using an initial Gaussian height of \SI{1}{kJ\per mol} and a temperature scaling factor of 10. \MPzbt{} only approximately identifies the features of the free energy landscape, whilst \MPzbt{}-FT identifies the positions of the minima in better qualitative agreement with the classical forcefield. \MPzbt{}-FT also more accurately predicts the relative population of the minima. However, neither model fully resolves the $\text{a}\beta$ and PPII minima around ($\phi=-100^{\circ}$, $\psi=150^{\circ}$). \MPzbt{} in particular overpopulates the $\alpha$-H(L) basin at $\phi=80^{\circ}$, $\psi=-100^{\circ}$. .

\subsubsection*{Similarity statement}
The MP dataset contains 99 structures containing exclusively the elements H, C, O and N. Based on UMAP analysis, we observe that all atomic environments in the example system are similar to environments in the training set. On closer inspection, we find that the most similar environments to the majority of the example configurations are clusters of water, ammonia and NO\textsubscript{2} molecules. Several examples contain oxygen-bearing molecules near water molecules, allowing for sampling of hydrogen bonding. We also find configurations containing clusters of large aromatic compounds containing carbon, nitrogen, oxygen and hydrogen. Interestingly, we also find a large cluster containing diverse configurations of small C,H,N,O-containing fragments, including carboxyl and amide fragments, \textit{e.g.} mp-997182.

\subsubsection*{Performance summary}
Stable MD at ambient conditions, but incorrect free energy of conformers.

\clearpage

\subsection{Molecule-Surface Interactions}\label{sec:molecule_surf_interactions}
\subsubsection{Adsorption energies - S24 dataset}

\begin{figure}[hbtp!]
    \centering\includegraphics[width=0.9\linewidth]{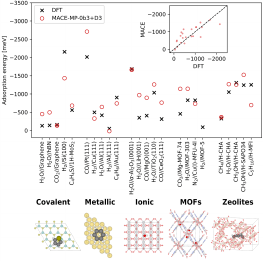}
    \caption{\label{fig:s24_dataset}Comparison between DFT (PBE+D3(BJ), red crosses) and \MPzbt{}+D3 (blue circles) adsorption energies calculated for a diverse set of surfaces. These surfaces consist of covalent, metallic, and ionic bonds as well as the porous material classes: MOFs and zeolites. Filled circles indicate surfaces where the MACE model reaches chemical accuracy ($43\,$meV) agreement to DFT. The inset shows that there is a good correlation between and \MPzbt{}+D3 and the DFT references.}
\end{figure}

Describing the interaction between a molecule and a surface with first-principles accuracy is central towards designing new and improved materials for heterogeneous catalysis, gas storage and separation, and many more~\cite{norskovComputationalDesignSolid2009a}.
Here, we test the accuracy of \MPz{} augmented with the Becke-Johnson D3 dispersion correction for a set of prototypical systems found in surface chemistry, encompassing metallic, covalent and ionic-bonded surfaces, together with porous metal-organic frameworks (MOFs) and zeolites.
The structures used within this work - dubbed the S24 dataset - were taken from an amalgamation of published~\cite{al-hamdaniPropertiesWaterBoron2017a,brandenburgPhysisorptionWaterGraphene2019,ehlertCO2GrapheneBenchmarking2023a,tsatsoulisReactionEnergeticsHydrogen2018a,tsatsoulisComparisonQuantumChemistry2017,yeInitioSurfaceChemistry2023,lustembergVibrationalFrequenciesCeriumOxideBound2020c,shiManyBodyMethodsSurface2023a,doi:10.1126/science.abj0890,doi:10.1021/acscatal.2c05493,https://doi.org/10.1002/anie.202013671} and unpublished works.
Fig.~\ref{fig:s24_dataset} summarises the computed PBE-D3(BJ) adsorption energies using \textsc{pymatgen} to generate VASP~\cite{VASP1,VASP2,VASP3} inputs with the \texttt{MPRelaxSet} settings.

\begin{table}
\centering
\caption{\label{tab:s24_mads}Mean absolute deviation in meV for the S24 dataset and its sub-categories of surfaces.}
\begin{tabular}{lr}
\hline
{} &  MACE-MP-0b3+D3 \\
\hline
Covalent &             308 \\
Ionic    &             361 \\
MOF      &             380 \\
Metal    &             265 \\
Overall  &             305 \\
Zeolite  &             229 \\
\hline
\end{tabular}
\end{table}

We summarise the overall MAD as well as the MAD within the sub-categories of surfaces for \MPzbt{}+D3 in Table~\ref{tab:s24_mads}.
Performance is moderately good, with an overall MAD of $305\,$meV, in line with the errors observed from a typical DFT functional.
Performance across the various sub-categories of surfaces is relatively uniform, with best performance for zeolites with an MAD of $229\,$meV.
It performs poorly for H$_2$ on Si(100), with errors exceeding $600\,$meV.

\subsubsection*{Similarity statement}

The S24 dataset contains 72 structures comprising 24 adsorbates, surfaces, and adsorbate-surface combinations. Based on the UMAP analysis, the elemental compositions of the adsorbates occur 6 to 27 times, the surfaces 0 to 9043 times and the adsorbate-surface combinations 0 to 4896 times on their own or along with other elements. Importantly, the training dataset does not contain any gas-phase molecules, surface-truncated models, or MOFs. For bare surfaces, on average, similar element compositions occur in the training data as follows: 3997 covalent, 5805 metallic, 2859 ionic, 46 MOFs, and 88 zeolites. For adsorbate-surface combinations, on average, the number of similar element compositions occurring in the training data are lowered to: 1492 covalent, 340 metallic, 499 ionic, 46 MOFs, and 59 zeolites. There is no clear correlation between the number of similar training data for a surface type and the accuracy of MACE-MP-0b3+D3.

\subsubsection*{Performance summary}
Moderately good agreement is generally achieved for adsorption energies, with improved performance for adsorption onto zeolites for MACE-MP-0b3+D3.
Performance appears to become poorer with dissociative chemisorption of H$_2$ on Si(100), albeit still reproducing the adsorption energy to the correct order of magnitude.

\clearpage

\subsubsection{Relative energies - OC157 dataset}

\begin{figure}[!ht]
    \includegraphics[width=6.69in]{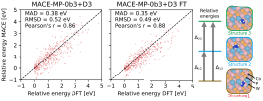}
    \caption{\label{fig:oc_dataset}Comparison of MACE-MP-0b3+D3 and its fine-tuned (FT) version (see text) against DFT (PBE-D3(BJ)) in predicting the relative energies between three structures across 157 molecule-surface combinations. This contains a highly diverse set of molecule-surface combinations which span 54 elements of the periodic table involving up to three elements per surface.}
\end{figure}

Identifying the most stable structures for a molecule-surface system is pivotal towards predicting the activity and selectivity of a catalyst, facilitating rational design of new catalyst materials~\cite{norskovDensityFunctionalTheory2011b}.
We compare MACE-MP-0b3+D3 and the model following fine-tuning against \texttt{MPRelaxSet} DFT at predicting the relative energies between 3 structures for 157 molecule-surface combinations.
The fine-tuned (FT) models were trained on a dataset containing one high-energy structure (not from the test set) for each of the 157 molecule-surface combination.
These surfaces were taken from the Open Catalyst Challenge 2023~\cite{chanussotOpenCatalyst20202021,tranOpenCatalyst20222023}, using structures generated by the baseline EquiformerV2 model trained on the \textsc{OC20-S2EF-2M} dataset.
While 200 molecule-surface combinations were originally provided, we have excluded those containing oxygen (O) in combination with several transition metals (Co, Cr, Fe, Mn, Mo, Ni, V and W) as this leads to complications with the Hubbard U correction (see main text) with \texttt{MPRelaxSet} settings.
Fig.~\ref{fig:oc_dataset} shows the 471 relative energies ($\Delta_{12}$, $\Delta_{13}$ and $\Delta_{23}$) for each of the remaining 157 molecule-surface systems, where $\Delta_{XY}$ is the relative energy between structure $X$ and $Y$, predicted with the MACE models and calculated with DFT.
Overall, we observe a strong correlation between MACE-MP-0b3+D3 and DFT on the relative energies, providing a Pearson correlation coefficient of 0.86 and an MAD of $0.38\,$eV.
We find that fine-tuning produces a small improvement, decreasing the MADs by $0.03\,$meV.
In particular, out of the 157 molecule-surface combinations, the lowest DFT energy configuration was correctly identified by MACE-MP-0b3+D3 for 126 of the surfaces. 
Fine-tuning further increases this number to 132 for MACE-MP-0b3+D3 FT.

\subsubsection*{Similarity statement}

The molecule-surface combinations in the OC157 dataset cover all elements up to and including Bi (atomic number 83), except He, Li, Be, B, F, Ne, Mg, Cl, Ar, Br, Kr, I, Xe, Ba and all lanthanoids. The UMAP analysis shows that 126 of the 157 molecule-surface combinations have no similarity to the training dataset, with a further 20 having less than 10 and only 3 having more than 20 similar compositions. 

\subsubsection*{Performance summary}
Good performance at predicting the absolute value of the relative energies between various molecule-surface configurations.
Lowest energy (DFT) structure correctly predicted in 80\% of the cases for MACE-MP-0b3+D3 with fine-tuning further improving performance to 84\%

\clearpage

\subsection{Computational efficiency (twenty-element alloy)}\label{sec:timings}
\begin{figure}[!ht]
  \centering
  \includegraphics[width=\textwidth,keepaspectratio]{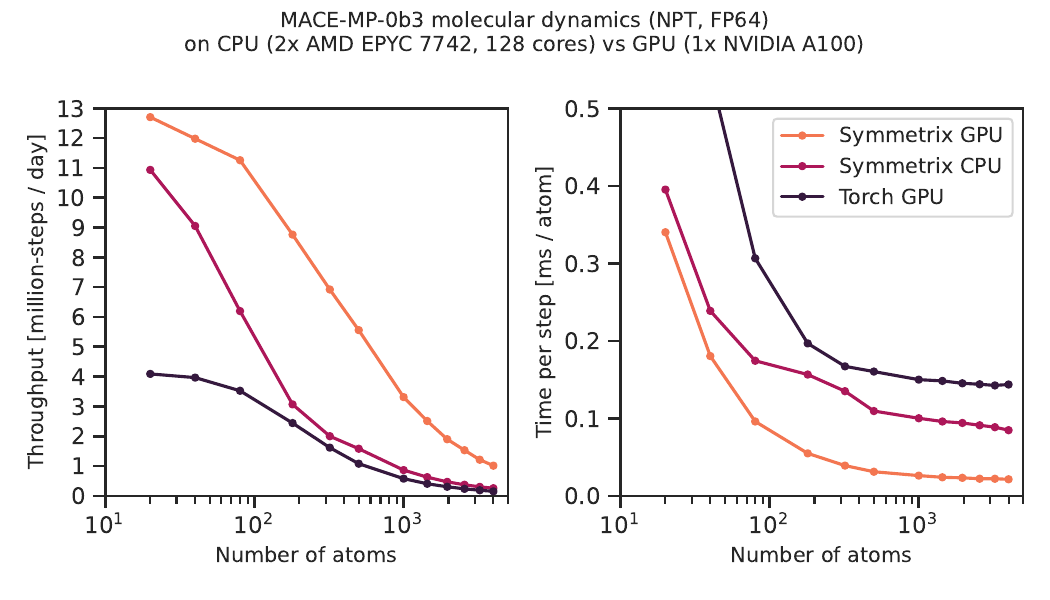}
  \includegraphics[width=\textwidth,keepaspectratio]{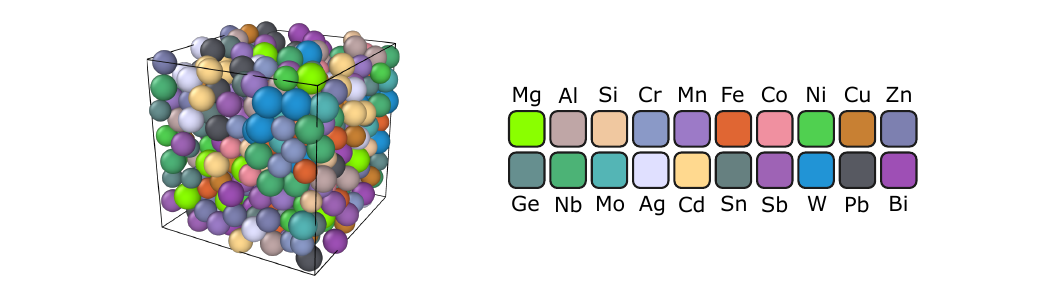}
  \caption{Double-precision computational performance for NPT dynamics in LAMMPS, measured for the 20-element Cantor alloy at ambient pressure and \SI{300}{\kelvin}. The two plots show the same data, highlighting throughput (left) and speed (right). Results are shown for a 128-core CPU node utilizing AMD EPYC 7742 CPUs as well as a single NVIDIA A100 GPU. Each plot has three curves: one for the Torch baseline on GPU and two for an optimized MACE evaluator available from the \texttt{symmetrix} package. The \texttt{symmetrix} evaluator uses Kokkos, a vendor-agnostic library, to enable fast CPU and GPU performance.}
  \label{fig:HEA-single}
\end{figure}
This section demonstrates the computational performance of the medium MACE-MP-0b3 model (without dispersion correction), while also illustrating stable dynamics for a large, diverse system. High entropy alloys are multicomponent mixtures with at least four or five distinct elements, where each component appears in non-trace proportions. Here, we consider the 20-component alloy investigated by Cantor\cite{cantor2004microstructural}, which contains equal amounts of Mn, Cr, Fe, Co, Ni, Cu, Ag, W, Mo, Nb, Al, Cd, Sn, Pb, Bi, Zn, Ge, Si, Sb, and Mg. [See also recent computational work by Ceriotti and co-workers \cite{lopanitsyna2023modeling, mazitov2023surface}.] Our simulations were performed with LAMMPS \cite{thompson2022lammps}, utilizing its Kokkos extensions\cite{trott2022kokkos}, on AMD CPUs and NVIDIA A100 GPUs.

\begin{figure}[!ht]
  \centering
  \includegraphics[width=\textwidth,keepaspectratio]{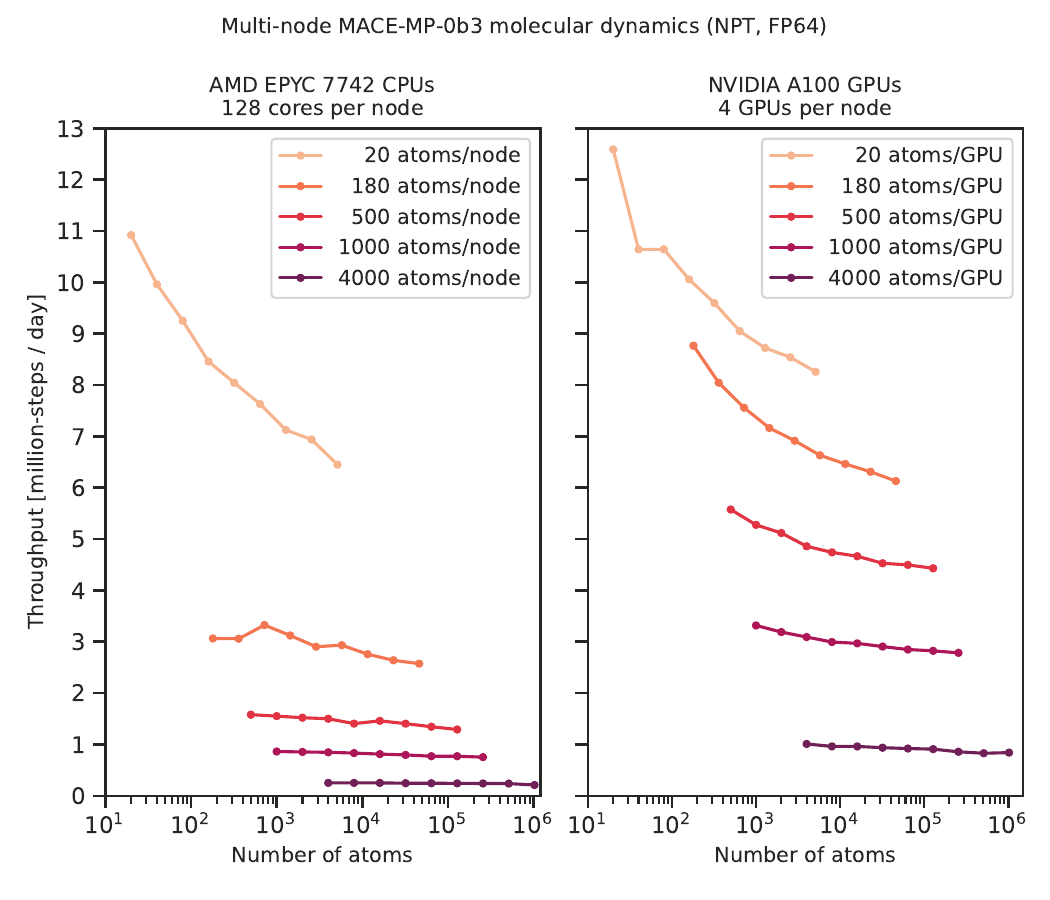}
  \caption{Double-precision multi-node performance for NPT dynamics in LAMMPS, measured for the 20-element Cantor alloy at ambient pressure and \SI{300}{\kelvin}. The left plot contains data for 1-256 CPU nodes, with 128 cores per node, and the right plot has data for 1-256 A100 GPUs, with four GPUs per node. Near-perfect weak scaling is achieved with as little as a few thousand atoms per CPU node or GPU.}
  \label{fig:HEA-multi}
\end{figure}

\Cref{fig:HEA-single} shows single-node computational performance and \Cref{fig:HEA-multi} demonstrates multi-node scaling. For each simulation, we began with a 20-atom FCC primitive cell and generated randomised supercells with sizes up to 1,024,000 atoms, measuring the performance over at least \SI{5}{\pico\second} of NPT dynamics at ambient pressure and \SI{300}{\kelvin}. In all cases, the primary aim was to measure performance for a reasonably well-mixed system, and we did not attempt to reach full equilibration.

\subsubsection*{Similarity statement}

Of the 150k MP structures, roughly 100k have at least one of the 20 elements used in this example, but only 3461 have compositions drawn exclusively from that set.
Of this latter group, 193 are single-component structures, 2079 are binaries, 1155 are ternaries, and 34 are quaternaries. Moreover, no structure in the entire database has five or more of the 20 elements considered here.
While it is encouraging that the dynamics appear stable for such a diverse, out-of-sample composition, we expect that incorporating dedicated datasets like that reported in \cite{lopanitsyna2023modeling} could enhance quantitative predictions.

\clearpage
\subsection{MPA-0: Scaling up with the Alexandria dataset}
\label{sec:alexandria}

To investigate the effect of training set size on model performance, we developed a MACE model, MPA-0, using identical hyperparameters to the MP-0b3 medium model but trained on an expanded dataset combining MPtraj and sAlex~\cite{BarrosoLuque2024Open}. The sAlex dataset comprises 10.5M structures extracted from the original Alexandria dataset, curated for compatibility with the Matbench Discovery benchmark. The Alexandria dataset consists only of randomly generated and relaxed crystal structures. Here, we demonstrate that increased data volume on just crystal structures significantly enhances model robustness across multiple applications. Notably, we observe strong enhancement in the accuracy of the MPA-0 model compared to MP-0b3 on applications for which coverage was not increased, namely amorphous phases and small molecules on surfaces.

\subsubsection{Metals}
\begin{figure}[h!]
\centering
\includegraphics[width=0.9\textwidth]{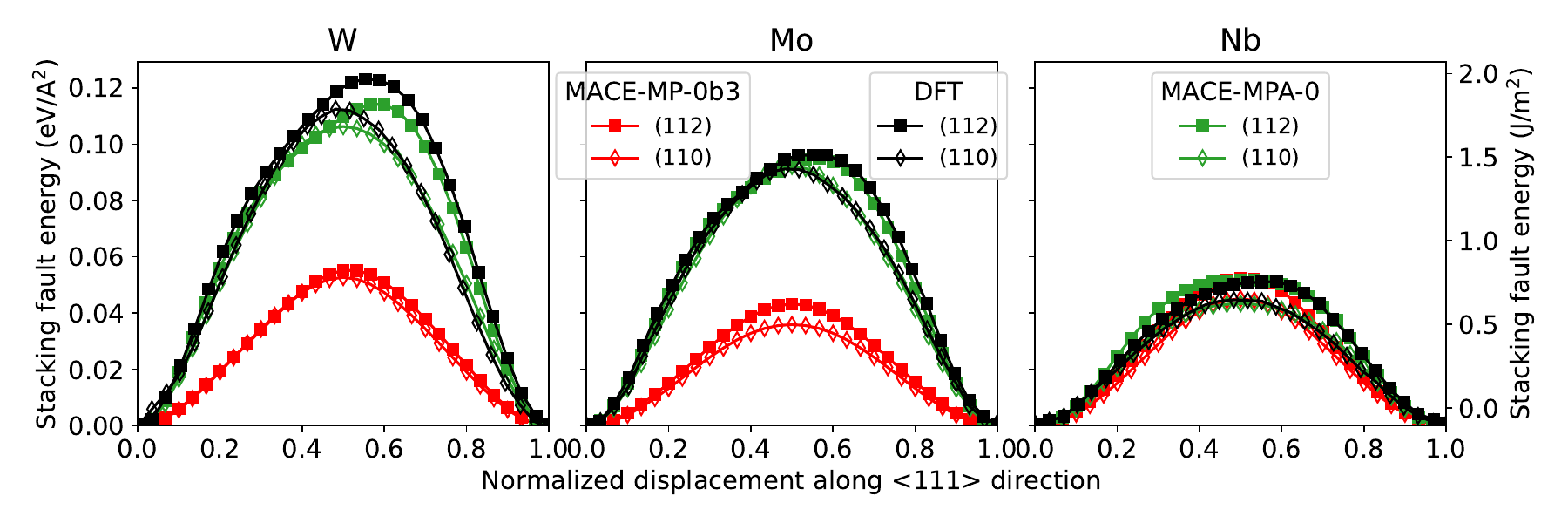}
\caption{Generalized stacking fault profiles for (112)[111] and (110)[111] $\Gamma$-surfaces predicted by MP-0b3 (red) and MPA-0 (green) compared to DFT reference data (black). MPA-0 demonstrates improved agreement across all three BCC metals, especially for Nb.}
\label{fig:stacking_fault}
\end{figure}

Figure~\ref{fig:stacking_fault} compares generalized stacking fault profiles for body-centered cubic (BCC) metals (W, Mo, Nb) along (112)[111] and (110)[111] $\Gamma$-surfaces. MPA-0 shows markedly improved agreement with DFT reference data compared to MP-0b3 for W and Mo where MP-0b3 substantially underestimates the fault energies. For W and Mo, MPA-0 correctly captures the peak heights and the overall energy landscape, while maintaining proper symmetry across the normalized displacement. This improved accuracy in stacking fault energetics is critical for predicting mechanical properties and plastic deformation mechanisms in structural metals. This enhancement is likely due to an increase in chemical coverage for these metals in the Alexandria dataset.

\subsubsection{Surfaces}
\begin{figure}[h!]
\centering
\includegraphics[width=0.9\textwidth]{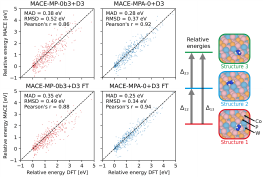}
\caption{Comparison of MP-0b3+D3, MPA-0+D3 and their fine-tuned (FT) versions against DFT (PBE-D3(BJ)) in predicting relative energies between three structures across 157 molecule-surface combinations. MPA-0 demonstrates consistently smaller errors and improved correlation with DFT values.}
\label{fig:surface_energies}
\end{figure}

The prediction of surface energetics represents a critical test for materials models. Figure~\ref{fig:surface_energies} shows parity plots comparing the performance of MP-0b3+D3 and MPA-0+D3 in predicting relative energies between three structures across 157 molecule-surface combinations. MPA-0+D3 achieves a significantly lower MAD (0.28 eV vs. 0.38 eV), RMSD (0.37 eV vs. 0.52 eV), and higher Pearson correlation (0.92 vs. 0.86) compared to MP-0b3+D3. Fine-tuning further improves these metrics for MPA-0+D3 FT, reducing MAD to 0.25 eV and RMSD to 0.34 eV while increasing correlation to 0.94. This enhanced accuracy spans a diverse range of molecule-surface interactions involving 54 elements, indicating improved transferability for heterogeneous catalysis and materials interface applications. It is notable that Alexandria does not contain any surfaces, nor isolated small molecules, meaning that better coverage of ly crystal structureson improves even not directly related downstream tasks.

\subsubsection{Amorphous carbon}
\begin{figure}[h]
\centering
\includegraphics[width=0.9\textwidth]{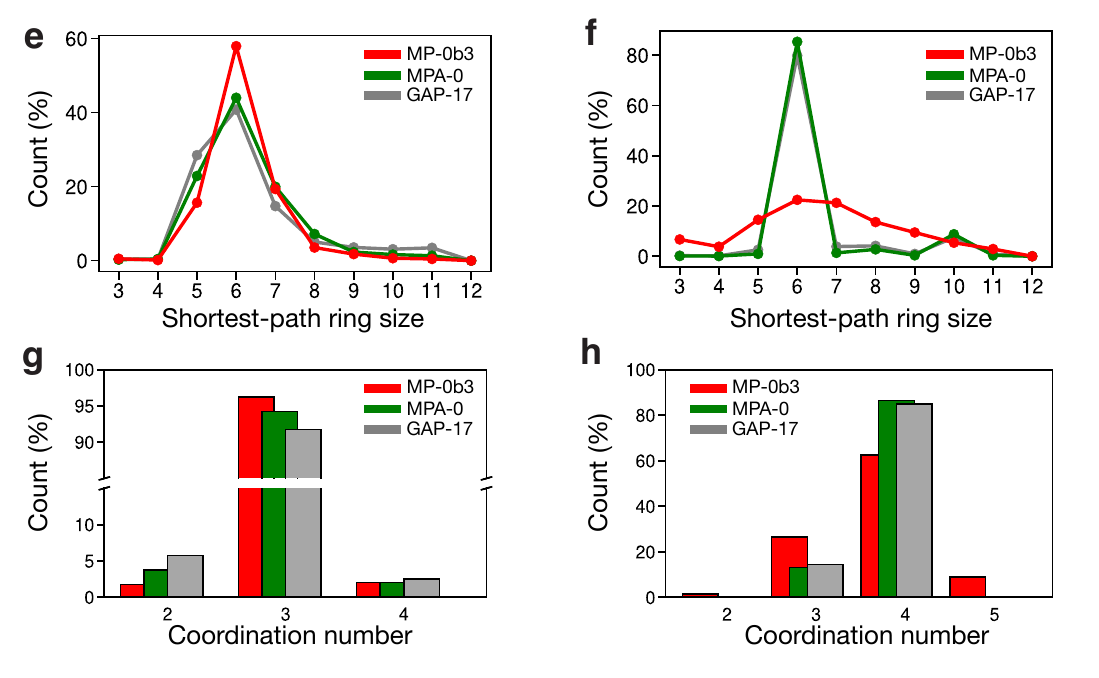}
\caption{Structural analysis of 4,096-atom amorphous carbon configurations. (e-f) Shortest-path ring size distributions and (g-h) coordination number distributions for MP-0b3 (red), MPA-0 (green), and GAP-17 reference model (gray). MPA-0 better captures the correct bonding environments and medium-range order.}
\label{fig:amorphous_carbon}
\end{figure}

Accurate modeling of amorphous materials requires precise representation of local atomic environments and bonding. Figure~\ref{fig:amorphous_carbon} compares the structural properties of amorphous carbon predicted by MP-0b3, MPA-0, and the reference GAP-17 model. While MP-0b3 performs qualitatively well at low density (panel e-g), MPA-0 significantly improves the distribution of shortest-path ring sizes for the high density phases, showing a sharper peak at 6-membered rings (panel f) that closely aligns with GAP-17, whereas MP-0b3 displays a broader, less defined distribution across ring sizes 6-8. For coordination numbers at high density (panel h), MPA-0 reproduces the correct sp$^3$:sp$^2$ ratio with approximately 85\% 4-coordinated and 15\% 3-coordinated carbon atoms, closely matching GAP-17. In contrast, MP-0b3 shows excess 3-coordinated (sp$^2$) and 5-coordinated carbon atoms, indicating structural deficiencies in the predicted amorphous networks. The Alexandria dataset does not contain any amorphous structures, demonstrating that an increase in broad chemical coverage of crystals helps in predicting amorphous phases.

\subsubsection{Metastable cell}
\begin{figure}[h]
\centering
\includegraphics[width=0.9\textwidth]{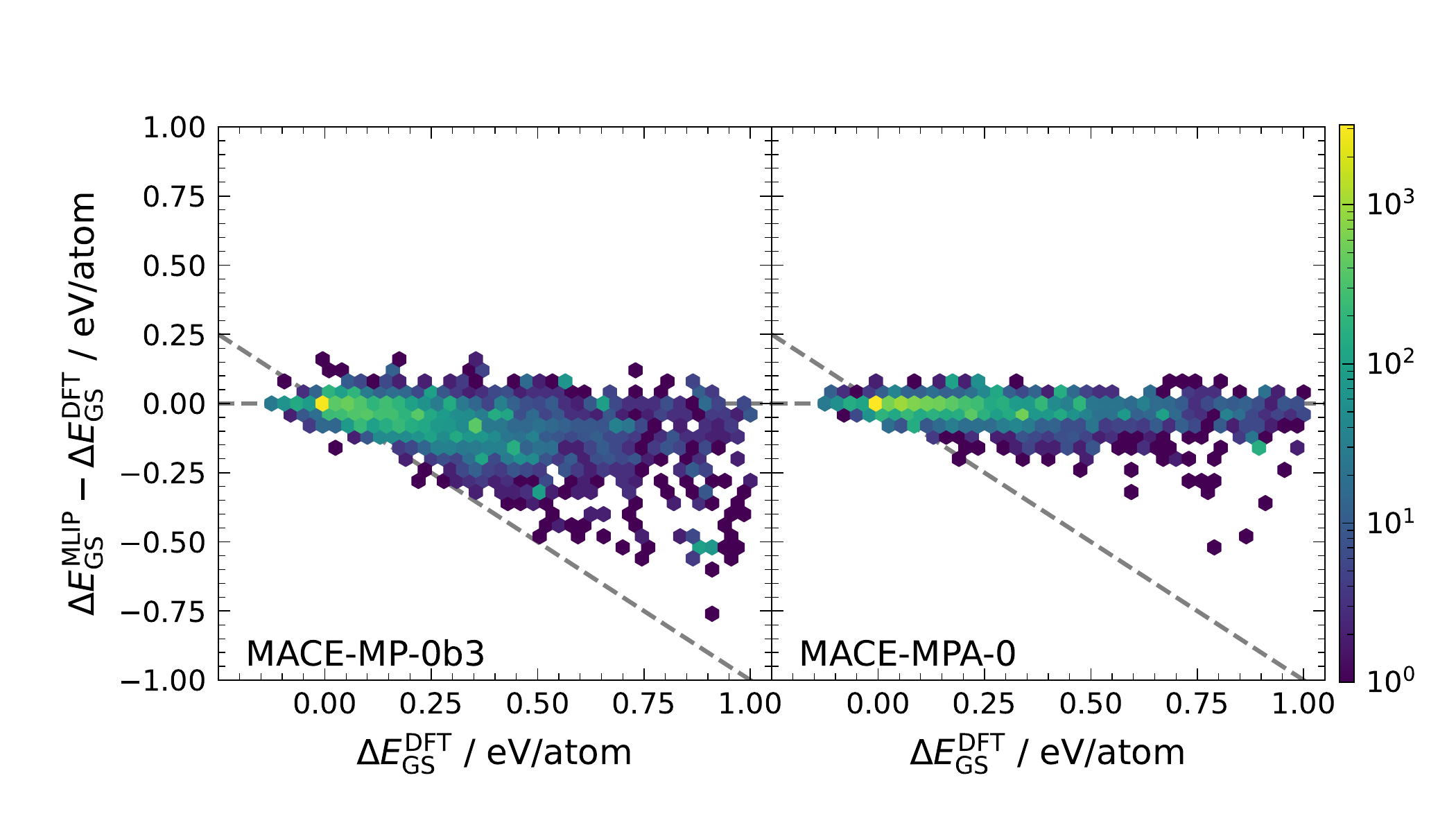}
\caption{Errors in prediction of energies above the ground state for MP-0b3 (left) and MPA-0 (right) across 16,126 relaxed structures from 100 different compositions. The color scale indicates point density. Dashed lines represent zero error (horizontal) and the boundary where structures would be erroneously predicted to have lower energy than the true ground state (diagonal). MPA-0 shows reduced error magnitude and spread, particularly for higher-energy structures.}
\label{fig:metastable_energies}
\end{figure}

A critical test for materials discovery applications is a model's ability to correctly rank metastable structures relative to the ground state. Figure~\ref{fig:metastable_energies} compares the error distributions of MP-0b3 and MPA-0 in predicting energy differences above the ground state ($\Delta E_{\text{MLIP}} - \Delta E_{\text{DFT}}$) across 16,126 relaxed structures from 100 different compositions. Both models exhibit a systematic tendency to overstabilize local minima, as indicated by negative errors that increase with distance from the ground state. However, MPA-0 shows a considerably narrower error distribution and significantly reduced overstabilization, particularly for structures with $\Delta E_{\text{DFT}} > 0.5$ eV/atom. This improved energy ranking enhances MPA-0's reliability for polymorph prediction, phase stability assessment, and materials discovery applications.

\clearpage
\section{Benchmarks}

\subsection{Phonons}\label{sec:phonons}
\begin{figure}[h]
    \centering
    \begin{minipage}{0.45\linewidth}
        \centering
        \includegraphics[width=\linewidth]{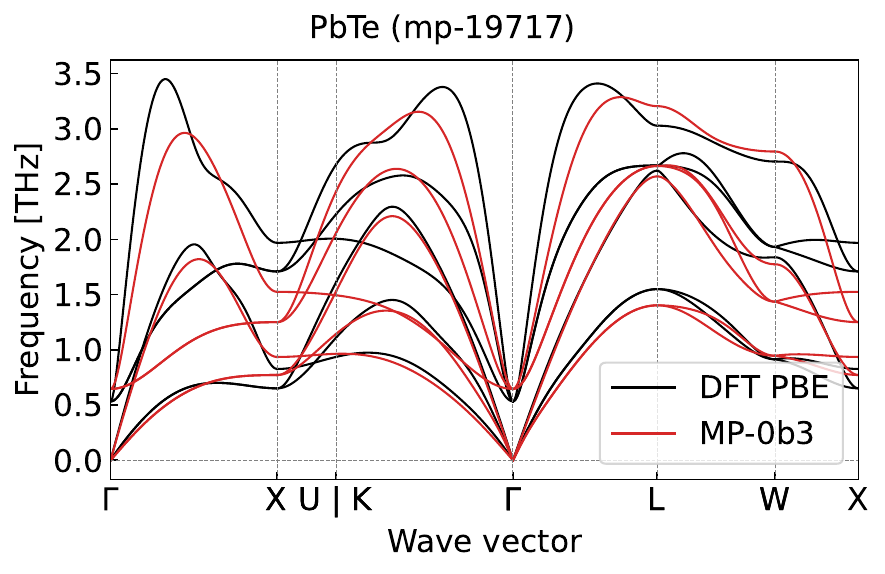}
        \caption*{(a) PbTe phonon band structure}
    \end{minipage}
    \hspace{0.05\linewidth} 
    \begin{minipage}{0.45\linewidth}
        \centering
        \includegraphics[width=\linewidth]{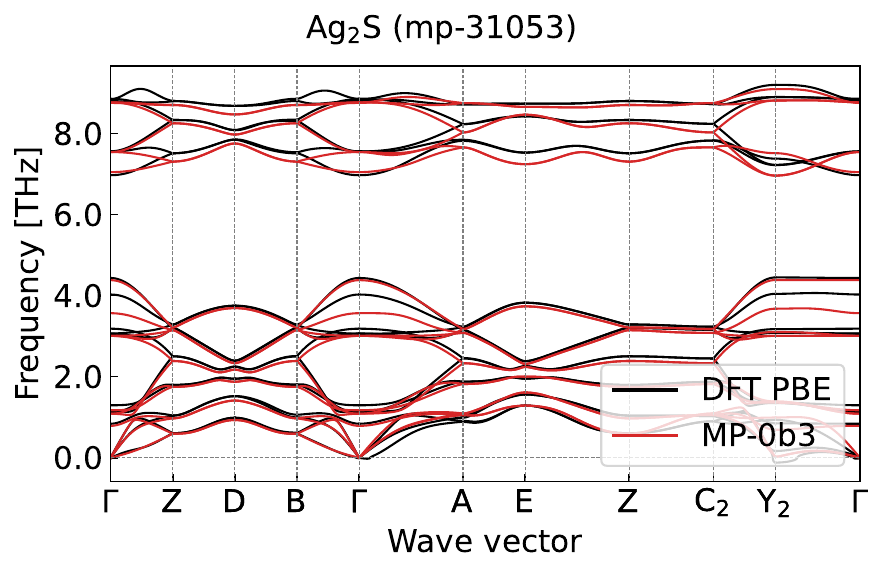}
        \caption*{(b) Ag$_2$S phonon band structure}
    \end{minipage}
    \begin{minipage}{0.45\linewidth}
        \centering
        \includegraphics[width=\linewidth]{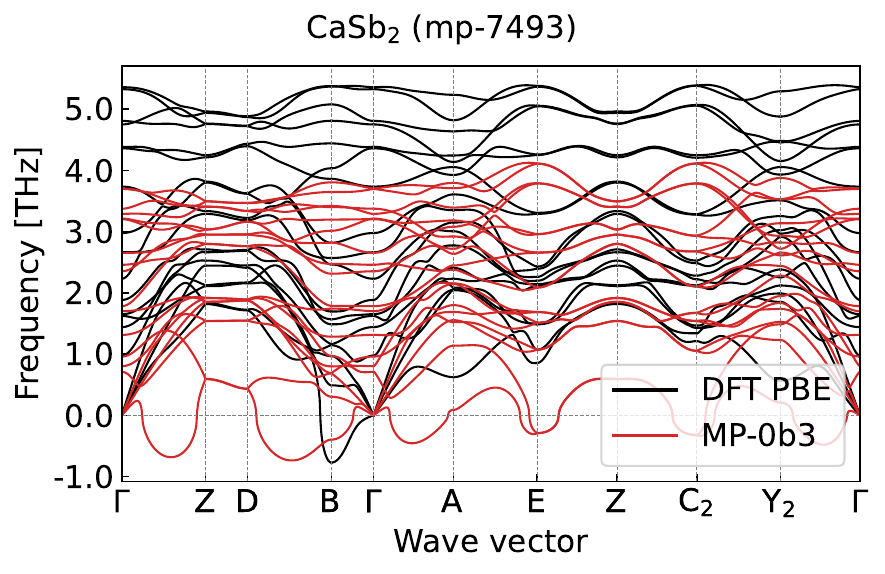}
        \caption*{(c) CaSb$_2$ phonon band structure}
    \end{minipage}
    \begin{minipage}{0.45\linewidth}
        \centering
        \vspace{0.5cm}
        \includegraphics[width=\linewidth]{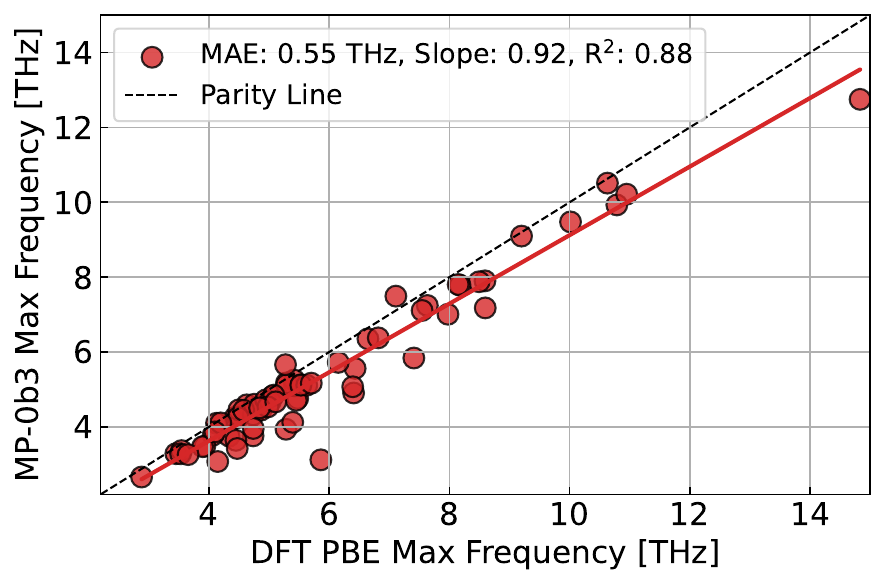}
        \caption*{(d) MACE vs PBE highest phonon frequency parity plot}
    \end{minipage}
    \caption{Comparison of DFT PBE and \MPzbt{} phonon band structures. Imaginary modes are plotted with a negative sign. a) and b) show examples of particularly good phonon bands. c) is a particularly bad example.  d) Parity plot of \MPzbt{} vs PBE highest phonon band frequency showing good agreement across the chosen materials.
showing excellent agreement across diverse materials.}
    \label{fig:phonon_bands}
\end{figure}

\begin{figure}[h]
    \centering
    \begin{minipage}{0.3\linewidth}
            \centering
            \includegraphics[width=\linewidth]{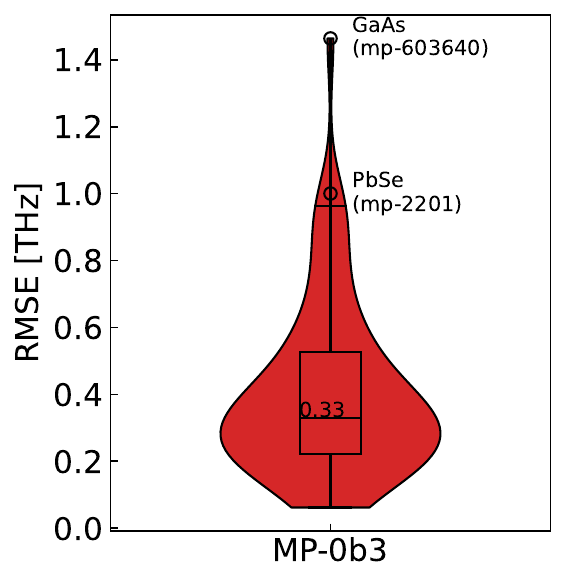}
            \caption*{(a) Violin/Box plot of phonon RMSE.}
    \end{minipage}
    \begin{minipage}{0.3\linewidth}
            \centering
            \includegraphics[width=\linewidth]{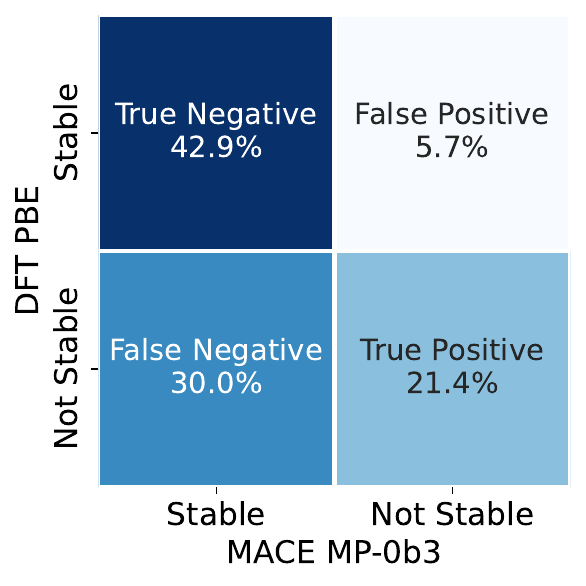}
            \caption*{(b) confusion matrix}
    \end{minipage}
    \begin{minipage}{0.3\linewidth}
            \centering
            \includegraphics[width=\linewidth]{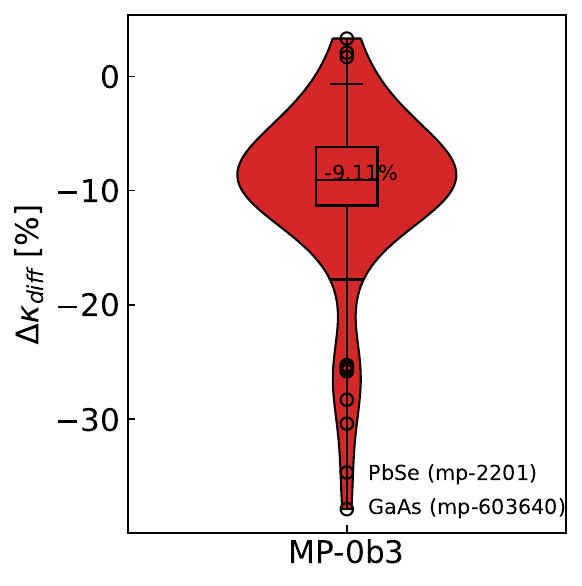}
            \caption*{(c) Relative error of diffusive thermal conductivity}
    \end{minipage}
    \caption{Comparison of DFT PBE and \MPzbt{} phonons. (a) shows the violin and box plot
    of the RMSE over the whole Brillouin zone. (b) shows the confusion matrix of occurrence of imaginary modes with a threshold of $|\omega_{imag}|$ < 0.05 THz. (c) Relative Error of the diffusive thermal conductivity}
    \label{fig:metrics_phonons}
\end{figure}

Accurate modeling of phonons is essential for determining dynamic stability of crystals, as well as thermoelectric properties and entropic contributions to the free energy \cite{stoffel_ab_2010, bartel_review_2022}. Those are important factors in the discovery of new materials. Harmonic phonons are typically calculated from the restoring force on atomic displacements and require highly accurate force predictions to be physical. An ML potential trained on PBE forces should be able to reproduce PBE lattice vibrations \cite{morrow_how_2023, george_combining_2020}. To assess the accuracy of \MPzbt{} restoring forces, we compare harmonic phonon modes predicted via the finite-displacement method as implemented in Phonopy \cite{togo_firstprinciples_2023,togo_implementation_2023} and atomate2 \cite{ganose2025atomate2, atomate2}. For this purpose, we use a benchmark set of 70 thermoelectric, phase-change, and chalcogenide-based materials. Furthermore, the reference set only contains materials in which magnetism does not play a role and U-correction is not applied in the MP. \\
For all materials in the benchmark set we used the MPRelaxMaker \cite{ganose2025atomate2} for structure optimization and chose a k-point density of 300 k-points per reciprocal volume and a cutoff energy of 520 eV (1.3*ENMAX of the highest ENMAX indicated in the pseudopotential files for the PAW method for each compound). We chose an electronic convergency criterion of 10$^{-7}$ eV and the ionic convergence criterion of 10$^{-6}$ eV.
We perform the calculations of single-atom-displaced supercells with the MPStaticMaker \cite{ganose2025atomate2}, setting the displacement amplitude to 0.01 \AA~ and setting the minimum edge length to 20 \AA, which allows us to perform the calculation with only one k-point. All data and scripts for reproducing the benchmark can be found under: https://doi.org/10.5281/zenodo.15462976\\
As examples, we show two compounds out of our set where DFT and \MPzbt{} agree well (Figs. \ref{fig:phonon_bands}a and b), and one of the compounds of largest qualitative disagreement (Fig \ref{fig:phonon_bands}c).
The noticeable drop in the optical modes around Gamma for PbTe arises from neglecting the non-analytic corrections derived from Born charges, which are unavailable from electronic structure-less ML potentials.
The parity plot of \MPzbt{} vs PBE highest-frequency phonon modes in Fig. \ref{fig:phonon_bands}d reveals that \MPzbt{} typically slightly underestimates the highest overall frequency in the phonon band structure (henceforth referred to as bandwidth, which correlates with the stiffness of the material’s strongest bond) by a mean absolute error of 0.55 THz in comparison to DFT. 
\MPzbt{} bandwidths have an excellent predictive power of R$^2$ = 0.88.
The least squares regression (red line), with a slope close to one, indicates that the error is largely independent of the actual bandwidth.
The largest absolute discrepancy between PBE vs \MPzbt{} maximal phonon frequency is around 2.7 THz (47\%) for orthorhombic GaAs (mp-603640). In contrast, 49 out of 70 materials show a maximal phonon frequency error of less than 10\%.
The violin and box plot of the root mean square errors (RMSEs) in Fig. \ref{fig:metrics_phonons}a indicate a median error of 0.33 THz. We calculate the RMSE over the entire Brillouin zone, as the density of states (DOS) is the only invariant quantity and the choice of the $k$-path is arbitrary.
Highly specialised models for one element can achieve much lower errors of 0.1–0.2 THz over the whole band structure \cite{george2020combining}. \MPzbt{} attains this level of accuracy for a few materials but does not achieve it on average. However, the deviations are better compared to results from models that specifically train only on the highest phonon peak on MatBench \cite{dunn_benchmarking_2020}. Here, the best models arrive at mean absolute errors of 0.8–1 THz for the last DOS peak of phonons, which is a similar but not identical quantity to the width of the band. \\
As shown in the confusion matrix inset to Fig. \ref{fig:metrics_phonons}, the model also provides a signal on the presence of imaginary modes. The presence of such lattice vibrations indicates that the structure is dynamically unstable, meaning it can transform into a lower-energy structure by displacing atoms along the corresponding vibrational eigenvector. \MPzbt{} achieves 64\% accuracy on binary dynamic stability classification with a
PBE unstable rate of 51.4\% (a nearly balanced data set). Although this outperforms the dummy accuracy of 50\%, it also leaves room for significant improvement. 
A set threshold of 0.05 THz determines how large the absolute value of the imaginary mode must be to count as imaginary. For example, a mode depicted at -0.04 THz in the phonon band structure would not be considered dynamically unstable while -0.06 THz would.
The false negative values are strikingly high. We assume there are three reasons for this. Firstly, a caveat to the above analysis is that the PBE unstable labels we use as ground truth may themselves contain false negatives due to potential convergence errors. Secondly, the data set contains mainly thermoelectric materials, 
which are more prone to the occurrence of imaginary modes. Thirdly, some materials in the dataset have exceptionally long unit cells (~20 \AA) in one direction. In these nearly amorphous structures in one direction, the energy landscape becomes more complex due to the increased degrees of freedom and leads faster to false positives or false negatives. \\
As is, \MPzbt{} can already serve as a useful pre-screening filter for dynamically unstable materials, especially given the lower false positive (5.7\%) than false negative rate (30.0\%). This means \MPzbt{} is less likely to predict materials as unstable that are stable than vice versa and, therefore, is biased toward keeping materials in the candidate pool. \\
We also emphasize the low computational cost of these predictions. Generating a complete phonon band structure
on an Apple M2 Max CPU takes approximately 30 s. Running \MPzbt{} on a supercomputer could easily generate approximate phonon bands for every material in MP or other large databases. However, the accuracy of harmonic phonons of \MPzbt{} is still insufficient to be a standard substitute for DFT. \\
However, \MPzbt{} is already well suited for estimating diffusive thermal conductivity, which represents a lower limit of thermal conductivity. Following the proposed model of Agne et al. \cite{agne2018minimum}, the diffusive thermal conductivity in the high-temperature limit only depends on the average phonon frequency. 
Figure \ref{fig:metrics_phonons} shows the violin and parity plot of the relative diffusive thermal conductivity error of the prediction by \MPzbt{}.
Since the phonon bandwidth is systematically underestimated (Fig.\ref{fig:phonon_bands}d), a consistent downshift is observed, with a median of -9.11\%.
The other differences in the phonon band shapes between DFT and \MPzbt{} are effectively averaged out. Imaginary modes were not included in the calculation.
The long tail in the violin plot arises from a few cases where the maximum phonon frequencies are significantly underestimated. This occurs due to a softening effect \cite{deng2025systematic}, which results in an effective underestimation of the PES and, consequently, an underestimation of the forces, leading to a reduced phonon mean frequency.

\subsubsection*{Similarity statement}
The structures whose phonon predictions we analyzed are all part of the training set.
However, the supercells, including single-atom perturbations performed by the finite-displacement method are not.

\clearpage
\subsection{Bulk and Shear Moduli}\label{sec:bulk_moduli}

\begin{figure}[htbp!]

  \centering  \includegraphics[width=\textwidth,keepaspectratio]{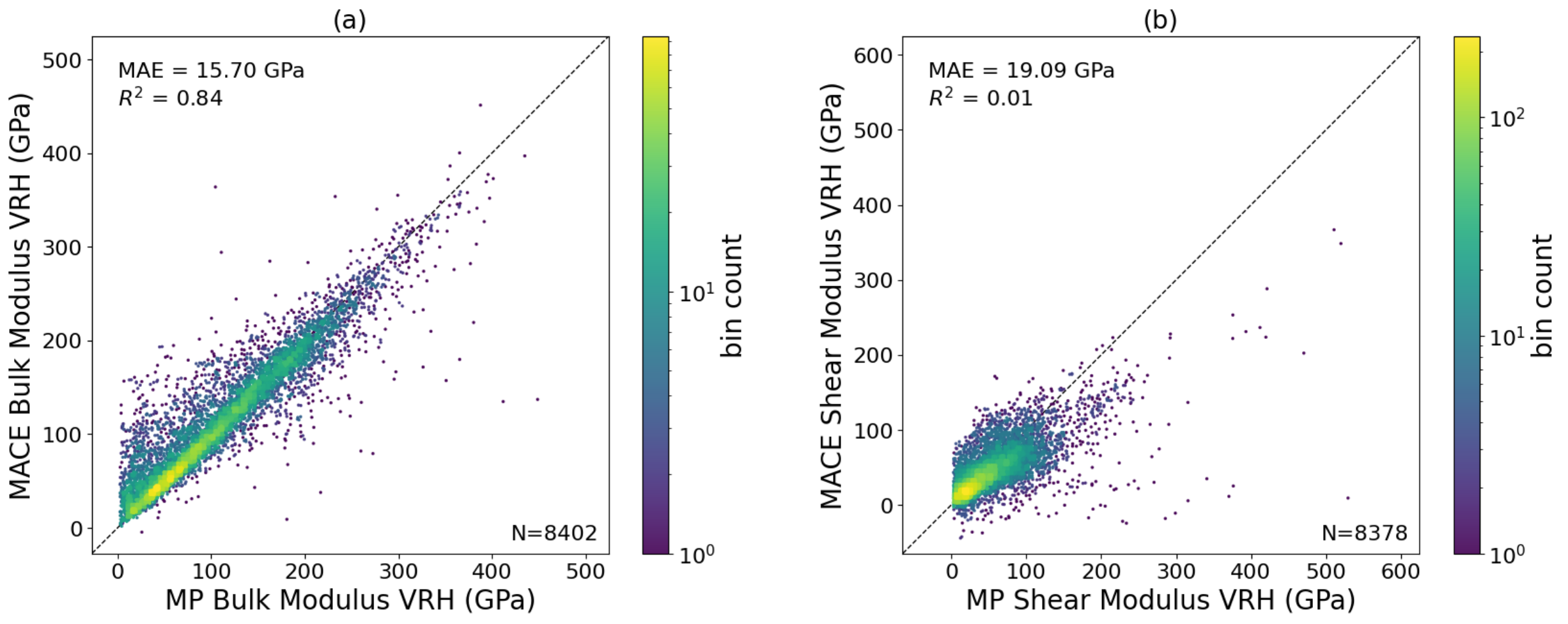}
  \caption{Comparison between \MPz{}- and MP-calculated (a) bulk moduli and (b) shear moduli for approximately 8,400 materials stored in the MP Database. The dashed line is a parity line (\textit{i.e.} the target distribution of data). For shear moduli, note that 10 points were excluded from this plot, as \MPz{} predicted an unphysically high (\SI{\geq 600}{\giga\pascal}) or low (\SI{\leq -50}{\giga\pascal}) shear modulus; this includes mp-\{1007819, 1008669, 1009019, 11478, 2458, 27954, 631377, 631633, 721759, 984628\}.}
  \label{fig:bulk-and-shear-moduli}
\end{figure}

\MPz{} was benchmarked against the elastic properties of over 8,000 materials stored in the MP database. Being able to capture elastic properties such as bulk and shear modulus - which depend on the second derivatives of the energy with respect to strain - demonstrates a more precise ability to capture the potential energy surface.

Specifically, \MPz{} was used to calculate the Voigt-Reuss-Hill average \cite{voigt_lehrbuch_1910, reuss_berechnung_1929, hill_elastic_1952} bulk modulus
and shear modulus
as derived from stress-strain relations. The initial structures used for these calculations were the relaxed PBE \cite{perdew1996generalized} structures from MP; these structures were then re-relaxed using the \MPz{} model, and then deformed. Specifically, a total of 4 strain magnitudes were used along 6 independent strain modes (in Voigt notation): $\boldsymbol{\epsilon} \in [\epsilon_{11}, \epsilon_{22}, \epsilon_{33}, \epsilon_{44}, \epsilon_{55}, \epsilon_{66}]$. For $\epsilon_{11}$, $\epsilon_{22}$, and $\epsilon_{33}$, the strain magnitudes were $\pm 0.01$ and $\pm 0.005$. For $\epsilon_{44}$, $\epsilon_{55}$, and $\epsilon_{66}$, the strain magnitudes were $\pm 0.06$ and $\pm 0.03$. These calculations were performed using the \texttt{elasticity} module from the \texttt{MatCalc} package \cite{Riebesell_MatCalc_2023}. Hence, all of these predictions are based on equilibrium, bulk crystals alone. Moreover, to filter out likely unphysical DFT predictions, elastic properties from MP were excluded from this analysis if the DFT VRH average bulk or shear modulus are less than \SI{-50}{\giga\pascal} or greater than \SI{600}{\giga\pascal}. Note that the data excluded due to unphysical DFT-based properties are distinct from the data not plotted in \cref{fig:bulk-and-shear-moduli} due to poor \MPz{} predictions.

Results comparing MP and \MPz{} bulk moduli with MAE of \SI{15.70}{\giga\pascal} and $R^2$ of 0.84 are shown in \cref{fig:bulk-and-shear-moduli}(a). This compares favorably to the $R^2$ value of 0.757 reported for M3GNet \cite{chen_universal_2022}. Similarly, results for shear moduli are shown in \cref{fig:bulk-and-shear-moduli}(b). \MPz{} struggles to predict shear properties, likely due to a lack of sheared structures in MP.

\subsubsection*{Similarity statement}
None of the DFT deformation calculations contained in MP are present in the MPtrj training set used for \MPz{}. 


\clearpage
\subsection{Cohesive energies} \label{sec:lattice-energies}

\begin{figure}[htbp!]
  \centering
  \includegraphics[width=\textwidth,keepaspectratio]{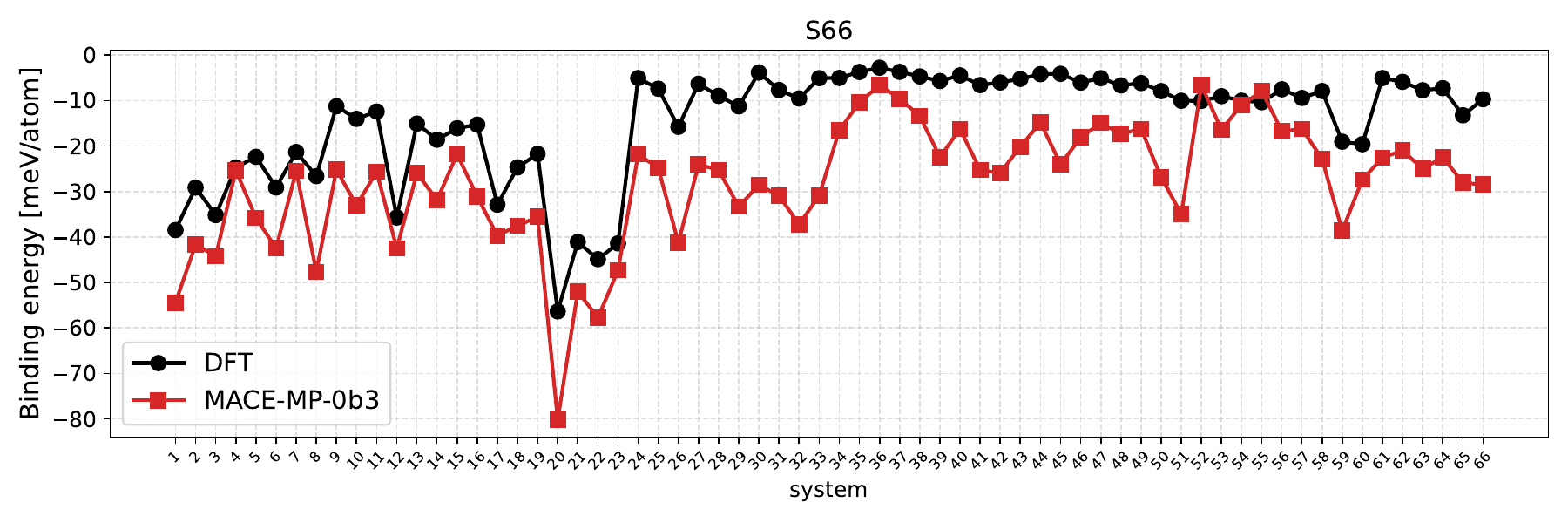}
  \caption{Comparison between DFT (black) and \MPz{} (red) calculated binding energies of the S66 dimers. The binding energies are divided by the number of atoms in each dimer. The lines are guides for the eye.
  }
  \label{fig:s66}
\end{figure}

\begin{figure}[htbp!]
  \centering
  \includegraphics[width=\textwidth,keepaspectratio]{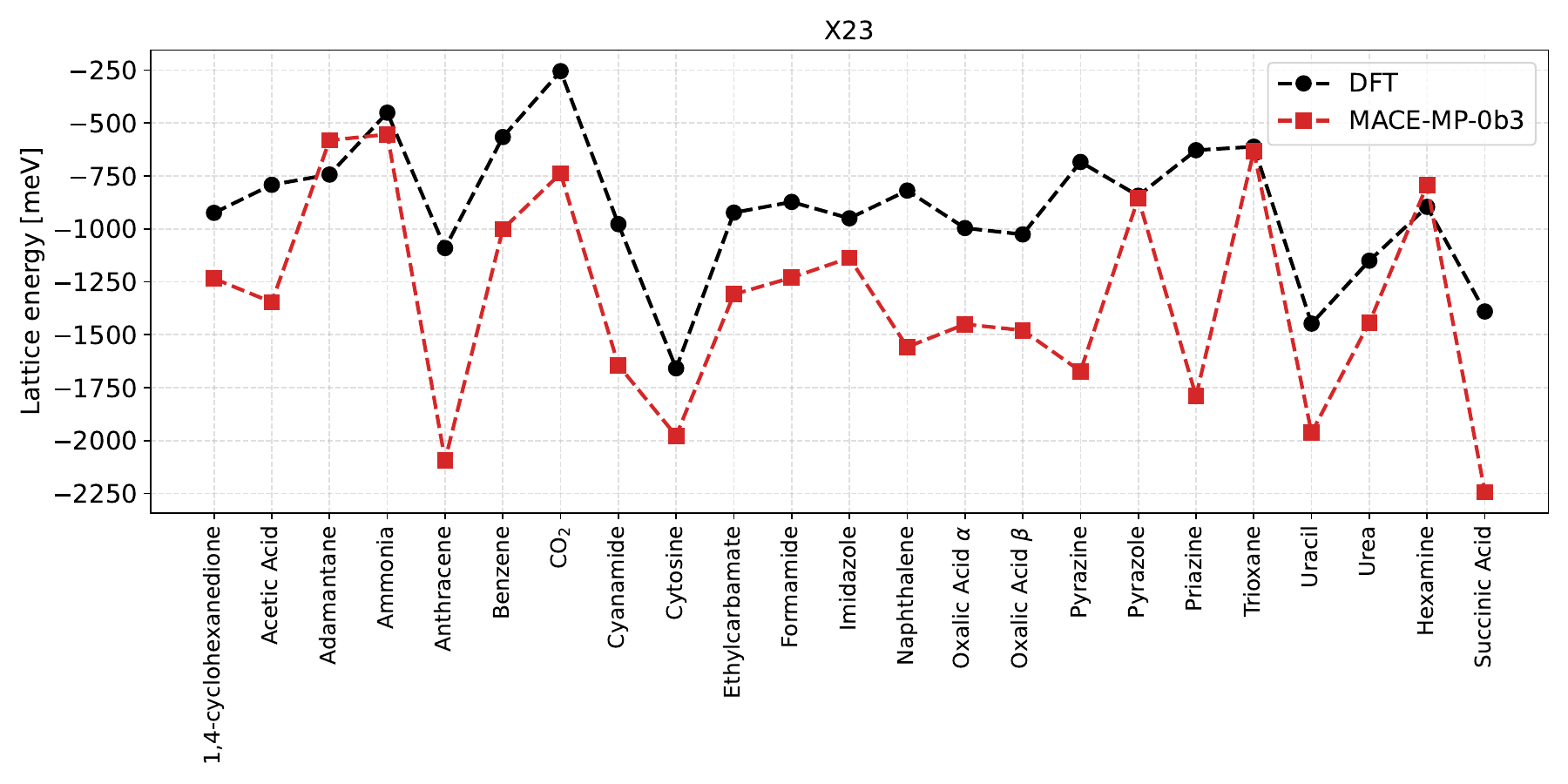}
  \caption{Comparison between DFT (black) and \MPz{}-calculated (red) lattice energies of the X23 dataset. The lines are guides for the eye.
  }
  \label{fig:x23}
\end{figure}

\begin{figure}[htbp!]
  \centering
  \includegraphics[width=\textwidth,keepaspectratio]{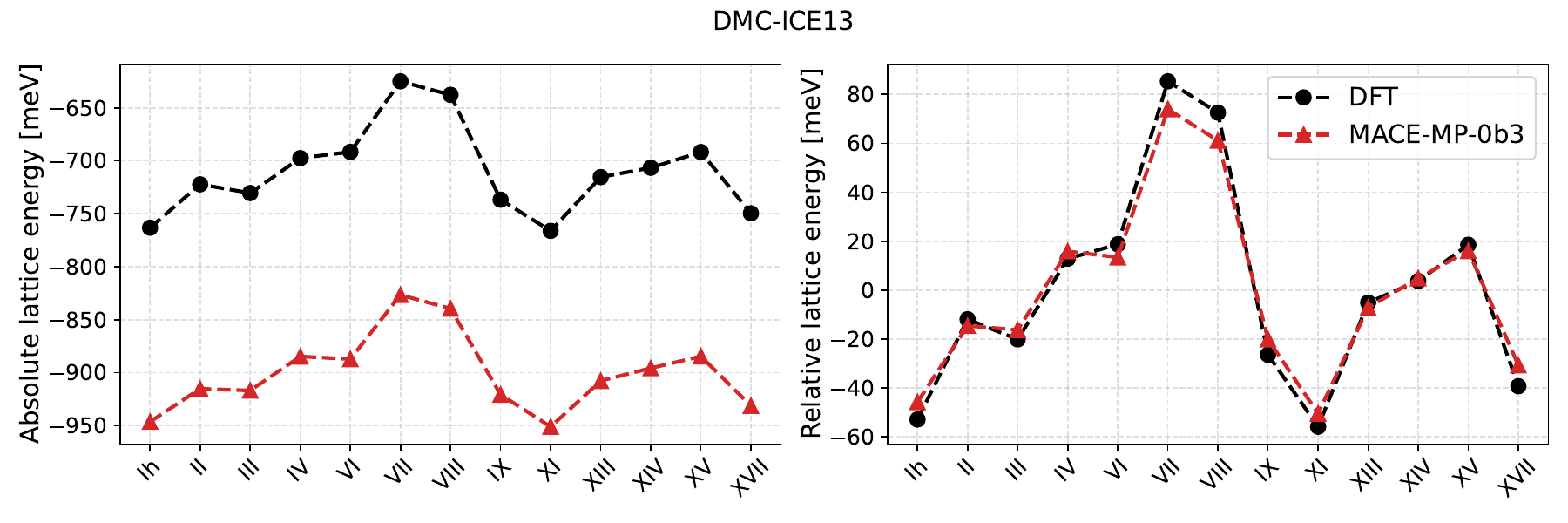}
  \caption{Comparison between DFT (black) and \MPzbt{}  (red) lattice energies of the DMC-ICE13 dataset. We report both the absolute lattice energies (left), \textit{i.e.} the energy per molecule of each crystalline phase with respect to the gas phase, and the relative lattice energies (right), \textit{i.e.} the lattice energy relative to mean value among the 13 polymorphs.
  }
  \label{fig:dmcice13}
\end{figure}

In this section, we benchmark \MPz{} against the cohesive energies of widely used data sets of molecules and molecular crystals, S66\cite{S66} and X23\cite{x23_rt}. S66 is a dataset comprising 66 molecular complexes at their reference equilibrium geometries, designed to cover the most common types of noncovalent interactions in biomolecules while keeping a balanced representation of dispersion and electrostatic contributions. X23 is a dataset of 23 organic molecular crystals. Furthermore, we analyze the relative stabilities of the ice polymorphs in DMC-ICE13\cite{dmcice13}. The DFT calculations were performed using VASP\cite{VASP1,VASP2,VASP3} with the PBE functional and D3 dispersion correction with Becke-Johnson damping. The energy cutoff is \SI{520}{eV}. Gas phase calculations were performed at the $\Gamma$ point in a \SI{25}{\Angstrom} cubic box. Solid phase calculations were performed with a $4\times 4 \times 4$ k-point grid. The results comparing \MPz{} and DFT binding energies of the S66 dimers, and the lattice energies of X23 and DMC-ICE13 are shown in \cref{fig:s66}, \cref{fig:x23}, and \cref{fig:dmcice13}. The MAE is approximately \SI{4}{meV/atom} for S66, \SI{459}{meV} for X23, and \SI{190}{meV} for the DMC-ICE13 absolute lattice energies. The relative stabilities of the ice polymorphs are correctly captured with an MAE of \SI{5}{meV} on the relative lattice energies.

\subsubsection*{Similarity statement} 
S66 and X23 comprise dimers and molecular crystals containing C, H, N, or O atoms. Ice polymorphs contain H and O atoms. The MP database contains 73799 structures with O atoms, 10312 structures with H atoms, 11356 structures with N atoms, and 9043 structures with C atoms. The database contains 6 structures matching an exact chemical formula in S66; these are \ch{H4O2}, \ch{C4HO8O4}, \ch{C8H8} and \ch{C4H4}. 16 structures match an exact chemical formula in X23; these are \ch{C8H16O8}, \ch{C20H32}, \ch{H12N4}, \ch{C4O8}, \ch{C8H16N16} and \ch{C2H8N4O2}. 9 structures match an exact chemical formula in DMC-ICE13; these are \ch{H24O12} and \ch{H16O8}. Overall, the database contains 630 structures with organic molecules and 1342 structures with water molecules. We provide \verb|s66_chemiscope_input.json|, \verb|x23_chemiscope_input.json|, and \verb|dmcice13_chemiscope_input.json| to help visualize the interactive UMAP on \url{chemiscope.org}.

\subsection{Atomization energies and lattice constants of solids}
\label{sec:atomization-energies}
In the following section, we benchmark \MPz{} against the atomization energies and lattice constants of a set of solids. The details of the DFT calculations are the same as in \cref{sec:lattice-energies}. Solid phase DFT total energies were computed with a $16\times 16 \times 16$ k-point grid. \MPz{} and DFT atomization energies (in eV/atom) are reported in \cref{tab:atomization-energies}. The MAE is \SI{0.03}{eV/atom}. Lattice constants are computed on equilibrium structures obtained by geometry relaxation with a force convergence threshold of \SI{0.03}{eV\per\Angstrom}. \MPz{} and DFT lattice constants (in \unit{\Angstrom})
are reported in \cref{tab:lattice-constants}. The MAE is \SI{0.03}{\unit{\Angstrom}}.

\subsubsection*{Similarity statement} 
All the tested solids are contained in the database, with an exact matching chemical formula for 71 structures. We provide \verb|solids_chemiscope_input.json| to help visualize the interactive UMAP on \url{chemiscope.org}.

\begin{table}[ht]
\centering
\begin{tabular}{ccccc}
    \multicolumn{5}{c}{Atomization energy of solids}    \\         
\hline \hline
 & DFT & \MPz{} & $\Delta$ & $\Delta$/DFT [\%] \\
\hline
Ag & 3.0980 & 3.0560 & 0.0420 & 1.4 \\ 
Pd & 4.3980 & 4.3560 & 0.0420 & 1.0 \\ 
Rh & 6.4040 & 6.4380 & -0.0340 & 0.5 \\ 
Li & 1.7820 & 1.7820 & 0.0000 & 0.0 \\ 
Na & 1.2440 & 1.2300 & 0.0140 & 1.1 \\ 
K & 0.9870 & 0.9700 & 0.0170 & 1.7 \\ 
Rb & 0.8810 & 0.8500 & 0.0310 & 3.5 \\ 
Cs & 0.8080 & 0.7770 & 0.0310 & 3.8 \\ 
Ca & 2.1380 & 2.1370 & 0.0010 & 0.0 \\ 
Sr & 1.8100 & 1.8120 & -0.0020 & 0.1 \\ 
Ba & 2.0790 & 2.0790 & 0.0000 & 0.0 \\ 
Al & 3.8910 & 3.8320 & 0.0590 & 1.5 \\ 
Cu & 4.0750 & 4.0840 & -0.0090 & 0.2 \\ 
Si & 4.9220 & 4.8680 & 0.0540 & 1.1 \\ 
Ge & 4.0360 & 4.0870 & -0.0510 & 1.3 \\ 
C & 8.0330 & 7.9910 & 0.0420 & 0.5 \\ 
LiF & 4.2670 & 4.1830 & 0.0840 & 2.0 \\ 
NaF & 3.7980 & 3.7320 & 0.0660 & 1.7 \\ 
NaCl & 3.1290 & 3.1080 & 0.0210 & 0.7 \\ 
MgO & 4.7100 & 4.7020 & 0.0080 & 0.2 \\ 
SiC & 5.7620 & 5.7130 & 0.0490 & 0.9 \\ 
GaAs & 2.5480 & 2.4720 & 0.0760 & 3.0 \\ 
LiCl & 3.3770 & 3.3540 & 0.0230 & 0.7 \\ 
\hline
 &  &  & MAE &  \\ 
 &  &  & 0.03 &  \\ 

\end{tabular}
\caption{Comparison between DFT and \MPzbt{}-calculated atomization energies of solids. The column
$\Delta$ reports the difference between DFT and \MPz{} values. Energies are reported in eV/atom.}
\label{tab:atomization-energies}
\end{table}

\begin{table}[ht]
\centering
\begin{tabular}{ccccc}
    \multicolumn{5}{c}{Lattice constants of solids}    \\         
\hline \hline
 & DFT & \MPzbt{} & $\Delta$ & $\Delta$/DFT [\%] \\
\hline
Ag & 4.0820 & 4.0820 & 0.0000 & 0 \\ 
Al & 4.0020 & 3.9880 & 0.0140 & 0 \\ 
Ba & 4.9760 & 4.9490 & 0.0270 & 1 \\ 
C & 3.5620 & 3.5510 & 0.0110 & 0 \\ 
Ca & 5.4630 & 5.4410 & 0.0220 & 0 \\ 
Cs & 6.1060 & 5.9220 & 0.1840 & 3 \\ 
Cu & 3.5680 & 3.5480 & 0.0200 & 1 \\ 
GaAs & 5.6900 & 5.6760 & 0.0140 & 0 \\ 
Ge & 5.7190 & 5.6770 & 0.0420 & 1 \\ 
K & 5.1910 & 5.1910 & 0.0000 & 0 \\ 
Li & 3.3520 & 3.2890 & 0.0630 & 2 \\ 
LiCl & 5.0560 & 5.0490 & 0.0070 & 0 \\ 
LiF & 3.9950 & 4.0210 & -0.0260 & 1 \\ 
MgO & 4.2030 & 4.2070 & -0.0040 & 0 \\ 
Na & 4.1070 & 4.1850 & -0.0780 & 2 \\ 
NaCl & 5.5850 & 5.5750 & 0.0100 & 0 \\ 
NaF & 4.6190 & 4.5980 & 0.0210 & 0 \\ 
Pd & 3.8910 & 3.9120 & -0.0210 & 1 \\ 
Rb & 5.5720 & 5.5720 & 0.0000 & 0 \\ 
Rh & 3.7600 & 3.8090 & -0.0490 & 1 \\ 
Si & 5.4340 & 5.4170 & 0.0170 & 0 \\ 
Sr & 5.9080 & 5.9660 & -0.0580 & 1 \\ 
SiC(a) & 3.0720 & 3.0780 & -0.0060 & 0 \\ 
SiC(c) & 5.0290 & 5.0310 & -0.0020 & 0 \\ 
\hline
 &  &  & MAE &  \\ 
 &  &  & 0.03 &  \\ 

\end{tabular}
\caption{Comparison between DFT and \MPzbt{} lattice constants of solids. The column $\Delta$
reports the difference between DFT and \MPzbt{} values. Lattice constants are in \unit{\Angstrom}.}
\label{tab:lattice-constants}
\end{table}

\clearpage
\subsection{Reaction barrier heights}
\label{sec:reaction-barrier}
In the following section, we benchmark \MPz{} against the reaction barrier heights of the CRBH20 database\cite{crbh20}, comprising 20 barrier heights for the cycloreversion of heterocyclic rings. The set-up of the DFT calculations is the same as in \cref{sec:lattice-energies}. \MPz{} and DFT barrier heights (in \unit{eV}) are reported in \cref{tab:barrier-heights}. The MAE is approximately \SI{0.3}{eV}.

\subsubsection*{Similarity statement} The CRBH20 dataset comprises barrier heights of organic molecules, containing H, O, C, N, S, and F atoms. We report the list of atoms with the number of structures in which they are contained in parentheses: O (73799), S (11972), H (10312),  N (11356),  F (11277), C (9043).
Overall, the database contains no structures matching an exact chemical formula in CRBH20.  We provide \verb|crbh20_chemiscope_input.json| to help visualize the interactive UMAP on \url{chemiscope.org}.

\begin{table}[ht]
\centering
\begin{tabular}{ccccc}
    \multicolumn{5}{c}{Reaction barrier heights}    \\         
\hline \hline
 & DFT & \MPzbt{} & $\Delta$ & $\Delta$/DFT [\%] \\
\hline
   1 & 1.7194 & 2.0431 & -0.3237 & 19 \\ 
   2 & 1.9241 & 1.9988 & -0.0747 & 4 \\ 
   3 & 1.7499 & 1.8505 & -0.1006 & 6 \\ 
   4 & 1.8238 & 1.8179 & 0.0059 & 0 \\ 
   5 & 1.7237 & 1.8621 & -0.1384 & 8 \\ 
   6 & 1.5653 & 1.1341 & 0.4312 & 28 \\ 
   7 & 1.0911 & 1.1134 & -0.0223 & 2 \\ 
   8 & 1.8983 & 1.6116 & 0.2867 & 15 \\ 
   9 & 1.5477 & 1.7161 & -0.1684 & 11 \\ 
  10 & 1.7115 & 1.1390 & 0.5725 & 33 \\ 
  11 & 1.7379 & 1.8005 & -0.0626 & 4 \\ 
  12 & 2.0361 & 1.6698 & 0.3663 & 18 \\ 
  13 & 1.8739 & 1.5611 & 0.3128 & 17 \\ 
  14 & 1.9760 & 1.6166 & 0.3594 & 18 \\ 
  15 & 1.8865 & 1.5719 & 0.3146 & 17 \\ 
  16 & 1.5741 & 0.7963 & 0.7778 & 49 \\ 
  17 & 1.2587 & 0.8127 & 0.4460 & 35 \\ 
  18 & 1.7497 & 1.3373 & 0.4124 & 24 \\ 
  19 & 1.6989 & 1.4281 & 0.2708 & 16 \\ 
  20 & 1.7654 & 0.8742 & 0.8912 & 50 \\ 
\hline
 &  &  & MAE &  \\ 
 &  &  & 0.3 &  \\ 

\end{tabular}
\caption{Comparison between DFT and \MPzbt{} barrier heights for CRBH20. The column
$\Delta$ reports the difference between DFT and \MPzbt{} values. Energies are in eV.}
\label{tab:barrier-heights}
\end{table}


\clearpage
\subsection{Homonuclear diatomics}\label{sec:dia}

\begin{figure*}[htbp!]
  \centering
  \includegraphics[width=\linewidth,keepaspectratio]{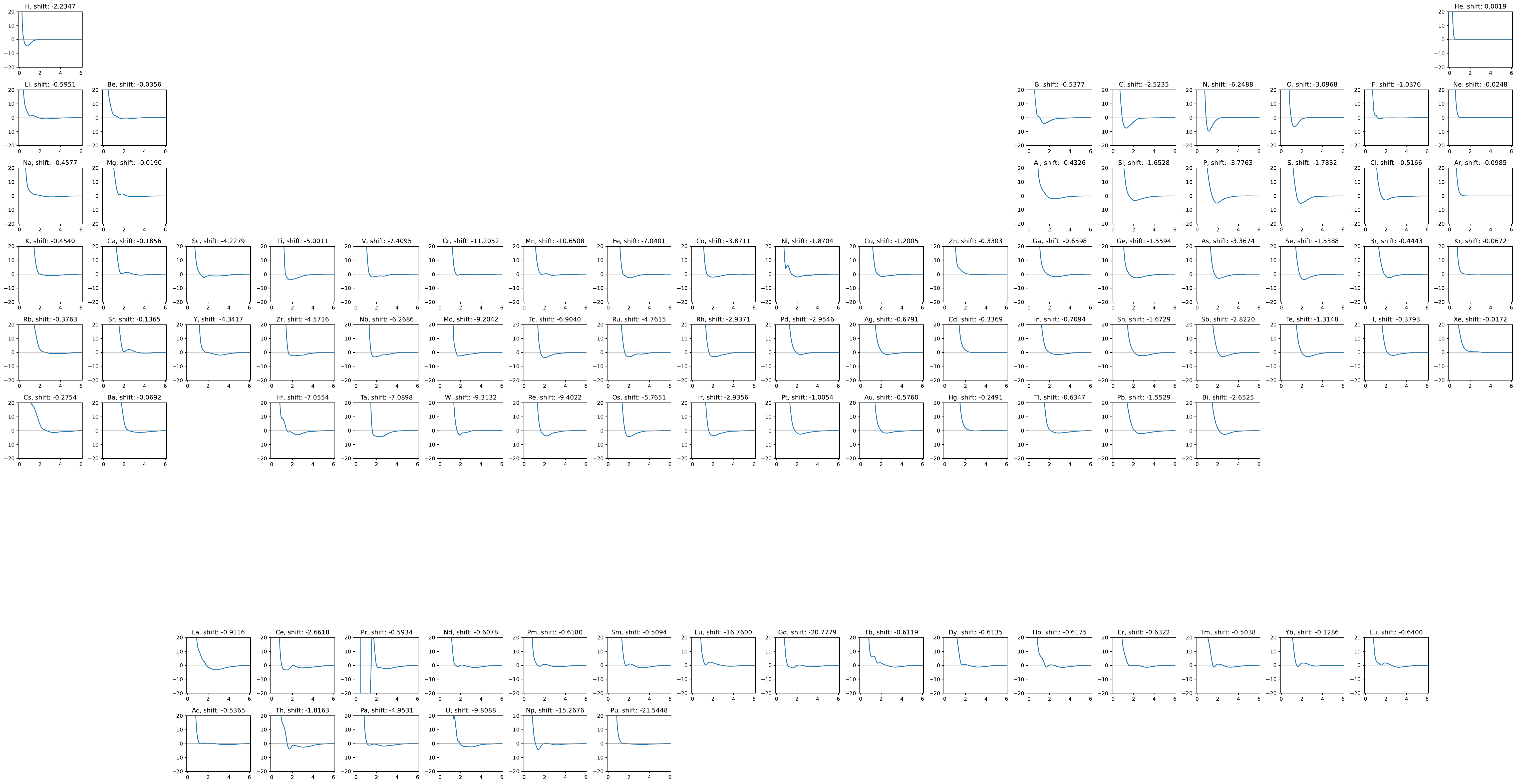}
  \caption{Energies of homonuclear diatomics in vacuum. Interactions are repulsive at small distances for the entire periodic table.}
  \label{fig:homonuclear-2b}
\end{figure*}

\begin{figure*}[htbp!]
  \centering
  \begin{subfigure}[b]{0.99\textwidth}
    \includegraphics[width=\textwidth,keepaspectratio]{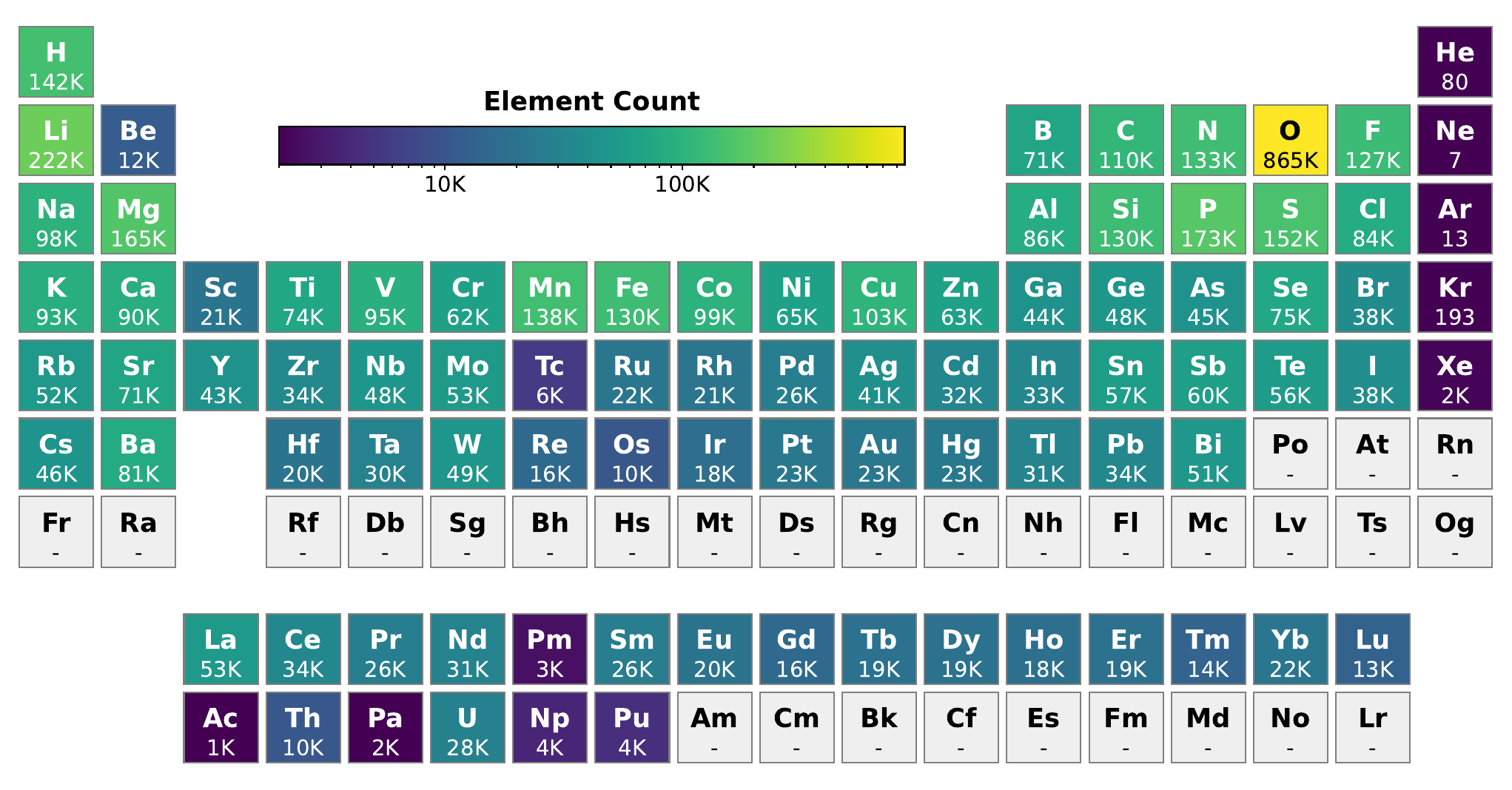}
    \caption{MPtrj training set element occurrence}
    \label{fig:mp-trj-element-counts-by-occurrence-log}
  \end{subfigure}
  \begin{subfigure}[b]{0.99\textwidth}
    \includegraphics[width=\textwidth,keepaspectratio]{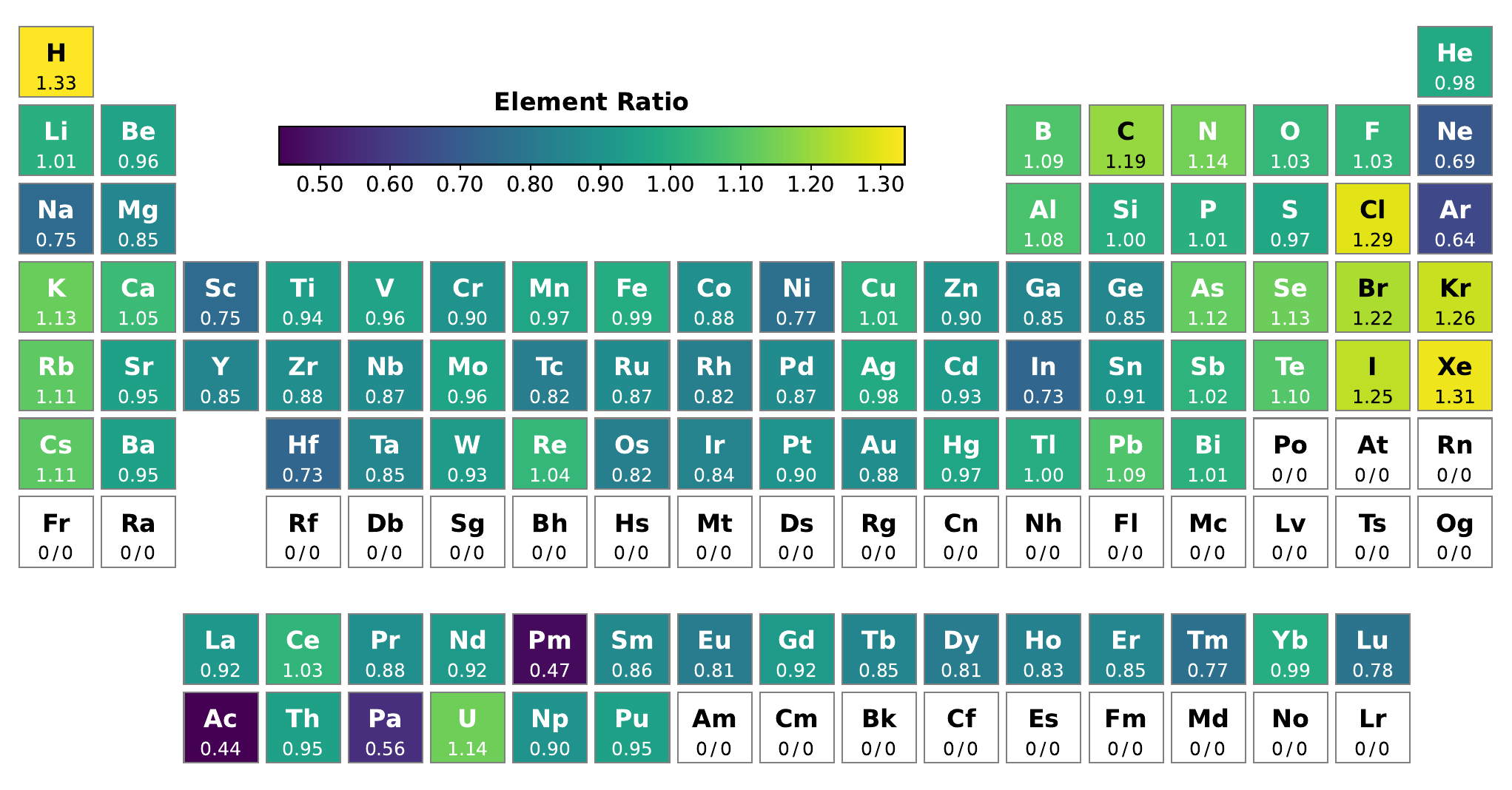}
    \caption{Normalized ratio of elements in MPtrj to MP, $\frac{\text{MPtrj / len(MPtrj)}}{\text{MP / len(MP)}}$}
    \label{fig:mp-trj-mp-ratio-element-counts-by-occurrence}
  \end{subfigure}
  \caption{
    The number of structures containing a given element in the MPtrj training set  \cite{deng_chgnet_2023}.
    MPtrj consists of multiple configurations from every relaxation trajectory in MP.
    Some elements can require more ionic steps to relax than others.
    To visualize this, \subref*{fig:mp-trj-mp-ratio-element-counts-by-occurrence} shows the overabundance of elements relative to the number of structures containing a given element in MP ground states (after normalizing by dividing each dataset by its number of structures).
    That is, the factor 1.33 for hydrogen in \subref*{fig:mp-trj-mp-ratio-element-counts-by-occurrence} indicates that structures containing hydrogen were selected 33\% more frequently than the base prevalence of hydrogen in MP ground states.
  }
  \label{fig:element-counts-ratio-by-occurrence}
\end{figure*}

Core repulsion is essential for stable modeling of atomic interactions, so atoms are prevented from coming too close together, especially when modeling high temperatures and pressures. The energies of all pairs of atoms were evaluated with the model in vacuum to test the 2-body interaction. The resultant curves for homonuclear diatomics are plotted in \cref{fig:homonuclear-2b}. All elements have repulsive potential at small distances, even elements with minimal presence in the training set (\cref{fig:element-counts-ratio-by-occurrence}). As an outlier, the Praseodymium pair shows an attractive non-physically large potential with an energy gain of over 450 eV for two \ch{Pr} atoms combining in vacuum.

\clearpage

\section{Training Methods and Data Exploration}

\subsection{Training protocol}\label{sec:trainingprotocol}

The pretrained \MPz{} interatomic potentials consist of many-body message passing (interaction) layers. In each layer, the message is encoded in the irreducible representation basis with $C$ channels up to an angular frequency of order $L$. This is specified as \verb|(128x0e+128x1o)| for 128 channels and $L=1$.

In each batch updating step, the weighted sum of Huber losses~\cite{Huber1992} of energy, forces, and stress incurred by all structures in a batch are averaged and back-propagated into the neural networks: \begin{equation}
  \begin{aligned}
    \mathcal{L} & = \frac{\lambda_E}{N_b} \sum_{b=1}^{N_b}\mathcal{L}_\text{Huber}\biggl(\frac{\hat{E}_b}{N_a}, \frac{E_b}{N_a}, \delta_E\biggr) + \frac{\lambda_F}{3\sum_{b=1}^{N_b}N_a} \sum_{b=1}^{N_b} \sum_{a=1}^{N_a}\sum_{i=1}^{3}\mathcal{L}^\star_\text{Huber}\biggl(-\frac{\partial\hat{E}_b}{\partial {r}_{b,a,i}}, F_{b,a,i}, \delta_F\biggr) \\
                & \hphantom{=}+ \frac{\lambda_\sigma}{9N_b} \sum_{b=1}^{N_b}\sum_{i=1}^3\sum_{j=1}^3\mathcal{L}_\text{Huber}\biggl(\frac{1}{V_b}\frac{\partial\hat{E}_b}{\partial {\varepsilon}_{b,ij}}, \sigma_{b,ij}, \delta_\sigma\biggr),
  \end{aligned}
\end{equation} where $\lambda_E, \lambda_F, \lambda_\sigma$ are predetermined weights of energy, forces, and stress losses. $(\lambda_E, \lambda_F, \lambda_\sigma) = (1, 10, 10)$ is adopted. 
Huber deltas $\delta_E = \delta_F = \delta_\sigma = 0.01$ are used. In particular, we use conditional Huber loss $\mathcal{L}^\star_\text{Huber}$ for forces, where the Huber delta $\delta_F$ is adaptive to the force magnitude on each atom. To be specific, the Huber delta $\delta_F$ decreases step-wise by a factor from 1.0 to 0.1 as the atomic force increases from \num{0} to \SI{300}{eV\per\Angstrom\per atom}. The Huber loss for forces can therefore be equivalently represented as: 

\begin{equation}
  \mathcal{L}^\star_\text{Huber}\left(\frac{\partial\hat{E}_b}{-\partial {r}_{b,a,i}}, F_{b,a,i}, \delta_F\right) = \left\lbrace\begin{aligned}
    \mathcal{L}_\text{Huber}(\dots, \delta_F)\,,    & \qquad F_{b,a,i} \textup{ \textless } \ 100         \\
    \mathcal{L}_\text{Huber}(\dots, 0.7\delta_F)\,, & \qquad 100 \le F_{b,a,i}  \textup{ \textless } \ 200 \\
    \mathcal{L}_\text{Huber}(\dots, 0.4\delta_F)\,, & \qquad 200 \le F_{b,a,i} \textup{ \textless } \ 300 \\
    \mathcal{L}_\text{Huber}(\dots, 0.1\delta_F)\,, & \qquad F_{b,a,i} \ge 300                 \\
  \end{aligned}\right.
\end{equation}

Standardization of target variables (here energies, forces and stresses) with different scales has been proven to be important for weight initialization and training stability \cite{bishop1995neural}, in that a large spread of input or output will result in large and uneven weight values and cause model instability. After each message passing layer $k$, the node energies $\epsilon_a$ are scaled and shifted before sum pooling. The energy prediction of each structure therefore reads: \begin{equation}
  \hat{E} = \sum_{a=1}^{N} \left[\sigma \left(\sum_{k=1}^K
    \epsilon_a^{(k)}\right) + \mu_{Z_a}\right],
\end{equation} where $K$ denotes the total number of message passing layers and $\epsilon_a^{(k)}$ is the node energy of atom $a$ at $k$-th layer. $\sigma$ is the root mean square of the atomic forces computed over the training dataset. In order to ensure the correct limit for the dissociation of atoms, we take $\mu_Z$ as the isolated atom energy computed with the MPtrj DFT. The predicted forces and stress are computed through PyTorch's automatic differentiation \texttt{torch.autograd} of total energy with respect to atomic positions and lattice strain tensor.

The original \MPz{}a models were trained for 200 epochs with \numrange{40}{80} NVIDIA A100 GPUs across \numrange{10}{20} nodes on HPE (Hewlett Packard Enterprise) Cray EX supercomputer Perlmutter, maintained by National Energy Research Scientific Computing Center (NERSC), a U.S. Department of Energy Office of Science User Facility located at Lawrence Berkeley National Laboratory (LBNL). 

The updated \MPzbt{} models is trained for 99 epochs with 32 NVIDIA H100 GPUs across 8 nodes on the Jean Zay cluster managed by the GENCI–IDRIS.

All the finetuned model were trained on a single NVIDIA H100 GPUs Jean Zay cluster managed by the GENCI–IDRIS.
\begin{figure}[htbp!]
  \centering
  \includegraphics[width=\textwidth,keepaspectratio]{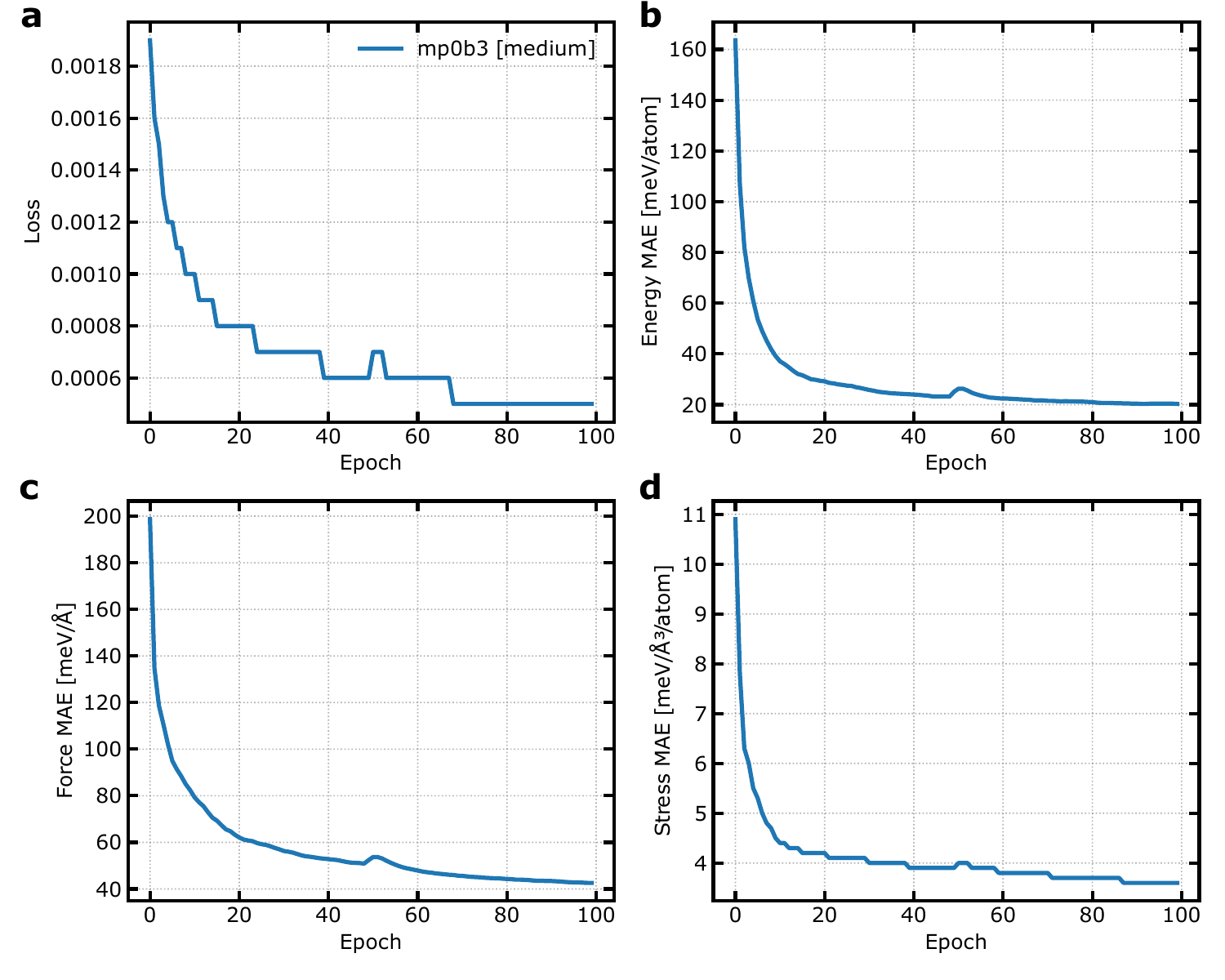}
  \caption{Training curves of \MPzbt{} models fitted to MPtrj data. (a) Loss (\cref{eq:loss}). (b-d) Root Mean Square Error (RMSE) of (b) energy per atom, (c) force, and (d) stress per atom. All curves are evaluated on the validation set.}
  \label{fig:metrics}
\end{figure}

\clearpage

\subsection{Fine-tuning protocol}\label{sec:finetuningprotocol}

As described in the previous section, MACE-MP-0 offers qualitatively good performance across a wide range of chemistry and materials at the PBE+U level of theory. 
However, for specific applications, it may be beneficial to fine-tune the model to improve its accuracy.
There are two main motivations to fine-tune a foundation model:
\begin{enumerate}
    \item To reach better quantitative accuracy for a specific application, in order to compute an observable that requires high precision.
    \item To increase the electronic-structure level of theory, in case a more accurate functional is required for a specific application.
\end{enumerate}
The fine-tuning process involves training the model on a new dataset, called the fine-tuning dataset, which contains a limited amount of data that are only relevant to the specific application.
The goals in designing a good fine-tuning protocol are threefold:
\begin{enumerate}
    \item To ensure that the fine-tuned model achieves the desired accuracy for the specific application (no under-fitting).
    \item To ensure that the fine-tuned model retains the robustness of the foundation model (no catastrophic forgetting).
    \item Ensure that the cost of the fine-tuning process is reasonable.
\end{enumerate}
Fine-tuning protocols have received significant attention in the machine learning community~\cite{dodge2020finetuningpretrainedlanguagemodels, liu2022fewshot}, especially in the context of computer vision and natural language processing.
The most naive protocols consist of just continuing training the model on the new dataset and restarting from the pre-trained model parameters. However, this approach is prone to catastrophic forgetting~\cite{McCloskey1989CatastrophicII}, where the model forgets the knowledge it has learned during the pre-training phase.
In the context of ML potentials, this can lead to a model that is not robust and that is more prone to explode during molecular dynamics simulations.
To mitigate catastrophic forgetting, several strategies have been proposed that can be decomposed into four main categories: (1) regularization, (2) architecture, (3) data, and (4) optimization.

Today, there exists a wide variety of fine-tuning strategies in the literature that are often specific to the application, the resource constraints, and the desired accuracy.
Most state-of-the-art fine-tuning protocols use a combination of regularization, architecture, data, and optimization techniques~\cite{lesort2019generative}.
One of the most widely used approaches is the replay buffer that consists in storing a subset of the pre-training dataset and replaying it during the fine-tuning phase.

In the context of ML potentials, the pre-training dataset is often very large, and efficient subsampling techniques are required to maximize the diversity of the replay buffer while keeping the computational cost reasonable.
Moreover, the model might be fine-tuned to new data that were generated using a different level of theory or electronic structure code from the one used to generate the pre-training dataset.
Therefore, the model needs to be able to learn from potentially inconsistent data effectively.

In the following, we present a fine-tuning protocol that we call the \textit{multi-head replay} that is specifically designed for ML potentials and that is used to fine-tune the \MPz{} model for different tasks.
We demonstrate that fine-tuned models using this protocol achieve the desired accuracy for the specific application using a few configurations, while retaining the robustness of the foundation model.

\subsubsection{Multi-head replay fine-tuning protocol}

The multihead replay protocol consists of two main steps: (1) the replay buffer construction and (2) the fine-tuning phase.
The construction of the replay buffer consists of selecting a subset of the pre-training dataset that will maximize the robustness of the fine-tuned model while keeping the computational cost reasonable.

Let $\mathcal{D}_\text{pre}$ be the pre-training dataset, and $\mathcal{D}_\text{fine}$ be the fine-tuning dataset. And let $Z_\text{pre}$ and $Z_\text{fine}$ be the set of atomic numbers present in $\mathcal{D}_\text{pre}$ and $\mathcal{D}_\text{fine}$, respectively.
We first select a subset of the pre-training dataset that has at least two elements that are present in the fine-tuning dataset,
\begin{equation}
    \mathcal{D}_{\text{Z}} = \{d \in \mathcal{D}_\text{pre} \mid \exists Z_1, Z_2 \in Z_\text{fine}, Z_1 \neq Z_2, Z_1, Z_2 \in d\}.
\end{equation}
If the size of $\mathcal{D}_{\text{Z}}$ is smaller than a predefined threshold, we add a random selection of configurations from $\mathcal{D}_\text{pre}$, $\mathcal{D}_{\text{Z}} = \mathcal{D}_{\text{Z}} \cup \text{random\_sample}(\mathcal{D}_\text{pre}$).
If the size of $\mathcal{D}_{\text{Z}}$ is larger than a predefined threshold, we subsample it to keep the computational cost reasonable by using a farthest point sampling algorithm. 
We embed the configurations in a high-dimensional space using the pre-trained model descriptors, and we select the configurations that are the farthest from each other. Note that other subsampling approaches might be more efficient and this is an ongoing research question.

A MACE model can be decomposed into two main parts, the descriptor part and the readout part. The descriptor part is the part of the model that is responsible for computing the atomic features that are invariant to rotations, and the readout maps the atomic features to the atomic energies.
In order to train the model on both the pre-training and fine-tuning datasets, we use a multi-head architecture, where the model has two readouts, one for the pre-training dataset and one for the fine-tuning dataset.
As the descriptor part is shared between the two readouts, the model can learn from both data sets simultaneously.

An important detail in MACE is the normalization of the total energies. 
When fitting the energies, we usually subtract the sum of the isolated atomic energies, $\mu_{Z_a}$, to the total energy. 
This is done to ensure that the model is able to predict the energy of the isolated atoms correctly. 
As the isolated atomic energies are different for the pre-training and fine-tuning datasets, we need to ensure that the model is able to predict the energy of the isolated atoms correctly for both datasets.

\subsubsection{Multi-head fine-tuning examples}

\paragraph{r2scan Materials Project dataset}
One application of fine-tuning is to increase the level of theory of the model.
To test the performance of our fine-tuning protocol, we use the r2scan dataset, which is a subset of the Materials Project data set that was recomputed using the r2scan functional.
It contains 17,000 relaxed configurations out of the 1.5M configurations in the MPtrj dataset.
\begin{table}[ht]
  \centering
  \begin{tabular}{l|c}
  \hline
  \textbf{Method} & \textbf{Energy MAE (eV)} \\ \hline
  Scratch ($n_{\text{r2scan}} = 17200$) & 0.5430 eV \\
  MP-0 Fine-tune No Replay ($n_{\text{r2scan}} = 17200$) & 0.1620 eV \\
  MP-0 Fine-tune Replay + Heads ($n_{\text{r2scan}} = 300$) & 0.077 eV \\ 
  MP-0 Fine-tune Replay + Heads ($n_{\text{r2scan}} = 3000$) & 0.039 eV \\ 
  MP-0 Fine-tune Replay + Heads ($n_{\text{r2scan}} = 17200$) & \textbf{0.032} eV \\ 
  \hline
  \end{tabular}
  \caption{Comparison of energy MAE for different methods.
  Models are trained on 17000 configurations from the r2scan Materials Project dataset
  and test on the 300,000 configurations from the newly discovered Gnome dataset \cite{merchant_scaling_2023}.
  "Scratch" refers to a model trained from scratch on the r2scan dataset.
  "Fine-tune No Replay" refers to a model fine-tuned just by restarting training.}
  \label{tab:energy_mae}
  \end{table}

We observe in Table~\ref{tab:energy_mae} that going from a model trained from scratch to a model fine-tuned even naively reduces the MAE energy by a factor of 3.
Using the multi-head replay protocol with randomly selected 100000 configurations fron MPtraj as the replay buffer, we are able to reduce the energy MAE by a factor of 17 compared to training from scratch, reaching an energy MAE of 0.032 eV.
The test set consists of the Gnome dataset, which is entirely composed of new crystals. 

\paragraph*{Fine-tuning on Applications}

To further demonstrate the effectiveness of our fine-tuning protocols, we pick configuration on a variety of examples of the present paper, and perform fine-tuning on them. For each selected examples, we select a hundred representative configurations and we recompute them using the MPtraj DFT. We then perform a multihead fine-tuning using both the MP-0 and MPA-0 models. We use randomly selected 100000 configurations fron MPtraj as the replay buffer, a learning rate of 0.0001 and exponential moving average decay of 0.9999, with Adam optimizer and a batch size of 16. We use a 10\% validation set for each task. In order to assess the performance of our fientuning protocal, we train models from scratch, using the same model sizes as the foundation models and the default optimization parameters.
\begin{figure}[!ht]
    \centering
    \includegraphics[width=0.9\linewidth]{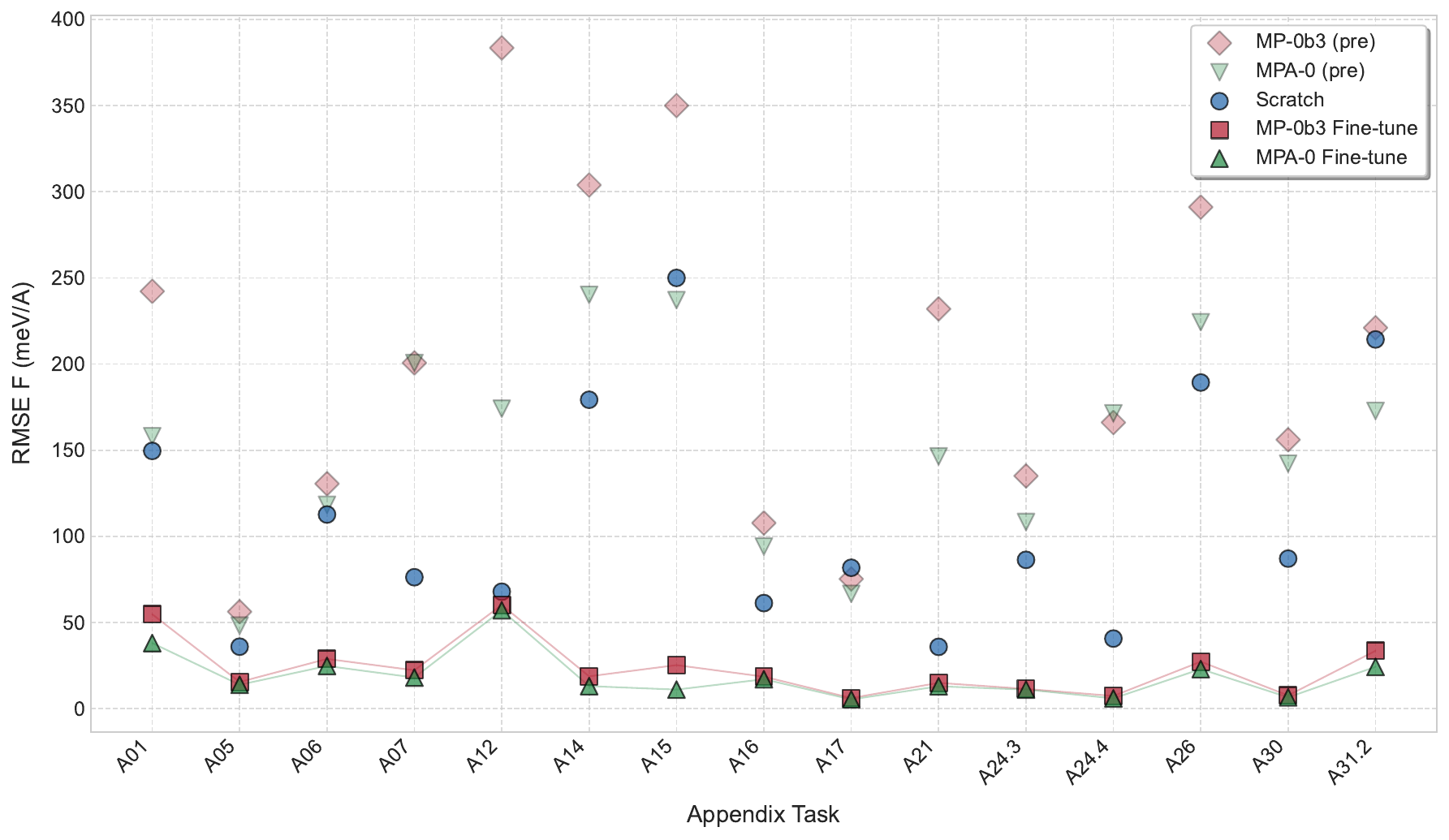}
    \caption{Comparison of validation RMSE on different applications for out of the box foundation models MP-0 and MPA-0, equivalent models trained from sratch and multihead fine-tuned models.}
    \label{fig:validationrsmse}
\end{figure}

In \ref{fig:validationrsmse}, we compare the RMSE forces errors on the validation set of each task, of (i) the two out of the box foundation models (ii) the equivalent models trained from scratch and (ii) the multihead replay fine-tuned models. The first remarkable result is that the foundation model often matches in performance the from scratch trained model showing the good out of the box performance of these models. The MACE-MPA-0 shows consistent out of the box improvement. Secondly, the fine-tuning shows an improvement of an x4 to x10 in performance compared to the models trained from scratch. Fine-tuning shows impressive data efficiency and reduces significantly the cost of building a system specific model even when the out of the box model is not the most accurate. We also observe only a slight improvement on accuracy for the fine-tuned MPA-0 model compared to MP-0, demonstrating that they are both good for fine-tuning purposes.

\begin{figure}[!ht]
    \centering
    \includegraphics[width=0.9\linewidth]{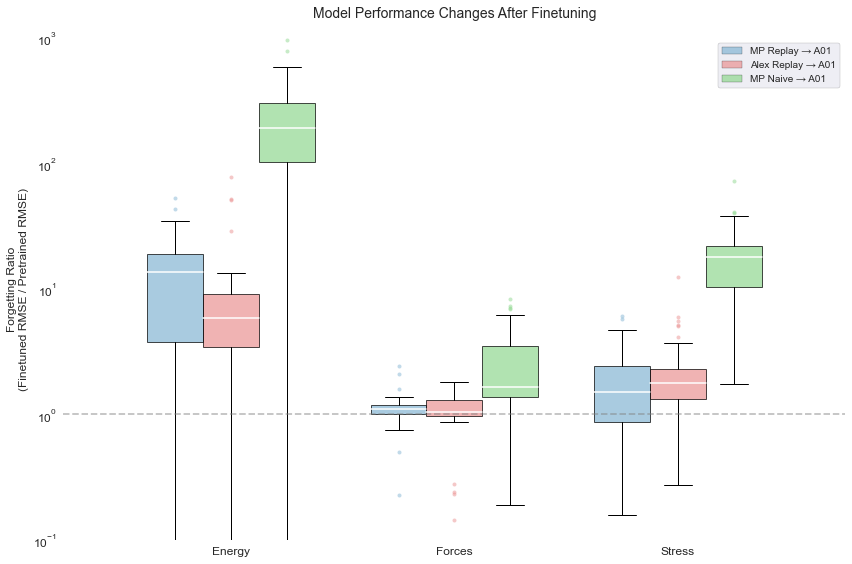}
    \caption{Comparison of performance degradation after fine-tuning across three model variants: MP with multihead replay (blue), Alexandria with multihead replay (pink), and MP with naive fine-tuning (green). The forgetting ratio (y-axis, log scale) represents the ratio of RMSE between fine-tuned and pretrained models for energy, forces, and stress predictions. A ratio of 1.0 (dashed line) indicates no degradation in performance. The box plots show the distribution of ratios across different configurations, with outliers represented as individual points.}
    \label{fig:forgetting-naive}
\end{figure}

To evaluate the amount of forgetting that occurs during fine-tuning, we assess the accuracy of a model fine-tuned on one task (A01) across configurations from other tasks. We then compute the ratio of fine-tuned model accuracy compared to the original model's accuracy. We call this metric the "Forgetting ratio." In Figure \ref{fig:forgetting-naive}, we compare the forgetting ratios of two multihead replay fine-tuned models and one naively fine-tuned model. All three models achieve similar validation RMSE on the A01 task and require similar GPU resources for the fine-tuning step.
We observe minimal force forgetting in the multihead replay fine-tuned model (forgetting ratios of approximately 1.1 for MPA-0 and 1.4 for MP-0), while the naively fine-tuned model (without replay) shows forgetting ratios around 4.0 for forces. For energy and stress predictions, the multihead replay fine-tuned models demonstrate an order of magnitude improvement in forgetting ratios compared to the naively fine-tuned model.
This dramatic reduction in forgetting significantly improves the stability and robustness of the fine-tuned model without compromising accuracy or performance, as all these models maintain similar performance metrics and GPU resource requirements.

\clearpage
\subsection{Exploration of the training data}
\label{sec:eda}

\begin{figure*}[htbp!]
  \centering
  \begin{subfigure}[b]{0.49\linewidth}
    \includegraphics[width=\linewidth,keepaspectratio]{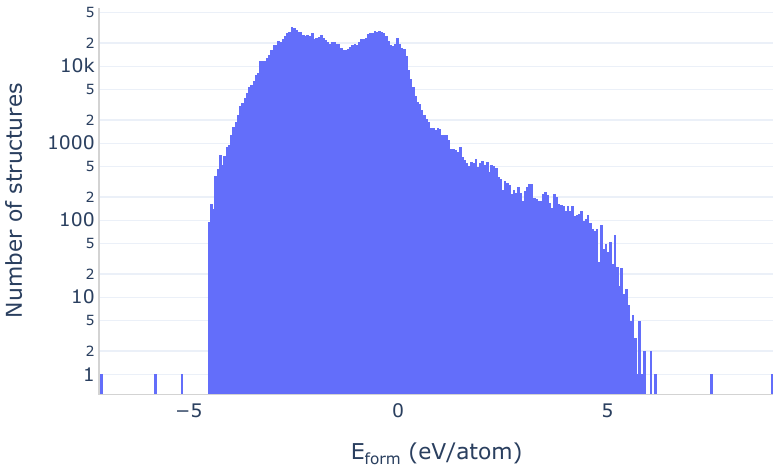}
    \caption{MPtrj formation energies}
    \label{fig:mp-trj-e-form-hist}
  \end{subfigure}
  \begin{subfigure}[b]{0.49\linewidth}
    \includegraphics[width=\linewidth,keepaspectratio]{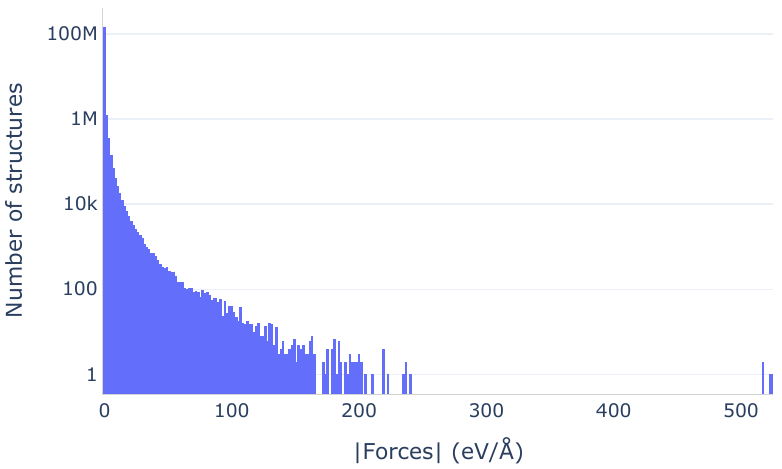}
    \caption{MPtrj forces}
    \label{fig:mp-trj-forces-hist}
  \end{subfigure}
  \begin{subfigure}[b]{0.49\linewidth}
    \includegraphics[width=\linewidth,keepaspectratio]{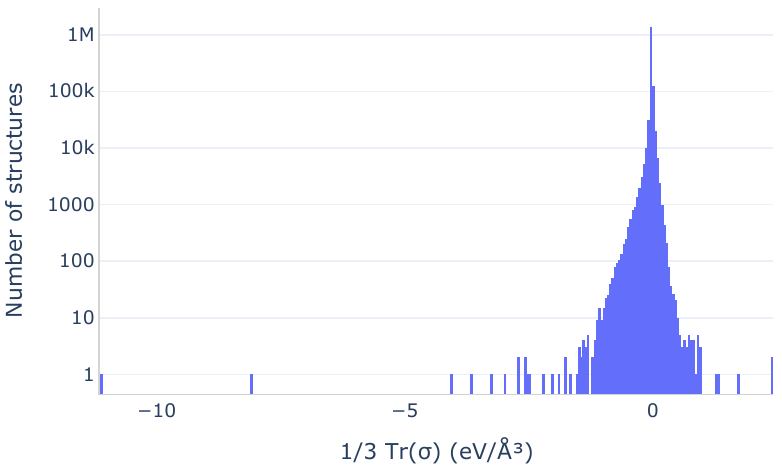}
    \caption{MPtrj stresses}
    \label{fig:mp-trj-stresses-hist}
  \end{subfigure}
  \begin{subfigure}[b]{0.49\linewidth}
    \includegraphics[width=\linewidth,keepaspectratio]{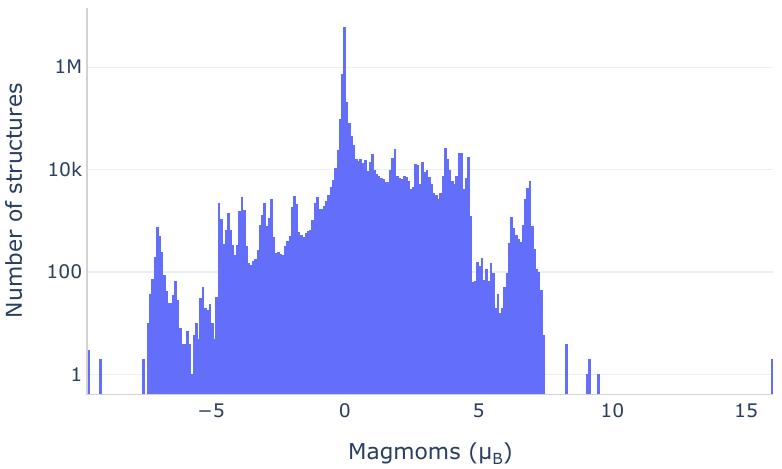}
    \caption{MPtrj magnetic moments}
    \label{fig:mp-trj-magmoms-hist}
  \end{subfigure}
  \caption{
    Distribution of energies, forces, stresses and magnetic moments in the MPtrj dataset \cite{deng_chgnet_2023,riebesell_matbench_2023}.
    The bimodality in the formation energy distribution is due to the MP anion correction scheme \cite{wang_framework_2021,kingsbury2022flexible} which significantly lowers oxide formation energies.
  }
  \label{fig:mp-trj-hists}
\end{figure*}

\begin{figure*}[htbp!]
  \centering
  \includegraphics[width=\linewidth,keepaspectratio]{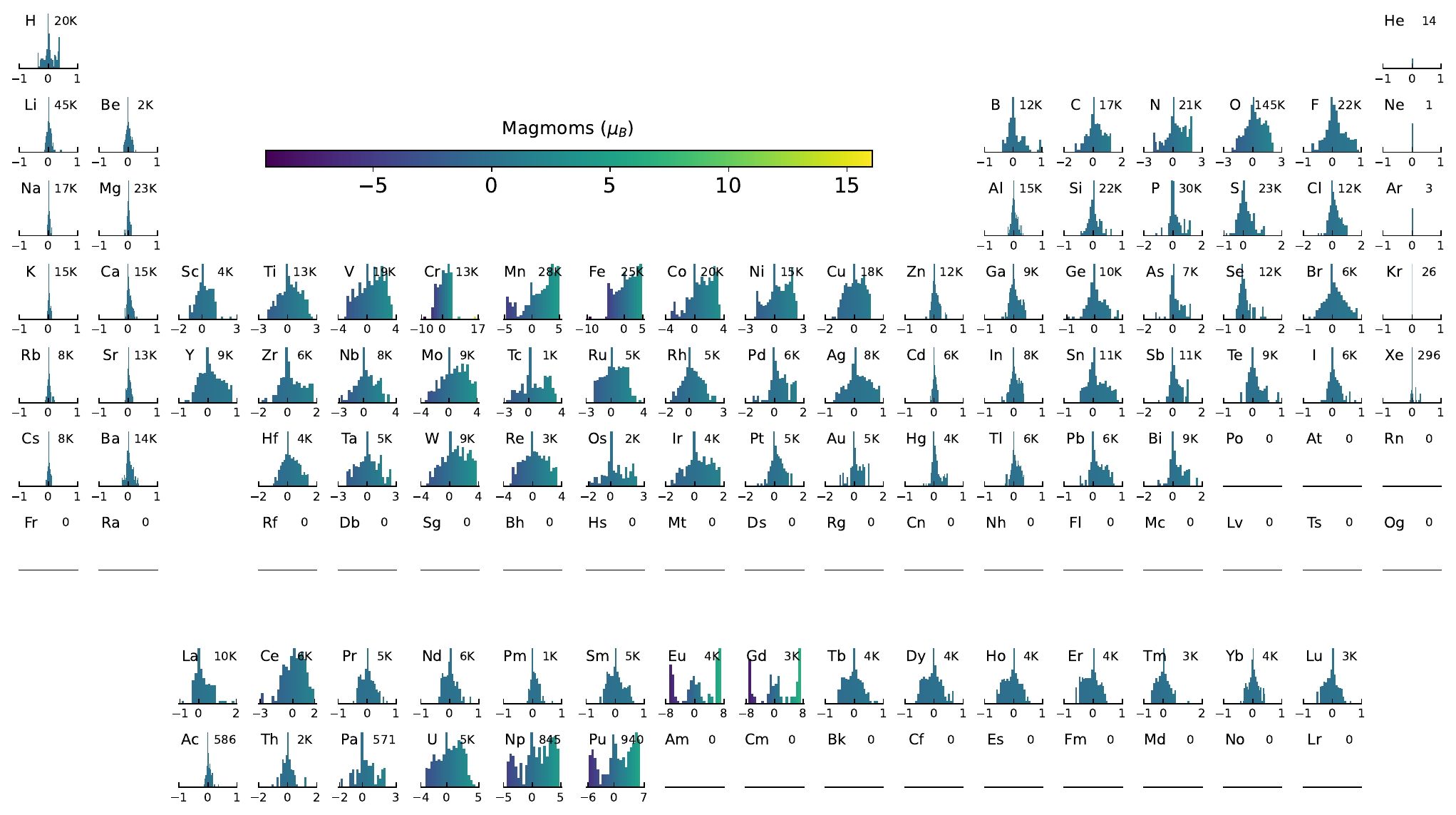}
  \caption{
    Distribution of magnetic moments for each element in the MPtrj dataset \cite{deng_chgnet_2023,riebesell_matbench_2023,riebesell_pymatviz_2022}.
    The $y$-axis is log-scaled to allow visualization of the tail of high magnetic moments in some elements with a sharp peak at 0.
    The number in the top right corner of each element tile counts magnetic moments for that element in the MPtrj dataset.
    This plot reveals rare erroneous data points in MPtrj.
    For instance, \ch{Cr} has a single-point calculation with a highly unphysical magnetic moment of \num{17}{$\mu_\text{B}$}.
  }
  \label{fig:mp-trj-magmoms-ptable-hists}
\end{figure*}

\begin{figure}[!htbp]
  \centering
  \includegraphics[width=0.7\columnwidth,keepaspectratio]{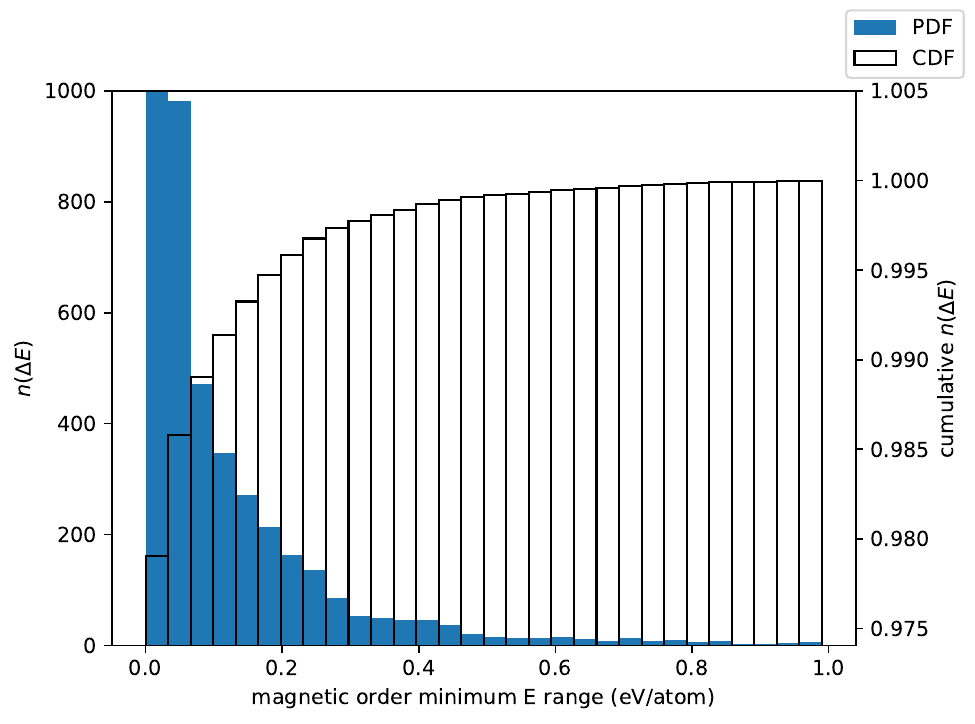}
  \caption{Distribution (PDF) of approximate deviation of energy within
    each material due to variation in magnetic order (non-magnetic calculation, calculation
    converges to moment zero, ferromagnetic, and other magnetic orders), with cumulative distribution (CDF) on right axis. The vast majority
    of materials have very small variation (y axis range does not show full extent of first PDF bar), but a few hundred include different magnetic orders with energies that vary by more than \SI{0.1}{eV/atoms}}
  \label{fig:E_magmom_range}
\end{figure}

\begin{figure}
  \centering
  \begin{subfigure}[b]{0.495\linewidth}
    \includegraphics[width=\linewidth,keepaspectratio]{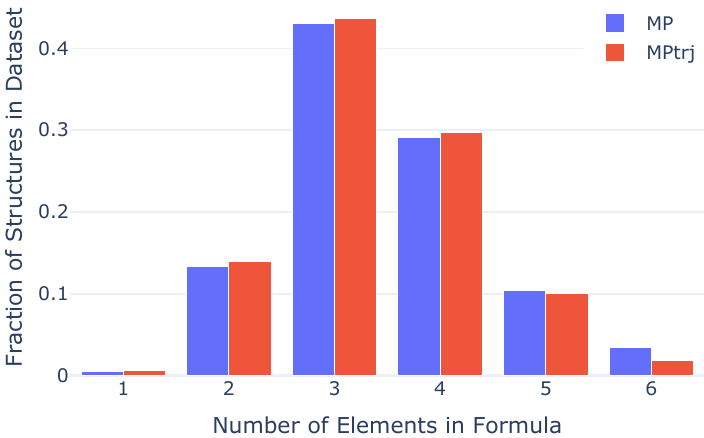}
    \caption{}
    \label{fig:mp-vs-mp-trj-arity-hist}
  \end{subfigure}
  \hfil
  \begin{subfigure}[b]{0.495\linewidth}
    \includegraphics[width=\linewidth,keepaspectratio]{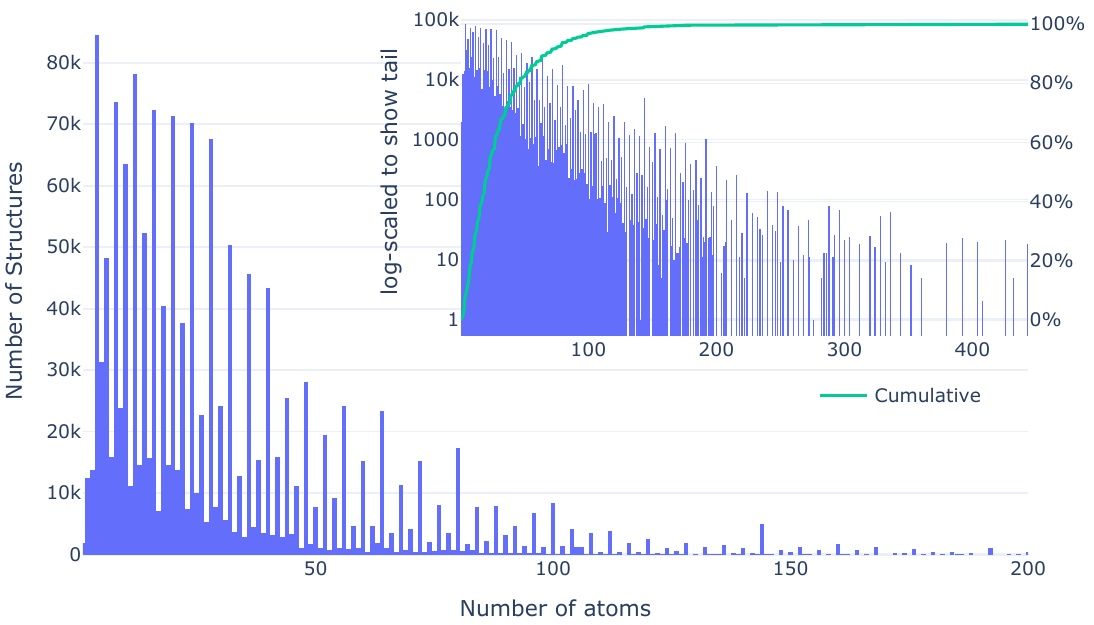}
    \caption{}
    \label{fig:mp-trj-n-sites-hist}
  \end{subfigure}
  \caption{
    \subref*{fig:mp-vs-mp-trj-arity-hist}) Distribution of the number of elements in the compositions of MP structures compared to MPtrj.
    We observe a slight overabundance of small numbers of elements in MPtrj relative to MP.
    \subref*{fig:mp-trj-n-sites-hist}) Distribution of a number of sites in MPtrj.
    The inset shows the same distribution log-scaled to visualize the tail of high site counts.
    The green cumulative line in the inset shows that 82\% have less than 50 sites and 97\% of structures in MPtrj have less than 100 atoms.
  }
  \label{fig:mp-trj-arity-and-n-sites-hist}
\end{figure}

\begin{figure*}[htbp!]
  \centering
  \includegraphics[width=\linewidth,keepaspectratio]{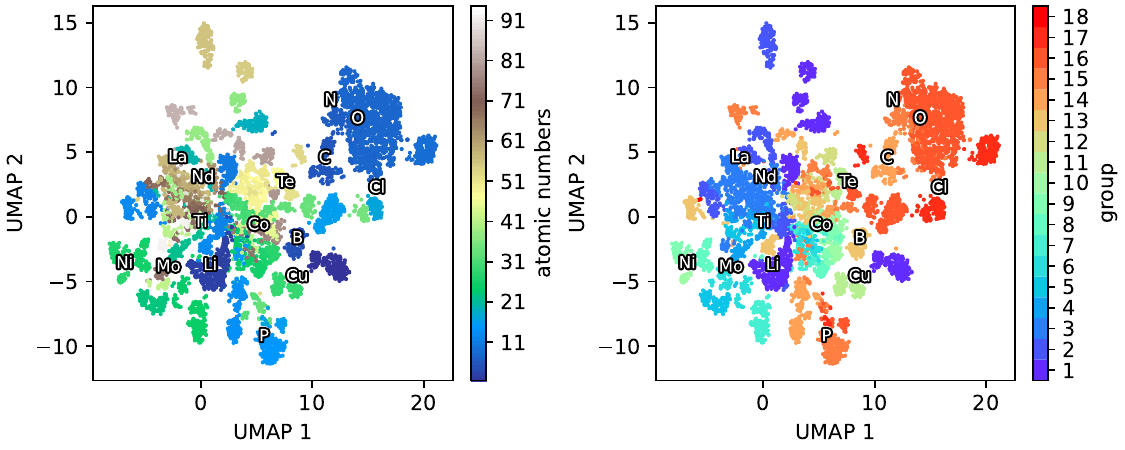}
  \caption{
    UMAP projection of MACE descriptors for atoms in MPtrj. Each point represents the averaged feature vector of a single element in one structure and is colored by atomic number (left) and group (right) in the periodic table. The features  of \MPz{} model are 256-dimensional vectors (concatenation from both first and second layer of 128 channels). Manhattan distance is used for the construction of a high-dimensional UMAP manifold.
  }
  \label{fig:mp-trj-umap}
\end{figure*}


\clearpage
\subsection{Similarity analysis}

In the examples above, we have shown that \MPz{}\ is capable of surprising degrees of extrapolation. The use of semi-local features (as a result of message passing)~\cite{batatia2022design} and element mixing~\cite{Darby2023Trace} within the MACE architecture are key components underlying \MPz{}'s capabilities. These components allow \MPz{}\ to extrapolate to systems that globally seem completely different from the training data but have close matches locally.

To quantify the similarity, we compare atomic environments from test systems to filtered portions of the training data using the following procedure:
\begin{enumerate}
  \item Filter training data to a subset with elemental compositions similar or exactly matching the test system.
  \item Use \MPz{}\ to extract invariant descriptors for all atoms in both the test system and the filtered training subset.
  \item Calculate the cosine similarity between the atoms in the test system and each filtered training structure. For each atom, we use the maximum cosine similarity found this way. This is essentially a best-match structure kernel~\cite{deComparingMoleculesSolids2016a} that allows many-to-one mappings.
  \item Average these maximum atomic similarities by element and then combine them by averaging again, yielding an element-stratified similarity.
\end{enumerate}

Through this procedure, we identify training set structures that contain the most similar local environments to those in any given test system. In addition, we create \texttt{chemiscope}\cite{frauxChemiscopeInteractiveStructureproperty2020} (\url{https://chemiscope.org/}) input files containing UMAP~\cite{mcinnesUMAPUniformManifold2020} projections of the atomic descriptors (fitted only on the training environments), allowing a more granular and interactive inspection of the environments in the test and training data.

The code for analysing the data and generating \texttt{chemiscope} inputs is available as a Python package~\cite{rokasel_2023_10426282}.

\section{Uncertainty quantification}
\label{sec:uq}

\subsection{Theory and implementation}

Predictive uncertainties for \MPzbt{} are obtained with the last-layer prediction rigidity (LLPR) method~\cite{Bigi2024}. This approach is well-suited for \MPzbt{} as it is simple, scalable, and allows to obtain uncertainty for neural networks that have already been trained. In the simplest form, LLPR uncertainties are given by 
\begin{equation}\label{eq:llpr}
    \sigma_i^2 = \alpha^2 \mathbf{f}_i^\top \mathbf{H}_o^{-1} \mathbf{f}_i,
\end{equation}
where $\mathbf{f}_i$ is the hidden features of structure $i$ in the last layer of the neural network (before entering the final linear transformation), $\mathbf{H}_o$ is the generalized Gauss-Newton pseudo-Hessian of the loss at the end of training, $\alpha^2$ is a calibration parameter, and $\sigma_i^2$ is the resulting LLPR uncertainty for structure $i$.

In calculating the $\mathbf{H}_o$ for \MPzbt{}, only the energy part of the loss function is taken into account, while the force and stress terms are omitted. This greatly simplifies the practical calculation of the pseudo-Hessian, and---as we show in Fig.~\ref{fig:validation}---still allows one to obtain high-quality error estimates on forces and stresses.
In simple cases, LLPR uncertainties can easily be analytically propagated to the derived targets. A simple example is given by the subtraction of two different energies, which is later used to generate uncertainties for several NEB experiments in Section~\ref{sec:case-studies}. In such cases, the uncertainty in the difference between the energies of structures $i$ and $j$ is computed as $\alpha^2 (\mathbf{f}_i-\mathbf{f}_j)^\top \mathbf{H}_o^{-1} (\mathbf{f}_i-\mathbf{f}_j)$. Other cases can be treated similarly by making use of the connection~\cite{Bigi2024} between the LLPR method and the Laplace approximation~\cite{laplace1774memoire, daxberger2021laplace}.

In terms of computational cost, the energy uncertainties obtained with~\eqref{eq:llpr} generate negligible overhead when compared to simple energy predictions of MACE architecture. 
Predicting the error on gradients (e.g. forces, stresses) is more demanding, as it requires performing multiple backpropagation steps. However, this is only needed when explicit estimates of the gradient error, or of properties that depends explicitly on the gradient values, are desired.
A LLPR-wrapped version of \MPzbt{} that can output uncertainties for energy, force, and stress predictions are made available alongside the original model.

\subsection{Distribution of errors in the validation set}

\begin{figure}[htb]
    \centering
    \includegraphics[width=\textwidth]{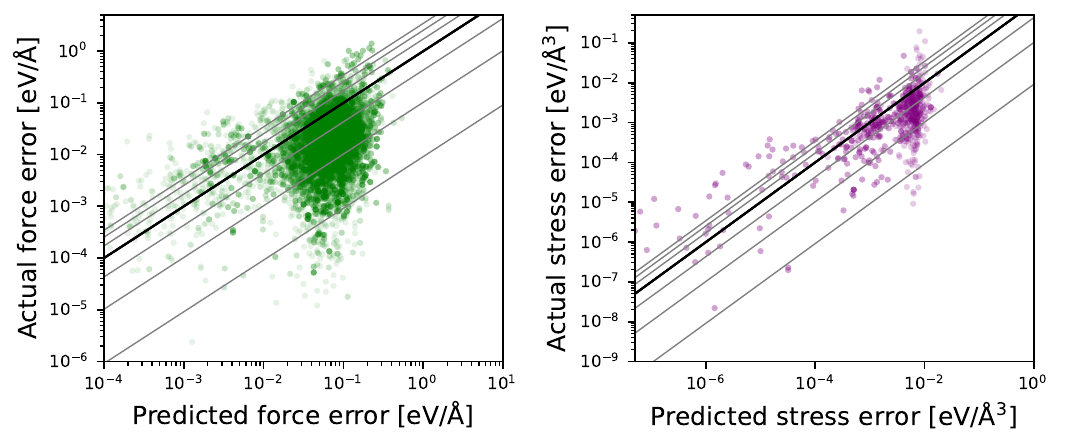}
    \caption{Force (left) and stress (right) errors on the validation structures for \MPzbt{}. The black line represents $y=x$, while the grey lines are pairs of isolines enclosing approximately 68\%, 95\% and 99\% of the expected distribution, respectively.
    Errors on very small predictions (which are present due to the symmetry of several validation structures) are cut out, but they are nonetheless well-predicted by the LLPR approach. See Section~\ref{sec:WBM} for more details on these plots and their expected distribution.}
    \label{fig:validation}
\end{figure}

Figure~\ref{fig:validation} shows the distributions of the predicted and actual errors on the forces and stresses of the validation set used in the training of \MPzbt{} potential. Energies are omitted as the limited size of the validation set (160 structures) is not sufficient to gather significant statistics. Nonetheless, it is clear that the LLPR method recovers good-quality uncertainty predictions in both cases, despite the exclusion of forces and stresses terms from the pseudo-Hessian calculations. See Section~\ref{sec:WBM} for results on out-of-domain energy predictions.

\subsection{Case studies}\label{sec:case-studies}

In this section, we show the LLPR uncertainties for a subset of the case studies found in section~\ref{sec:applications}. When interpreting these results, it is important to keep in mind that only the WBM dataset (section~\ref{sec:WBM}) provides reference DFT energies that are consistent with the dataset \MPzbt{} has been trained on.
This is not the case for the other case studies considered, which makes it difficult to make any quantitative assessments, and in particular to pinpoint whether the discrepancy between the DFT reference data and the \MPzbt{} predictions stems from the error in model prediction or the inconsistencies in the DFT methods employed.
Nonetheless, the following subsections demonstrate that performing uncertainty quantification can still be insightful in detecting where the model succeeds or fails, even in the absence of consistent DFT calculations. Note that uncertainties for the NEB case studies are demonstrated using the NEB frames acquired with a previous generation of the model (\MPz{}a).

\subsubsection{Uncertainty quantification for the WBM dataset}\label{sec:WBM}

\begin{figure}[htb]
    \centering
    \includegraphics[width=15cm]{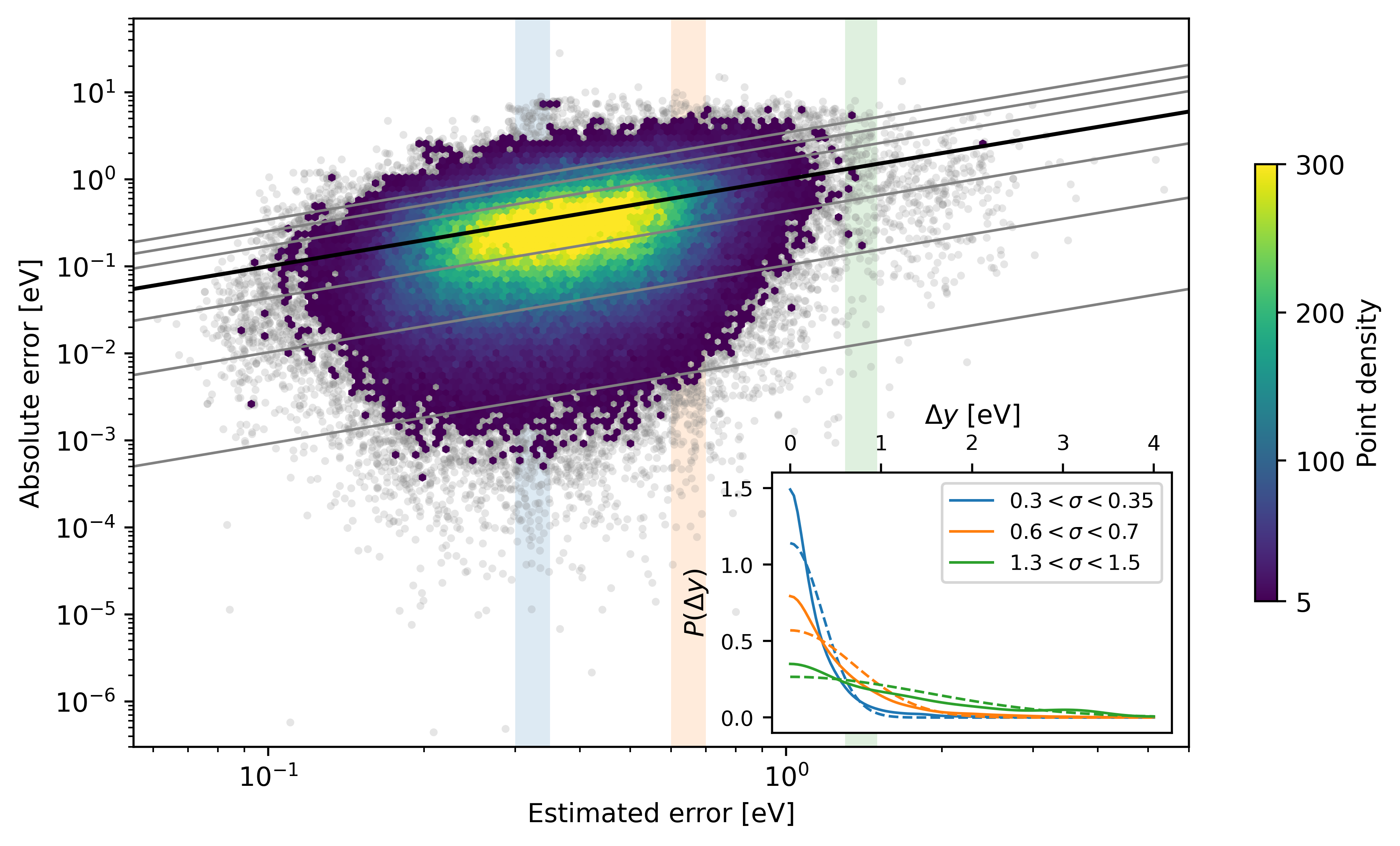}
    \caption{Scatter plot that shows the absolute error vs. estimated error in the total energies for the WBM dataset of Wang et al.~\cite{wang2021predicting} $y = x$ is shown in black, and isolines that successively bound 68\%, 95\%, and 99\% of the distribution are shown in gray. Scatter plot for the entire dataset (light gray) is shown together with its 2D histogram. The inset shows the probability densities of points within three ``slices'' (highlighted in the main panel) of estimated errors. Solid line shows the probability density within a given slice estimated by Gaussian kernel density estimation. Dashed line shows the actual Gaussian distribution using the mean estimated error of the slice.}
    \label{fig:WBM}
\end{figure}

As discussed in section~\ref{sec:materials discovery}, the WBM dataset~\cite{wang2021predicting} can be regarded as an out-of-distribution test set for \MPzbt{}. The WBM dataset also provides reference DFT energies that are consistent with that of the MPtrj dataset which the \MPzbt{} has been trained on. We hence demonstrate the efficacy of the LLPR-based error estimates by performing uncertainty quantification for the total energies of the structures in the WBM dataset. Figure~\ref{fig:WBM} shows the distribution of the absolute errors versus the estimated errors. 

It should be understood that the estimated error corresponds to the width of the distribution of the prediction errors, and that it does not directly correlate with the actual absolute error.
This is apparent from the probability densities of the vertical ``slices'' (the inset of Figure~\ref{fig:WBM}) in the distribution, where the observed probability densities closely match that of the expected Gaussian distributions with $\sigma$ values corresponding to each slice. One can see that a majority of the empirical errors lie within $3\sigma$ of the estimated error, and there is a limited number of statistically significant underestimations.

\subsubsection{Uncertainty quantification for elemental defects in alumina}

\begin{figure}[ht]
    \centering
    \includegraphics[width=15cm]{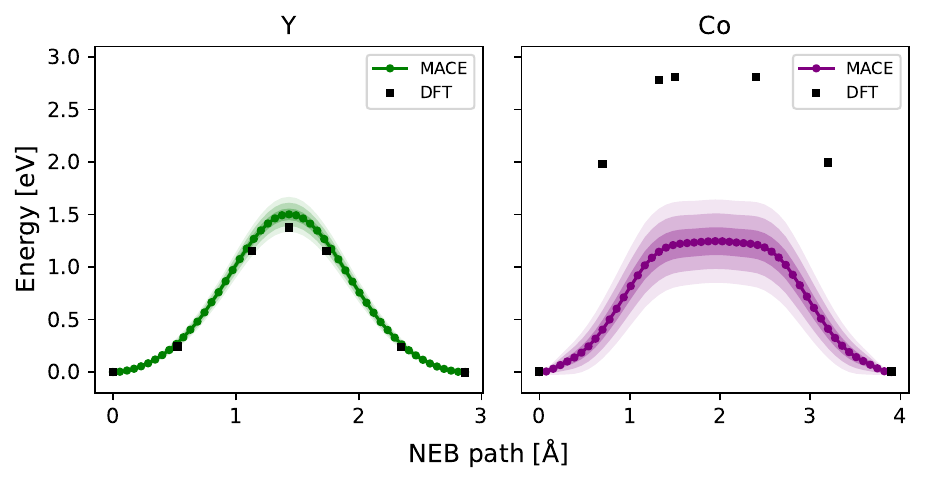}
    \caption{\MPzbt{}-predicted relative energies along the NEB barrier paths of elemental defect diffusion in alumina, for Y (left) and Co (right). Estimated error bounds are shaded along the plot, where $1\sigma$, $2\sigma$, and $3\sigma$ bounds are shown in successively lighter shades. Available DFT data (that is not entirely consistent with the MPtrj dataset) is shown with black square markers.}
    \label{fig:alum_defect}
\end{figure}

In section~\ref{sec:alumina-elemental} (Figure \ref{fig:al2o3}d-e), it is shown that \MPzbt{} accurately recovers the reference NEB barrier path of an elemental defect diffusion for Y, whilst failing to do so for Co. In Figure~\ref{fig:alum_defect}, we show that LLPR-based uncertainty quantification can correctly discern the model accuracy between these two cases. The uncertainties for the case of Y remain small in the beginning and end of the NEB path, and only shows notable uncertainty in the middle of the path. On the contrary, the estimated uncertainties for the Co NEB barrier path is considerably large throughout the entire NEB path, even close to the beginning and end. In fact, it is generally larger by a factor of 4 when compared with that of Y.

\subsubsection{Uncertainty quantification for CO oxidation on Cu}

\begin{figure}[htb]
    \centering
    \includegraphics[width=10cm]{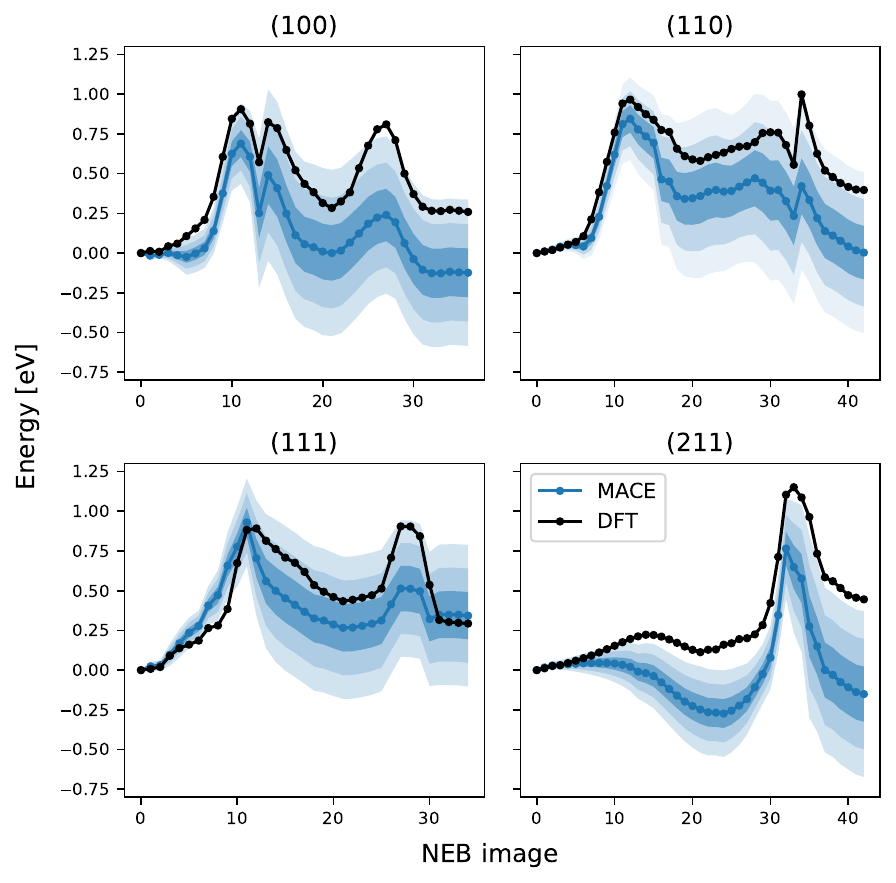}
    \caption{\MPzbt{}-predicted relative energies along the NEB barrier paths of CO oxidation on four different surfaces of Cu. Estimated error bounds are shaded along the plot, where $1\sigma$, $2\sigma$, and $3\sigma$ bounds are shown in successively lighter shades. The DFT reference data (that is not entirely consistent with the MPtrj dataset) is shown in black.}
    \label{fig:hetcat-uq}
\end{figure}

Section~\ref{sec:catalysis} and section~\ref{sec:hetcat} demonstrates the application of \MPzbt{} for CO oxidation on several different surfaces of Cu. Figure~\ref{fig:hetcat-uq} shows the results for the first half of the pathway accompanied by the LLPR-based uncertainties. In all cases, LLPR uncertainties generally remain small for the reaction pathway up to coupling at (i). Despite inconsistencies in the DFT details, this is indeed where the \MPzbt{} predictions remain close to the available DFT data. Past this range in the reaction coordinate, the uncertainties become larger in all cases, and the actual errors with respect to the available DFT data also become more severe.

\subsubsection{Uncertainty quantification for carborane rearrangement}

\begin{figure}[htb]
    \centering
    \includegraphics[width=15cm]{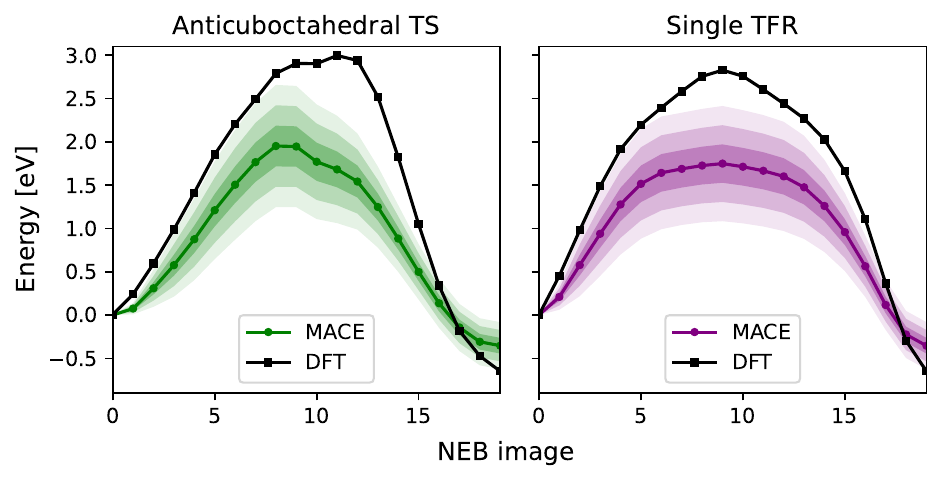}
    \caption{Relative energies predicted by \MPzbt{} along the NEB barriers obtained for the anticuboctahedral transition state and single triangular face rotation pathways suggested for carborane rearrangement. Estimated error bounds are shaded along the plot, where $1\sigma$, $2\sigma$, and $3\sigma$ bounds are shown in successively lighter shades. DFT reference data (that is not entirely consistent with the MPtrj dataset) is shown in black.}
    \label{fig:carborane-uq}
\end{figure}

Section \ref{sec:carborane} discusses the capability of \MPzbt{} to qualitatively capture the energetics of carborane rearrangement. Figure~\ref{fig:carborane-uq} shows that LLPR-based uncertainty quantification can further shed light on the accuracy of the model along the NEB pathways. In both cases, the estimated uncertainties grow quickly along the NEB path especially towards the transition state, but diminsh towards the end when the system reaches the \textit{meta} isomer products. 
These uncertainties are in good alignment with the actual discrepancies between the \MPzbt{} predictions and the available DFT reference data, where the \MPzbt{} predictions are found to notably underestimate the energies in the middle of the NEB paths, but becomes comparable when predicting the overall reaction energy.

\subsection{Uncertainty Calibration}

Post-processing techniques offer a way to calibrate inaccurate uncertainty estimates. A common approach involves introducing a calibration set, $\mathcal{D}_{\rm cal} := \big\{(\mathbf{X}_n, \mathbf{Y}_n)\big\}_{n=1}^{N}$, to calibrate the original uncertainty. Among these techniques, conformal prediction (CP) has gained popularity for its simplicity and effectiveness in both classification and regression tasks~\cite{angelopoulos2021gentle, bai2022efficient, hu2022robust}. 

To calibrate a heuristic uncertainty $\sigma(\mathbf{X})$ using CP, we first define a score function: $s(\mathbf{X}, \mathbf{Y}) := \|\widetilde{\mathbf{Y}} - \mathbf{Y}\|/\sigma(\mathbf{X})$, where $\widetilde{\mathbf{Y}}$ represents the model's prediction (e.g., energy, forces, or stress). For each data point in the calibration set $\mathcal{D}_{\rm cal}$, we compute the corresponding score $s_i = s(\mathbf{X}_i, \mathbf{Y}_i)$. Next, let $\alpha \in (0, 1)$, and define $q_\alpha$ as the empirical quantile of $\{s_i\}_{i=1}^{N}$, given by
\[
q_{\alpha} := \text{quantile}\left(\{s_i\}_{i=1}^{N}, \frac{\lceil (N+1)(1-\alpha) \rceil}{N}\right).
\]  
The calibrated uncertainty is then obtained as $q_\alpha\sigma(\mathbf{X})$. This adjustment guarantees that the calibrated uncertainty provides a probabilistic bound on the true error, $\|\widetilde{\mathbf{Y}} - \mathbf{Y}\|$, with at least $1-\alpha$ confidence~\cite{angelopoulos2021gentle, shafer2008tutorial}. 

A recent extension of the standard CP method is to incorporate the dependence of the quantile value on the local atomic environment (LAE)~\cite{forceuq}. Specifically, the quantile forms a step function, denoted as $\hat{q}_{\alpha}\big(\mathcal{G}(\mathbf{X}_i)\big)$, where $\mathcal{G}$ is a classification model that assigns each LAE $\mathbf{X}_i$ to a distinct class. The calibration $\hat{q}_{\alpha}$ is then applied independently within each class, a technique referred to as class-based CP. 

The hyperparameter $\alpha$ is a crucial component of this framework and must be carefully fined. In our numerical experiments, we set $\alpha = 0.5$ by default unless stated otherwise, as it as it consistently yields stable results across different scenarios. Alternatively, cross-validation on various calibration sets can be employed to determine the optimal $\alpha$. The classification model $\mathcal{G}$ is constructed by clustering invariant descriptors generated by MACE-MP-0B3 using a Bayesian variant of the Gaussian Mixture Model (GMM)~\cite{bishop2006pattern}. 

Next, we examine the feasibility of uncertainty calibration by applying CP and its classification-based extensions to three case studies: (1) validation error on the MP dataset, (2) NEB transition paths for CO oxidation on Cu with different surface orientations, and (3) NEB transitions in carborane rearrangement. In this section, force errors are measured using the magnitude of the force on each atom rather than considering force components, as done in the previous section.

\subsubsection{Calibrating uncertainty for the MP validation set}

\begin{figure}
    \centering
    \includegraphics[width=0.8\linewidth]{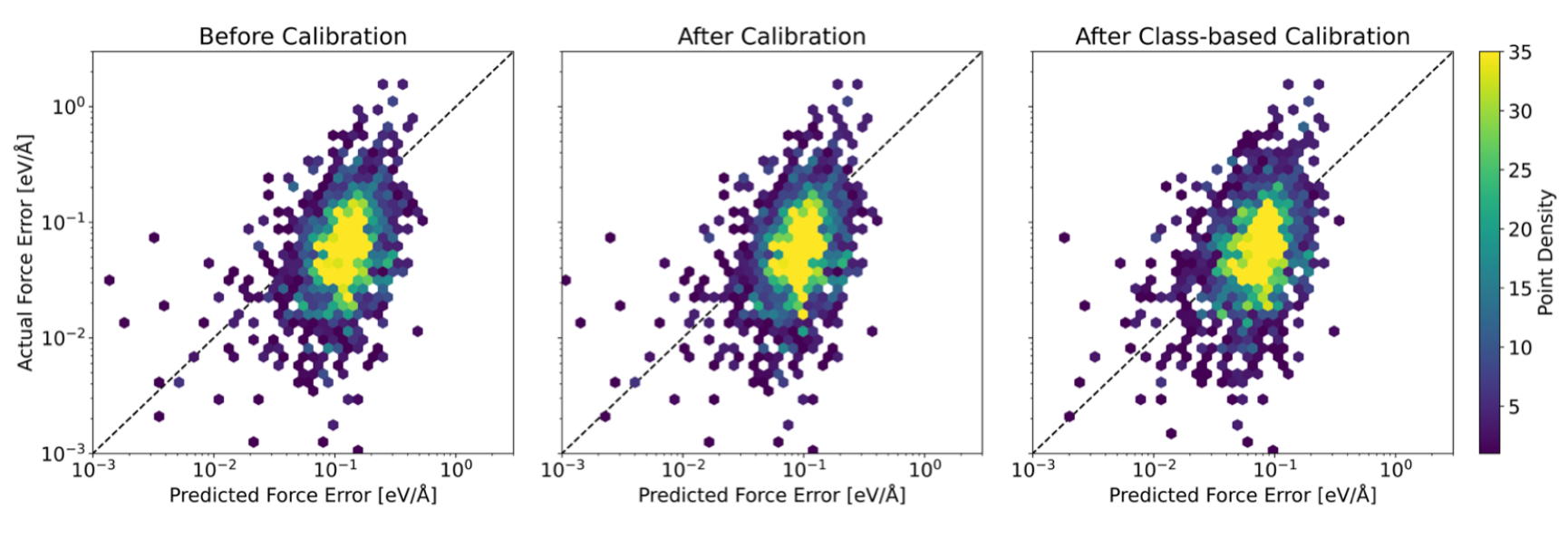}
    \caption{Force errors on the validation set for Large MACE-MP0B3. Quantile in simple CP is $\hat{q} = 0.76$. Quantile for class-based CP with $6$ classes ranges from $0.5$ to $2.0$ and are of similar range with increasing number of clusters.}
    \label{fig:MP_valid_calibration}
\end{figure}

We begin by demonstrating the improvements in predicted error using CP and class-based CP. In this example, 10\% of the data was randomly selected for calibration. As shown in Figure~\ref{fig:MP_valid_calibration}, the predicted force error from LLPR (left panel) overestimates the true force error. The original CP without classification (middle panel) provides a modest improvement over LLPR. Notably, applying class-based CP (right panel) results in a further slight enhancement, primarily in terms of point density. However, increasing the number of classes in GMM clustering does not lead to any significant improvement. These findings suggest that while CP effectively improves uncertainty calibration, the additional benefit of class-based CP is marginal for large, diverse datasets. This outcome is expected, given the inherent difficulty of classifying complex atomic environments.

\subsubsection{Calibrating uncertainty for CO oxidation on Cu}

To highlight the potential of classification on calibrating force uncertainty in specific cases, we evaluate the uncertainty of forces and energy differences along the NEB paths for CO oxidation on the (111) and (211) surfaces, where LLPR provides only qualitative uncertainty estimates. For calibration, we select 10 uniformly spaced images along each NEB path rather than using random sampling.

\begin{figure}[htbp]
    \centering
    \begin{subfigure}[b]{0.7\linewidth}
        \centering
        \includegraphics[width=1.0\linewidth]{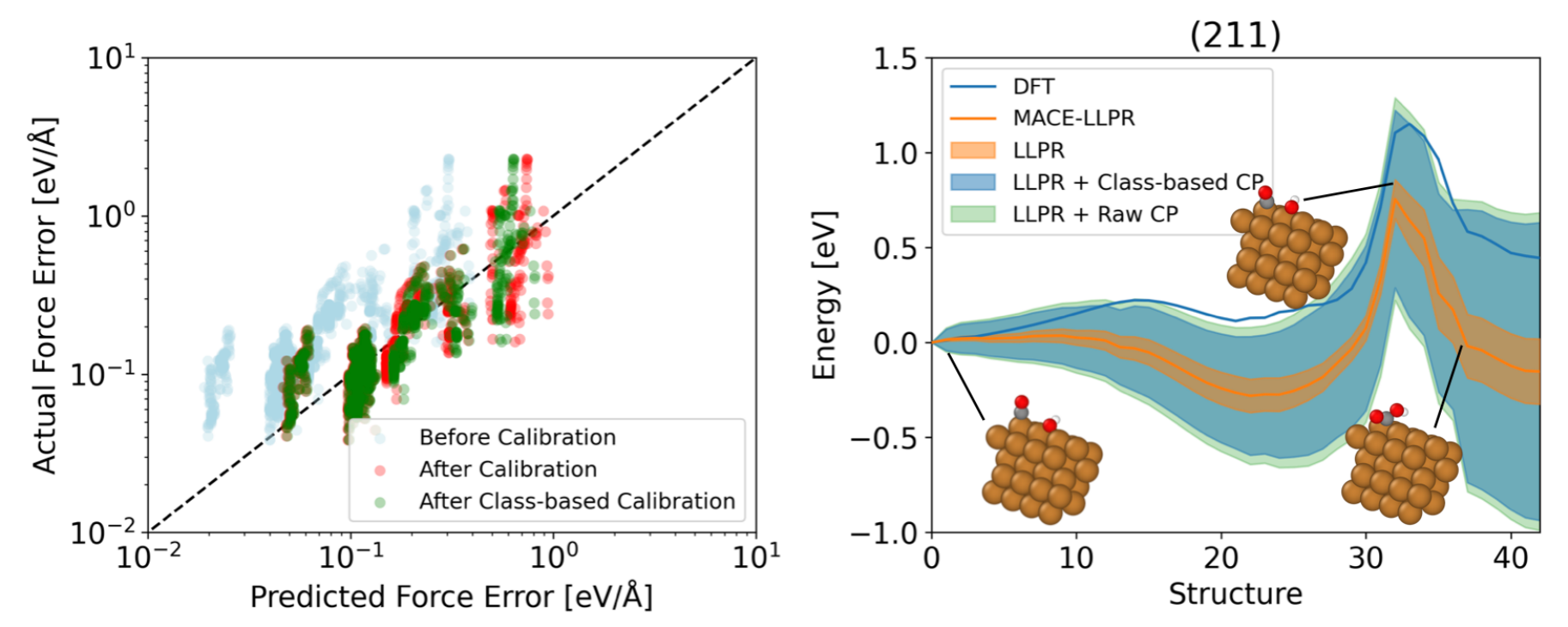}
        \vspace{-0.5cm}
        \label{fig:MP_valid_calibration_211}
    \end{subfigure}
    \begin{subfigure}[b]{0.7\linewidth}
        \centering
        \includegraphics[width=1.0\linewidth]{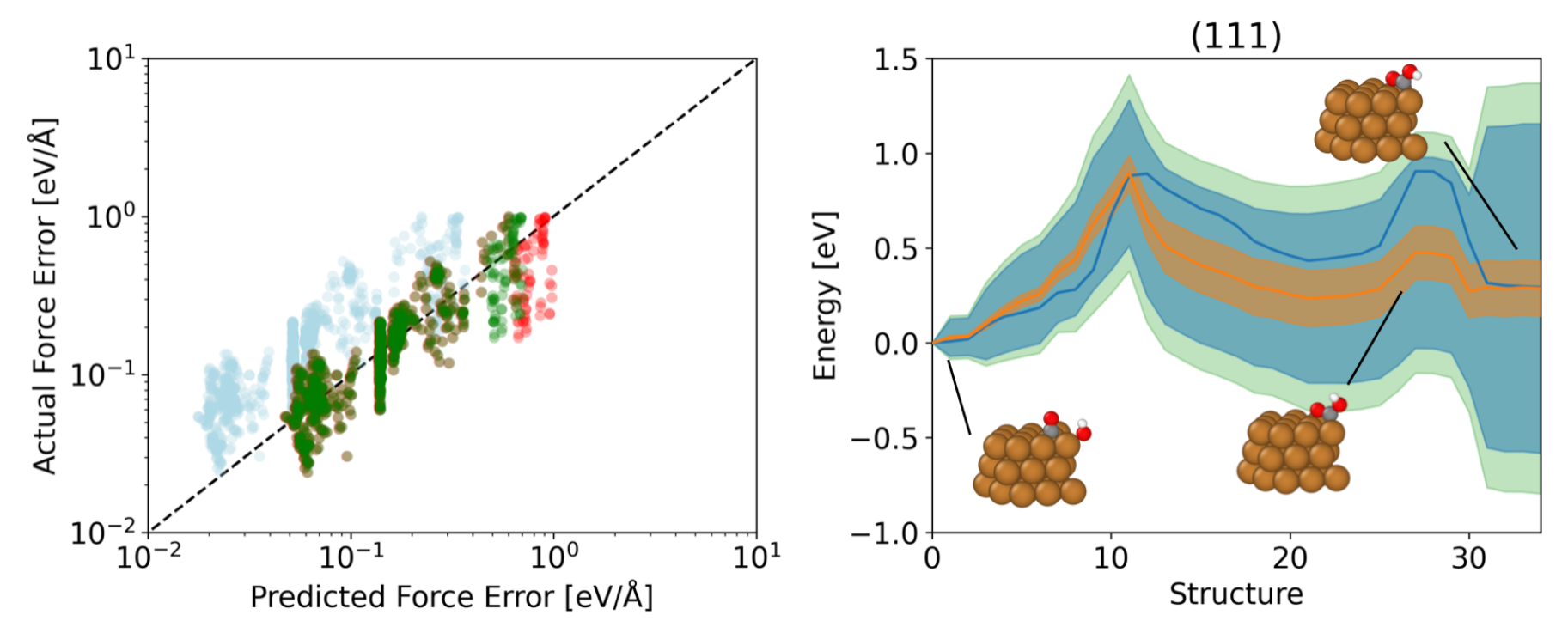}
        \vspace{-0.5cm}
        \label{fig:MP_valid_calibration_111}
    \end{subfigure}
    \caption{MACE-predicted forces and relative energies along the NEB barrier paths for CO oxidation on the Cu(211) (top) and Cu(111) (bottom) surfaces. The estimated force and NEB barrier path errors from MACE-LLPR uncertainty, after calibration and class-based calibration, are shown in different colors. The DFT reference data for the NEB barrier path is shown in blue.
    }
    \label{fig:combined-figure-nebs}
\end{figure}

Figure~\ref{fig:combined-figure-nebs} compares the predicted force errors against DFT references and illustrates the NEB barrier paths for two different surfaces, with error bars derived from LLPR uncertainty, original CP, and class-based CP.

After calibration, the predicted force errors align more closely with the ground truth for both surfaces. However, when applying original CP to all sites on the (111) surface, force uncertainties remain inaccurate due to the predominance of “bulk” copper surface sites in the calibration set, as indicated by the brown atoms in the right panels of Figure~\ref{fig:combined-figure-nebs}. This imbalance leads to overestimated quantiles and inflated errors for high-error sites. By incorporating classification, the predicted errors for outlier sites better match the true errors, as shown in the left panel of the figure.

For uncertainties in energy differences along the NEB paths (right panel of Figure~\ref{fig:combined-figure-nebs}), original CP provides accurate calibration based on LLPR uncertainties. Class-based CP offers a slightly more refined calibration, particularly for the (111) surface.

\subsubsection{Calibrating uncertainty for carborane rearrangement}

Next, we further evaluate the robustness of the proposed method on a more challenging system: carborane rearrangement. We selected Anticuboctahedral TS and single TFR cases as their results are representative. Calibration samples are selected using the same scheme as in the previous example. We set $\alpha = 0.2$, as this choice enables meaningful variation in quantiles across distinct classes during class-based CP calibration.

\begin{figure}[htbp]
    \centering
    \begin{subfigure}[b]{0.7\linewidth}
        \centering
        \includegraphics[width=1.0\linewidth]{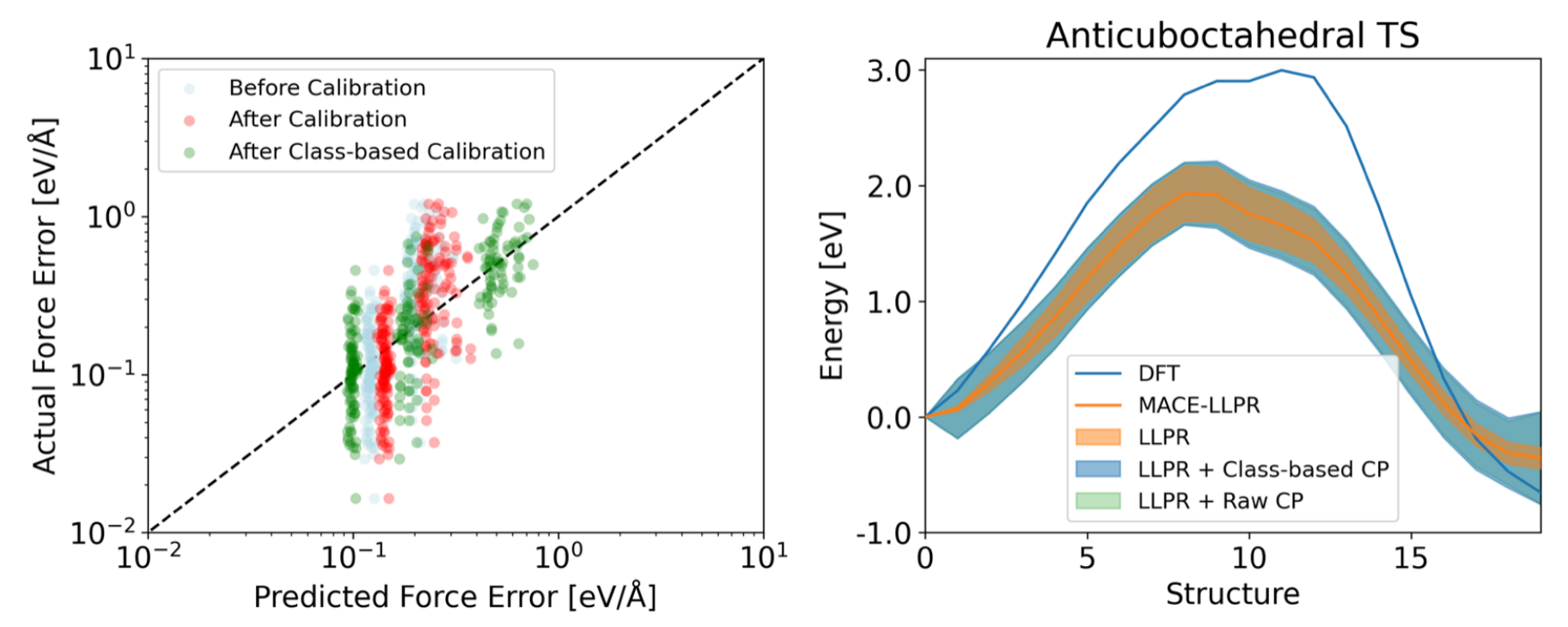}
        \vspace{-0.5cm}
        \label{fig:MP_valid_calibration_ts}
    \end{subfigure}
    \begin{subfigure}[b]{0.7\linewidth}
        \centering
        \includegraphics[width=1.0\linewidth]{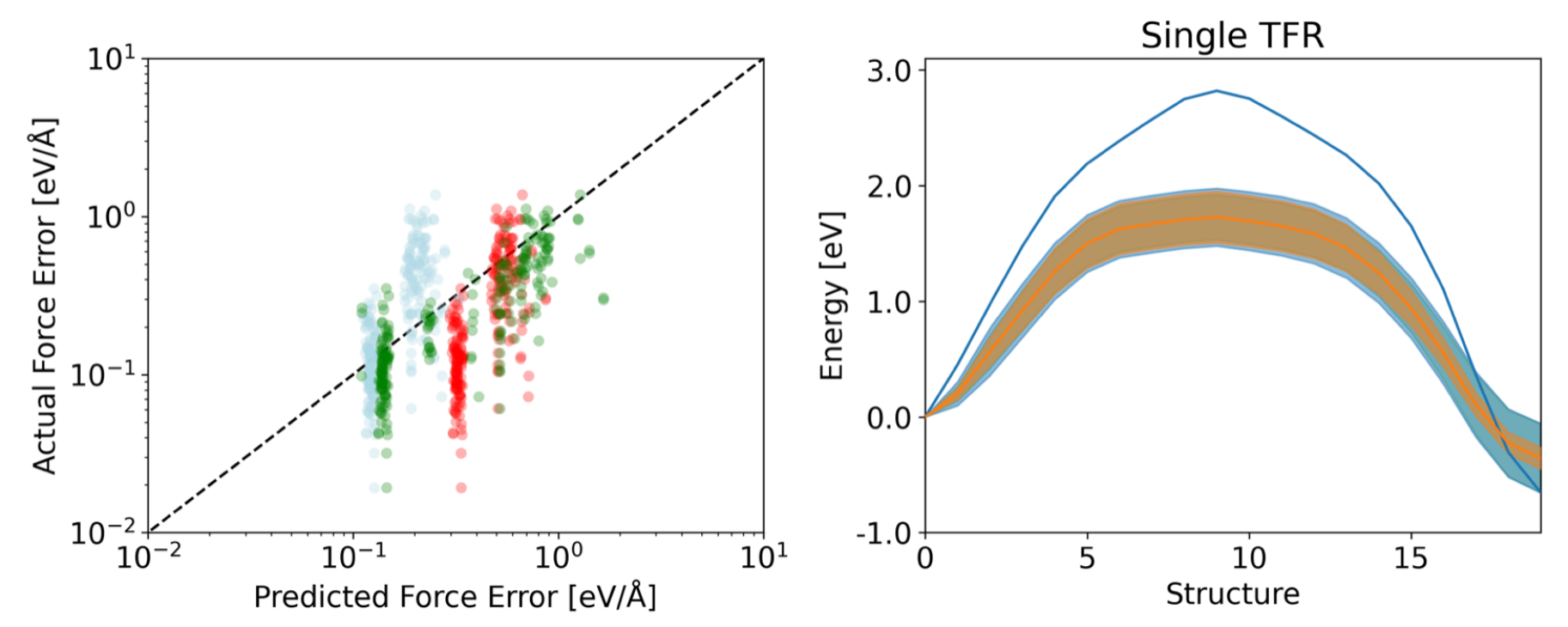}
        \vspace{-0.5cm}
        \label{fig:MP_valid_calibration_tfr}
    \end{subfigure}
    \caption{MACE-predicted forces and relative energies along the NEB barrier paths for carborane 
    rearrangement in Anticuboctahedral TS and single TFR. The estimated force and NEB barrier path errors from MACE-LLPR uncertainty, after calibration and class-based calibration, are shown in different colors. The DFT reference data for the NEB barrier path is shown in blue.
    }
    \label{fig:combined-figure-carbonrane}
\end{figure}

As shown in the left panels of Figure~\ref{fig:combined-figure-carbonrane}, the LLPR method slightly underestimates the actual force error in both cases, while calibration improves alignment with the true error. Notably, classification plays a crucial role in achieving robust calibration, as it accounts for the coexistence of both accurate and underestimated LLPR uncertainties within the system. This effect is particularly evident in the isomerization reaction of TFR (bottom), where separate calibration for different subsets of observations is necessary. This underscores the potential application of class-based calibration in active learning for chemically diverse systems, where both severe under- and overestimation of force uncertainties occur during iterative fine-tuning.

However, as shown in the right panels of Figure~\ref{fig:combined-figure-carbonrane}, while calibration significantly improves force error estimates and classification enhances robustness, this effect does not directly carry over to the calibration of energy differences along the NEB path. This discrepancy arises from the strong symmetry in system movement, which leads to error cancellation when summing uncertainty contributions from individual atoms. We speculate that this issue could be mitigated by performing calibration directly on energy rather than force, particularly for energy-related properties.

\end{document}